\documentclass[preprint,12pt]{elsarticle}

\usepackage{amssymb}
\usepackage{amsthm}
\theoremstyle{plain}
\newtheorem{theorem}{Theorem}[section]

\newtheorem{lemma}[theorem]{Lemma}
\newtheorem{corollary}[theorem]{Corollary}
\newtheorem{remark}[theorem]{Remark}
\usepackage{amsmath}

\numberwithin{equation}{section}
\usepackage[super,compress]{cite}
\usepackage{dcolumn}% Align table columns on decimal point
\usepackage{bm}% bold math
\usepackage{xcolor}
\usepackage{graphicx}
\usepackage{adjustbox}
\usepackage[utf8]{inputenc} % allow utf-8 input
\usepackage[T1]{fontenc}    % use 8-bit T1 fonts
\usepackage{hyperref}       % hyperlinks
\usepackage{url}            % simple URL typesetting
\usepackage{booktabs}       % professional-quality tables% blackboard math symbols
\usepackage{nicefrac}       % compact symbols for 1/2, etc.
\usepackage{microtype}      % microtypography
\usepackage{mathtools}
\usepackage{commath}
\usepackage[sc,osf]{mathpazo}

\usepackage{float}
\usepackage{varioref}
\usepackage[sans]{dsfont}
\usepackage{tikz}
\usepackage{esint}

\makeatletter
\@for\next:={int,iint,iiint,iiiint,dotsint,oint,oiint,sqint,sqiint,
  ointctrclockwise,ointclockwise,varointclockwise,varointctrclockwise,
  fint,varoiint,landupint,landdownint}\do{%
    \expandafter\edef\csname\next\endcsname{%
      \noexpand\DOTSI
      \expandafter\noexpand\csname\next op\endcsname
      \noexpand\ilimits@
    }%
  }
\makeatother

\usepackage[numbers]{natbib}
\usepackage[makeroom]{cancel}

\usepackage{xspace}
\DeclareMathOperator{\sign}{sign}
\newcommand{\Mathematica}{\textit{Mathematica\textsuperscript{\resizebox{!}{0.8ex}{\textregistered}}}}
\def\8{\infty}
\def\oh{\frac{1}{2}}
\def\ot{\frac{1}{3}}
\def\oq{\frac{1}{4}}
\def\tt{\frac{2}{3}}
\def\ft{\frac{4}{3}}
\def\tq{\frac{3}{4}}
\def\ofi{\frac{1}{5}}

\def\d{\partial}
\def\i{\imath\,}
\def\alp{\alpha}
\def\bet{\beta}
\def\gam{\gamma}
\def\lam{\lambda}
\def\eps{\epsilon}

\def\del{\delta}
\def\dal{\partial_{\alpha}}
\def\dbe{\partial_{\beta}}
\def\dga{\partial_{\gamma}}
\def\dla{\partial_{\lambda}}

\def\undertext#1{\vtop{\hbox{#1}\kern 1pt \hrule}}
\def\ra{\rightarrow}
\def\Ra{\Rightarrow}
\def\lra{\longrightarrow}
\def\O#1{O\left(#1\right)}
\def\VEV#1{\left\langle #1\right\rangle}

\def\tr{\hbox{tr}\,}
\def\diag#1{\hbox{diag}\left(#1\right)}
\def\dd#1{\frac{d}{d#1}}
\def\dbyd#1#2{\frac{d#1}{d#2}}
\def\pp#1{\frac{\partial}{\partial#1}}
\def\pbyp#1#2{\frac{\partial#1}{\partial#2}}

\def\fbyf#1#2{\frac{\delta#1}{\delta#2}}

\def\br{\\ \nonumber & &}
\def\inv#1{\frac{1}{#1}}
\def\bea{\begin{eqnarray} & &}
\def\eea{\end{eqnarray}}
\def\ct#1{\cite{#1}}
\def\rf#1{(\ref{#1})}
\let\oldexp\exp
\renewcommand{\exp}[1]{\oldexp\left(#1\right)}
\def\INT#1#2{\int_{#1}^{#2}}
\def\INT#1#2{\int_{#1}^{#2}}

\def\PBR#1{\left[#1\right]}

\def\C{\oint_C d }
\def\Y{\Psi_C(\gamma)}

\def\et{{\mathcal E}}

\def\CL{Clebsch}

\def\PB{Poisson brackets}
\def\KO{Kolmogorov}
\def\NS{Navier-Stokes}
\def\KH{Kelvin-Helmholtz}
\def\BS{Biot-Savart}

\def\BL {Beltrami}
\def \BT{Burgers-Townsend}

\def\SYM {symplectomorphisms}
\def\GBF {$\mbox{GBF}$}

\def\DS {discontinuity surface}
\def\CVS{\textit{CVS}}
\def\CFE{Clebsch Flow Equation}

\def\val{v_{\alpha}}
\def\vbe{v_{\beta}}

\def\ral{r_{\alpha}}
\def\rbe{r_{\beta}}
\def\rga{r_{\gamma}}

\def\oal{\omega_{\alpha}}
\def\obe{\omega_{\beta}}

\def\omu{\omega_{\mu}}

\def\Xint#1{\mathchoice
   {\XXint\displaystyle\textstyle{#1}}%
   {\XXint\textstyle\scriptstyle{#1}}%
   {\XXint\scriptstyle\scriptscriptstyle{#1}}%
   {\XXint\scriptscriptstyle\scriptscriptstyle{#1}}%
   \!\int}
\def\XXint#1#2#3{{\setbox0=\hbox{$#1{#2#3}{\int}$}
     \vcenter{\hbox{$#2#3$}}\kern-.5\wd0}}

\def\dashint{\Xint-}

\newcommand{\Z}{\mathbb{Z}}
\DeclareMathOperator*{\argmax}{arg\,max}
\DeclareMathOperator*{\argmin}{arg\,min}

\DeclareMathOperator{\erf}{erf}
 
\renewcommand{\Re}{\textbf{Re }}
\renewcommand{\Im}{\textbf{Im }}

\newcommand{\pct}[2]{
\begin{figure}
    \centering
    \includegraphics[width=\textwidth]{#1}
    \caption{#2}
    \label{fig::#1}
\end{figure}
}

\journal{Physics Reports}

\begin{document}

%\frontmatter

\title{Statistical Equilibrium of Circulating Fluids}

\author[inst1]{Alexander Migdal}

\affiliation[inst1]{organization={Department of Physics, New York University Abu Dhabi},%Department and Organization
            addressline={Saadiyat Island}, 
            city={Abu Dhabi},
            postcode={PO Box 129188}, 
            state={Abu Dhabi},
            country={United Arab Emirates}}
\begin{abstract}
We are investigating the inviscid limit of the Navier-Stokes equation, and we find previously unknown anomalous terms in Hamiltonian, Dissipation, and Helicity, which survive this limit and define the turbulent statistics.

We find various topologically nontrivial configurations of the confined Clebsch field responsible for vortex sheets and lines. In particular, a stable vortex sheet family is discovered, but its anomalous dissipation vanishes as $\sqrt{\nu}$.

Topologically stable stationary singular flows, which we call Kelvinons, are introduced. They have a conserved velocity circulation $\Gamma_\alpha$ around the loop $C$ and another one $\Gamma_\beta$ for an infinitesimal closed loop $\tilde C$ encircling $C$, leading to a finite helicity. The anomalous dissipation has a finite limit, which we computed analytically.

The Kelvinon is responsible for asymptotic PDF tails of velocity circulation, \textbf{perfectly matching numerical simulations}.

The loop equation for circulation PDF as functional of the loop shape is derived and studied. This equation is \textbf{exactly} equivalent to the Schrödinger equation in loop space, with viscosity $\nu$ playing the role of Planck's constant. 

Kelvinons are fixed points of the loop equation at WKB limit $\nu \rightarrow 0$. The anomalous Hamiltonian for the Kelvinons contains a large parameter $\log \frac{|\Gamma_\beta|}{\nu}$. The leading powers of this parameter can be summed up, leading to familiar asymptotic freedom, like in QCD. In particular, the so-called multifractal scaling laws are, as in QCD, modified by the powers of the logarithm.

\end{abstract}

\maketitle

\tableofcontents

%\mainmatter
\newpage
\section{Prologue}
\subsection{Paths to the summit}
  The unsolved problem of classical turbulence has challenged us for centuries, like a shining Himalaya peak. %

Burgers, Onzager, Heisenberg, Landau, Kolmogorov, Feynman, and many other great scientists attempted and failed to reach the summit but blazed the trail for the next generations.

Richard Feynman wrote half a century ago in his famous "Lectures in Physics"

\begin{quote}
\textit{
"there is a physical problem that is common to many fields, that is very old, and that has not been solved. It is not the problem of finding new fundamental particles, but something left over from a long time ago—over a hundred years. Nobody in physics has really been able to analyze it mathematically satisfactorily in spite of its importance to the sister sciences. It is the analysis of circulating or turbulent fluids."}
\end{quote}

  This analysis would aim to properly redefine and solve the \NS{} equations in a limit of vanishing viscosity at fixed energy flow.

  In doing so, we must answer some obvious questions.
  \begin{itemize}

\item \textit{Is there an analogy between turbulent velocity and vector potential in the gauge field theory?}

  \item \textit{Are some "fundamental particles" hidden inside as it was alluded to by Feynman?}
  
   \item \textit{How is the inviscid limit of the \NS{} dynamics different from the Euler dynamics?}
   
 \item \textit{What is the origin of the randomness of the circulating fluid? }

  \item \textit{Is it spontaneous, and what makes it irreversible?}

  \item \textit{What are the properties of the fixed manifold covered by this random motion? }

\item \textit{What is the microscopic mechanism of the observed multifractal scaling laws?}

\end{itemize}
    The modern Turbulence theory with ad hoc random forcing at low wavelengths does not even try to answer these questions.
    
    It resembles the Strong Interaction Theory 60 years ago when I started my scientific career.

    Werner Heisenberg promoted an S-matrix dogma, which stated that there was nothing inside the protons, neutrons, and mesons. 

    All particles consisted of each other, and it was FORBIDDEN to ask unobservable questions like "what is inside"?

    "Build phenomenological models instead of trying to define and solve the microscopic theory"  dictated to us the S-matrix dogma.

    The Yang-Mills theory already existed but was discarded by Heisenberg and his follower Lev Landau, who declared that \begin{quote} \textit{"the Lagrangian is dead and should be buried with all due honors."}\end{quote}
     Another outstanding mathematical physicist, Sasha Zamolodchikov, recently noted
     \begin{quote}
    \textit{"they buried the Lagrangian but forgot to drive the stake through its heart."}\end{quote}

    And then came quarks, gluons, and asymptotic freedom! The rest is history.
    
    We learned that gluon strings confine quarks inside protons and neutrons. We cannot directly observe quarks and gluons, but we can study their almost free motion at small distances inside hadrons.

    Quarks and gluons are true fundamental particles, not protons, neutrons, and mesons, as Heisenberg and Landau thought.

\subsection{Ideal gas of vortex structures vs. uniform rough velocity field}
Few general introductory words before we proceed.

The big picture of Turbulent incompressible flow, which we accept, is similar to statistical mechanics.

There is an infinite volume, filled with various compact vortex structures, moving in a common velocity field generated by the sum of the \BS{} formulas from all these structures. This \BS{} law has long tails; therefore, at any given point, all the structures contribute to this common velocity, inversely proportional to some power of distance.

There is a finite but small density of these structures, which means that their number is proportional to the volume going to infinity.
This picture is qualitatively the same as the ideal gas of point vortexes advocated by Onsager, except our vortexes have finite sizes.

Our picture also resembles the Richardson cascade, with a hierarchy of eddies of various sizes, passing the energy flow from large spatial scales to the smallest one, where the energy dissipates due to viscosity.

Our picture has an ensemble of eddies, though they do not obey any hierarchical pattern in coordinate space. As for the wavelength space, it is inadequate to describe our singular vortex lines and sheets. It would be like describing the smile of Mona-Lisa by spectral analysis of the paints used by Leonardo. Or, to use a more direct analogy, this would be like describing the gluon strings confining quarks as a collection of plane waves (gluons).

Mathematicians often use another picture of the flow in a unit cube, which corresponds to an infinite shrinking of our picture of a large number of finite vortex structures. In this mathematical picture, the infinitesimal vortex structures fill the unit volume. Therefore the velocity field becomes rough, with a singularity at every point.
This singularity resembles fractal geometry where $\Delta v \sim \Delta r^{\ot}$.

Further investigation of the experimental data and the numerical simulations of the forced \NS{} equations for the last 25 years revealed that higher powers $\Delta \vec v^n$ do not scale as $ \Delta r^{n/3}$. To explain this paradox, Parisi and Frisch suggested the multifractal models where each power of $\Delta v$ scales with its index $\Delta v^n \sim \Delta r^{\zeta(n)}$. These models describe the large-scale DNS \cite{S19, S21} at the phenomenological level, but so far, led us nowhere close to a microscopic theory of turbulence.

We insist on our picture, as it allows us to resolve the singularities involved and opens the way for a microscopic theory. In our approach, these rough velocity fields have at least two more intrinsic scales: the size of each vortex structure plus a much smaller thickness of vortex lines and surfaces inside these structures. We use a microscope to build a microscopic theory.

These singular lines and surfaces at even larger magnification turn out to be some well-known (but almost forgotten)  exact solutions of the \NS{} equations.
The thickness of these singular lines and the width of these vortex sheets go to zero as a square root of viscosity when it tends to zero in a strong turbulence limit.

Thus, instead of mysterious multifractal, we have statistics of Euler vortex structures built around the \NS{} vortex lines and sheets.

Instead of an infinite Richardson cascade, we have just three hierarchy levels: the box size, the Kelvinon size, and the thickness of the viscous lines and surfaces, where the energy finally dissipates.

In the limit of the small density of these vortex structures, we can single one out and study its statistical equilibrium with the rest of the structures, acting as a thermostat.

The irreversibility of the \NS{} dynamics makes this equilibrium different from the Boltzmann-Gibbs distribution, but, as we shall see, 
simpler in some aspects once we get used to the nontrivial topology of the flow.

Finally, note that this approximation of a small density of vortex structures is optional. In a more general approach, we use the so-called Loop Equation, which makes no approximations. It exactly relates the equilibrium statistics of velocity circulation to the ground state problem of a quantum system in loop space, with the Action being the velocity circulation around a fixed loop in space.

The Loop Equation can be solved at large loops, leading to the Area law, matching the recent DNS.

\subsection{Confined fields of Turbulence}

In our recent works\cite{M20a, M20b, M20c, M21b, M21c, M21d, M22}, we conjectured that there were confined gauge fields in turbulence, 
hidden inside the velocity and vorticity fields.

    These so-called spherical Clebsch variables are mechanically equivalent to rigid rotators. They are the canonical Hamiltonian variables in the Euler dynamics but cannot propagate as wave excitations.

    They play the same role as quarks and gluons in the QCD. They are "fundamental particles" inside the turbulent fluid, unobservable in pure form.
    These spherical Clebsch variables are subject to gauge transformations, which leave vorticity and velocity invariant. 

    These gauge transformations are global in physical space but local in target space ($S_2$ in this case). These are area-preserving diffeomorphisms or canonical transformations.
    
    This unbroken invariance is the mathematical reason for the Clebsch confinement.

    In the same way, as gluons collapse to surfaces bounded by the closed world lines of quarks, these confined variables in turbulent flow collapse into vortex sheets bounded by vortex lines (Kelvinons).
     
      We have found the Hamiltonian for these Kelvinons, which equals the Euler Hamiltonian plus previously unknown anomalous terms, coming from the core of the Burgers vortex. A normalization condition is added to this minimization problem to balance energy pumping with energy dissipation.

      In addition to the singular vortex sheet bounded by a singular vortex line, there is a soliton field around these singularities, mapping the remaining physical space into the target space $S_2$. Two winding numbers $n, m$ describe this mapping and provide topological stability of a soliton. It minimizes this Hamiltonian within a given topological class $n,m$.
    
    The fluctuations of these variables die out in the strong turbulence, echoing the asymptotic freedom. Only three "zero modes" are left, dramatically simplifying the property of the turbulent fixed point.

    We cannot rigorously prove our conjectures and claim that we have reached the shining summit of Turbulence mountain, but we have found a new trail climbing new heights.
    
    Let us give a flyover of this trail before tedious walking step by step.
    
\section{Introduction}

\subsection{Historical Notes. Field-Geometry Duality}
The potential role of vortex sheets and lines as dominant structures in fully developed turbulence was first suggested by Burgers in 1948 \cite{BURGERS1948} with his famous cylindrical vortex. This solution was then extended to a planar vortex sheet by Townsend \cite{TW51}, but nothing was done with this solution until recently.

In the late '80s, we found that the vortex sheet dynamics represent a gauge-invariant 2D Hamiltonian system with multivalued Action \cite{M88}. The steady-state corresponded to the minimum of the Hamiltonian of this system. 

This correspondence was the first manifestation of Duality between the fluctuating vector field and fluctuating geometry. The ADS/CFT Duality in quantum field theory discovered by Juan Maldacena in 1999 became one of the most important developments in modern theoretical physics. 

The velocity-vortex sheet duality is still developing, with some important advances in recent years.

We proposed in \cite{AM89} a discrete system version with a triangulated surface with conserved variables $\Gamma$ at vertices.

There were multiple works by applied mathematicians studying this time evolution numerically (Kaneda \cite{Kaneda90} and some others). The consensus was that this evolution suffered from the \KH{} instability, leading to the loss of smooth shape and possibly a finite time roll-up singularity.

Recently we made some more progress in vortex sheet theory as Lagrangian dynamics and studied solutions with various topologies.

Various fascinating regimes preceded the fully developed isotropic turbulence, such as shear flows \cite{Cvitanovi__2010}. There are complex patterns of periodic motions in these intermediate regimes, suggesting even more complex patterns in extreme turbulence.

We suggested \cite{M93, M20a, M21b} that turbulence is a 2D statistical field theory of random vortex sheets, using some elements of the string theory.
The problem with that approach is that random surfaces are unstable as a statistical field theory in 3 dimensions.

The vortex surfaces are also unstable due to \KH{}, so this idea is questionable.

The \NS{}  Hopf equation for statistical distribution was reduced to the loop equation in \cite{M93}. Mathematically, this reduced dynamical variables from vector field $\vec v(\vec r)$ in three dimensions to the vector loop field $\vec C(\theta)$ in one dimension.

We have found some asymptotic relations from this equation, including Area law.

This loop equation is equivalent to a Schrödinger equation in loop space with complex Hamiltonian and viscosity in place of Planck's constant.

This analogy with quantum mechanics is not a poetic metaphor nor some model approximation: this is an \textbf{exact} mathematical equivalence.
There are observable "quantum" effects in classical turbulence rooted in this amazing analogy.

The Area law was derived from the loop equation and later confirmed in remarkable DNS by Sreenivasan with collaborators \cite{S19, S21}.

This inspired the further development of the geometric theory of turbulence\cite{M19a,M19b,M19c,M19d,M19e,M20a,M20b,M20c,M21a,M21b,M21c,M21d,M22}.

In particular, recently, it was discovered \cite{M21a, KS21} that there are new generic solutions of the \NS{}  equations with arbitrary constant background strain near the planar vortex sheet.

The stable vortex sheets can be curved in the general case as long as they satisfy the local nonlinear equation $\hat S(\vec r) \cdot \Delta \vec v(\vec r)=0$ where $\hat S$ is a strain at every point $\vec r$ of the vortex sheet. 

These equations restrict the shape of the vortex sheet, and we have solved them exactly for cylindrical geometry.
This solution represents a stable fixed point of the Euler dynamics.

Unfortunately, compact \CVS{} (confined vorticity surfaces) were proven not to exist by De Lellis and Ru\'e (see \ref{Camillo}). Our solution \eqref{velocityField} was non-compact, which made it less relevant for the turbulence problem.

Recently we found relevant compact solutions of Euler dynamics (Kelvinons), with conserved velocity circulation around a static loop due to the nontrivial topology of the \CL{} field \cite{M22}. Similar topological defects (KLS domain walls bounded by monopole strings) were observed in real experiments in the superfluid helium \cite{Vol15, ZH20}.

These singular solutions perfectly match the cylindrical Burgers vortex at the core of the loop with the local tangent axis. This matching plays a crucial role in the theory, letting us compute anomalous dissipation. This dissipation is proportional to the line integral of the tangent strain times the square of circulation $\Gamma_\beta$ around the infinitesimal loop encircling the monopole loop $C$. There is also the anomalous term in the Euler Hamiltonian, proportional to the line integral of the logarithm of this tangent strain.

Thus we have found a true turbulent limit of the \NS{} theory: the effective Euler Hamiltonian $\oh \dashint\vec v^2$ plus some universal anomalous terms depending upon the Euler flow outside the singular loop.
The principal value of the volume integral here corresponds to the exclusion of an infinitesimal tube around the loop $C$, where the energy integral logarithmically diverges. Inside the tube, it is regularized by a matching Burgers vortex.
The stationary solutions, minimizing this effective Hamiltonian, are responsible for the PDF tails of velocity circulation $\Gamma = \oint_C \val d \ral$ in the extreme turbulent limit $|\Gamma| \gg \nu$. 

Compact motion is associated with quantized winding numbers, which label Kelvinons. The PDF of velocity circulation represents the sum of exponential terms with quantized slopes, like the black body radiation. 

In the case of radiation, the energy quantization follows from the compactness of the motions of electrons in atoms. In our case, quantization of the winding numbers, which define the decrements of exponential PDF decay, is a consequence of the compactness of the target space $S_2$ for \CL{} variables.

This quantization is one of the quantum effects that all follow from the loop equation's equivalence to quantum mechanics in loop space.

As we recently found, Kelvinon also explains and correctly describes the asymmetry of the circulation PDF (i.e., time-reversal breaking).

The perfect fit of the observed circulation PDF\cite{S19} by the Kelvinon prediction confirms the Duality of the turbulence theory and the Kelvinon mechanism of the intermittency of velocity circulation.

On a deeper level, it provides a strong argument in favor of the quantum description of the classical turbulence by the loop equation.

\subsection{The statement of the problem and basic results}
The ultimate goal of turbulence studies is to properly define and solve the inviscid \NS{} equations and determine why and how the solution covers some manifold rather than staying unique given initial data and boundary conditions. 

We need to understand why it is irreversible at zero viscosity when the \NS{}  equations formally become the Euler equation corresponding to the reversible Hamiltonian system with conserved energy. This limit is nontrivial -- some anomalous terms remain at $\nu \ra 0$ and cause energy dissipation.

Once we know why and how the solution covers some manifold -- a degenerate fixed point of the \NS{}  equations -- we would like to know the parameters and the invariant measure on this manifold.

We know (or at least assume) that the vortex structures in this extreme turbulent flow collapse into thin clusters in physical space. 
Snapshots of vorticity in numerical simulations  \cite{VTubes, Tubes19} show a collection of tube-like structures relatively sparsely distributed in space.

The large vorticity domains and the large strain domain lead to anomalous dissipation: the enstrophy integral is dominated by these domains so that the viscosity factor in front of this integral is compensated, leading to a finite dissipation rate. It was observed years ago and studied in the DNS\cite{SreeniDissipation}.

There was an excellent recent DNS \cite{S19, S21} studying statistics of vorticity structures in isotropic turbulence with a high Reynolds number. Their distribution of velocity circulation appeared compatible with circular vortex structures (which we now call Kelvinons) suggested in \cite{M19d, M22} following our early suggestions about area law for the velocity circulation \cite{M93}.

The authors of \cite{S19, S21} confirmed the area law and compared it with the tensor area law, which would correspond to a constant uniform vortex, irrelevant to the turbulence. These two laws are indistinguishable for simple loops like a square, and they both are solutions to the loop equation but are very different for twisted or non-planar loops.

Their data supported the area law, which inspired our search for the relevant vorticity structures behind that law.

Some recent works also modeled sparse vortex structures in classical \cite{VortexGasCirculation20} and quantum \cite{QuantumCirculation21} turbulence.

We know such singular structures in Euler dynamics: vortex surfaces and lines. Vorticity collapses into a thin boundary layer around the surface, 
 or the core surrounding the vortex line, moving in a self-generated velocity field.
 
The vortex surface motion is known to be unstable against the \KH{} instabilities, which undermines the whole idea of random vortex surfaces.

However, the exact solutions of the \NS{}  equations discovered in the previous century by Burgers\cite{BURGERS1948, BurgersVortex} and Townsend \cite{TW51} show stable planar sheets with Gaussian profile of the vorticity in the normal direction, peaked at the plane. 

Thus, the viscosity effects in certain cases suppress the \KH{} instabilities leading to stable, steady vortex surfaces.

The same applies to the Burgers vortex, corresponding to a cylindrical core, regularizing a singular vortex line.

The Burgers cylindrical vortex has a constant strain, which is not the most general; there is only one independent eigenvalue instead of two.
The general asymmetric solution approaches the symmetric Burgers vortex with the potential background flow as it was subsequently found in\cite{BMO84} in the turbulent limit $|\Gamma| \gg \nu$ with asymmetric strain.

Let us define the conditions and assumptions we take.

We adopt Einstein's implied summation over repeated Greek indexes $\alpha,\beta,\gamma,\dots$. We also use the conventional vector notation for coordinates $\vec r$, velocity $\vec v$, and vorticity $\vec \omega = \vec \nabla \times \vec v$.
The Greek indexes $\alpha,\beta,\dots $  run from $1$ to $3$ and correspond to physical space $R_3$, and the lower case Latin indexes $a,b,\dots $ take values $1,2$ and correspond to the internal parameters on the surface. 
We shall also use the Kronecker delta $\delta_{\alpha\beta}, \delta_{a b}$ in three and two dimensions as well as the antisymmetric tensors $e_{\alpha\beta\gamma}, e_{a b}$.

\begin{enumerate}
    \item We address the turbulence problem from the first principle, the Navier-Stokes equation, without any random forces.
    \item  We study an infinite isotropic flow in the extreme turbulent limit (viscosity going to zero at a fixed dissipation rate). 
    \item We start with vortex sheet solutions and investigate the stability of these solutions, leading to new boundary conditions (\CVS{}) of the Euler equations on vortex surfaces.
    \item
    New mathematical theorems eliminate stable, compact vortex sheets, leaving only unbounded surfaces. We find exact analytical solutions for the family of such surfaces and related flow.
    \item
    We compute the dissipation on vortex sheets/lines in the turbulent limit as a surface/line integral and prove that it is conserved. In the case of a vortex sheet, dissipation goes to zero as $\sqrt{\nu}$, and in the case of vortex lines, it tends to a constant, providing the desired anomalous dissipation.
    \item
    We find a new class of singular stationary flows (Kelvinon) of the \NS{} theory in the limit of $\nu \ra 0$. This flow has a nontrivial topology with two winding numbers corresponding to two cycles of the infinitesimal torus around a stationary loop $C \in R_3$. The helicity of a singular loop is proportional to the product of these two winding numbers.
    \item
    We also compute the PDF tails for the velocity circulations, dominated, as we argue, by a Kelvinon. This PDF perfectly matches the exponential laws observed in DNS.
    The time-odd topological winding number explains the time-reversal breaking in the DNS.
    \item
    We derive and systematically investigate the loop equations for the circulation PDF and express the turbulent limit as the WKB limit.
    \item
    We elaborate a mystifying mathematical equivalence of the loop equation to the Schrödinger equation in loop space with complex Hamiltonian, leading to quantization of classical circulation on the topological solutions.
    \item
   Another surprising consequence of this equivalence is the superposition principle. The general solution for circulation PDF involves a sum over topological classes of Kelvinons.
    \item
    We do not take any model simplifications in our theory, but certain parameters remain phenomenological constants, which we cannot yet compute.
\end{enumerate}

\section{Canonical Clebsch Variables}
\subsection{ \CL{} variables}

Before going into details of the vortex singularity dynamics, let us mention the relation between vortex structures and so-called \CL{} variables.
These variables play an important role in solving the Euler equations with nontrivial topology.

We parameterize the vorticity by two-component \CL{} field $\phi = (\phi_1,\phi_2) \in R_2$:
\begin{equation}
    \vec \omega= \vec \nabla \phi_2 \times \vec \nabla \phi_1
\end{equation}
The Euler equations are then equivalent to passive convection of the \CL{} field by the velocity field (modulo gauge transformations, as we argue below):
\begin{align}\label{CLEq}
     &\d_t \phi_a = -\vec v \cdot \vec \nabla \phi_a\\
     &\vec v =\left(\phi_2 \vec \nabla \phi_1\right)^\perp \label{Vperp}
\end{align}
Here $V^\perp$ denotes projection to the transverse direction in Fourier space, or:

\begin{equation}
    V^\perp_\alpha(r) = V_\alpha(r) + \dal \dbe \int d^3 r' \frac{V_\beta(r')}{4 \pi |r-r'|}
\end{equation}
One may check that projection (\ref{Vperp}) is equivalent to the Biot-Savart law (\ref{BS}).

The conventional Euler equations for vorticity:
\begin{equation}
    \d_t \vec \omega = \vec \omega \cdot \vec \nabla \vec v - \vec v \cdot \vec \nabla \vec \omega \label{EulerO}
\end{equation}
follow from these equations.

The \CL{} field maps $R_3$ to $R_2$ and the velocity circulation around the loop $C \in R_3$:
\begin{equation}
    \Gamma(C) = \oint_{C}  d \vec r \cdot \vec v = \oint_{\gamma_2} \phi_2 d \phi_1 = \mbox{Area}(\gamma_2)
\end{equation}
becomes the oriented area inside the planar loop $\gamma_2 = \phi(C)$. We will discuss this later when we build the Kelvinon.

The most important property of the \CL{} fields is that they represent a $p,q$ (canonical) pair in this generalized Hamiltonian dynamics. The phase-space volume element $D \phi = \prod_x d \phi_1(x) d \phi_2(x)$ is time-independent, as the Liouville theorem requires.

The generalized \BL{} flow (\GBF{}) corresponding to steady vorticity is described by $\vec G(x)=0$ where:
\begin{equation}\label{GBFConst}
    \vec G \overset{\mbox{def}}{=} \vec \omega \cdot \vec \nabla \vec v - \vec v \cdot \vec \nabla \vec \omega 
\end{equation}
These three conditions are, in fact, degenerate, as $\vec \nabla \cdot\vec G =0$. So, there are only two independent conditions, the same as the number of local \CL{} degrees of freedom. However, as we see below, the relation between vorticity and \CL{} field is not invertible. 

Moreover, there are singular flows with discontinuous \CL{} variables, which do not belong to the Beltrami flow class. These singular (weak) solutions of the Euler equations, after proper regularization by the \NS{} equation in the singular layers and cores, will play a primary role in our theory.
\subsection{Gauge invariance}\label{GaugeInv}

Our system possesses some gauge invariance (canonical transformation in the Hamiltonian dynamics or area preserving diffeomorphisms geometrically).

\begin{align}\label{SDiff}
    &\phi_a(r) \Ra M_a(\phi(r))\\
    &\det \pbyp{M_a}{\phi_b} = \frac{\d(M_1,M_2)}{\d(\phi_1,\phi_2)} = 1.
\end{align}
These transformations manifestly preserve vorticity and, therefore, velocity. \footnote{These variables and their ambiguity were known for centuries \cite{Lamb45}, but they were not utilized within hydrodynamics until the pioneering work of Khalatnikov \cite{khalat52} and subsequent publications of Kuznetzov and Mikhailov \cite{KM80} and Levich \cite{L81} in early 80-ties. The modern mathematical formulation in terms of \SYM{} was initiated in \cite{M83}. Yakhot and Zakharov \cite{YZ93} derived the K41 spectrum in weak turbulence using some approximation to the kinetic equations in \CL{} variables. 
}

In terms of field theory, this is an exact gauge invariance, much like color gauge symmetry in QCD, which is why back in the early 90-ties, we referred to \CL{} fields as "quarks of turbulence." To be more precise, they are both quarks and gauge fields simultaneously. 

It may not be very clear that there is another gauge invariance in fluid dynamics, namely the $\bf{volume}$ preserving diffeomorphisms of Lagrange dynamics. Due to incompressibility, the element of the fluid, while moved by the velocity field, preserved its volume. However, these diffeomorphisms are not the symmetry of the Euler dynamics, unlike the $\bf{area}$ preserving diffeomorphisms of the Euler dynamics in \CL{} variables.

One could introduce gauge fixing, which we will discuss later. The main idea is to specify the Clebsch space metric $g_{ i j}$ as a gauge condition. The \SYM{} varies the metric while preserving its determinant. 

At the same time, the topology of \CL{} space is specified by this gauge condition. We are going to choose the simplest topology of $S_2$ without handles. 
This gauge fixing is investigated in  \ref{SphericalGauge}. It does not require Faddeev-Popov ghosts. 

As we argue in the later sections, the topology of the \CL{} space is not a matter of taste. There are two different phases of the turbulent flow-- weak and strong turbulence. The weak turbulence corresponds to $R_2$  as the \CL{} target space, and strong turbulence, which we study here -- to $S_2$.

The $R_2$ \CL{} field can exist in the form of the waves in physical space $R_3$ with some conserved charge corresponding to the $O(2) = U(1) $ rotation group in the target space\cite{YZ93}. There is also an invariance to the translation of \CL{} field in $R_2$.

The $S_2$ \CL{} field is confined to a sphere and cannot exist as a wave in physical space. It obeys a non-Abelian symmetry group $SO(2)$ of rotations of the sphere.

As we argue later, this compactification phase transition $R_2 \Ra S_2$ is the origin of the intermittency phenomena in strong turbulence.

Let us come back to the \CL{} field on a sphere.

Note that the \GBF{} condition comes from the Poisson bracket with Hamiltonian $H =\int d^3 r \oh \val^2$
\begin{align}\label{PBforGBF}
    &G_\alpha(r) = \PBR{\oal,H} = \\
    &\int d^3 r' \fbyf{\oal(r)}{\phi_i(r')} e_{i j} \fbyf{H}{\phi_j(r')}=\\
    &-\int d^3 r' \fbyf{\oal(r)}{\phi_i(r')} v_\lambda(r') \dla{} \phi_i(r')
\end{align}
In the \GBF{}, we only demand that this integral vanish. The steady solution for \CL{} would mean that the integrand vanishes locally, which may be too strong.

The Euler evolution of vorticity does not mean the passive advection of the \CL{} field: the solution of a more general equation
\begin{eqnarray}
    \d_t \phi_i = -\val \dal \phi_i + e_{i j} \pbyp{h(\phi)}{\phi_j}  \label{GaugeTranClebsch}
\end{eqnarray}
with some unknown function $h(\phi)$ would still provide the same Euler evolution of vorticity:
\begin{eqnarray}
 \d_t \oal = \int d^3 r' \fbyf{\oal(r)}{\phi_i(r')} \d_t \phi_i(r) = \PBR{\oal,H}
\end{eqnarray}

The last term in $\d_t \phi_i$ drops from here in virtue of infinitesimal gauge transformation $\delta \phi_a = \epsilon e_{a b} \pbyp{h(\phi)}{\phi_b}$ which leave vorticity invariant.

This invariance means that the \CL{} field is being gauge transformed while carried by the flow. 

This motion's topology will play a central role in our theory of intermittency.

\subsection{Spherical parametrization of Clebsch field}

To find configurations with nontrivial topology, we have to assume that at least one of the two independent \CL{} fields, say, $\phi_2$ is multivalued. The most natural representation was first suggested by Ludwig Faddeev and was used in subsequent works of Kuznetzov and Mikhailov \cite{KM80} and Levich \cite{L81} in the early '80s.

\begin{eqnarray}\label{CLS2}
    &&\vec \omega = \oh Z e_{a b c}  S_a  \vec \nabla S_b \times \vec\nabla S_c;\\
    &&S_1^2 + S_2^2 + S_3^2= 1;\\
    &&\phi_2 = \arg{ (S_1 + \i S_2)}; \; \phi_1 = ZS_3; 
\end{eqnarray}
 where $Z$ is some parameter with the dimension of viscosity, staying finite when $\nu \ra 0$. 
 
 It becomes an \textbf{integral of motion} in Euler dynamics. It plays a very important role in our theory: it matches the anomalous dissipation with energy flow into the Kelvinon.

 The ratio $\textbf{Re} =\frac{Z}{\nu}$ represents the Reynolds number of our theory.

The field $\phi_2$ is defined modulo $2 \pi$. We shall assume it is discontinuous across a vortex sheet shaped as a disk $\mathcal D$.

Note also that this parametrization of vorticity is invariant to the shift $\phi_2 \Ra \phi_2 + \pi$, corresponding to the reflection of the complex field $\Psi = S_1 + \i S_2$
\begin{eqnarray}
&&\Psi \Ra -\Psi;\\
&&\vec \omega = \mbox{inv}
\end{eqnarray}

This reflection symmetry allows us to look for the \CL{} field configurations with $\Psi$ changing sign when the coordinate goes around some loop, like the Fermion wave function in quantum mechanics.

Let us now look at the helicity integral
\begin{equation}\label{Hel}
    H = \int_{R_3 \setminus \mathcal D}  d^3 r \vec v \vec \omega
\end{equation}

Note that velocity
\begin{equation}\label{velCl}
    v_i = \phi_2 \d_i \phi_1 + \d_i \phi_3
\end{equation}
will have discontinuity $\Delta \vec v = 2 \pi n \vec \nabla \phi_1$

The $\phi_3$ can be written as a solution of the Poisson equation
\begin{equation}\label{TildePhi3}
   \phi_3(r) =  \Phi(\vec r) + \dbe \int d^3 r' \frac{\phi_2(r') \dbe \phi_1(r')}{4 \pi |r-r'|}
\end{equation}
where $\Phi(\vec r)$ is a solution of the Laplace equation.

We will treat $\phi_3$ as an independent field while computing helicity as a functional of the \CL{} field. At the end of the computation, we can set $\phi_3$ to \eqref{TildePhi3}, thus projecting back the \CL{} space to two dimensions $\phi_1, \phi_2$.

If there is a discontinuity of the phase $\phi_2$ and the corresponding discontinuity of velocity field on some surface $\mathcal S_{disc}$, the solution of the Poisson equation for $\phi_3$ must satisfy Neumann boundary conditions (vanishing normal velocity at this surface).

In that case, this discontinuity surface will be steady in the Euler dynamics, as our theory requires (see later).

The  helicity integral could now written as  a map $R_3 \mapsto (\phi_1,\phi_2,\phi_3)$
\begin{align}
    &\mathcal H = \int_{R_3 \setminus \mathcal D} d^3 r \left(\phi_1 \d_i \phi_2+ \d_i \phi_3\right) e_{i j k } \d_j \phi_1 \d_k \phi_2 =\\
   & \int_{R_3 \setminus \mathcal D} d\phi_1 \wedge d\phi_2 \wedge d\phi_3 \label{fullHel}
\end{align}

Here is the most important point. There is a following surgery performed in three-dimensional \CL{}  space. An incision is made along the surface $\phi\left(\mathcal D\right)$ and $n$ more copies of the same space are glued to this incision like sheets of a Riemann surface of $ f(z) = z^{k + \oh}$, except this time the three-dimensional spaces are glued at the two-dimensional boundary, rather then the two-dimensional Riemann sheets glued at a one-dimensional cut in the complex plane.

Best analogy: this disk is a portal to other \CL{} universes like in science fiction movies.
The winding number of $\phi_2$ counts these parallel universes when passing through that surface.

Integrating over $\phi_2 $  in \eqref{fullHel}, using discontinuity $\Delta \phi_2\left(\mathcal D\right) = 2 \pi n$ and then integrating  $\int_{\mathcal D} d \phi_3 \wedge d \phi_1$ we find a simple formula
\begin{equation}\label{Helicity}
    H =2\pi n \oint_C \phi_3 d \phi_1
\end{equation}
This helicity is $2 \pi $ times the number of parallel universes times the area of the portal.

These singular flows also have a monopole singularity at the loop $C$, leading to anomalous contribution to helicity.

\subsection{Euler flow as Clebsch flow in product space}

In virtue of gauge invariance of the \CL{} parametrization of vorticity \eqref{CLS2}, the most general \CL{} dynamics involve time-dependent gauge transformation, as we have seen in Section \ref{GaugeInv}.

The Euler-Lagrange flow passively carries the Clebsch field and its gauge transformation. For the vector $\vec S \in S_2$, this flow becomes:
\begin{eqnarray}\label{CLFlow}
&&\d_t S_i +(\vec v \cdot \vec \nabla) S_i +  [h,S_i]=0;\; i = 1,2,3;\\
&& [A,B] = \pbyp{ A}{\phi_1} \pbyp{ B}{\phi_2} -\pbyp{ A}{\phi_2} \pbyp{ B}{\phi_1}
\end{eqnarray}

As discussed above, this equation, with arbitrary gauge function $h(\phi_1, \phi_2)$, leads to the usual Euler equations for vorticity and velocity.
We call this equation \CFE{}.
The $h-$ terms all cancel in the Euler equations for $\vec \omega, \vec v$.

Geometrically, this equation describes the passive flow in direct product space $ R_3 \otimes S_2$ with velocity 
\begin{eqnarray}
V = \left(v_1,v_2,v_3, -\pbyp{ h}{\phi_2},\pbyp{ h}{\phi_1}\right)
\end{eqnarray}
This flow conserves the volume element in each of the two spaces.

\begin{eqnarray}
div  V = \d_i v_i + \d_a e_{a b} \d_b h =0 + 0 =0
\end{eqnarray}

 This flow corresponds to the vector $\vec S(\vec r)$ being rotated while transported along the flow line in $R_3$, thereby covering the target space $S_2$.

This rotation is nonlocal in the target space, as it involves derivatives $\pbyp{ S_i}{\phi_a}$.
Geometrically, the point $\vec S$ moves on a sphere while being carried by the Euler flow in physical space.

This gauge transformation accompanying the flow is a significant modification of the traditional view of the passive advection of the \CL{} field.

\section{Vortex sheets}\label{VortexSheets}
\subsection{Topological Invariants for closed surface}

Let us now consider the case of a closed vortex sheet.

In that case, the first \CL{} field $\phi_2$ takes constant values: $-\Delta \phi_2$ inside the surface and zero outside would lead to velocity gap  $\Delta \vec v = \Delta \phi_2 \vec \nabla\phi_1$, corresponding to the potential gap $\Gamma = \phi_1 \Delta \phi_2 $. 

In short, the closed vortex surface (if it exists) is the bubble of the \CL{} field.

The values of the \CL{} field $\phi_1$ outside the surface drop from the equation. One can take any smooth interpolating field $\phi_1$ between closed vortex surfaces, and the vorticity field will stay zero outside and inside the surfaces.

Furthermore, the velocity circulation around any contractible loop at each surface vanishes because there is no normal vorticity on these surfaces.

The \CL{} topology plays no role in the case of closed surfaces with the spherical topology, unlike the case with open surfaces with edges $C_n = \d S_n$ we considered above (see also \cite{M20c}).
There, the normal component of vorticity at the surface was present (and finite).

Still, our collection of closed vortex sheets in the general case has some nontrivial helicity \eqref{Helicity}. This helicity is pseudoscalar, but it preserves time-reversal symmetry. 
The higher genus surfaces, say the torus, will have nontrivial circulation around its cycles, corresponding to conserved vorticity flux through the skin of the handle.

The parity transformation $P$ changes the velocity sign, keeping vorticity invariant, whereas the time-reversal $T$ changes the signs of both velocity and vorticity.

The helicity integral is $T$-even and $P$-odd. It measures the knotting of vortex lines between these surfaces. 
\pct{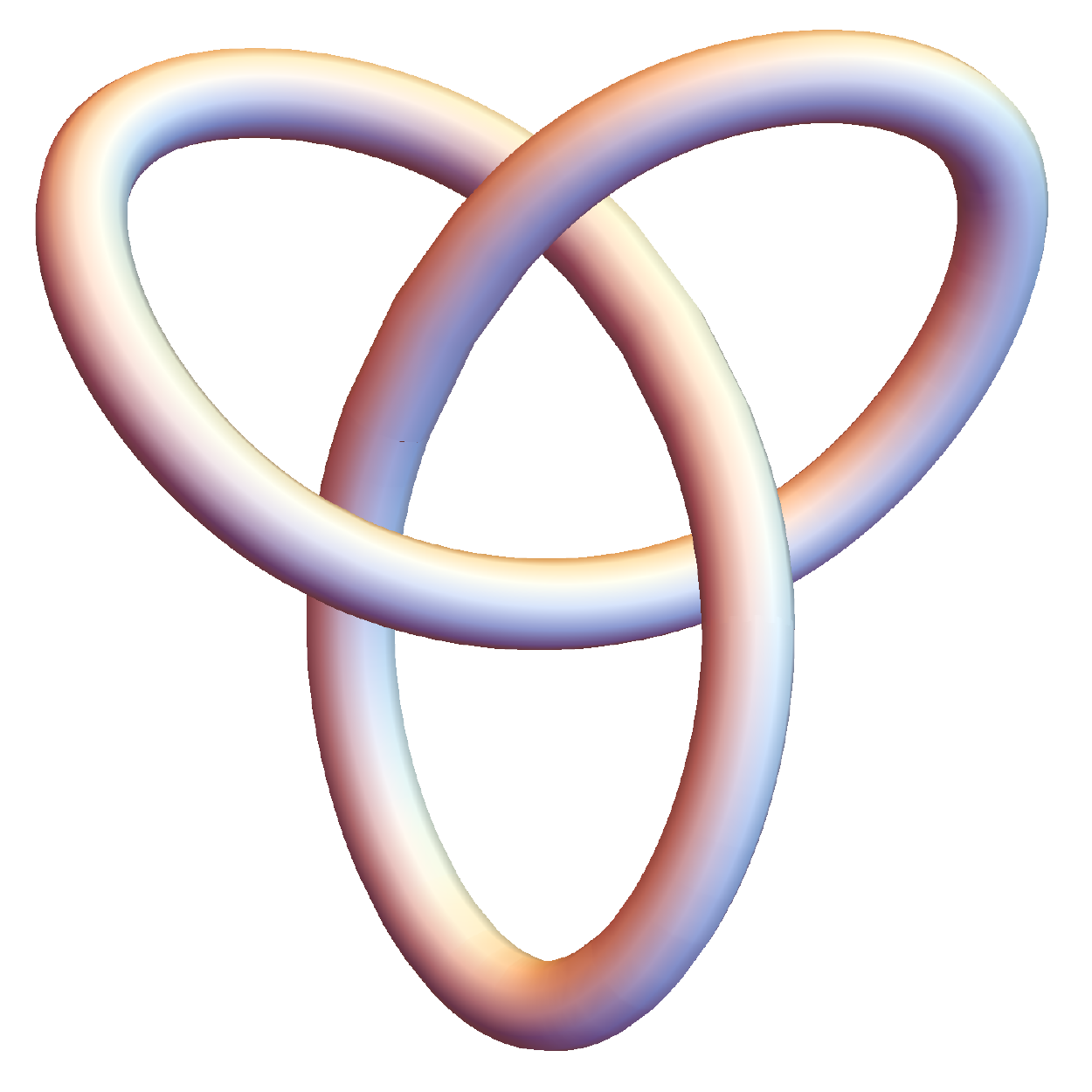}{The torus is knotted to produce nontrivial helicity.}

As noted in \cite{AM89}, the surfaces avoid each other and themselves, so these are not just random surfaces. This property was studied in their time evolution and recently in their statistics \cite{M20c}.

The circulation around each contractible loop on the surface will still be zero. However, the loop winding around a handle would produce a topologically invariant circulation  $n \Delta \Gamma$ for any loop winding $n$ times.

This $\Delta \Gamma$ is the period of $\Gamma(\xi)$ when the point $\xi$ goes around this handle\cite{M88}. This period depends upon the surface's size and shape, but it does not change when the path varies along the surface as long as it winds around the handle. 

There is a flux through the handle related to tangential vorticity inside the skin of the surface. This flux through any surface intersecting the handle is topologically invariant and equals $\Delta \Gamma$.

Considering the circulation around some fixed loop in space will reduce to an algebraic sum of the circulations around closed surfaces' handles encircled by this loop. It will be topologically invariant when the loop moves in space without crossing any surfaces. 

In particular, the closed vortex tube (topological torus) encircled by a fixed loop $C$ in space would produce the circulation $\oint_C \vec v \cdot d \vec r = \Delta \Gamma$, which is equal to the period of $\Gamma$ around the $\alpha$ cycle, corresponding to shrinking of $C$.
\pct{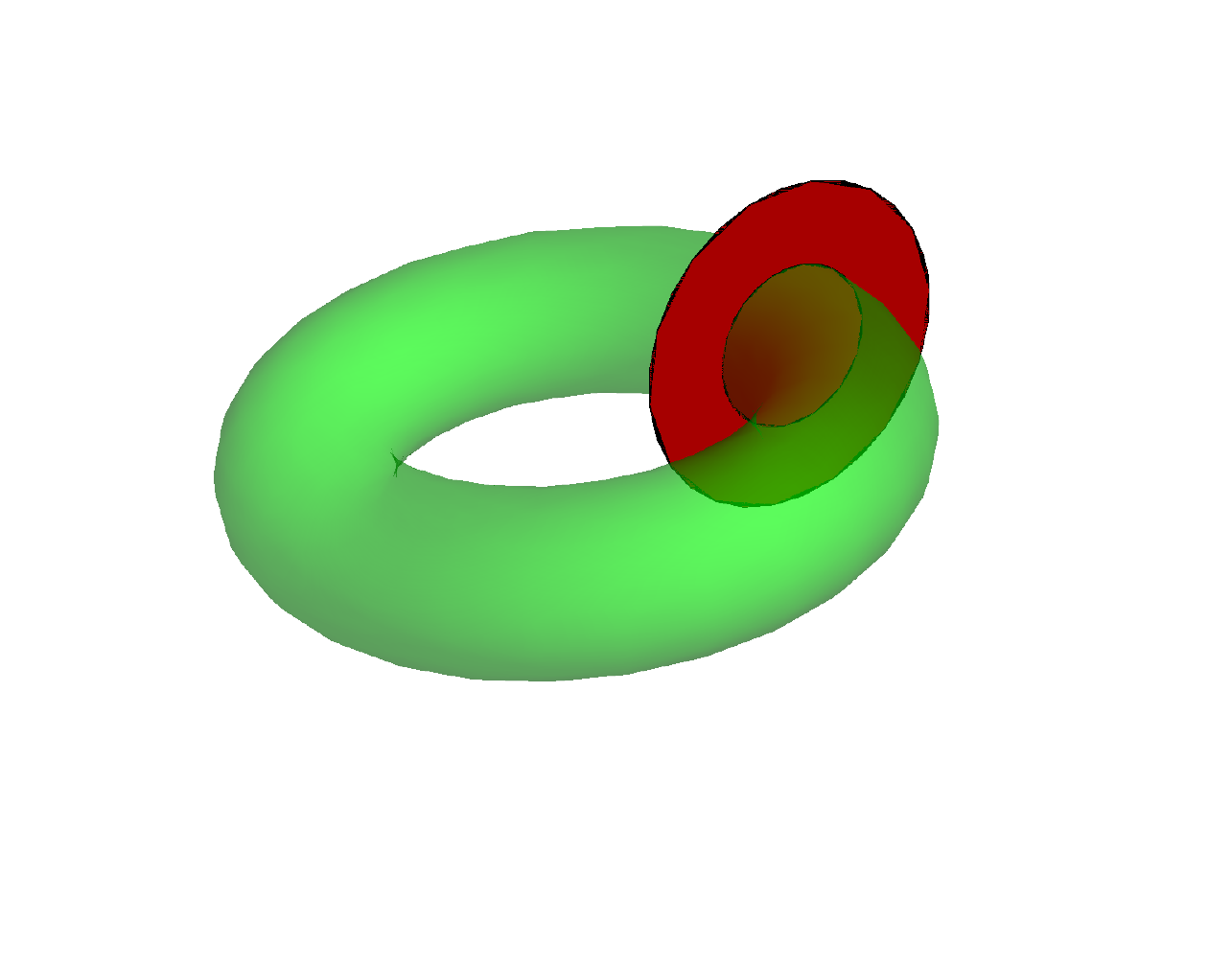}{The green vortex tube $T$ cut  vertically by a red disk $D_C$. The vorticity flux through the disk reduces to the integral of velocity discontinuity over the $\alpha$ cycle.}

Another cross-section of the same vortex tube leads to circulation around another cycle of the torus.
\pct{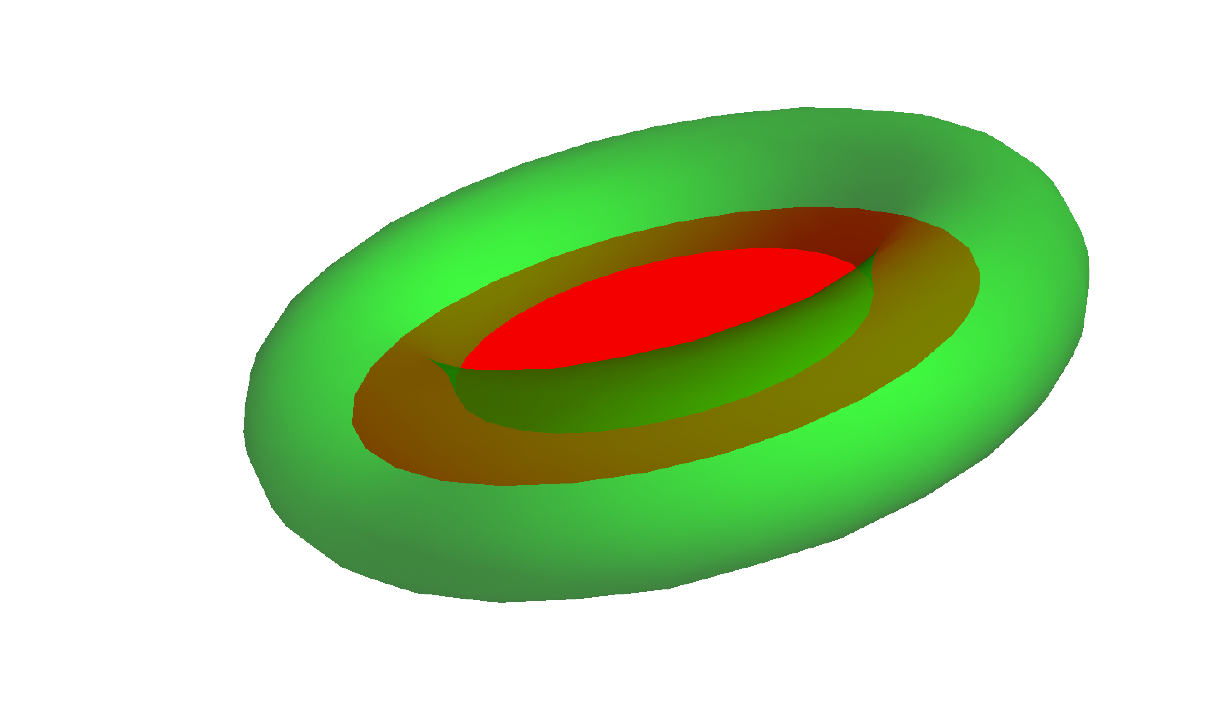}{The green vortex tube $T$ cut horizontally by a red disk $D_C$.The vorticity flux through the disk reduces to the integral of velocity discontinuity over the $\beta$ cycle.}

To prove the relation between circulation over the disk edge $C$ and $\Delta \Gamma$ over the corresponding cycle of the handle, let us consider the vorticity flux through the disk $D_C$ bounded by $C$ intersecting this torus $T$.

By Stokes' theorem, this flux equals the circulation on the external side of $T$ minus the internal side. On the other hand, it is equal to the circulation around the edge $C$ of this disk
\begin{eqnarray}
    \oint_C \vec v \cdot \vec d r = \oint_{\gamma\in T}  \Delta \vec v d \vec r = \oint d\Gamma = \Delta \Gamma
\end{eqnarray}

\subsection{Stable vortex sheets and irreversibility of turbulence}
The recent research \cite{M21a, KS21} revealed that the \BT{} regime required certain restrictions on the eigenvalues and eigenvectors of the background strain\footnote{The alignment of vorticity with strain tensor in the turbulent flow was observed long ago \cite{VortStrain87} and studied in the DNS since then.} for the planar vortex sheet solution.

This research is summarized and reviewed in our paper \cite{M21c}. The reader can find the exact solutions and the plots illustrating the vorticity leaks in the case of the non-degenerate background strain.

Abstracting from these observations, we conjectured in that paper that the local stability condition of the vortex surface with the local normal vector $\vec \sigma$ and local mean boundary value of the strain tensor $\hat S$ reduces to three equations (with $\vec v_\pm$ being the boundary values of velocity on each side)

\begin{eqnarray}\label{CVSEQ}
&& \vec v_\pm \cdot \vec \sigma =0;\\
&&\hat S \cdot (\vec v_+- \vec v_-) =0;\\
&& \hat S_{n n} =\vec \sigma \cdot \hat S \cdot \vec \sigma < 0
\end{eqnarray}

We call these equations the Confined Vortex Surface or \CVS{} equations. They are supposed to hold at every point of the vortex surface, thus imposing extra boundary conditions on the Euler equations. 

As conventional Neumann boundary conditions for potential flow on each side of for fixed vortex surface (the first \CVS{} equation) uniquely determine the local strain tensor,  the two additional \CVS{} boundary conditions restrict the allowed shapes of vortex surfaces.

The first \CVS{} equation is simply a statement that the surface is steady. Only the normal velocity must vanish for such a steady surface-- the tangent flow around the surface reparametrizes its equation but does not move it in the Euler-Lagrange dynamics.

The second \CVS{} equation is an enhanced version of the vanishing eigenvalue requirement. The velocity gap $\Delta \vec v = \vec v_+- \vec v_-$ should be a null vector for this zero eigenvalue. This requirement is stronger than $\det \hat S  =0$.

The third equation demands that the strain move the fluid towards both sides of the surface.  

With negative normal strain and zero strain in the direction of the velocity gap, the flow effortlessly slides along the surface on both sides without leakage and pile-up.

Here is an intuitive explanation of the \CVS{} conditions-- they provide permanent tangential flow around the surface, confining vorticity inside the boundary layer.

Once these requirements are satisfied in the local tangent plane, one could solve the \NS{}  equation in the (flat) boundary layer and obtain the Gaussian \BT{} solution, vindicating the stability hypothesis. The error function of the normal coordinate will replace the velocity gap, and the delta function of tangent vorticity will become the Gaussian profile with viscous width.

The flow in the boundary layer surrounding the local tangent plane has the form \cite{MB16} (with $\Phi_\pm(r)$ being velocity potentials on both sides )
\begin{eqnarray}\label{BTSol}
&& \vec v\left(\vec r_0 + \vec \sigma_0 \zeta\right) = \nonumber\\
&&\oh (\vec v_+(\vec r_0) +\vec v_-(\vec r_0))+
\oh \left(\vec v_+(\vec r_0) -\vec v_-(\vec r_0)\right)\erf{ \left(\frac{\zeta}{h \sqrt 2}\right)}; \\
&& \vec \sigma_0 = \vec \sigma(\vec r_0);\\
&& \vec v_\pm(\vec r_0) = \vec \nabla \Phi_\pm(\vec r_0);\\
&& \vec \omega\left(\vec r_0 + \vec \sigma_0 \zeta\right) = \frac{\sqrt  2}{h \sqrt \pi }\vec \sigma_0 \times (\vec v_+(\vec r_0) -\vec v_-(\vec r_0))\exp{-\frac{\zeta^2}{2 h^2}};\\
&& h = \sqrt{\frac{\nu}{-\vec \sigma_0 \cdot \hat S(\vec r_0) \cdot \vec \sigma_0}};\\
&& \hat S_{\alpha\beta}(\vec r_0) = \oh \dal \dbe \Phi_+(\vec r_0) + \oh \dal \dbe \Phi_-(\vec r_0)
\end{eqnarray}
With the inequality $ S_{ n n } = \vec \sigma_0 \cdot \hat S(\vec r_0) \cdot \vec \sigma_0 < 0$ satisfied, this width  $h$ will be real positive.

It is important to understand that this inequality breaks the reversibility of the Euler equation: the strain is an odd variable for time reflection, although it is even for space reflection.

Thus, we have found a dynamic mechanism for the irreversibility of the turbulent flow: the vorticity can collapse only to the surface with negative normal strain; otherwise, it is unstable. \footnote{Later, we see that the same applies to the vortex lines; their stability also requires a negative trace of the strain in the local normal plane to the vortex line.}

Another way of saying this is as follows. The strain tensor has three local eigenvalues, adding up to zero. We sort them in decreasing order.
The first one is positive, the last one is negative, and the second one lies in between and can, in principle, have any sign. 

The stable vortex sheet orients itself locally normal to the last eigenvector and distorts the flow so that the middle eigenvalue vanishes. The velocity gap in this flow is aligned with the middle eigenvector, the one with the vanishing eigenvalue.

Let us systematically describe this theory, starting from the vortex sheet dynamics.

\subsection{Hamiltonian Dynamics of Vortex Sheets}

The vortex sheets (velocity gaps) have a long history, with various famous scientists contributing to the theory.
The vortex surfaces recently came to our attention after it was argued \cite{M20c, M21b} that they provide the basic fluctuating variables in turbulent statistics. 

Let us define here the necessary equations.

\NS{}  equations 
\begin{subequations}
\begin{eqnarray}\label{NSv}
   && \d_t \vec v + (\vec v \cdot \vec \nabla) \vec v + \vec \nabla p =  \nu \vec \nabla^2\vec v;\\
   && \vec \nabla \cdot \vec v =0;
\end{eqnarray}
\end{subequations}
can be rewritten as the equation for vorticity
\begin{subequations}
\begin{eqnarray}
&&\vec \omega = \vec \nabla \times \vec v ;\\
    &&\d_t \vec \omega + (\vec v \cdot \vec \nabla) \vec \omega - (\vec \omega \cdot \vec \nabla) \vec v =  \nu \vec \nabla^2\vec \omega;\label{NSE}
\end{eqnarray}
\end{subequations}

As for the velocity, it is given by a \BS{} integral
\begin{equation}\label{BS}
    \vec v(r)= -\vec \nabla \times \int d^3 r' \frac{\vec \omega(r')}{4\pi | r - r'|}
\end{equation}
which is a linear functional of the instant value of vorticity. There is, in general, a solution of the Laplace equation to be added to the \BS{} integral.

The energy dissipation is related to the enstrophy (up to total derivative terms, vanishing by the Stokes theorem)
\begin{eqnarray}\label{Diss}
    -\d_t H = \mathcal{E} = \nu \int d^3 r \vec \omega^2 
\end{eqnarray}

The Euler equation corresponds to setting $\nu =0$ in \eqref{NSv}, \eqref{NSE}. This limit is known not to be smooth, leading to a statistical distribution of vortex structures, which is the whole turbulence problem. 

Within the Euler-Lagrange equations, the shape $S$ of the vortex surface is arbitrary, as well as the density $\Gamma(\vec r \in S) $ parametrizing the velocity discontinuity $ \Delta \vec v = \vec \nabla \Gamma$. The corresponding vortex surface dynamics\cite{M88, AM89} represents a special case of the Hamiltonian dynamics in 2 dimensions with parametric invariance similar to the string theory.

 \subsection{ \KH{} instability}

 In conventional Euler-Lagrange dynamics, the vortex surface's shape is evolving, subject to \KH{} instability, while $\Gamma$ stays constant due to the Kelvin theorem.

The steady solution for the vortex surface $\mathcal S$, as we recently argued in  \cite{M20c, M21b} would correspond to a particular potential gap $\Gamma(\mathcal S) = \Phi_+(\mathcal S) - \Phi_-(\mathcal S) $  minimizing the fluid Hamiltonian.

This minimization is equivalent to the Neumann boundary condition for the potential flow $ \vec v(\vec r) = \vec \nabla \Phi_\pm(\vec r)$ inside and outside the vortex surface
\begin{eqnarray}
   \d_n \Phi_+(\mathcal S)= \d_n \Phi_-(\mathcal S) =0
\end{eqnarray}

The local normal displacement $z$ of the surface\footnote{the tangent displacement is equivalent to reparametrization of the surface and, as such, can be discounted \cite{M21a}.} satisfies the Lagrange equation
\begin{eqnarray}
  && \d_t z = v_n  = S_{ n n} z + O(z^2)\\
  && S_{ n n} = \oh \left(\d_n^2 \Phi_+(\mathcal S) + \d_n^2 \Phi_-(\mathcal S)\right)
\end{eqnarray}

The positive normal strain $\hat S_{n n} $ would lead to the (trivial version of) \KH{} instability, but in case
\begin{eqnarray}
   \hat S_{n n}(\mathcal S) <0,
\end{eqnarray}
this steady vortex surface would be stable against smooth variations in a normal direction.

The \KH{} instability remains when we allow high wavelength variations, but in certain cases, it is stabilized by viscosity (see below).

In the conventional analysis of the \KH{} instability \cite{BAT00}, Section 7, the variation of the displacement in the tangent direction is taken into account, which leads to a more general equation
\begin{eqnarray}
   \d_t z = -a z -\hat b z
\end{eqnarray}
where $\hat b$ is a linear differential operator. The normal strain $-a z$ was ignored in that book.

The equation is linear and can be solved in Fourier space where $\hat b z \Ra \omega(k) \tilde z(k)$
\begin{eqnarray}
  \tilde z(k) = c_1 \exp{-(a + \omega(k)) t }
\end{eqnarray}

This $\omega(k)$ has two eigenvalues $\pm k \Delta v_t$, of opposite sign, and the negative eigenvalue leads to the \KH{} instability.
This instability happens at large enough wave vectors $k \sim \frac{a}{\Delta v_t}$, invalidating the Euler vortex sheets.

However, the smearing of the vortex sheet in the boundary layer by viscosity stabilizes it under certain conditions, which we discuss below.
 \subsection{ The \BT{} vortex sheet}
We know some exact solutions of the \NS{}  equations where a boundary layer replaces the vortex sheet with the width $h \sim \sqrt{\frac{\nu}{a}}$. 

The \BT{} solution reads
\begin{subequations}
\begin{eqnarray}
  &&\vec v = \{ a x, b F_h(z), -a z\};\\
  && S_{\alpha\beta} = \diag{a,0,-a};\\
  &&\vec \omega =\{-b F_h'(z),0,0\} ;\\
  &&F_h'(z) =\frac{1}{h \sqrt{2 \pi} } \exp{- \frac{z^2}{2 h^2}};\\
  \label{ErrorFunc}
  && F_h(z) =\erf \left( \frac{z}{h \sqrt 2}\right);\\
  && h = \sqrt{\frac{\nu}{a}};\label{aheq}
\end{eqnarray}
\end{subequations}

We verified this solution analytically in various coordinate systems in a \Mathematica notebook \cite{MB} for the reader's convenience. 
In the Euler limit of $\nu \ra 0$, the vorticity reduces to $\delta(z)$, and the velocity gap reduces to $\sign(z)$.

The dissipation in this limit is proportional to the area of the sheet
\begin{eqnarray}
  &&\mathcal E =  b\sqrt{\frac{2 a \nu}{ \pi} } \int d S;
\end{eqnarray}

This result disappointed Burgers, who was looking for a finite limit of anomalous dissipation at $\nu \ra 0$. In this solution, that finite limit would require $a b^2$ to grow as $1/\nu$, for which there were no physical reasons.

\subsection{The vortex sheet dynamics and parametric invariance}
As we know, since the times of Euler and Lagrange, the Euler equations are equivalent to the statement that every element of any vortex structure is moving with the local velocity.

However, the mathematical meaning of this remarkable equivalency was not elaborated initially. In particular, how does this statement apply to the singular vortex structures, where the local velocity is not defined?

In the case of the vortex line, the Euler velocity diverges at every point at the line; therefore, the \NS{}  equation must be used. It produces some exact solutions with an infinite velocity at the line in the vanishing viscosity limit.

We will study these solutions in the next Sections: our current goal is the vortex sheet, where the Euler velocity is finite on both sides of the surface, with the tangent velocity gap.

In this case, the velocity driving each point of the vortex surface is an arithmetic mean of the velocities on both sides, in other words, the principal value of the Biot-Savart integral.

As we discussed in\cite{M88, AM89}, the velocity at the surface 
\begin{eqnarray}
    \vec v^S(\vec r) = \oh \left(\vec v\left(\vec X^+(\xi)\right) + \vec v\left(\vec X^-(\xi)\right)\right)
\end{eqnarray}

The normal component $v_n$ of velocity describes the surface's genuine change– its global motion or the change of its shape. This normal component must vanish at every surface point in a steady flow.

On the other hand, the remaining two tangent components $\vec v_t$ of the velocity move points along the surface without changing the surface position or shape. To see that, we rewrite the tangent part $\vec v^S_t $ of the surface velocity $\vec v^S(\vec r)$ as time-dependent re-parametrization $\xi \Ra \xi(t)$:
\begin{eqnarray}\label{ParInv}
    &&\vec v^S_t  = \d_t \vec X(\xi(t)) = \d_a \vec X \d_t\xi^a ;\\
    &&\d_t\xi^a =  g^{a b} \d_b \vec X \cdot \vec v^S_t;\\
    && g_{a b} = \d_a \vec X \cdot \d_b \vec  X;
\end{eqnarray}
Here $g_{a b} $ is an induced metric, and $g^{a b}$ is its inverse. To verify this identity, one has to expand $\vec v^S_t $ in the two local plane vectors $\d_1 \vec X, \d_2 \vec X$ and use two identities 
\begin{eqnarray}
    &&\d_b \vec X \cdot \d_c \vec X = g_{b c};\\
    && g^{a b} g_{b c} = \delta_{a c}
\end{eqnarray}

For a closed surface, these tangent motions will never leave the surface. For the surface with the fixed edge $C$, there is a boundary condition that velocity normal to the edge vanishes $ \vec v_t\times d \vec r =0; \forall \vec r \in C$. In this case, the fluid will slide along $C$ leading to its reparametrization, but it never leaves the surface.

We will restrict ourselves to the parametric invariant functionals, which do not depend on this tangent flow and will stay steady. 
\subsection{Multivalued Action}\label{ActionSurface}

We introduced the Lagrange action of vortex sheets in the old paper\cite{M88}. In that paper, we also conjectured the relation of turbulence to Random Surfaces.\footnote{I could not find \cite{M88} anywhere online, but it exists in university libraries such as Princeton University Library and U.C. Berkeley Library.}

In the next paper\cite{AM89}, this Lagrange vortex dynamics was simulated using a triangulated surface. We calculated each triangle's contributions to the velocity field in terms of elliptic integrals. The positions of the triangle vertices served as dynamical degrees of freedom. 

There were conserved variables related to the velocity gap as a function of a point at the surface. These points passively moved with the surface by the mean velocity on the surface's two sides. 

Later, our equations were recognized and reiterated in traditional terms of fluid dynamics\cite{Kaneda90} and simulated with a larger number of triangles \cite{Brady98} with similar results regarding the \KH{} instability of the vortex sheet. There were dozens of publications using various versions of the discretization of the surface and various simulation methods.

Let us reproduce this theory for the reader's convenience before advancing it further. 

The following ansatz describes the vortex sheet vorticity: 
\begin{equation}\label{omegaInv}
   \vec \omega(\vec r) = \int_\Sigma d\vec \Omega \delta^3\left(\vec X-\vec r\right)
\end{equation}
where the 2-form
\begin{eqnarray}
    d \vec \Omega \equiv d \Gamma\wedge d\vec X = d\xi_1 d \xi_2 e_{a b} \pbyp{\Gamma}{\xi_a} \pbyp{\vec X}{\xi_b} ;
\end{eqnarray}

This vorticity is zero everywhere in space except the surface, where it is infinite. This ansatz must satisfy the divergence equation (the conservation of the "current" $\vec \omega$ in the language of statistical field theory) to describe the physical vorticity of the fluid, 
\begin{eqnarray}
    \vec \nabla \cdot \vec \omega =0;
\end{eqnarray}
This relation is built into this ansatz for arbitrary $\Gamma(\xi)$, as can be verified by direct calculation. In virtue of the singular behavior of the Dirac delta function, it may be easier to understand this calculation in Fourier space
\begin{eqnarray}
&& \vec \omega^F(\vec k) = \int d^3 r e^{\i \vec k \cdot \vec r} \vec \omega(\vec r) = \int_\Sigma d \vec \Omega e^{\i \vec k \cdot \vec X};\\
    &&\i \vec k \cdot \vec\omega^F(\vec k) =\int_\Sigma  d \Gamma \wedge d \vec X \cdot (\i \vec k) e^{\i \vec k \cdot \vec X}=\int_{\d \Sigma} d\Gamma e^{\i \vec k \cdot \vec X};
\end{eqnarray}
If there is a surface boundary, this $\Gamma(\xi)$ must be a constant at this boundary for the identity $\vec k \cdot  \omega^F(\vec k)=0$ to hold. Transforming to $R_3$ from Fourier space, we confirm the desired relation $\vec \nabla \cdot \vec \omega =0$ in the sense of distribution, like the delta function itself.

It may be instructive to write down an explicit formula for the tangent components of vorticity in the local frame, where $x,y$ is a local tangent plane and $z$ is a normal direction
\begin{eqnarray}
   && \omega_j(x,y,z) =  \d_i\Gamma  e_{i j}  \delta(z);\\
   && \omega_z(x,y,z) =0;
\end{eqnarray}
In particular, outside the surface, $\vec \omega =0$, so that its divergence vanishes trivially.

The divergence is manifestly zero in this coordinate frame 
\begin{eqnarray}
    \vec \nabla \cdot \vec \omega = \delta(z)\d_j \d_i\Gamma  e_{i j} =0;
\end{eqnarray}
Let us compare this with the \CL{} representation
\begin{equation}
    \oal =  e_{\alpha\beta\gamma} \dbe \phi_1 \dga\phi_2,
\end{equation}
We see that in case $\phi_2$ takes one space-independent value $\phi_2^{in}$ inside the surface and another space-independent value $\phi_2^{out}$ outside, the vorticity will have the same form, with 
\begin{eqnarray}
    \Gamma = \phi_1 (\phi_2^{in} - \phi_2^{out})
\end{eqnarray}

Neither \cite{AM89} nor any subsequent papers noticed this relation between \CL{} field discontinuity and vortex sheets.

As we already noted in \cite{M88,AM89}, the function $\Gamma(\xi_1, \xi_2)$ is defined modulo diffeomorphisms $\xi \Ra \eta(\xi); \det {\d_i \eta_j} >0$ and is conserved in Lagrange dynamics:
\begin{equation}
    \d_t \Gamma =0;
\end{equation}
This function is related to the $1$-form of velocity discontinuity
\begin{equation}\label{DiscGamma}
    d \Gamma  = \Delta \vec v \cdot  d \vec X
\end{equation}
where the velocity gap
\begin{equation}
    \Delta \vec v(\xi) = \vec v\left(X_+(\xi)\right) -\vec v\left(X_-(\xi)\right)
\end{equation}

Another way of writing the relation for $\Gamma$ is to note that the velocity field is purely potential outside the surface because there is no vorticity.

However, the potential $\Phi^-$ inside is different from the potential $\Phi^+ $ outside.

In this case, the velocity discontinuity equals the difference between the gradients of these two potentials or, equivalently
\begin{eqnarray}
    \Gamma(\vec r) = \Phi^+(\vec r) - \Phi^-(\vec r); \forall \vec r \in S
\end{eqnarray}

The surface is driven by the self-generated velocity field (mean of velocity above and below the surface). Let us substitute our ansatz for vorticity  into the Biot-Savart integral for the velocity field  and change the order of integration
\begin{eqnarray}\label{BSGamma}
    &&\vec v(\vec r) = -\frac{1}{4 \pi}  \vec \nabla \times\int d^3 r' \frac{1}{|\vec r - \vec r'|}\int d \vec \Omega   \delta^3(X - \vec r')= \nonumber \\
    && \frac{1}{4 \pi}\int d \vec \Omega \times \vec \nabla \frac{1}{|\vec r - \vec X|}
\end{eqnarray}

\subsection{Steady State as a Minimum of the Hamiltonian}

The Lagrange equations of motion for the surface
\begin{equation}
    \d_t \vec X(\xi) = \vec v\left(\vec X(\xi)\right)
\end{equation}
were shown in\cite{M88, AM89} to follow from the action
\begin{eqnarray}\label{action}
   && S = \int \Gamma d  V- \int H d t;\\
   && d V = d \xi_1 d \xi_2 d t  \pbyp{\vec X}{\xi_1} \times \pbyp{\vec X}{\xi_2} \cdot \d_t \vec X ;\\
   && H = \oh \int d^3 r \vec v^2 = \frac{1}{2} \int_{S}\int_{S}  \frac{ d \vec \Omega\cdot d \vec \Omega' }{4 \pi |\vec X - \vec X'|}; 
\end{eqnarray}
This $d V$ is the 3-volume  swept by the surface area element in its movement for the time $ d t$.

The easiest way to derive the vortex sheet representation for the Hamiltonian is to go in Fourier space where the convolution becomes just multiplication and use the incompressibility condition $\vec k \cdot \vec v^F(\vec k)=0$
\begin{eqnarray}
    && \vec \omega^F(\vec k)  = \i \vec k \times \vec v^F(\vec k);\\
    &&\vec v^F(\vec k) \cdot \vec v^F(-\vec k) = \frac{\vec \omega^F(\vec k) \cdot \vec \omega^F(-\vec k)}{\vec k^2}
\end{eqnarray}
In the case of the handle on a surface, $\Gamma$ acquires extra term $\Delta \Gamma = \oint_{\gamma} \Delta \vec v \cdot d \vec r$ when the point goes around one of the cycles $\gamma = \{\alpha, \beta\}$ of the handle. 

This $\Delta \Gamma$ does not depend on the path shape because there is no normal vorticity at the surface, and thus there is no flux through the surface. This topologically invariant $\Delta \Gamma$ represents the flux through the handle cross-section.

This ambiguity in $\Gamma$ makes our action multivalued as well.

Let us check the equations of motion emerging from the variation of the surface at fixed $\Gamma$:
\begin{eqnarray}
    &&\delta \int H d t = \int d \vec \Omega \times \delta \vec X\cdot \vec v(\vec X) d t;\\\label{deltaH}
    &&\delta \int \Gamma d V = \int d \vec \Omega \times \delta \vec X\cdot \d_t \vec X d t\label{deltaS}
\end{eqnarray}

As we discussed above, the tangent components of velocity at the surface create tangent motion, resulting in the surface's reparametrization.

One of the two tangent components of the velocity (along the line of constant $\Gamma(\xi)$) does not contribute to the variation of the action so that the correct Lagrange equation of motion following from our action reads
\begin{equation}
    \d_t \vec X(\xi) = \vec v(\vec X(\xi)) \mod{e^{i j} \d_i\Gamma  \d_j \vec X}
\end{equation}

We noticed this gauge invariance before in\cite{M88}. Now we see that both tangent components of the velocity could be absorbed into the reparametrization of a surface;  therefore, these components do not represent an observable change.

However, the normal component of the velocity must vanish in a steady solution, which provides a linear integral equation for the conserved function $\Gamma(\xi)$. \footnote{To be more precise, in the case of a uniform global motion of a fluid, the normal component of the velocity field must coincide with the normal component of this uniform global velocity $\vec v^G$.}

In the general case, when there is an ensemble of such surfaces $S_n, n =1,\dots  N$ each has its discontinuity function $\Gamma_n(\xi)$. At each point  on each surface $\vec r \in  S_n$, the net normal velocity adding up from  all surfaces, including this one in the \BS{} integral, must vanish:
\begin{eqnarray}
   && \vec \Sigma_n(\xi) \cdot \vec v(\vec X_n(\xi))=0;\\
   && \vec \Sigma_n(\xi) = e^{ i j}\d_i \vec X_n \times \d_j\vec X_n ;\\
   && \vec v(\vec r)  = \frac{1}{4 \pi} \sum_m\int d \vec \Omega_m (\xi)\times \vec \nabla\frac{1}{|\vec r - \vec X_m(\xi)|}
\end{eqnarray}
 
 This requirement provides a linear set of $N$ linear integral equations (called Master Equation in \cite{M20c}) relating $N$ independent surface functions $\Gamma_1\dots\Gamma_n$.
 With this set of equations satisfied, the collection of surfaces $S_1 \dots S_N$ will remain steady up to reparametrization.
 
Here is an essential new observation we have found in that paper. 

\textbf{The Master Equation is equivalent to the minimization of our Hamiltonian by $\Gamma_n, n=1\dots N$}
 \begin{eqnarray}\label{Ham}
   &&H[\Gamma,\vec X]= \frac{1}{2}\sum_{n,m} \int_{S_n}\int_{S_m}  \frac{ d \vec \Omega_n(\xi)\cdot d \vec \Omega_m(\xi) }{4 \pi | \vec X_n(\xi) - \vec X_m(\xi')|}; \\
   && \fbyf{H[\Gamma,\vec X]}{\Gamma_n(\xi)} = \vec \Sigma_n(\xi) \cdot \vec v\left(\vec X_n(\xi)\right);
\end{eqnarray}

The tangential components of velocity $\vec v_t$ are included in the parametric transformations, as noted above. They are equivalent to variations of the Hamiltonian by the parametrization of $\Gamma, \vec X$ and are, therefore, the tangent components of Lagrange equations of motion satisfied by parametric invariance. 

Therefore, the normal surface velocity in the general case is equal to the Hamiltonian variation by $\Gamma$, as if $\Gamma$ is the conjugate momentum corresponding to the surface's normal displacement. To be more precise, $\Gamma$ in our action \eqref{action} is a conjugate momentum to the volume, which is locally equivalent -- the variation of volume equals the area element times the normal displacement. 

In other words, we can consider an extended dynamical system with the same Hamiltonian \eqref{Ham} but a wider phase space $\Gamma,\vec X \mod{\textbf{Diff}}$. We can introduce an extended Hamiltonian dynamics with our action \eqref{action}. 

This system is degenerate because for an arbitrary evolution of $\vec X$ providing an extremum of the action, the evolution for $\Gamma$ is absent, i.e., $\Gamma$ is constant. It is a conserved momentum in our Hamiltonian dynamics with volume as a coordinate.

This conservation of $\Gamma$ is a consequence of Kelvin's theorem. To see this relation\cite{AM89}, we rewrite this $\Gamma$ as a circulation over the loop $C$ puncturing the surface in two points $A, B$ and going along some curve $\gamma_{A B}$ on one side, then back on the same curve $\gamma_{B A}$ on another side. The circulation does not depend upon the shape of $\gamma_{ A B}$ because there is no normal vorticity at the surface.

Another way to arrive at the conservation of $\Gamma$ is to notice that it is related to the \CL{} field on the \DS{}, as we mentioned in the introduction.

The steady solution for $\vec X \mod \mbox{Diff}$ corresponds to the Hamiltonian minimum as a (quadratic) functional of $\Gamma$. 

\subsection{Does Steady Surface mean Steady Flow?}

There is a subtle difference between the steady \DS{} and steady flow. After all, the flow around a steady object is not necessarily steady. There could be time-dependent motions in the bulk while the flow's normal component vanishes at the solid surface (as it always does).

This logic applies to the generic flow around steady solid objects but does not apply here. The big difference is that, by our assumption, there is no vorticity outside these discontinuity surfaces. 

The  Biot-Savart integral for the velocity field \eqref{BSGamma} is manifestly parametric invariant, if we transform \textbf{both} $\Gamma, \vec X$
\begin{eqnarray}
    &&\Gamma(\xi) \Ra \Gamma(\eta(t,\xi));\\
    && \vec X(\xi) \Ra \vec X(\eta(t,\xi));\\
    && \d_t \eta_a = u_a(\eta);
\end{eqnarray}

This transformation describes the flux of coordinates $\eta$ in parametric space with the velocity field $u_a(\eta)$.
The tangent flow around the surface is equivalent to the transformation of $\vec X$, as demonstrated in \eqref{ParInv}. 

However, in the Lagrange dynamics of vortex sheets, the function $\Gamma(\xi)$ remains constant, not constant up to reparametrization. However, an absolute constant so that $\d_t \Gamma(\xi)=0$ where time derivative goes at fixed $\xi$.

Therefore, the velocity field $\vec v(\vec r)$ generally changes when the surface gets re-parametrized, but $\Gamma$ does not. Naturally, one could not get a steady solution without solving some equations first :).

However, in our steady manifold, $\Gamma^*(\xi)$ is related to the surface by the Master Equation. Let us write it down once again
\begin{eqnarray}
    &&0=\vec \Sigma_n \cdot \sum_m \int_{S_m} d \Gamma^*_m \wedge d \vec X_m \times \vec \nabla_n \frac{1}{|\vec X_n - \vec X_m|} ;\\
    && \vec \Sigma_n = e^{a b} \d_a \vec X_n \times \d_b \vec X_n
\end{eqnarray}
This equation is invariant under the parametric transformation of both variables $\Gamma^*, \vec X$. This invariance means that the solution of this equation for $\Gamma^*$ would come out as a parametric invariant functional of $\vec X$, also invariant by translations of $ \vec X$.

As we have seen, this Master Equation
leads to a vanishing normal velocity $\vec \Sigma_n\cdot \d_t \vec X_n=0$. The remaining tangent velocity leaves $\vec X$ steady up to reparametrization. Therefore, in virtue of this Master Equation, the velocity field will also be steady.

We introduced a family of steady solutions of Lagrange equations, parametrized by an arbitrary set of discontinuity surfaces $\vec X_n(\xi)$ with discontinuities $\Gamma^*_n(\xi)$ determined by the minimization of the Hamiltonian.

The surfaces are steady up to time-dependent reparametrization (diffeomorphism). The equivalence of Lagrange and Euler dynamics suggests that these are steady solutions to the Euler equations.

The above arguments are too formal to accept our steady solution of the Euler equation.
These are weak solutions with tangent discontinuities of velocity, and thus they require some care to investigate the equivalence between the Lagrange and Euler solutions.

We will investigate these solutions' stability in the following sections.
\subsection{CVS Equation for Cylindrical Geometry}

The motivation for cylindrical geometry is as follows. 
\begin{itemize}
    \item This is the only known case of exactly solvable stability equations. 
    \item Several numerical simulations \cite{Tubes19, VTubes} indicate vortex structures shaped as long tubes corresponding to the cylindrical geometry.
    \item Three theorems by De Lellis and Bru\'e severely limit the space of possible \CVS{}. There is no compact closed \CVS{}; moreover, there is no cylindrical \CVS{} with a bounded cross-section.
\end{itemize}

Our cylindrical unbounded solutions are (post-factum) allowed by these three theorems. 
A uniform strain tensor on one side allows the overdetermined system of equations to have a solution.

This solution heavily relies on complex analysis, thanks to the cylindrical geometry. There is no known analog of this analysis in full three-dimensional geometry.

Given the exceptional nature of this solution, we put forward the following conjecture:

\textbf{Conjecture 1}.The only \CVS{}  are unbounded cylindrical ones, with uniform strain tensor on one side.

We cannot prove this conjecture, so we leave it for further study.
With this conjecture proven, we would have the complete solution to the vortex sheet stability problem.

Before studying the associated turbulent statistics, let us describe, generalize and solve the cylindrical \CVS{} equations. 

In cylindrical geometry, the piecewise harmonic potential has the following form:
\begin{eqnarray}\label{complexPot}
&&\Phi_\pm(x,y,z) = \oh a x^2 + \oh b y^2  + \oh c z^2 + \Re \phi_\pm(\eta);\\
&& a + b + c =0,\, a \le b \le c;\; c >0, a <0;\\
&& \eta = x+ \i y;\\
&& \phi_\pm(\eta) =  \frac{1}{2 \pi \i}\int d \Gamma_\pm(\theta) \log \left(\eta - C(\theta) \right);\\\label{Vgamma}
&&v_z = c z;\\
&& V_\pm(x,y) = v_x - \i v_y = a x - \i b y + F_\pm(x+ \i y);\\
&& F_\pm(\eta) = \phi'_\pm(\eta);
\end{eqnarray}

Here $C(\theta)$ is a complex periodic function of the angle $\theta \in (0, 2 \pi)$ parametrizing the closed loop $C$.
We simplified the notations compared to \cite{M21c}, and now we denote the ordered eigenvalues of the background strain tensor as $a,b,c$.

The geometry is as follows. The vortex surface corresponds to the parallel transport of the $x y $ loop $C$ in the third dimension $z$. Thus, at some point on the surface, the local tangent frame is
\begin{eqnarray}
&&\vec E_1 =  \left(\frac{\Re C'}{|C'|}, \frac{\Im C'}{|C'|}, 0\right);\\
&&\vec E_2 =  \left(0,0, 1\right);\\
&& \vec E_3 = \left(\frac{\Im C'}{|C'|}, -\frac{\Re C'}{|C'|}, 0\right)
\end{eqnarray}
The first two orthogonal vectors define the tangent plane, and the third one $\vec E_3 = \vec E_1 \times \vec E_2$ corresponds to the normal $\vec \sigma$ to the surface. Do not confuse the $z$ direction with the normal -- this is one of the two tangent directions.

\pct{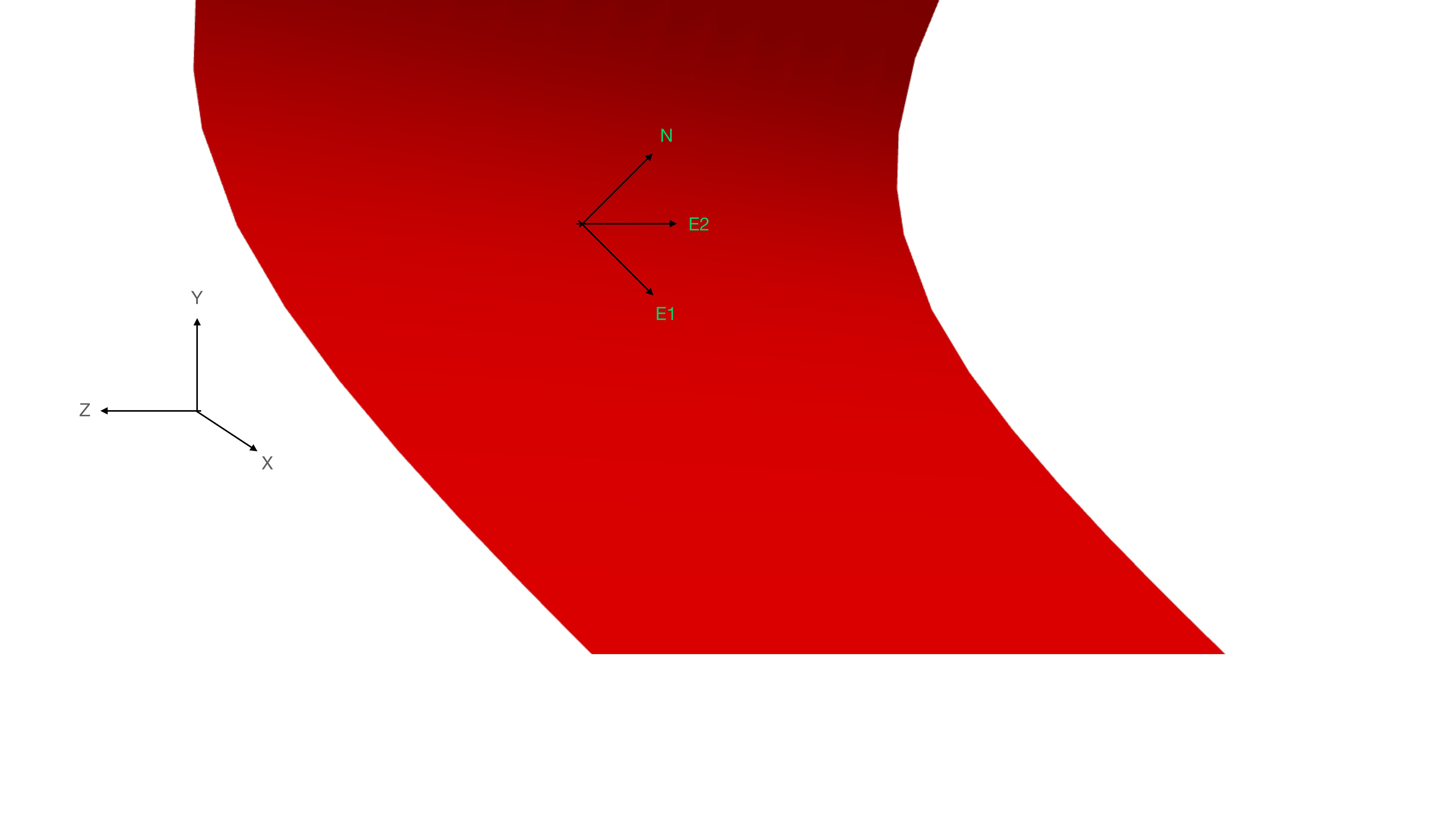}{The  local tangent plane $E_1, E_2, N$ and the global Cartesian frame $X,Y,Z$.}

In this solution, the functions $\Gamma_\pm(\theta)$ can be complex, but the circulation will depend on the real part.

The double-layer potential studied in the previous paper \cite{M21c}, corresponds to the real $\Gamma_+(\theta)= \Gamma_-(\theta)$. In that case, the reparametrization of the curve would eliminate this function so that the solution would be parametrized by the loop $C(\theta)$ modulo diffeomorphisms.

As we see it now, the condition of real and equal $\Gamma_\pm(\theta)$ is unnecessary. We shall find simple analytic solutions by dropping this restriction.

One can reduce the  condition of vanishing normal velocity at each side of the steady surface to two real equations 
\begin{eqnarray}\label{shapeEq}
&&\Im C'(\theta) V_+\left(C(\theta)\right) =0;\\
&&\Im C'(\theta) V_-\left(C(\theta)\right) =0;
\end{eqnarray}
with $V_\pm$ denoting the boundary values at each side of the surface.

The second \CVS{} equation was reduced in \cite{M21c} to the complex equation 
\begin{eqnarray}\label{CVSCyl}
V_+\left(C(\theta)\right) + V_-\left(C(\theta)\right) =0;
\end{eqnarray}

These two equations were derived in \cite{M21c} under assumptions of real $d\Gamma_\pm(\theta)$  in \eqref{Vgamma}, which we find unnecessary and drop now.

In Appendix A, we re-derive these two equations without extra assumptions.

It follows from equation \eqref{CVSCyl} that one of the two eigenvalues of the local tangent strain at the surface vanishes. This vanishing eigenvalue corresponds to the velocity gap $\Delta V =V_+ - V_- = - 2 V_-$ as an eigenvector.

The way to prove this is to note that differentiation of \eqref{CVSCyl} by $\theta$ reduces to projecting the complex derivatives of $V_+(C)+ V_-(C)$  on the complex tangent vector $C'(\theta)$. As the $\theta$ derivative of $V_+(C)+ V_-(C)$ is zero in virtue of \eqref{CVSCyl}, so is the projection of the strain in the direction of the curve tangent vector $ C'(\theta)$. 

There is no normal velocity gap (as both normal velocities are zero), nor is there any gap in the $z$ component of velocity $v_z = c z$. Thus, the velocity gap aligns with the curve tangent vector $C'(\theta)$, and the strain along the velocity gap vanishes. Direct calculation in Appendix A supports this argument.

The other eigenvalue $\hat S_{n n}$, corresponding to the eigenvector $ \vec \sigma =\vec E_3$  normal to the surface, is then uniquely fixed by the condition of a vanishing trace of the local strain tensor.

The third eigenvalue $c$  corresponds to our cylindrical tube's eigenvector $\vec E_2$.

Therefore, the normal component of the strain tensor is 
 \begin{eqnarray}
 \hat S_{n n} = -c
 \end{eqnarray}
 
 \subsection{Spontaneous breaking of time-reversal symmetry and Normal strain}
 
 As the largest $c$ of the three ordered numbers $a,b,c $ with the sum equal to $0$ is always positive (unless all three are zero), we have a negative normal strain as required by the stability of the \NS{}  equation in the local tangent plane.
 
 Contrary to our previous statements, this condition does not restrict the background strain tensor. The irreversibility of turbulence manifests itself in the negative sign of the normal strain on the surface, which is true for arbitrary finite eigenvalues $a,b,c$. 
 
 The stability conditions restrict the vortex tube shape: its axis aligns with the leading eigenvector of the background strain, and its profile loop adjusts to the local strain to annihilate its projection on the velocity gap.
 
 Unlike the general shape of the vortex tube, the cylindrical shape guarantees the negative sign of the normal strain. We may be dealing with the spontaneous emergence of cylindrical symmetry at the expense of breaking the time-reversal symmetry from the stability condition. 
 
 The maximal value of the normal strain on the closed vortex surface represents a functional of its shape. So, the problem is to find under which conditions this maximal value is negative. The potential being harmonic, this normal strain equals minus the surface Laplacian of the boundary value of potential.
 
 Thus, we are dealing with the minimal value of the surface Laplacian of a boundary value of harmonic potential on a closed surface. Under which conditions is this minimal value positive? 
 
 The third theorem by De Lellis and Bru\'e answers this question. This minimal value cannot be positive for any bounded smooth surface of an arbitrary genus because the average over such surface vanishes by the Gauss theorem.
 
 So, this leaves us with unbounded surfaces. We suspect these are only cylindrical ones with uniform strain on one side. This condition is sufficient for \CVS{}, as we shall see below, but we cannot prove it is necessary.
 
 The Euler equation is invariant under time reversal, changing the sign of the strain. Without the \CVS{} boundary conditions, both signs of the normal strain would satisfy the steady Euler equation. Therefore, this \CVS{} vortex sheet represents a dynamical breaking of the time-reversal symmetry.
 
 Out of the two time-reflected solutions of the Euler equation, only the one with the negative normal strain survives. If virtually created as a metastable phase, the other dissolves in the turbulent flow, but this remains stable. 
 
 Technically this instability displays itself in the lack of the real solutions of the steady \NS{}  equation for positive $\hat S_{n n}$. The Gaussian profile of vorticity as a function of normal coordinate formally becomes complex at positive $\hat S_{n n}$, which means instability or decay in the time-dependent equation.

In \cite{KS21}, the authors verified this decay/instability process. They solved the time-dependent \NS{}  equation numerically in the vicinity of the steady solution with arbitrary background strain.
 Only the \BT{} solution corresponding to our \CVS{} conditions on the strain was stable.

 No external forces drive the breaking of the time-reversal symmetry of the Euler dynamics; it rather emerges spontaneously by the stability requirement of internal \NS{}  dynamics. The result of this microscopic stability mechanism of the \NS{}  dynamics is the \CVS{} boundary conditions added to the ambiguous Euler dynamics of the vortex sheets.
 
 \subsection{Complex Curves}
 Let us proceed with the exact solution of the \CVS{} equations.
 
 The \CVS{} equations for the cylindrical geometry reduce to the equation  \eqref{shapeEq} for the boundary loop $C(\theta)$ plus a complex equation $V_+(x,y) +V_-(x,y) =0$ at the loop $ (x,y) \in C$.
 
Consider two holomorphic functions involved in this equation: $F_-(\eta)$ inside the loop, $F_+(\eta)$  outside.
The boundary condition  relates these two functions 
\begin{eqnarray}\label{Feq}
F_+(\eta) + F_-(\eta) + (a -b) \eta - c \bar \eta =0 ;\forall \eta \in C;
\end{eqnarray}

Let us look for the linear solution of the Laplace equation inside (constant strain):
\begin{eqnarray}
F_-(\eta) = (p + \i q) \eta
\end{eqnarray}
with some real $p,q$.

The vanishing normal velocity from the inside leads to the differential equation for $C(\theta)  $. 
\begin{eqnarray}\label{formEq}
&&\Im \left( (2 p + a - b + 2\i q  ) C(\theta) -c \bar C(\theta)\right)C'(\theta) =0
\end{eqnarray}

This equation is integrable for arbitrary parameters. 
We shall use two dimensionless parameters $\gamma, \beta$  to parametrize $p,q$ as follows:
\begin{eqnarray}\label{params}
&&p + \i q =\frac{b-a}{2} +\frac{c e^{-2\i\beta }}{2 (2 \gamma +1) };
\end{eqnarray}

The general solution of \eqref{formEq} in polar coordinates reads \cite{MB14} (up to an arbitrary normalization constant)
\begin{eqnarray}\label{gensol}
&&C(\theta) =  e^{\i\beta}(1 + \i \tau) \tau^{\gamma};\\
&& \tau = \tan \theta
\end{eqnarray}
This solution applies when $\tau^\gamma$ is real.

\subsection{Parabolic curves}

Let us consider positive $\gamma$ first (the parabolic curves).

This solution passes through the origin and will have no singularity at $\tau=0$ in case $\gamma = n, n \in \Z, n \ge 0$. 

The case $n=0$ corresponds to a tilted straight line.
\pct{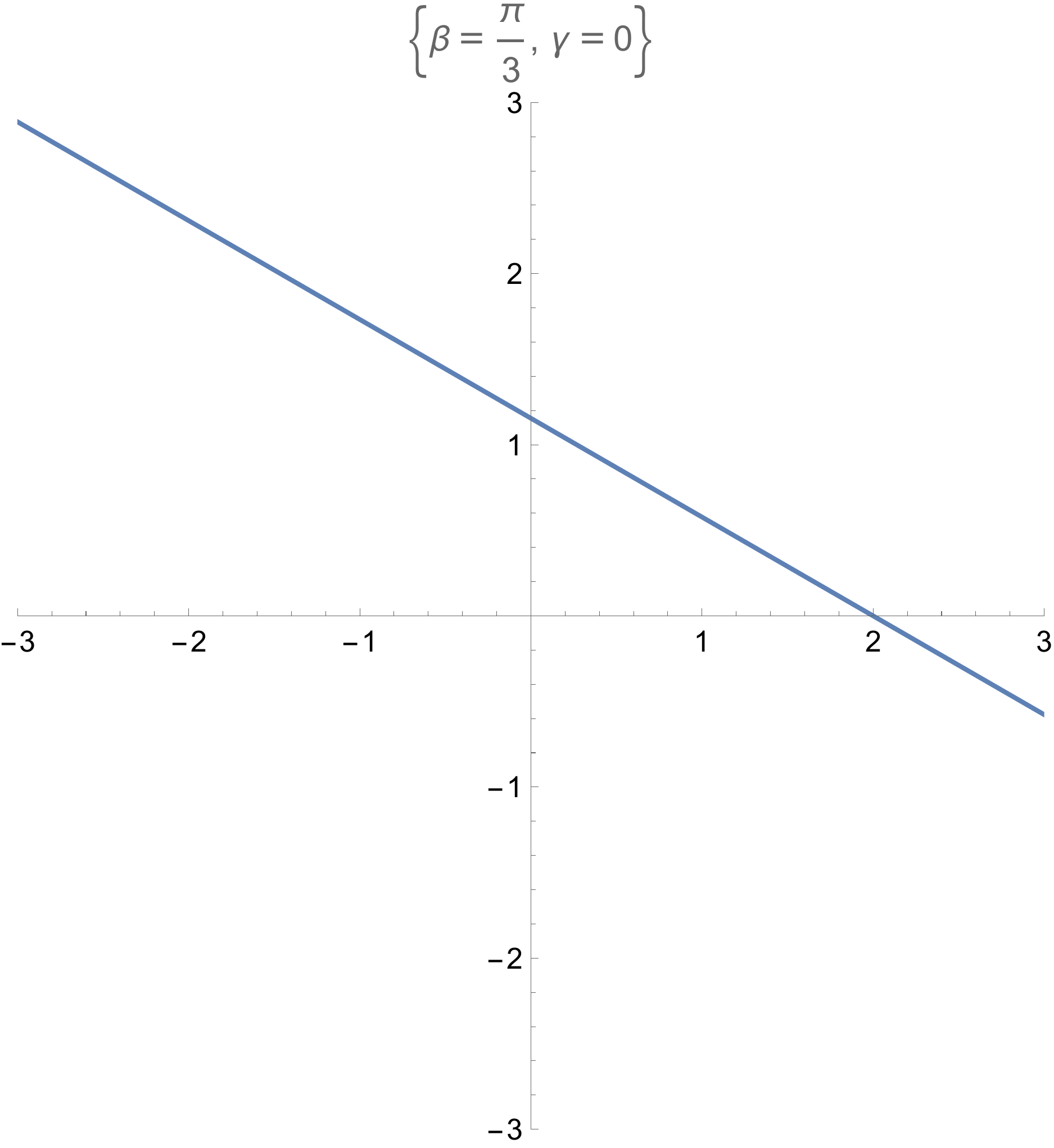}{The  \BT{} case, $n=0$.}
This case is just a \BT{} planar vortex sheet.
The next cases are already nontrivial.
\pct{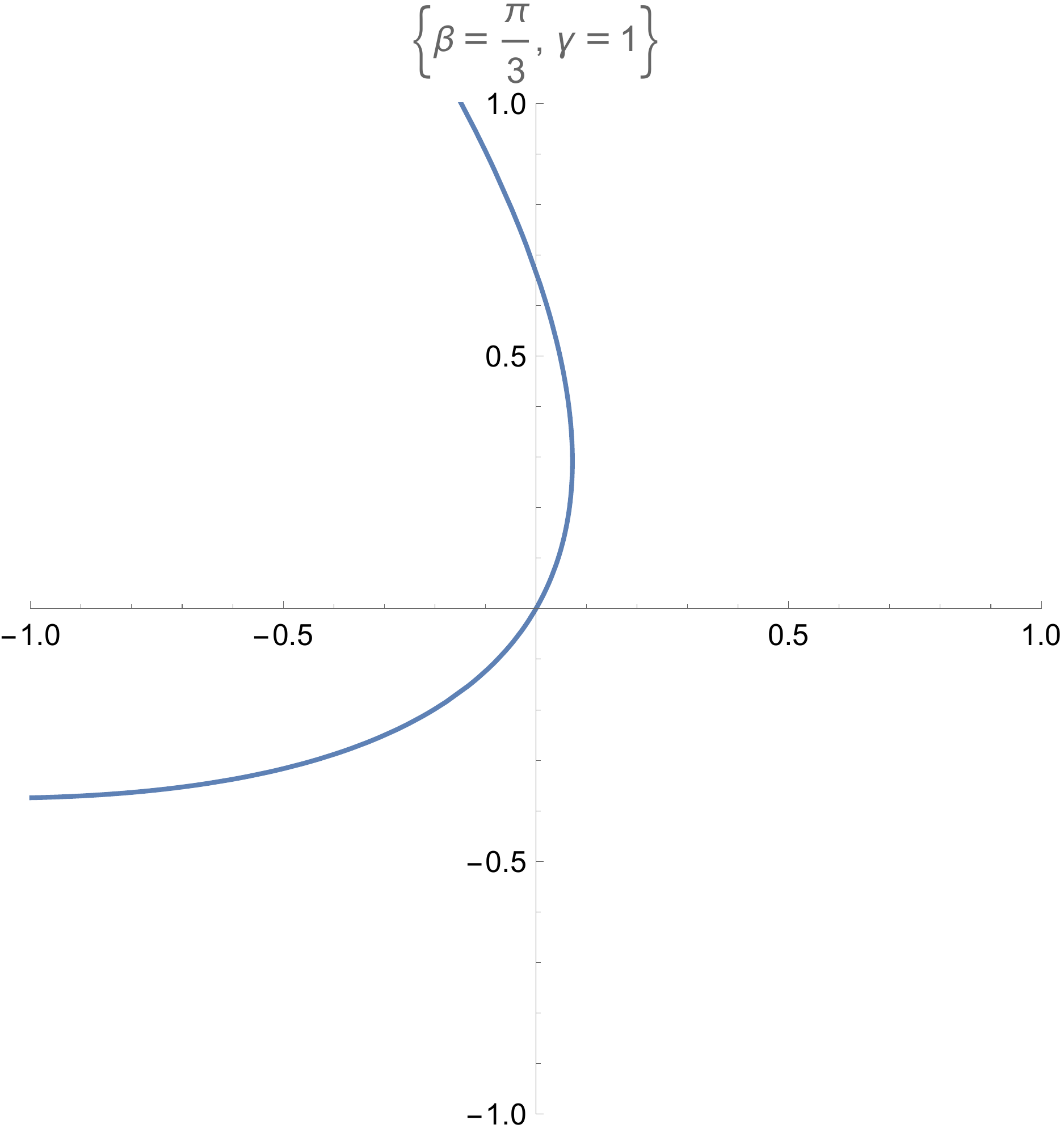}{The \CVS{} for $n=1$.}
\pct{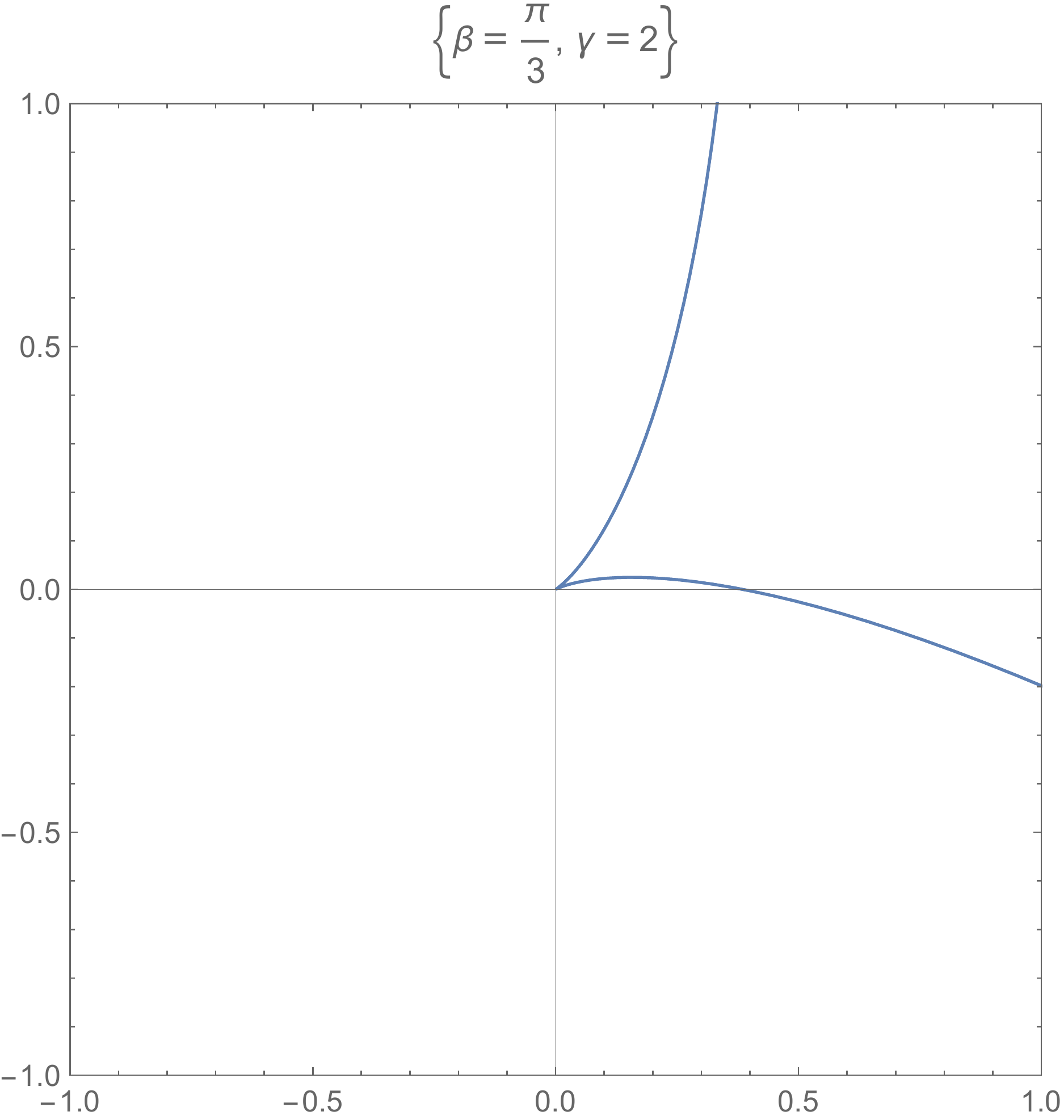}{The \CVS{} for $n=2$.}
\pct{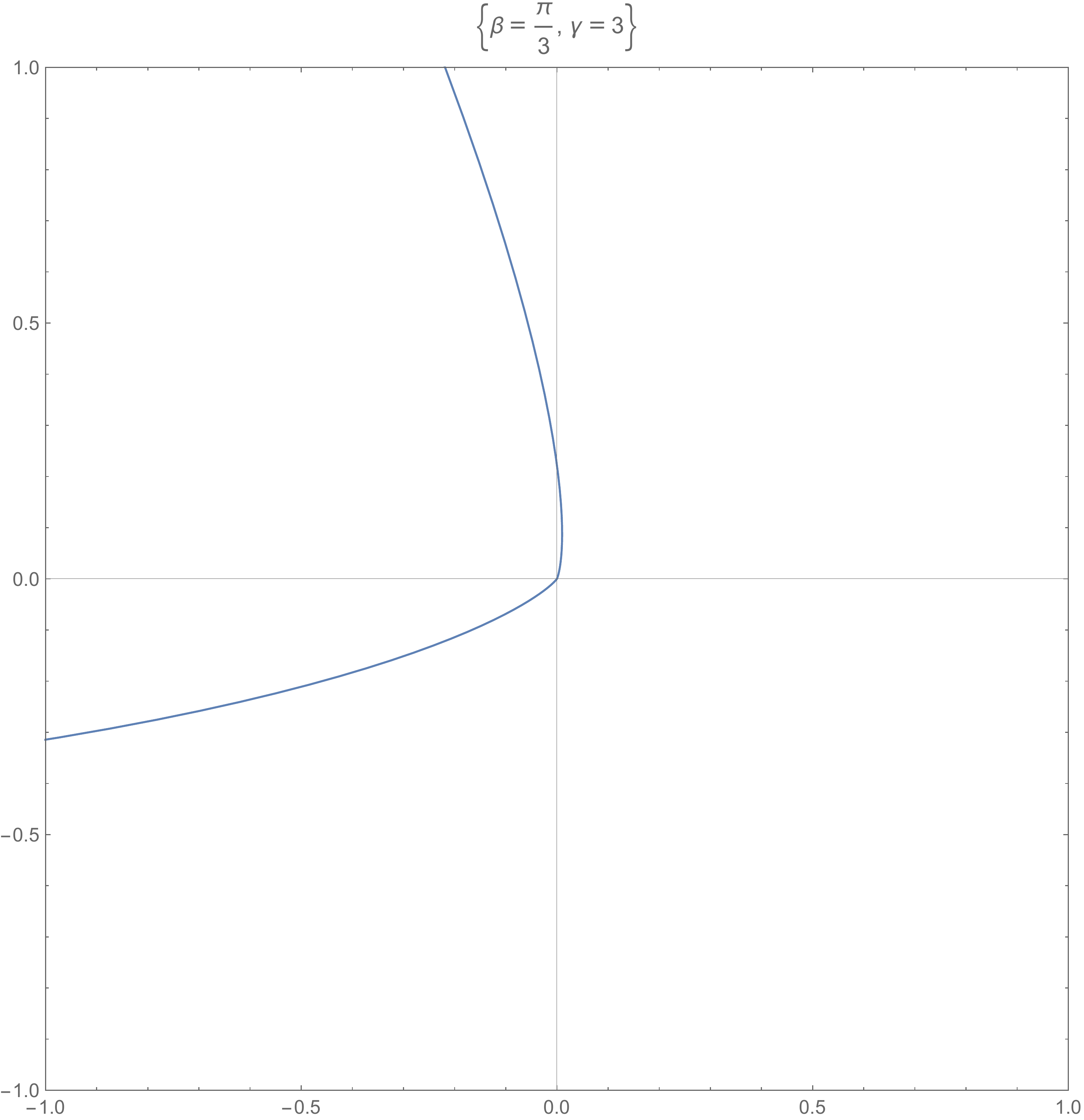}{The \CVS{} for $n=3$.}

However, not all positive $n$ are acceptable. For even positive $n$ we have a cusp in a curve because only positive $r(\theta)$ exist at $\theta\ra 0$. This cusp is visible at the $n=2$ curve.

For odd positive $n= 2 k+1$, there is a linear relation between $x,y$ at small $\tau$ provided $\beta \neq 0$.
The slope of the curve $y(x)$  at $x= \pm 0$ is the same.

Higher derivatives are singular, as we have 
\begin{eqnarray}
&&x \sim \tau^{2 k +1}  \cos \left(\beta\right) (1 - \tan\left(\beta\right) \tau) ;\\
&&y \sim \tau^{2 k +1}  \sin \left(\beta\right) (1 + \cot\left(\beta\right) \tau) ;\\
&& y \ra \tan \left(\beta\right) x \left( 1 + \frac{2\sign x }{\sin\beta} \left(\frac{|x|}{\cos \left(\beta\right)}\right)^{\frac{1}{2 k +1}} \right);\\
&& \dbyd{y}{x} \ra \tan \left(\beta\right) \text{ at } x \ra \pm 0
\end{eqnarray}

As the slopes $\dbyd{y}{x}$  are the same at $x = \pm 0$, the vortex sheet reduces to a plane up to higher-order terms. Thus, the \BT{} solution, assuming the infinitesimal thickness of the vorticity layer, will still apply here. Ergo, the sheet will be stable at this point, and the points on the rest of the surface.

Thus, we have found the discrete parametric family of \CVS{} shapes. It simplifies in terms of $\tau = \tan\theta$.
\begin{eqnarray}\label{Acurve}
&&\eta(\tau) = e^{\i \beta} \left(1 + \i \tau\right) \tau^{2 k + 1};\\
&&k \in \Z, k \ge 0;\\
&& -\infty < \tau < \infty;
\end{eqnarray}

We have a conformal map from the $\tau $ plane to the plane of $\eta = x + \i y$, with our curve $C$ corresponding to the real axis. The physical region outside the loop corresponds to the sector in the lower semiplane, $ -\frac{ \pi}{( k +1)}< \arg \tau < 0$.

The infinity in the physical space corresponds to infinity in the $\tau$ plane.

Two singular points correspond to the vanishing derivative $\eta'(\tau) =0$. The singularity happens at
\begin{eqnarray}
&&\tau = 0;\\
&& \tau =\frac{\i (2 k+1)}{2 (k+1)};
\end{eqnarray}

The first one maps to the origin in the $\eta$ plane. The second one is in the upper $\tau$ semiplane, so it does not affect the inverse function $\tau(\eta)$ in the physical sector of the lower semiplane.

The next step is to find the holomorphic function $F_+(\eta)$, which defines the complex velocity on the external side of the sheet (the internal side has linear velocity $F_-(\eta) = (p + \i q) \eta$ ).

From the \CVS{} equation, we get the boundary values, which we express in terms of $\eta, \eta^*$
\begin{eqnarray}\label{boundaryF}
 &&F_+(\eta) = -(p + a - b+ \i q)\eta  + c \eta^* ;\; \forall \eta \in C
\end{eqnarray}
We have to continue this function from the curve \eqref{Acurve} into the part of the complex plane outside this curve.

Let us express the right side in terms of $\tau$
\begin{eqnarray}\label{param}
&&f(\tau) = -(p + a - b+ \i q)\eta(\tau)  + c \tilde \eta(\tau)
\end{eqnarray}
Here $\tilde \eta(\tau)$ is the complex conjugate expression for $\eta$
\begin{eqnarray}
\tilde\eta(\tau) = e^{-\i \beta} \left(1 - \i \tau\right) \tau^{2 k + 1};\\
\end{eqnarray}
At the real axis of $\tau$ the variables $\eta(\tau),\tilde\eta(\tau)$ will be complex conjugates, and the boundary value
\begin{eqnarray}
f(\tau) = F_+(\eta(\tau)); \forall \Im \tau =0
\end{eqnarray}

Now we have an obvious analytic continuation to the lower semiplane of $\tau$. After some algebra, we get \cite{MB14}
\begin{eqnarray}
&&\eta(\tau) = e^{\i \beta} \left(1 + \i \tau\right) \tau^{2 k + 1};\\
&&f(\tau) = \frac{1}{2} e^{-i \beta } \tau ^{2 k+1}\nonumber\\
&&\left(\frac{c (-i (8 k+7) \tau +8 k+5)}{4 k+3}-i e^{2 i \beta } (\tau -i) (2 a+c)\right)
\end{eqnarray}

This parametric function in the sector $ -\frac{ \pi}{( k +1)}< \arg \tau < 0$ of the complex $\tau$ plane represents an implicit solution of the \CVS{} equation.

Now, let us consider the limit $\tau \ra \infty$ corresponding to $\eta \ra \infty$. In this limit, the function $F_+(\eta(\tau))$ linearly grows 
\begin{eqnarray}
&& F_+(\eta) \ra  B \eta;\\
&& B =-a+\frac{1}{2} c \left(-1-\frac{e^{-2 i \beta } (8 k+7)}{4 k+3}\right)
\end{eqnarray}

This linear growth means that the true value of the $x y$ components of strain at infinity is not  $\diag{a,-c-a}$ we thought it was. It is instead
\begin{eqnarray}
\hat S = \left(
\begin{array}{cc}
 -\frac{c ((8 k+7) \cos (2 \beta )+4 k+3)}{8 k+6} & -\frac{c (8 k+7) \sin (2 \beta )}{8 k+6} \\
 -\frac{c (8 k+7) \sin (2 \beta )}{8 k+6} & \frac{c ((8 k+7) \cos (2 \beta )-4 k-3)}{8 k+6} \\
\end{array}
\right)
\end{eqnarray}

The eigenvalues of this strain are 
\begin{eqnarray}
\left\{-\frac{c (6 k+5)}{4 k+3},\frac{2 c (k+1)}{4 k+3}\right\}
\end{eqnarray}

The angle $\beta $ dropped from the eigenvalues because it can be eliminated by the $U(1)$ transformation $\eta \ra \eta e^{-\i \beta}$, which leaves the eigenvalues invariant.

The parabolic solution does not decrease at infinity, which makes it unphysical.

\subsection{Hyperbolic curves}

Let us now consider the hyperbolic curves with negative $\gamma< 0$, which do not need to be half-integer. 
We redefine our parameters as follows:
\begin{eqnarray}
&&\eta(\xi) = \pm e^{\i \beta} \left(\xi +\i\xi^{-\mu}\right);\\
&& \xi = \tau^{\gamma};\\
&& \mu = -1 -\frac{1}{\gamma }; \mu >0;\\
&& 0 < \xi < \infty
\end{eqnarray}

These two curves correspond to various $\mu, \beta$. 

\pct{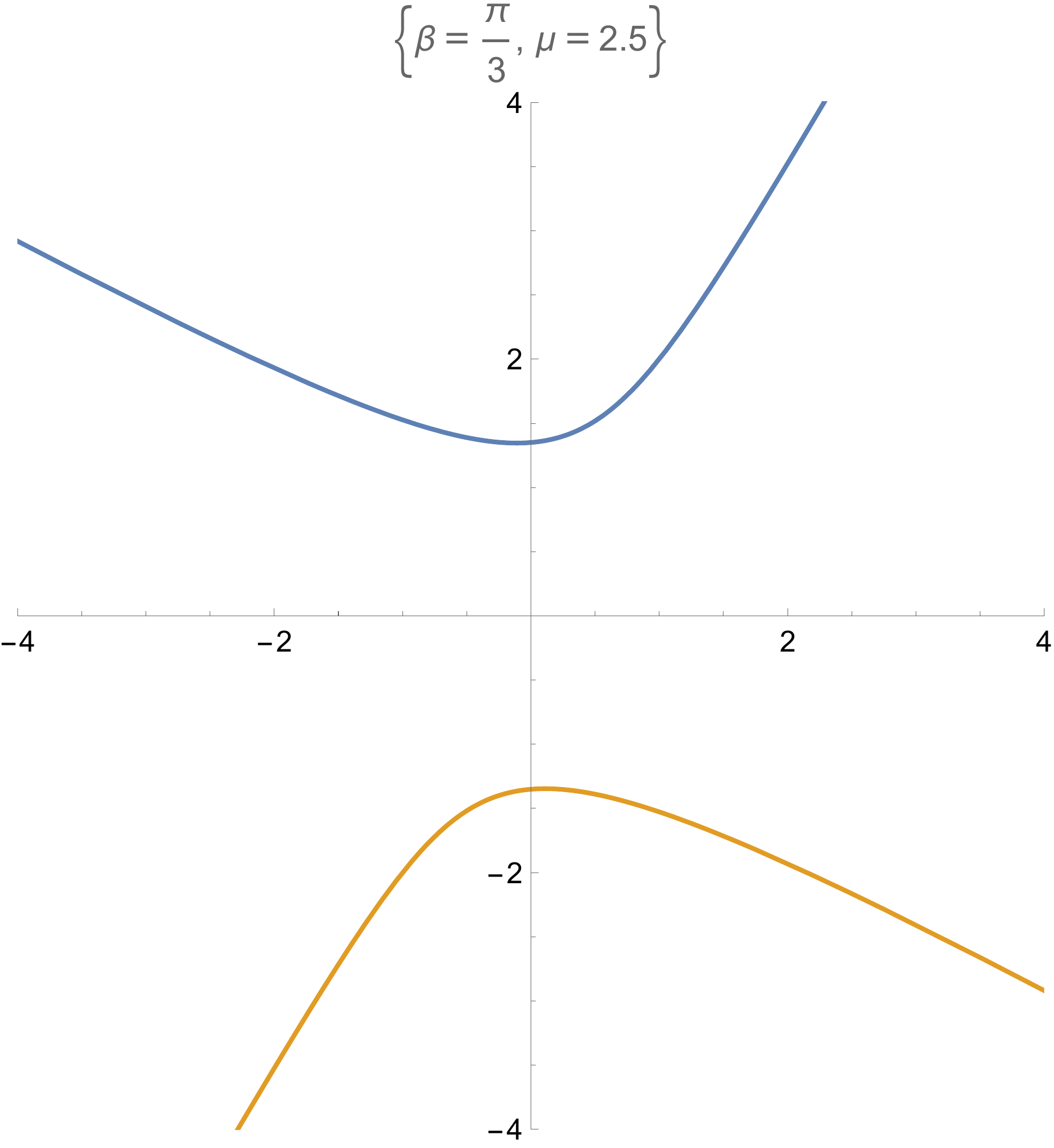}{The hyperbolic \CVS{} for $\beta = \pi/3, \mu = 2.5$.}
\pct{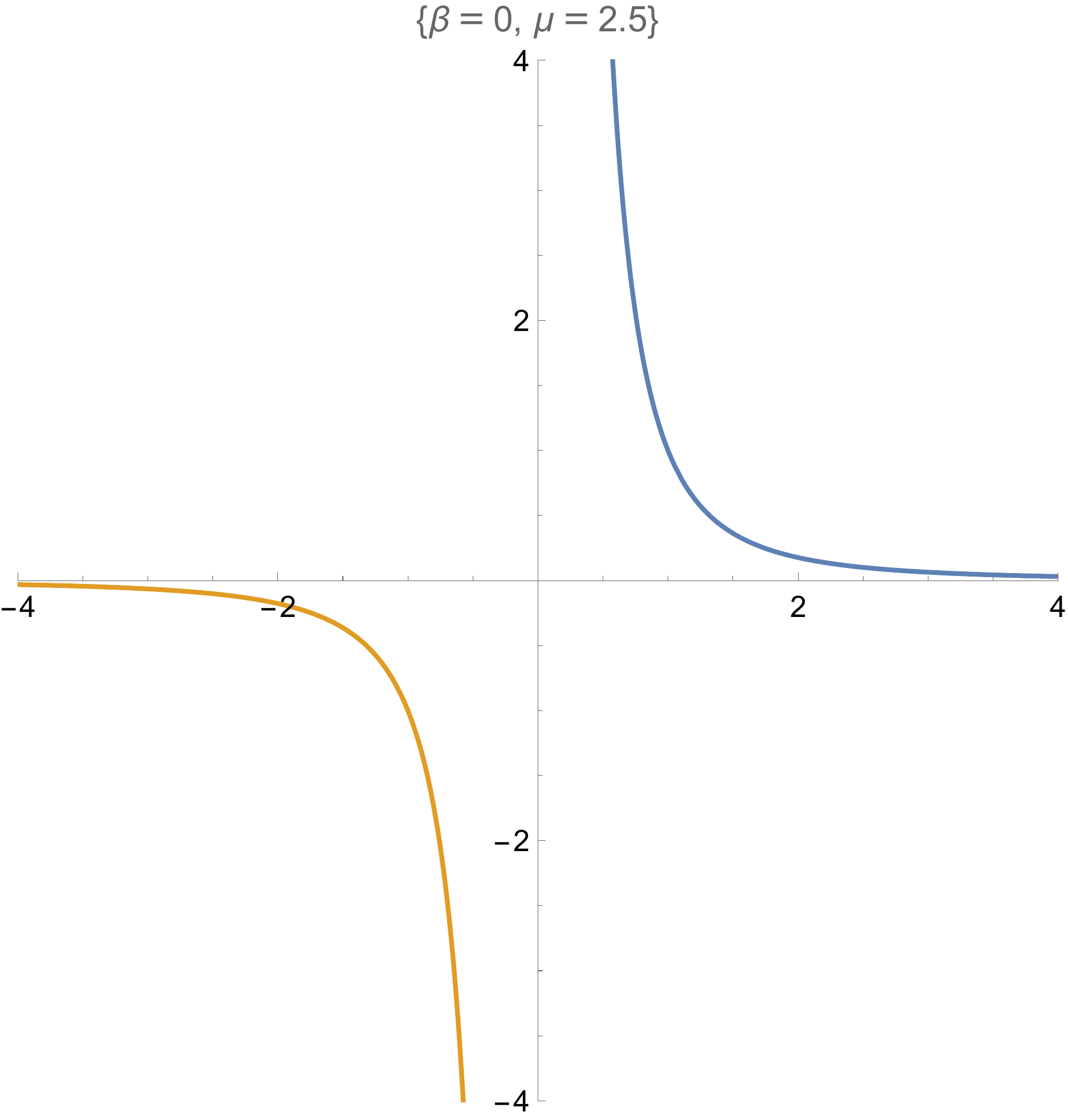}{The hyperbolic \CVS{} for $\beta = 0, \mu = 2.5$} 

Note that there are no singularities at the origin, as both curves are going around it.

The inverse function is singular at the point $\xi_0$  in the upper semiplane
where $\eta'(\xi_0)=0$:
\begin{eqnarray}
\xi_0 = (i \mu )^{\frac{1}{\mu +1}}
\end{eqnarray}
Therefore, we need to use the lower semiplane as a physical domain for $f(\xi ) =F_+(\eta(\xi))$.

The holomorphic function $f(\xi)$ is reconstructed employing the same steps as in the parabolic case. We get
\begin{eqnarray}
&&f(\xi) =\frac{1}{2} e^{-i \beta } \xi  \left(\frac{c (\mu -3)}{\mu -1}-e^{2 i \beta } (2 a+c)\right)-\nonumber\\
&&\frac{i e^{-i \beta } \xi ^{-\mu } \left(c (3 \mu -1)+e^{2 i \beta } (\mu -1) (2 a+c)\right)}{2 (\mu -1)}
\end{eqnarray}

Let us now find the limit at $\xi\ra \infty$, when $\eta$ and $f$ both linearly grow
\begin{eqnarray}
&&f \ra \eta B;\\
&& B = -a+\frac{1}{2} c \left(-1+\frac{e^{-2 i \beta } (\mu -3)}{\mu -1}\right)
\end{eqnarray}

We want this coefficient to be zero so that $f(\eta)$ decreases at infinity.

The solution of this complex equation for $\beta, \mu$ is
\begin{eqnarray}
\begin{cases}
\beta =0; &\mu =1 -\frac{c}{a};\\
\beta = \oh \pi; & \mu = 1-\frac{ c}{b};\\
\end{cases}
\end{eqnarray}
From the inequalities between the three eigenvalues $a<b<c$ adding to zero, we get
\begin{eqnarray}
 &&-2 c < a < -\frac{c}{2};\\
 && -\frac{c}{2} < b < c;
\end{eqnarray}
which makes this index $\mu$ limited to the intervals
\begin{eqnarray}
\begin{cases}
\beta =0; &\frac{3}{2} < \mu < 3;\\
\beta =\oh \pi; &-\infty < \mu < \infty;
\end{cases}
\end{eqnarray}

We choose the first solution, as in this case, $\mu$ varies in the positive region where there are no singularities.

Now, we observe that at $\beta =0,\oh\pi$, which means $q =0$ there is an extra symmetry in the equation \eqref{formEq}
\begin{eqnarray}
&&C \Ra C^*
\end{eqnarray}

This symmetry means that now all four branches of the hyperbola
\begin{eqnarray}
|y| |x|^\mu = \mbox{const};
\end{eqnarray}
are the parts of a single periodic (unbounded) solution $C(\theta)$ (Fig \ref{fig::IntegrationLoop}).

These four branches define the domain inside the loop $C$  where the holomorphic velocity
\begin{eqnarray}
F_-(\eta) = -2 a \eta;
\end{eqnarray} 

As for the function $F_+$ it is just a negative power
\begin{eqnarray}
&&F_+\left(x \pm\i |x|^{-\mu}\right) = + 2\i c |x|^{-\mu};
\end{eqnarray}
The loop $C(\theta)$ defined by these four branches is a periodic function with singularities at $ \theta = k \pi/2$.

Topologically, on a Riemann sphere $S_2$, these four branches divide the sphere into five regions.
There is an inside region, with the boundary touching Riemann infinity four times. 

Each of the four remaining outside regions is bounded by a hyperbola,  starting and ending at infinity.
\pct{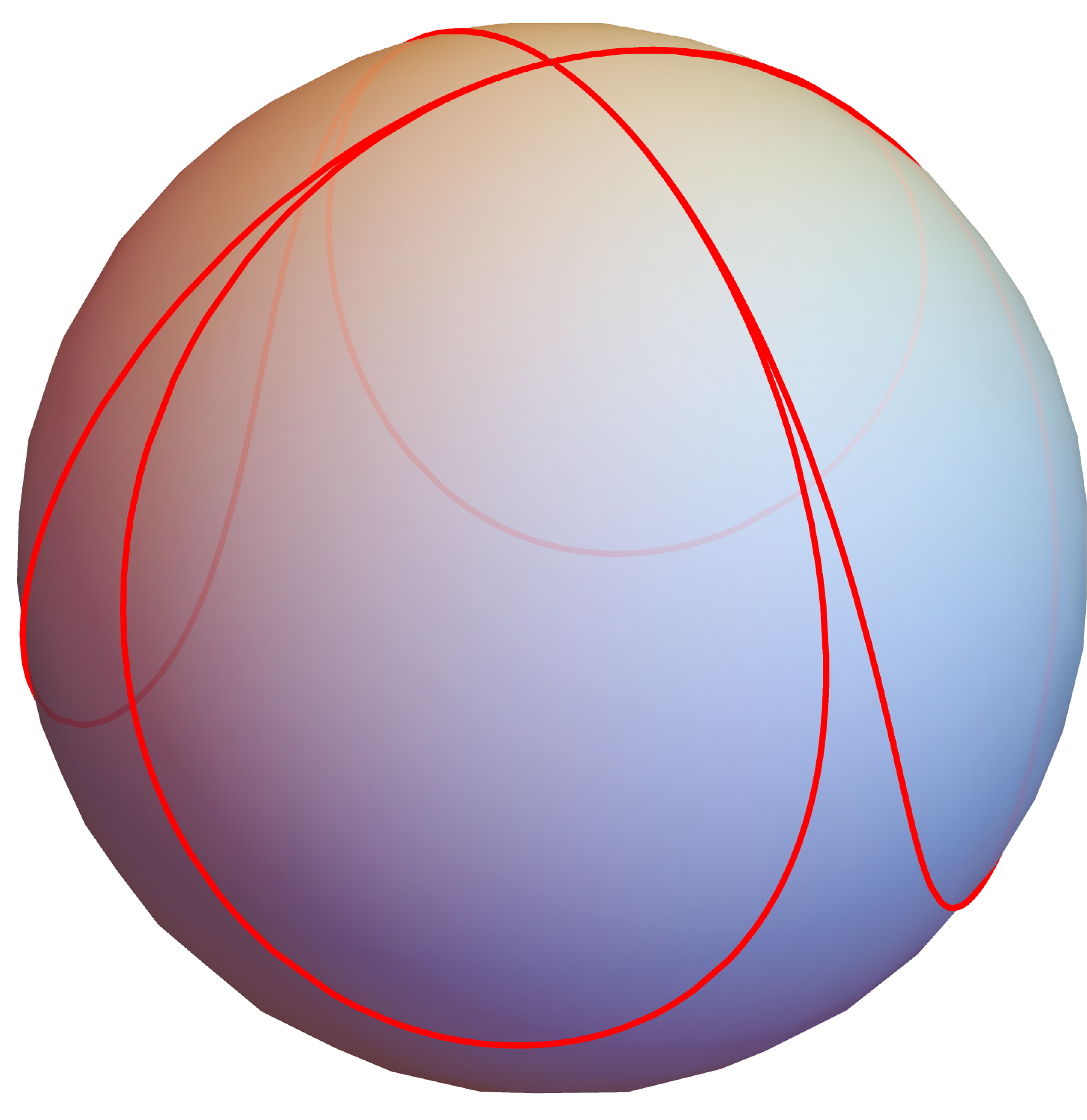}{The stereographic projection of four hyperbola branches on the Riemann sphere.}

Let us consider the lower right hyperbola with $y = -x^{-\mu}, x >0$

The values of $F_+(\eta)$ in three other external regions will be obtained from this one by reflection against the $x$ or $y$ axis.
\begin{eqnarray}
&&F_+(-\eta) = - F_+(\eta);\\
&&F_+(\eta^*) = F_+^*(\eta);
\end{eqnarray}

We can continue the above parametric equation for the function $f$ to the complex plane for the variable $w$, with the cut from $ -\infty $ to $0$. The phase of the multivalued function $w^{-\mu}$ is set to $\arg w^{-\mu} =0, w >0$.
\begin{eqnarray}
&&\eta = w - \i w^{-\mu};\\
&& F_+=  2\i c w^{-\mu};
\end{eqnarray}

This parametric function has a square root singularity at the point where derivative $\d_w \eta = 1 + \i \mu w^{-\mu-1}$ vanishes:
\begin{eqnarray}
w_c = (-\i \mu)^{\frac{1}{\mu+1}}
\end{eqnarray}

For positive $\mu$, this singularity is located in the lower semiplane of $w$, leaving the upper semiplane as a physical domain.

We have drawn the complex maps\footnote{Complex map in \Mathematica{} is a 3D surface of $x,y,|f(x+\i y)|$ for complex function $f(z)$ with its phase $\arg{f(x+\i y)}$ as color of the point at this surface.} of the derivative $\d_w \eta = 1 + \i \mu w^{-\mu-1}$ for  $\mu=1.51,2,2.99$.
\pct{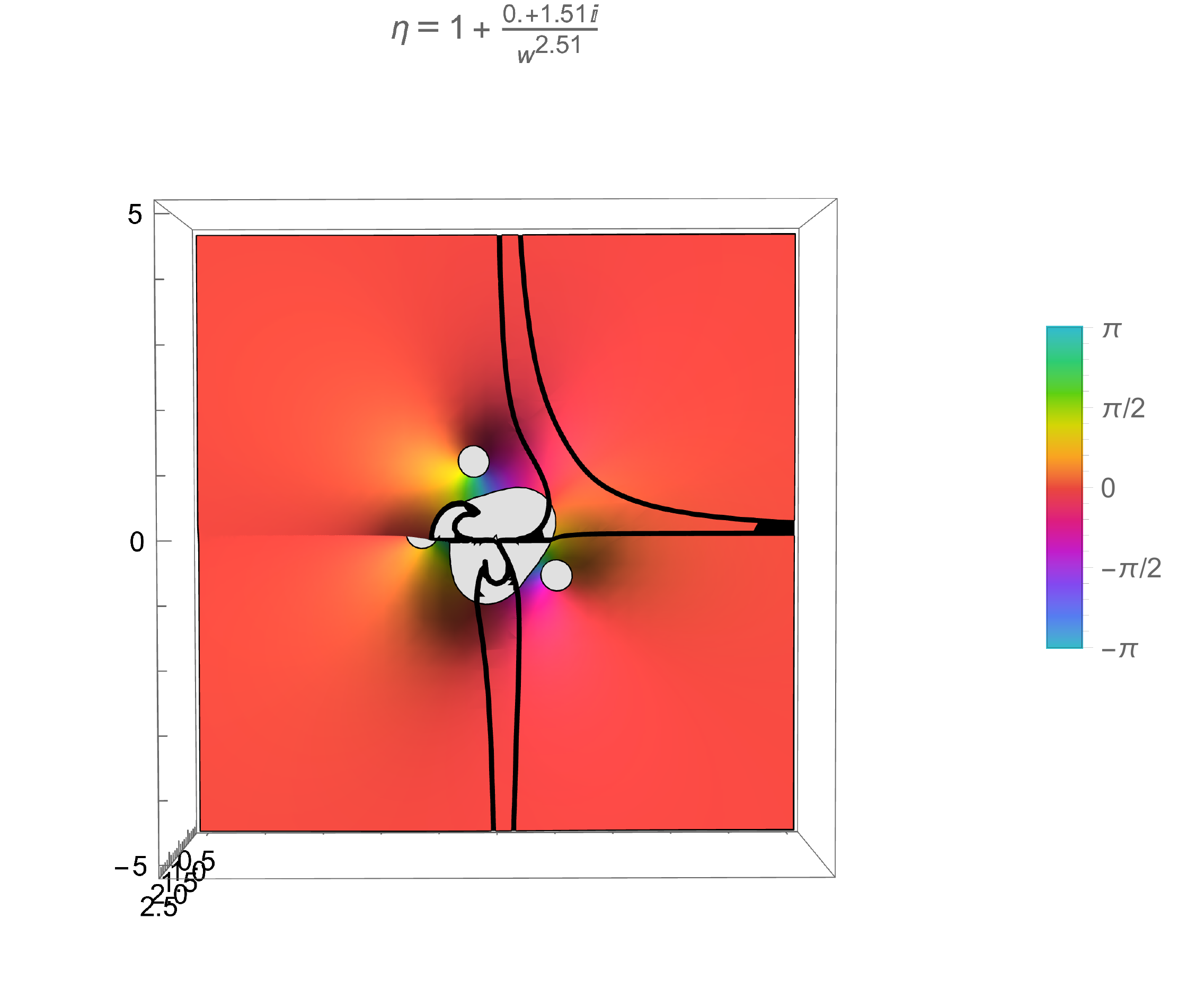}{The complex Map3D of the holomorphic function $\eta'(w)= 1 + \i \mu w^{\mu-1}$ for $\mu = 1.51$. The height is $|\eta'|$, the color is $\arg \eta'$. The square root singularities of the inverse function $\eta(w)$ lie at the points where $\eta'(w)=0$. We indicate them as holes on the surface (white circles). The black lines are  described by $(\Re\eta(w))^\mu \Im \eta(w) =\pm1$. The physical region is outside the black line in the first quadrant, and there are no singularities.}

\pct{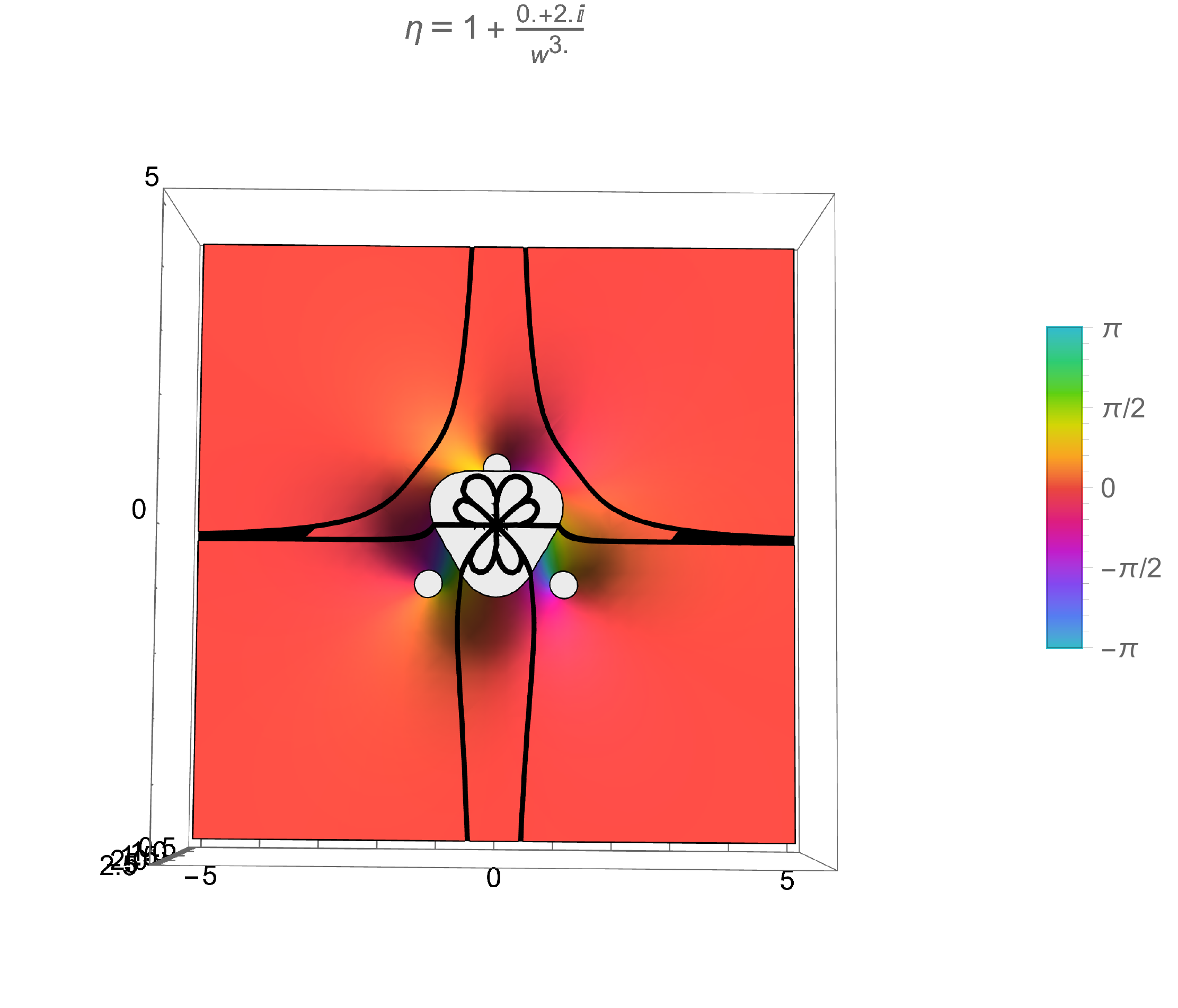}{The complex Map3D of the holomorphic function $\eta'(w)= 1 + \i \mu w^{-\mu-1}$ for $\mu = 2$. The height is $|\eta'|$, the color is $\arg \eta'$. The square root singularities of the inverse function $\eta(w)$ lie at the points where $\eta'(w)=0$. We indicate them as holes on the surface (white circles). The black lines are  described by $(\Re\eta(w))^\mu \Im \eta(w) =\pm1$. The physical region is outside the black line in the first quadrant, and there are no singularities.}
\pct{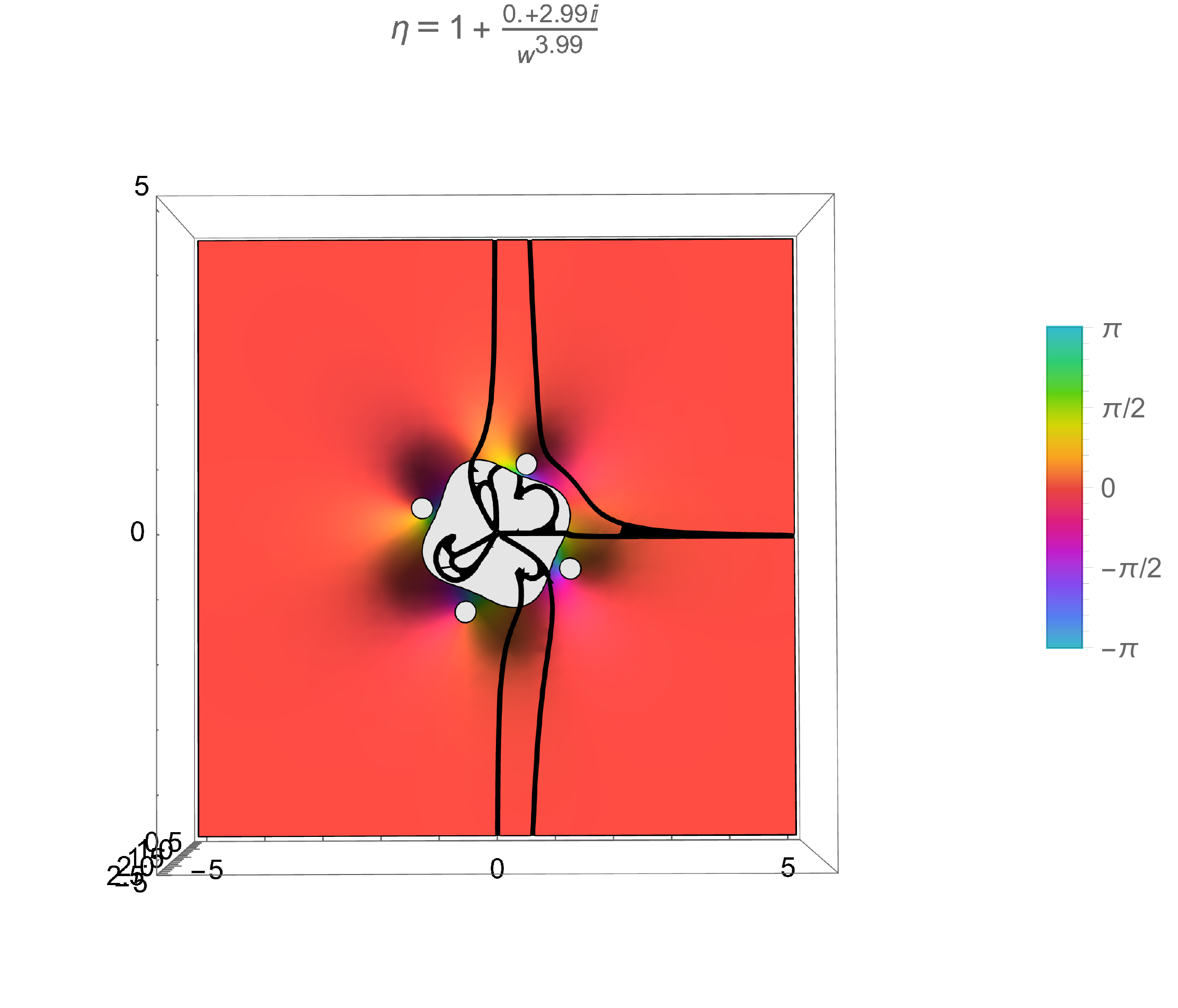}{The complex Map3D of the holomorphic function $\eta'(w)= 1 + \i \mu w^{-\mu-1}$ for $\mu = 2.99$. The height is $|\eta'|$, the color is $\arg \eta'$. The square root singularities of inverse function $\eta(w)$  lie at the points where $\eta'(w)=0$. We indicate them as holes on the surface (white circles). The black lines are  described by $(\Re\eta(w))^\mu \Im \eta(w) =\pm1$. The physical region is outside the black line in the first quadrant, and there are no singularities.}

We indicate the zeros of $\d_w \eta$  as holes on the surface (white circles). 

To check whether these singularities penetrate the physical region, we have drawn in black the boundaries of the physical regions
\begin{eqnarray}
\Re\eta^\mu \Im \eta =\pm1
\end{eqnarray}

We observe that these square root singularities lie outside the physical region. This physical region is above the black line in the first quadrant, and there are no singular points there.

Inside the tube, the velocity is linear, and an extra linear term $ F_-(\eta) = -2 a \eta;$ is a potential flow, which preserves the incompressibility
\begin{eqnarray}
V^- = a x - \i b y -2 a (x + \i y) = -a x + \i (c-a) y
\end{eqnarray}

To summarize, the velocity field $\vec v(x,y,z)$ and complex coordinates $\eta = x + \i y$ are parametrized as a function of a complex variable $w$ as follows
\begin{subequations}\label{velocityField}
\begin{eqnarray}
&&v_z = c z;\\
&& v_x - \i v_y = a x - \i b y +  F_\pm\left(\frac{x + \i y}{R_0}\right);\\
&& F_-(\eta) = - 2 a  \eta;\\
&& F_+(\eta)=  2\i c  w(\eta)^{-\mu};\; \Re \eta >0, \Im \eta < 0;\\
&& F_+(-\eta) = -F_+(\eta); F_+(\eta^*) = F_+^*(\eta);\\
&&w(\eta): \eta = w - \i w^{-\mu};\\
&&C:  |x|^\mu |y| = R_0^{1 + \mu};\\
&& \mu = 1 - \frac{c}{a};\; \frac{3}{2} < \mu < 3;\\
&& \Delta \Gamma = \oint_C \vec v(\vec r) d \vec r  =0;
\end{eqnarray}
\end{subequations}
We show the flow at  Figs.\ref{fig::Flow}, \ref{fig::Flow}.
\pct{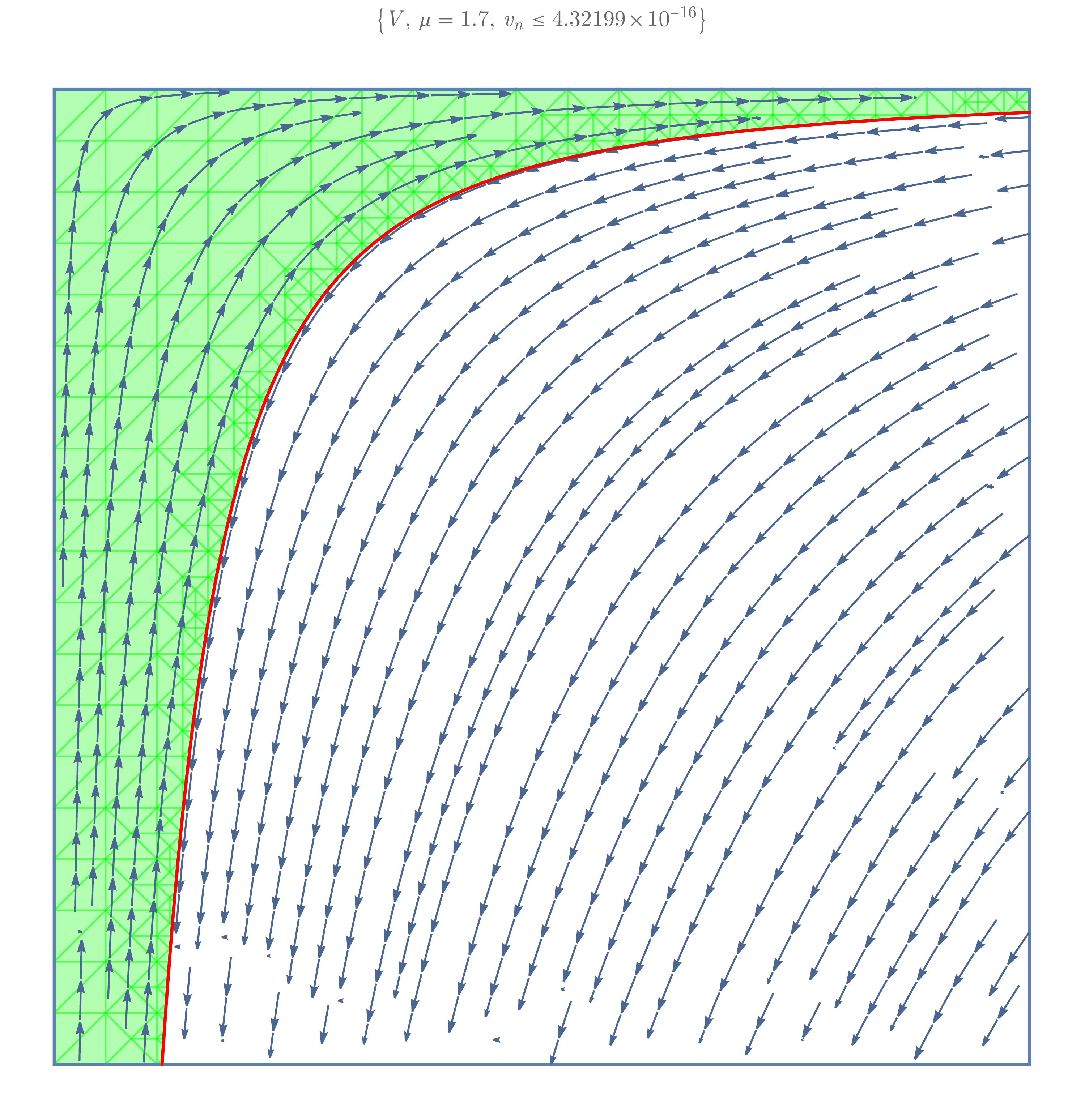}{The stream  plot in $IV$ quadrant of $x y $ plane for $\mu = 1.7$. The green fluid is inside; the clear fluid is outside the vortex surface (red).
    One can obtain the flow in other quadrants by reflection against the $x$ and $y$ axes.
    The normal velocity vanishes at the vortex sheet on both sides, with accuracy $\sim 10^{-16}$.} 
\pct{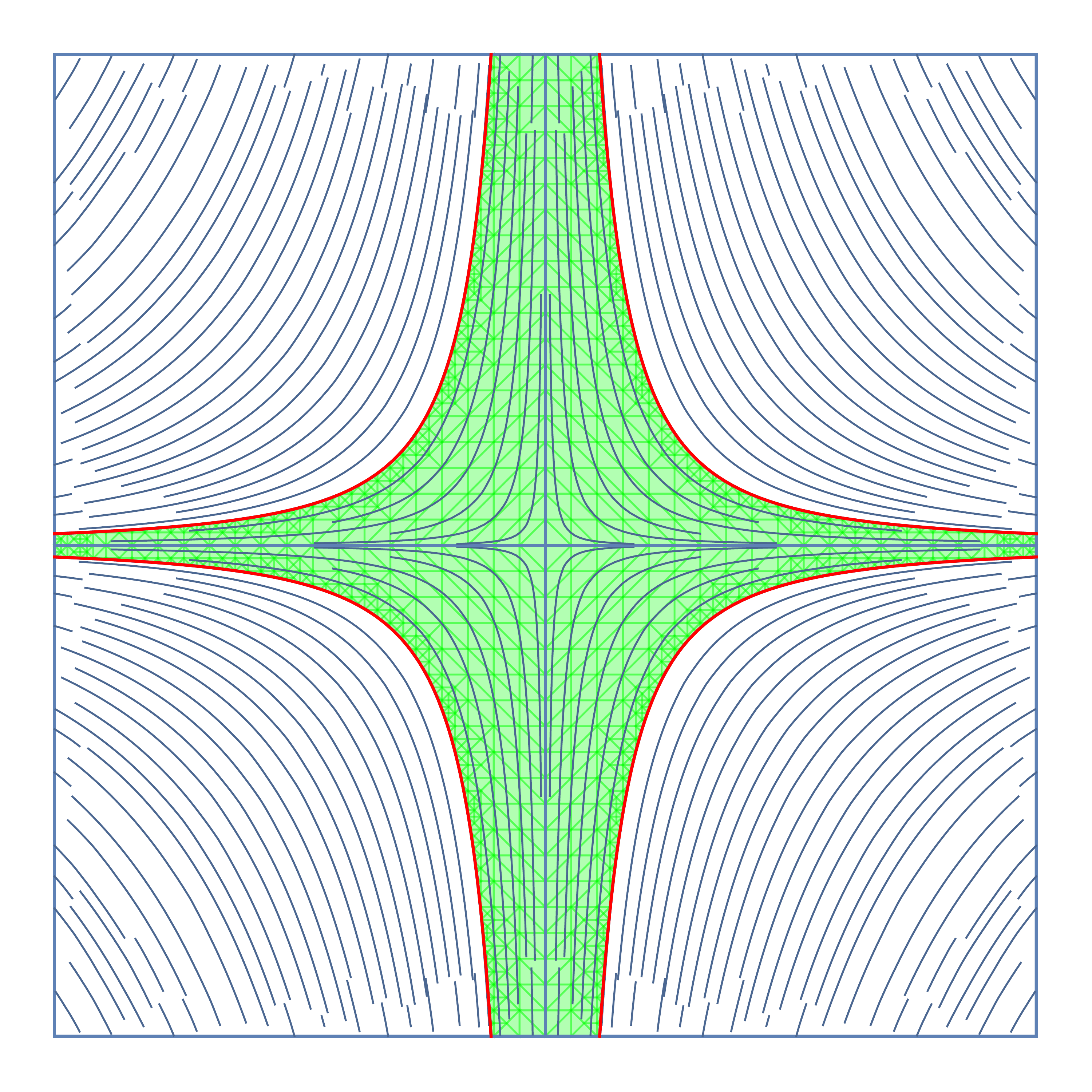}{The stream  plot in  $x y $ plane for $\mu = 1.7$. The green fluid is inside; the clear fluid is outside the vortex surface (red)}.
    The stream plot of this flow is projected in the $x y $ plane.
    
We performed analytical and numerical computations in \cite{MB15}.

While computing the complex velocity and coordinates using parametric equations, we restricted $w$ to the physical region and populated this region with an irregular grid in angular variables in the $w$ plane. 

After that, we used \textit{ListStreamPlot} method of \Mathematica.

The grid resolution does not allow tracking the boundary's immediate vicinity, but we computed the normal velocity numerically at the boundary. It was less than $10^{-15}$ on both sides of the sheet.

\subsection{Clipping the Cusps}

Let us consider the hyperbolic cylindrical CVS $ |x|^\mu |z| = 1, \mu = 2 + \frac{b}{a}$ (in proper units, with cylinder axis, parallel to $y$, and decreasing eigenvalues of background strain being $(a,b,-a- b)$).

With finite viscosity, in the \NS{}  system, there is a thickness $h = \sqrt{\nu/(a+b)} $ of the vortex sheet, which goes to zero at $\nu \ra 0$. We are not assuming that $a+b$ is finite at $\nu \ra 0$, only that $h\ra 0$.

There are three domains on the positive $z$ axis\footnote{and, likewise on any other axis}:
\begin{eqnarray}
&& I : 0 < z < z_0;\\
&& II :  z_0 < z < z_1;\\
&& III :  z_1 < z < \infty;
\end{eqnarray}
where
\begin{eqnarray}
&&  z_0 \ll h \ll z_1; 
\end{eqnarray}

In the third region, the two opposite velocity gaps $\pm \Delta v_y$ at the mirror branches of hyperbola $ |x|^\mu |z| = 1$ are present. In the first region, there is no gap, and the opposite gaps annihilated each other.

The solution in the intermediate (second) region has the same geometry as the Burgers sheet, but the velocity has no gap. In fact, up to the higher correction terms, the hyperbola is horizontal, with the CVS normals aligned at $ z >0, z <0$, so the same Anzats $v_y = f(z/h)$  applies, but this time the velocity is an even solution of the same equation
\begin{eqnarray}
&&f''(\xi)+ \xi f'(\xi)-\alpha f(\xi) =0;\\
&& \xi = \frac{z}{h};\\
&& \alpha =  \frac{b}{a+b};\\
&& f(\pm\infty) = 0;
\end{eqnarray}
With these boundary conditions, the only restriction on parameters is an inequality $\alpha <0$.

Hypergeometric functions give this solution for velocity and vorticity
\begin{eqnarray}
&&v_y = \, _1F_1\left(\frac{\alpha }{2};\frac{1}{2};-\frac{\xi^2}{2}\right);\\
&&\omega_x = -\d_z v_y = \frac{\alpha}{h}  \xi \, _1F_1\left(\frac{\alpha }{2}+1;\frac{3}{2};-\frac{\xi^2}{2}\right)
\end{eqnarray}

\pct{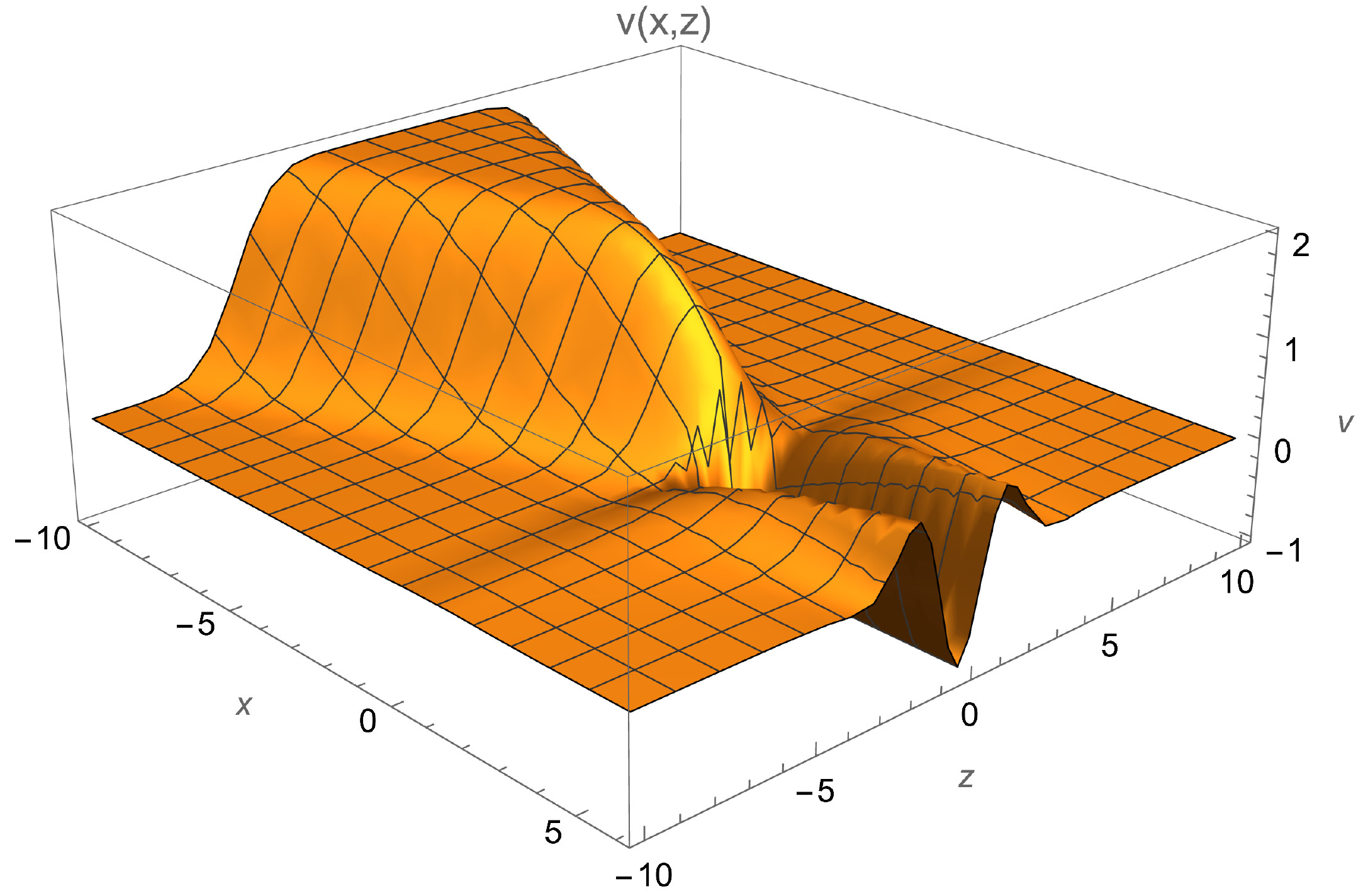}{The velocity profile $v_y(x,z)$}

The positions $\xi = \pm\Xi(\alpha)$ of the extrema of vorticity are given by the position of the extremum
\begin{eqnarray}
\Xi(\alpha)= \argmax_\xi {\left(\xi \, _1F_1\left(\frac{\alpha }{2}+1;\frac{3}{2};-\frac{\xi^2}{2}\right)\right)}
\end{eqnarray}

Now, matching the position of these symmetric extrema with the positions of the  extrema of the vorticity for the Burgers solution for two separated vortex sheets, we find
\begin{eqnarray}
 z_0 = h|\Xi(\alpha)|
\end{eqnarray}

There is a nontrivial velocity field structure in the region $ z_0 < z < z_1$, and the boundary layer proportional to $h$. However, at the distances $z < z_0$, the gaps are mutually annihilated, and velocity and its derivatives are all finite in the Euler limit.

This analysis is only a partial solution in the whole plane $x, z$. We just studied two regions out of the three.
The intermediate region where the two-gap solution deforms into the one with no gaps is yet to be studied.

The curve has an infinite length, and it encircles an infinite area because of the infinite cusps at $x =\pm 0, y = \pm 0$
(see Fig.\ref{fig::IntegrationLoop}).

At a large upper limit $x_{max} \ra \infty$ and small lower limit $x_{min}\ra 0$ the perimeter of the loop goes as
\begin{eqnarray}
&&P =  4 R_0 \int_{x_{min}}^{x_{max}} d x \sqrt{1 + \mu^2 x^{-2(\mu+1)}} \ra 4\left(x_{min}^{-\mu} + x_{max}\right) R_0;
\end{eqnarray}

Of course, these infinities will not occur in the viscous fluid with finite $\nu$. Our solution applies only as long as the spacing between the two branches of the hyperbola
is much larger than the viscous thickness of the vortex sheet.
\pct{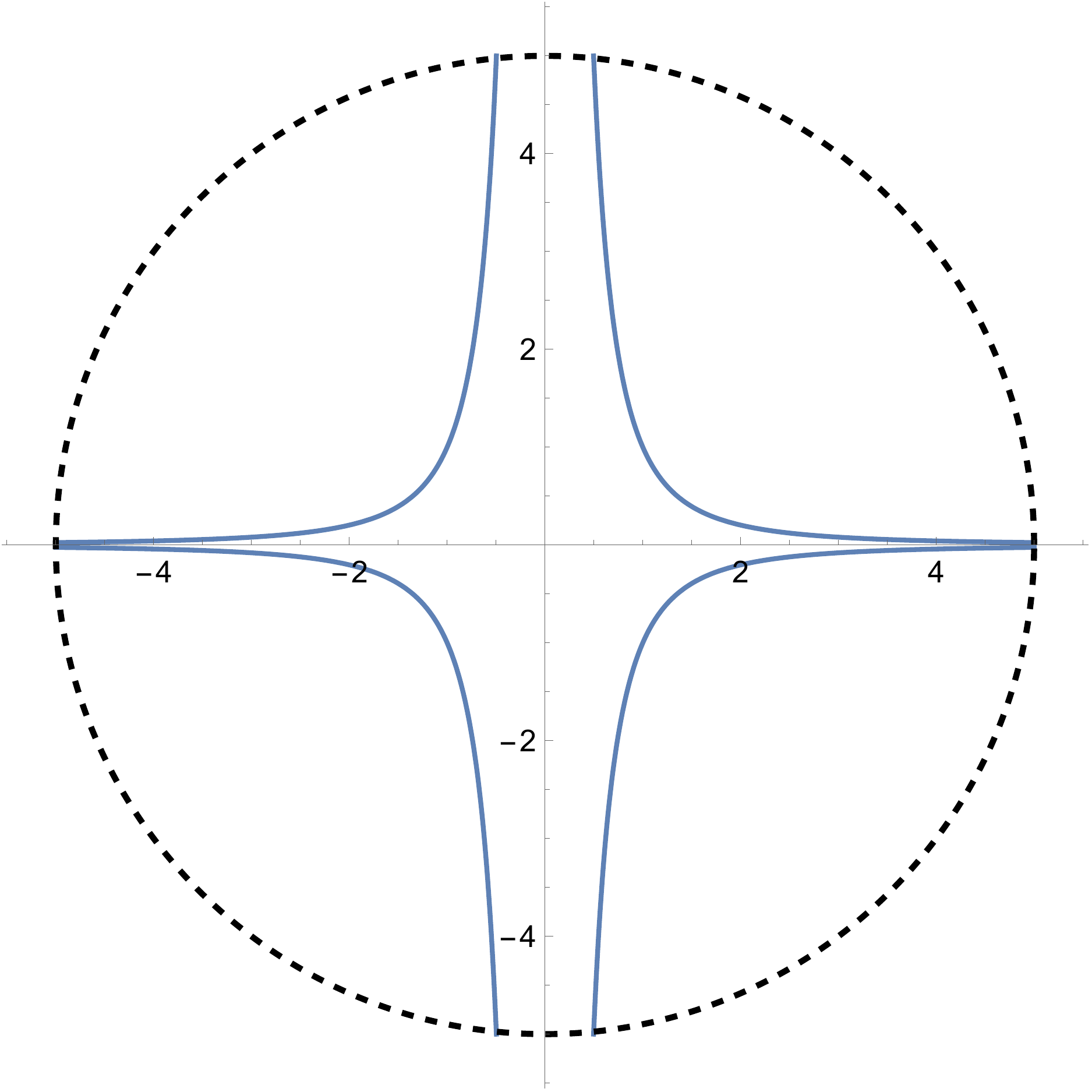}{The loop is made of four hyperbola branches.} 
\begin{eqnarray}\label{PR0}
&& h = \sqrt{\frac{\nu}{c}};\\
&& x_{min}  = \frac{h}{R_0};\\
&& x_{max} = \left(\frac{R_0}{h}\right)^{\frac{1}{\mu}};\\
&& P \ra  4 R_0\left(\frac{R_0}{h}\right)^\mu + 4R_0\left(\frac{R_0}{h}\right)^{\frac{1}{\mu}}  \ra 4 R_0^{\mu+1}h^{-\mu}
\end{eqnarray}

To keep the perimeter fixed in the extreme turbulent limit, we have to tend the parameter $R_0$ to zero as
\begin{eqnarray}\label{R0Scaling}
&&R_0 = \left(\frac{P}{4}\right)^{\frac{1}{\mu+1}} \left(\frac{\nu}{c}\right)^{\frac{\mu}{2(\mu+1)}} \ra 0;
\end{eqnarray}

The cross-section area of the tube
\begin{eqnarray}\label{Area}
\mbox{Area} = 4 R_0^2\int_{x_{min}}^\infty d x x^{-\mu} = \frac{4}{\mu-1} R_0^2 x_{min}^{1-\mu} \propto P h \sim P \left(\frac{\nu}{c}\right)^{\oh}
\end{eqnarray}

\subsection{Velocity Gap and Circulation}
Once the equation is solved, the parametric solution for $\Gamma$ is straightforward (with factors of $R_0$ restored from dimensional counting)
\begin{eqnarray}\label{eqGamma}
 &&\Gamma = \int (-2 a X X'+2 (a-c) Y Y') d x = \nonumber\\
 &&-a R_0^2\left(x^{2}-\mu |x|^{-2 \mu }\right);\\
 &&\Delta V = \frac{\Gamma'}{C'} = -2 a R_0 x\left(1 + \i \mu |x|^{-\mu-1 }\right)
\end{eqnarray}

We have not expected to find such a singular vortex tube, but it satisfies all requirements and must be accepted.

In \cite{M21c}, we appealed to the Brouwer theorem \cite{BrT} to advocate the existence of solutions of the \CVS{} equations. This theorem does not tell us how many fixed points are on a sphere made of the normalized Fourier coefficients when their number goes to infinity. 

De Lellis and Bru\'e have proven that there are no compact solutions.

Our non-compact solution is not normalizable, and it has no circulation 
\begin{eqnarray}
\Delta \Gamma =  \frac{c R_0^2 }{\mu -1}\left(x^{2}-\mu |x|^{-2 \mu }\right)_{x = -\infty}^{x = \infty} = 0
\end{eqnarray}

We assumed finite circulation in \cite{M21c}; thus, the Brouwer theorem does not apply. On the other hand, we no longer need an existence theorem once we have found an analytic solution.

Note that the net circulation would be infinite unless we combine all four branches of our hyperbola into a single closed loop (with infinite wings at the real and imaginary axes, but still closed).

\subsection{Anomalous dissipation }
 
 The role of stable, steady solutions of the \NS{}  equations is to provide the subspace of attractors in phase space. We expect the evolution of the solution to cover this subspace with a uniform measure like Newton's dynamics covers the surface of constant energy.
 
 The situation is more complex in turbulence. The energy $E$ of the fluid is not conserved; it rather dissipates. This dissipation is proportional to the enstrophy
 \begin{eqnarray}
 \d_t E = -\mathcal E = -\nu \int d^3 r \oal^2 
 \end{eqnarray}
 
 We have a steady state of constant energy dissipation in the turbulent limit, with the constant energy supply from external forces on the system's boundary, like the submarine moving in the ocean or the water pumped into the pipe.
 
 Although ultimately related to the viscous effects, one can study this dissipation in the turbulent limit within the framework of the Euler-Lagrange dynamics.
 
 Large vorticity in certain regions of space where dissipation predominantly occurs will offset the vanishing factor $\nu$ in front of the enstrophy. 
 
 Our theory starts with the conjecture that dissipation occurs in vortex surfaces, like the \BT{} sheet. In that case, the Euler-Lagrange equations describe the dynamics everywhere except the turbulent layer surrounding the vortex sheet, as discussed above.

Using the \BT{} vortex sheet solution in the linear vicinity of the local tangent plane to the surface, as in \cite{M20c, M21b}, we get the following integral for the dissipation
\begin{eqnarray}
\mathcal E  \propto \nu  \int_D d^2 \xi \sqrt{g} (\vec \nabla \Gamma)^2 \int_{-\infty}^{\infty} d \eta(\delta_h(\eta))^2;
\end{eqnarray}
  
Here $\eta$ is a local normal direction to the surface and
\begin{eqnarray}
  \delta_h(\eta) = \frac{1}{h \sqrt{2 \pi}}\exp{- \frac{\eta^2}{2 h^2}}
\end{eqnarray}
is the normal distribution.
The (local!) width $h$ is determined by
\begin{eqnarray}
  h = \sqrt{\frac{\nu}{ \lambda}}
\end{eqnarray}

Here $\lambda = - \hat S_{n n} = S^i_i$. 

Naturally, we assume that the local normal $\vec \sigma$ points on the third axis of the strain, where the finite eigenvalue is negative.

An important new phenomenon is the possibility of the variation of this eigenvalue along the surface.

The square of the Gaussian is also a Gaussian:
\begin{eqnarray}
  &&\delta_h(\eta)^2 = \frac{1}{ 2h \sqrt{\pi} }\delta_{\tilde h}(\eta);\\
  && \tilde h = h/\sqrt{2}.
\end{eqnarray}
In the limit $\tilde h \ra 0$, we are left with the surface integral
\begin{eqnarray}\label{DissipationStrain}
  \mathcal E =  \frac{\sqrt{\nu}}{ 2\sqrt{\pi} } \int_D d^2 \xi \sqrt{g}\sqrt{-\hat S_{n n} } (\vec \nabla \Gamma)^2;
\end{eqnarray}

In the paper \cite{M21b}, we also found an expression for the time derivative of the viscosity anomaly for the Euler equation. 

Repeating these steps with our local width $h$, we find (fixing wrong sign in \cite{M21b})
\begin{eqnarray}
    &&(\d_t \mathcal{E})_{Euler} =  2\nu \int d^3 r \left(-\dbe \left(\oh \vbe \oal^2\right) + \oal  \obe  \dbe \val \right) \ra\nonumber\\
    && +\frac{\sqrt{\nu}}{ 2\sqrt{\pi} } \int_D d^2 \xi \sqrt{g}\sqrt{S^i_i}  \tilde \d^i \Gamma S_{i j}\tilde \d^j \Gamma ;\\
    &&  \tilde \d^i = \epsilon^i_k \d_k
\end{eqnarray}

We did not add the viscous term of the \NS{}  equation in that paper, which was a mistake. Adding it now, we get, after a  simple algebra
\begin{eqnarray}
    &&(\d_t \mathcal{E})_{NS} = (\d_t \mathcal{E})_{Euler} + 2 \nu^2 \int d^3 r \; \vec \omega \cdot\vec \nabla^2 \vec \omega \ra \nonumber\\
    && +\frac{\sqrt{\nu}}{ 2\sqrt{\pi} } \int_D d^2 \xi \sqrt{g} \sqrt{S^k_k} \tilde \d^i \Gamma \tilde \d^j \Gamma \left(S_{i j}- S^k_k g_{i j}\right)
\end{eqnarray}
We transform this expression into a simpler one using the properties of two-dimensional tensors
\begin{eqnarray}
(\d_t \mathcal{E})_{NS} =-\frac{\sqrt{\nu}}{2 \sqrt{\pi} } \int_D d^2 \xi \sqrt{g} \sqrt{S^k_k}\d^i \Gamma S_{i j}\d^j \Gamma
\end{eqnarray}
In the general case, this expression is not zero and could have an arbitrary sign because the tangent eigenvalues must be both signs, given that their sum is positive.

However, in the \CVS{} case, the Euler and \NS{}  terms in the time derivative of the enstrophy cancel so that the surface dissipation integral is conserved!

\subsection{The  Surface enstrophy conservation}

The \CVS{} equation, derived from the microscopic stability of the Euler dynamics in the turbulent limit, also provides a new turbulent motion integral: the surface dissipation \eqref{DissipationStrain}. 

Unlike the energy functional, this is an \textit{integral of motion of the \NS{}  dynamics}, but not the Euler dynamics.

The \CVS{} surface is unique in having these two effects: vortex stretching and diffusion cancel each other, making the surface enstrophy an integral on motion of the \NS{}  dynamics in the turbulent limit.

Naturally, the enstrophy is constant on every steady solution of the \NS{}  equation, simply by the definition of a steady solution.
What is not trivial and is quite fortunate for the vortex sheet dynamics is that the \textit{ surface} enstrophy is conserved.

The energy could be dissipated all over the space, not just on the vortex surface. This phenomenon happens for every vortex sheet except the \CVS{}. The time derivative of the enstrophy is either negative on these surfaces, which means that vorticity is leaking outside, or it is positive, meaning an instability of the surface.

The total enstrophy integral, in that case, would not be dominated by a surface, which would make the theory incomplete.

The simple idea that the turbulent statistics is a Gibbs distribution with surface dissipation in the effective Hamiltonian was advocated in our recent paper \cite{M21b}. The conservation of the surface dissipation was unknown at that time, which was the problem.

Now we have this problem solved in \CVS{} dynamics. However, the Gibbs distribution with surface enstrophy in place of Gibbs energy was not confirmed. 

As we shall see in the last section, an extreme turbulence statistical distribution is even simpler. It is three-dimensional statistical distribution as opposed to the functional integral of the Gibbs theory.

The \CVS{} constraints reduce the arbitrary surface and arbitrary potential gap $\Gamma$ to some family of solutions with a finite number of parameters. This family of solutions is no longer the functional phase space of the general vortex sheet dynamics \cite{M88, AM89}.

A finite-dimensional attractor could be the Holy Grail of the turbulence quest, but first, we must thoroughly investigate this hypothesis.

We have found only non-compact \CVS{} surfaces. Our surfaces extend to infinity in the Euler limit, and only a finite viscosity cuts them off. 
De Lellis and Bru\'e subsequently proved that no compact \CVS{} surfaces can exist.

In the next chapters, we study Euler solutions with vorticity spreading outside the vortex sheet. This sheet forms a minimal surface $S_C$ bounded by a 3D loop $C$ (the \CL{} domain wall). The singular vorticity, located at the loop, dominates the enstrophy, and this time, there is a \textbf{finite} limit at $\nu \ra 0$.

These solutions have a nontrivial topology of the \CL{} field, leading to the circulation conservation for the loop bounding the vortex sheet. As we argue, these solutions (Kelvinons) have finite anomalous dissipation and dominate the tails of the PDF of velocity circulation. 

Are they the Holy Grail? Time will tell.

\subsection{Minimizing Euler Energy}

Let us now compute the energy of the vortex surface as a Hamiltonian system \cite{M88, AM89}. 
There is a regular part related to the background strain. This part is not involved in the minimization we are interested in; it depends on $a,b,c$, which are external parameters for our problem.

The internal part of the Hamiltonian is directly related to the potential gap $\Gamma$ we have computed in the previous section.

As shown in the previous sections, the extremum equations for the fluid Hamiltonian as a functional of $\Gamma$ are locally equivalent to the Laplace equation for the external and internal potentials with Neumann boundary conditions.

\begin{eqnarray}
H_{int} = \int_{\vec r_1,\vec r_2 \in \mathcal S}  d \Gamma(\vec r_1) \wedge d \vec r_1  \cdot d \Gamma(\vec r_1) \wedge d \vec r_1 \frac{1}{8 \pi |\vec r_1 - \vec r_2| }
\end{eqnarray}

One determines the remaining parameters in the solution from the global minimum condition.

In our case, we have a cylindrical surface
\begin{eqnarray}
d \Gamma(\vec r_1) \wedge d \vec r_1  = d \Gamma(\theta) \left\{0,0,d z \right\}
\end{eqnarray}
and a separation of variables $\theta, z$. 

The integration over $ z_1, z_2 $ provides the total length $L \ra \infty$ of the cylinder times logarithmically divergent integral over $z_1-z_2$. We limit this integral to the interval  $(-L, L)$ and compute it exactly 
\begin{eqnarray}
\int_{-L}^L \frac{1}{8 \pi  \sqrt{\eta ^2+z^2}} \, dz = \frac{\log \left(\frac{2 L \left(\sqrt{\eta ^2+L^2}+L\right)}{\eta ^2}+1\right)}{8 \pi }
\end{eqnarray}

Then we expand it for large $L$
\begin{eqnarray}
\int_{-L}^L \frac{1}{8 \pi  \sqrt{\eta ^2+z^2}} \, dz  \ra \frac{\log( 2 L) -\log |\eta|}{4 \pi} + O\left(\frac{\eta^2}{L^2}\right)
\end{eqnarray}

Thus we get in our case, with  $\Gamma'(w)$ from \eqref{eqGamma}
\begin{eqnarray}
&&\frac{H_{int}}{L} = \log (2 L) \frac{(\Delta \Gamma)^2 }{4 \pi}-\nonumber\\
&&\frac{1}{8 \pi} \dashint_{-\infty}^{\infty} d w_1\Gamma'(w_1) \dashint_{-\infty}^{\infty} d w_2\Gamma'(w_2) \log \abs{\xi_1 -\xi_2};\\
&& \xi_{1,2} = w_{1,2} - \i |w_{1,2}|^{-\mu};\\
&& \Gamma'(w) = \frac{2 c R_0^2 w }{\mu -1}\left(1 +  \mu^2 |w|^{-2 \mu -2}\right)
\end{eqnarray}

The first term is the leading one at $L \ra \infty$. Minimization of the energy leads to the condition
\begin{eqnarray}
\Delta \Gamma =0
\end{eqnarray}

Therefore, our solution with zero circulation is singled out among other combinations of hyperbolic vortex sheets by the requirement of minimization of energy.

Once the divergent term vanishes, one can compute the rest of the energy. 

The remaining principal value integral over $x_1, x_2$ converges at infinity. In the local limit, when $R_0 \ra 0$, it goes to zero.

We do not see a point in this computation compared to observable quantities such as the energy dissipation and the loop functional.

\subsection{The induced background strain}\label{InducedStrain}

What is the physical origin of the constant background strain $W_{\alpha\beta}$ which we used in our solution?

Usually, the external Gaussian random forces are added to the Navier-Stokes equation to trigger the effects of the unknown inner randomness.

In the above theory, the random forces come from many remote vortex structures, contributing to the background velocity field via the Biot-Savart law.

There is no contradiction: in the case of spontaneous magnetization of ferromagnetic, one can introduce an infinitesimal external magnetic field, which is enhanced by infinite susceptibility to produce finite magnetization.

One may also write the self-consistent equation for the mean field, which is analogous to our background strain. 

Ultimately, the randomness and the energy flow originate from the external forces at the boundary. These forces trigger internal randomness to be enhanced by the large vorticity structures in bulk.

Let us assume that the space is occupied by some localized vortex structures (not necessarily the \CVS{} surfaces) far from each other. In other words, let us consider an ideal gas of vortex bubbles. 

We shall see below that the mean size $R_0$ of the surface is small compared to the mean distance $\bar R$ between them in the turbulent limit $\nu \ra 0, \mathcal E = \mbox {const} $. This vanishing size vindicates the assumptions of the low-density ideal gas.

In such an ideal gas, we can neglect the collision of these extended particles but not the long-range effect of the strain they impose on each other.

The Biot-Savart formula \eqref{BS} for the velocity field induced by the set of remote localized vorticity bubbles $B$
\begin{eqnarray}
    \vec v(\vec r) = \sum_B\int_B d^3 r' \frac{\vec \omega(\vec r') \times (\vec r' - \vec r)}{4 \pi | \vec r - \vec r'|^3}
\end{eqnarray}\label{BubblesBS}
falls off as $1/r^2$ for each bubble, like an electric field from the charged body.

All vortex structures in our infinite volume contribute to this background velocity field, adding up to many small terms at every point in space.

While the \NS{}  equation is nonlinear, this relation between the local strain and contributions from each vortex tube is \textbf{perfectly linear}, as it follows from the linear Poisson equation relating velocity to vorticity.

Therefore, the interaction between bubbles decreases with distance by the power law, which justifies the ideal gas picture in the case of sparsely distributed vortex tubes.

\begin{center}
    \begin{tikzpicture}
\draw (0, 0) circle (3);
\draw[->, red] (3.0, 0.0) -- (3.18989763692, 0.206882004178);
\draw[->, red] (2.99817248106, 0.104698490108) -- (2.86187640313, 0.444418012577);
\draw[->, red] (2.99269215078, 0.209269421232) -- (2.78392904122, 0.0658099999994);
\draw[->, red] (2.9835656861, 0.313585389803) -- (3.02070562755, 0.837037261956);
\draw[->, red] (2.97080420622, 0.41751930288) -- (3.06995364287, 1.18536928931);
\draw[->, red] (2.95442325904, 0.520944533001) -- (3.0879480934, 0.493598974164);
\draw[->, red] (2.9344428022, 0.623735072453) -- (2.83891271278, 1.06033398555);
\draw[->, red] (2.91088717883, 0.725765686799) -- (3.10473557841, 0.510627909767);
\draw[->, red] (2.88378508781, 0.826912067451) -- (2.5060573585, 0.754345348178);
\draw[->, red] (2.85316954889, 0.927050983125) -- (2.78075252939, 0.901143347442);
\draw[->, red] (2.81907786236, 1.02606042998) -- (2.37399730859, 0.781558686335);
\draw[->, red] (2.7815515637, 1.12381978025) -- (3.09508229995, 1.19150164342);
\draw[->, red] (2.74063637293, 1.22020992923) -- (2.81299485895, 1.3564352197);
\draw[->, red] (2.6963821389, 1.31511344037) -- (2.49941942261, 1.26615212382);
\draw[->, red] (2.64884277858, 1.40841468836) -- (3.36903230154, 1.73842863018);
\draw[->, red] (2.59807621135, 1.5) -- (2.50870498738, 1.17045855207);
\draw[->, red] (2.54414428847, 1.5897577927) -- (2.71827641138, 1.58464664852);
\draw[->, red] (2.48711271767, 1.67757871041) -- (2.18813718767, 1.54686708672);
\draw[->, red] (2.42705098312, 1.76335575688) -- (2.09405794429, 1.66037100993);
\draw[->, red] (2.36403226082, 1.84698442598) -- (1.93252639477, 2.21807804904);
\draw[->, red] (2.29813332936, 1.92836282906) -- (2.43423272507, 1.51347700514);
\draw[->, red] (2.22943447643, 2.00739181908) -- (2.06854921552, 1.97359095614);
\draw[->, red] (2.15801940102, 2.08397511138) -- (1.9258335215, 1.7263424932);
\draw[->, red] (2.08397511138, 2.15801940102) -- (1.75565565496, 1.91907655062);
\draw[->, red] (2.00739181908, 2.22943447643) -- (1.66531344206, 1.99636529281);
\draw[->, red] (1.92836282906, 2.29813332936) -- (1.96566980134, 1.94379809258);
\draw[->, red] (1.84698442598, 2.36403226082) -- (1.70487875929, 2.37829869);
\draw[->, red] (1.76335575688, 2.42705098312) -- (1.62182528237, 2.21848916138);
\draw[->, red] (1.67757871041, 2.48711271767) -- (1.70293332919, 2.85993204194);
\draw[->, red] (1.5897577927, 2.54414428847) -- (1.63460823423, 2.25213441541);
\draw[->, red] (1.5, 2.59807621135) -- (1.51237421172, 2.36254630876);
\draw[->, red] (1.40841468836, 2.64884277858) -- (1.56044917181, 2.62350283287);
\draw[->, red] (1.31511344037, 2.6963821389) -- (1.4326193314, 2.58649487031);
\draw[->, red] (1.22020992923, 2.74063637293) -- (1.29146381578, 2.63264373772);
\draw[->, red] (1.12381978025, 2.7815515637) -- (0.613253182305, 2.8114841836);
\draw[->, red] (1.02606042998, 2.81907786236) -- (1.03497921441, 2.67324867851);
\draw[->, red] (0.927050983125, 2.85316954889) -- (0.551482030987, 3.2386645256);
\draw[->, red] (0.826912067451, 2.88378508781) -- (0.580443305142, 3.09117113964);
\draw[->, red] (0.725765686799, 2.91088717883) -- (0.863825020463, 2.69147350839);
\draw[->, red] (0.623735072453, 2.9344428022) -- (0.709160938294, 2.568562005);
\draw[->, red] (0.520944533001, 2.95442325904) -- (0.185063664054, 3.15727266888);
\draw[->, red] (0.41751930288, 2.97080420622) -- (0.477598819342, 3.01834760387);
\draw[->, red] (0.313585389803, 2.9835656861) -- (0.0527268101921, 2.69929557543);
\draw[->, red] (0.209269421232, 2.99269215078) -- (-0.145300820942, 3.36346044384);
\draw[->, red] (0.104698490108, 2.99817248106) -- (0.304621265325, 2.71379047122);
\draw[->, red] (1.83697019872e-16, 3.0) -- (-0.0732179374315, 3.62894708093);
\draw[->, red] (-0.104698490108, 2.99817248106) -- (-0.298047562323, 2.85671684955);
\draw[->, red] (-0.209269421232, 2.99269215078) -- (-0.24374159757, 3.00338661766);
\draw[->, red] (-0.313585389803, 2.9835656861) -- (-0.0184572014415, 3.41722939486);
\draw[->, red] (-0.41751930288, 2.97080420622) -- (-0.352661474172, 2.73506271443);
\draw[->, red] (-0.520944533001, 2.95442325904) -- (-1.67682304378, 2.60410098578);
\draw[->, red] (-0.623735072453, 2.9344428022) -- (-0.226344352811, 2.7172102314);
\draw[->, red] (-0.725765686799, 2.91088717883) -- (-0.874320159866, 2.67885694643);
\draw[->, red] (-0.826912067451, 2.88378508781) -- (-0.930387802353, 2.53269518768);
\draw[->, red] (-0.927050983125, 2.85316954889) -- (-1.13239640231, 2.79620127612);
\draw[->, red] (-1.02606042998, 2.81907786236) -- (-0.948619372619, 2.61302134644);
\draw[->, red] (-1.12381978025, 2.7815515637) -- (-1.21845112514, 2.9238346055);
\draw[->, red] (-1.22020992923, 2.74063637293) -- (-1.21402359492, 2.86791433052);
\draw[->, red] (-1.31511344037, 2.6963821389) -- (-1.59018092296, 2.71650729288);
\draw[->, red] (-1.40841468836, 2.64884277858) -- (-1.40797553653, 2.65572636607);
\draw[->, red] (-1.5, 2.59807621135) -- (-1.23848197872, 2.63974165954);
\draw[->, red] (-1.5897577927, 2.54414428847) -- (-1.57343114712, 2.91323923811);
\draw[->, red] (-1.67757871041, 2.48711271767) -- (-1.51732473682, 2.34852873138);
\draw[->, red] (-1.76335575688, 2.42705098312) -- (-2.06223149649, 2.76621785129);
\draw[->, red] (-1.84698442598, 2.36403226082) -- (-1.96648420569, 2.72176017255);
\draw[->, red] (-1.92836282906, 2.29813332936) -- (-1.68116668044, 2.38792019914);
\draw[->, red] (-2.00739181908, 2.22943447643) -- (-1.98427700146, 2.207824065);
\draw[->, red] (-2.08397511138, 2.15801940102) -- (-2.27486540816, 2.23729426669);
\draw[->, red] (-2.15801940102, 2.08397511138) -- (-2.58464805536, 1.67631550285);
\draw[->, red] (-2.22943447643, 2.00739181908) -- (-2.14390493528, 2.12149980891);
\draw[->, red] (-2.29813332936, 1.92836282906) -- (-2.62011809966, 1.92867779322);
\draw[->, red] (-2.36403226082, 1.84698442598) -- (-2.03101397202, 1.4033297666);
\draw[->, red] (-2.42705098312, 1.76335575688) -- (-2.49172131703, 1.64731954638);
\draw[->, red] (-2.48711271767, 1.67757871041) -- (-2.68711199588, 1.21852721685);
\draw[->, red] (-2.54414428847, 1.5897577927) -- (-2.86847556942, 1.34749171723);
\draw[->, red] (-2.59807621135, 1.5) -- (-2.69346300154, 1.4660419061);
\draw[->, red] (-2.64884277858, 1.40841468836) -- (-2.67130130732, 1.38862466366);
\draw[->, red] (-2.6963821389, 1.31511344037) -- (-3.16709170952, 1.45366002775);
\draw[->, red] (-2.74063637293, 1.22020992923) -- (-2.31560474605, 1.14453440601);
\draw[->, red] (-2.7815515637, 1.12381978025) -- (-2.94191152978, 1.43498732059);
\draw[->, red] (-2.81907786236, 1.02606042998) -- (-2.74434057758, 1.07006003882);
\draw[->, red] (-2.85316954889, 0.927050983125) -- (-2.69263106898, 1.08978634707);
\draw[->, red] (-2.88378508781, 0.826912067451) -- (-2.72734777215, 0.478442701806);
\draw[->, red] (-2.91088717883, 0.725765686799) -- (-2.84168325018, 0.830974929636);
\draw[->, red] (-2.9344428022, 0.623735072453) -- (-2.83489568062, 0.444299979843);
\draw[->, red] (-2.95442325904, 0.520944533001) -- (-2.92014840509, 0.428199685013);
\draw[->, red] (-2.97080420622, 0.41751930288) -- (-2.66383622494, 0.406924869236);
\draw[->, red] (-2.9835656861, 0.313585389803) -- (-2.79707618979, 0.109554742508);
\draw[->, red] (-2.99269215078, 0.209269421232) -- (-3.03547972658, 0.037666203143);
\draw[->, red] (-2.99817248106, 0.104698490108) -- (-2.64949550051, -0.0347696612005);
\draw[->, red] (-3.0, 3.67394039744e-16) -- (-3.04347529899, 0.0247195154345);
\draw[->, red] (-2.99817248106, -0.104698490108) -- (-3.14802135357, 0.136428384273);
\draw[->, red] (-2.99269215078, -0.209269421232) -- (-2.8633154347, -0.114792426255);
\draw[->, red] (-2.9835656861, -0.313585389803) -- (-2.85779839474, -0.299630149963);
\draw[->, red] (-2.97080420622, -0.41751930288) -- (-3.13178657115, -0.574847155054);
\draw[->, red] (-2.95442325904, -0.520944533001) -- (-2.91876244475, -0.405922871455);
\draw[->, red] (-2.9344428022, -0.623735072453) -- (-3.34436009073, -0.671144984914);
\draw[->, red] (-2.91088717883, -0.725765686799) -- (-3.22652305699, -0.594522096622);
\draw[->, red] (-2.88378508781, -0.826912067451) -- (-2.88965284355, -0.791280875084);
\draw[->, red] (-2.85316954889, -0.927050983125) -- (-2.57329714466, -1.18768088164);
\draw[->, red] (-2.81907786236, -1.02606042998) -- (-2.31548993148, -0.955560814495);
\draw[->, red] (-2.7815515637, -1.12381978025) -- (-2.64306314606, -1.43357356318);
\draw[->, red] (-2.74063637293, -1.22020992923) -- (-2.79556284294, -1.02477172835);
\draw[->, red] (-2.6963821389, -1.31511344037) -- (-2.86837394899, -0.929545490032);
\draw[->, red] (-2.64884277858, -1.40841468836) -- (-2.2928484287, -1.21347068465);
\draw[->, red] (-2.59807621135, -1.5) -- (-2.9452681636, -2.33491134724);
\draw[->, red] (-2.54414428847, -1.5897577927) -- (-2.38485580346, -1.38392513667);
\draw[->, red] (-2.48711271767, -1.67757871041) -- (-2.42693193622, -1.71005883878);
\draw[->, red] (-2.42705098312, -1.76335575688) -- (-2.58021736489, -2.05362842271);
\draw[->, red] (-2.36403226082, -1.84698442598) -- (-2.57753029198, -2.01808049085);
\draw[->, red] (-2.29813332936, -1.92836282906) -- (-2.05897015018, -1.71848590207);
\draw[->, red] (-2.22943447643, -2.00739181908) -- (-2.44570276606, -2.25970871455);
\draw[->, red] (-2.15801940102, -2.08397511138) -- (-2.40522641721, -2.57239496883);
\draw[->, red] (-2.08397511138, -2.15801940102) -- (-2.22804190616, -2.15804774161);
\draw[->, red] (-2.00739181908, -2.22943447643) -- (-2.26571530382, -2.75787491001);
\draw[->, red] (-1.92836282906, -2.29813332936) -- (-2.05874326567, -2.76354745826);
\draw[->, red] (-1.84698442598, -2.36403226082) -- (-1.87992422083, -2.72079287842);
\draw[->, red] (-1.76335575688, -2.42705098312) -- (-1.9130533174, -2.41974275229);
\draw[->, red] (-1.67757871041, -2.48711271767) -- (-1.99541526503, -2.9918263612);
\draw[->, red] (-1.5897577927, -2.54414428847) -- (-1.54597838458, -2.17331898586);
\draw[->, red] (-1.5, -2.59807621135) -- (-1.53108938192, -2.24666773134);
\draw[->, red] (-1.40841468836, -2.64884277858) -- (-0.853249123831, -2.6063190776);
\draw[->, red] (-1.31511344037, -2.6963821389) -- (-1.18136858564, -2.81789828092);
\draw[->, red] (-1.22020992923, -2.74063637293) -- (-1.43055933258, -2.70353724505);
\draw[->, red] (-1.12381978025, -2.7815515637) -- (-0.798275008208, -2.85745596864);
\draw[->, red] (-1.02606042998, -2.81907786236) -- (-0.984965307961, -2.14911041462);
\draw[->, red] (-0.927050983125, -2.85316954889) -- (-0.857307365463, -2.70898721676);
\draw[->, red] (-0.826912067451, -2.88378508781) -- (-0.836690279356, -3.13049458975);
\draw[->, red] (-0.725765686799, -2.91088717883) -- (-0.641349468948, -2.66763377287);
\draw[->, red] (-0.623735072453, -2.9344428022) -- (-1.1357114996, -2.99716954073);
\draw[->, red] (-0.520944533001, -2.95442325904) -- (-0.803722923223, -3.10453424292);
\draw[->, red] (-0.41751930288, -2.97080420622) -- (-0.386091178005, -2.86062802752);
\draw[->, red] (-0.313585389803, -2.9835656861) -- (-0.28277270597, -3.09864383986);
\draw[->, red] (-0.209269421232, -2.99269215078) -- (-0.516166908905, -2.85363952263);
\draw[->, red] (-0.104698490108, -2.99817248106) -- (0.0210513862726, -2.97429127975);
\draw[->, red] (-5.51091059616e-16, -3.0) -- (0.0242713301868, -3.52815135671);
\draw[->, red] (0.104698490108, -2.99817248106) -- (0.433559917904, -2.87756960072);
\draw[->, red] (0.209269421232, -2.99269215078) -- (0.0832811303765, -2.75274296472);
\draw[->, red] (0.313585389803, -2.9835656861) -- (0.625462350184, -3.47791128908);
\draw[->, red] (0.41751930288, -2.97080420622) -- (-0.215427846601, -2.91674753389);
\draw[->, red] (0.520944533001, -2.95442325904) -- (0.698746135555, -3.03995003002);
\draw[->, red] (0.623735072453, -2.9344428022) -- (0.381825012443, -3.45095631867);
\draw[->, red] (0.725765686799, -2.91088717883) -- (0.753334033268, -3.01887033499);
\draw[->, red] (0.826912067451, -2.88378508781) -- (0.705564136556, -2.92941724919);
\draw[->, red] (0.927050983125, -2.85316954889) -- (1.41113455115, -3.40534200591);
\draw[->, red] (1.02606042998, -2.81907786236) -- (1.07451301987, -3.03563601835);
\draw[->, red] (1.12381978025, -2.7815515637) -- (0.957630503643, -2.71400528226);
\draw[->, red] (1.22020992923, -2.74063637293) -- (1.00278291385, -2.15590056526);
\draw[->, red] (1.31511344037, -2.6963821389) -- (1.11584795112, -2.48156714269);
\draw[->, red] (1.40841468836, -2.64884277858) -- (1.47229470996, -2.42997355565);
\draw[->, red] (1.5, -2.59807621135) -- (1.36543984678, -2.9415074817);
\draw[->, red] (1.5897577927, -2.54414428847) -- (1.92005553603, -2.0135324779);
\draw[->, red] (1.67757871041, -2.48711271767) -- (1.60455737283, -2.49062004187);
\draw[->, red] (1.76335575688, -2.42705098312) -- (1.5606572359, -2.46252841489);
\draw[->, red] (1.84698442598, -2.36403226082) -- (1.44807230237, -2.00682935572);
\draw[->, red] (1.92836282906, -2.29813332936) -- (2.18815057107, -2.21852192352);
\draw[->, red] (2.00739181908, -2.22943447643) -- (2.15059591715, -2.43092969499);
\draw[->, red] (2.08397511138, -2.15801940102) -- (2.37654352353, -1.40822174014);
\draw[->, red] (2.15801940102, -2.08397511138) -- (1.95867211726, -2.51284844011);
\draw[->, red] (2.22943447643, -2.00739181908) -- (1.84163202633, -2.31766478478);
\draw[->, red] (2.29813332936, -1.92836282906) -- (2.85993170839, -1.87707818054);
\draw[->, red] (2.36403226082, -1.84698442598) -- (1.99752151071, -1.72229460185);
\draw[->, red] (2.42705098312, -1.76335575688) -- (1.8965835219, -1.62996390383);
\draw[->, red] (2.48711271767, -1.67757871041) -- (2.80647737019, -1.45189383519);
\draw[->, red] (2.54414428847, -1.5897577927) -- (2.5273239967, -1.66408286705);
\draw[->, red] (2.59807621135, -1.5) -- (2.43947066427, -1.72825089551);
\draw[->, red] (2.64884277858, -1.40841468836) -- (2.42336328609, -1.0644063492);
\draw[->, red] (2.6963821389, -1.31511344037) -- (2.77666951601, -1.35593722929);
\draw[->, red] (2.74063637293, -1.22020992923) -- (2.28728115697, -1.86575140416);
\draw[->, red] (2.7815515637, -1.12381978025) -- (2.85856133106, -0.824183760798);
\draw[->, red] (2.81907786236, -1.02606042998) -- (2.64034459611, -1.40192744459);
\draw[->, red] (2.85316954889, -0.927050983125) -- (2.52983147664, -1.1933025616);
\draw[->, red] (2.88378508781, -0.826912067451) -- (3.3150543001, -0.537321731144);
\draw[->, red] (2.91088717883, -0.725765686799) -- (3.20664407828, -1.07084249221);
\draw[->, red] (2.9344428022, -0.623735072453) -- (2.37786593194, -0.508453913022);
\draw[->, red] (2.95442325904, -0.520944533001) -- (3.36526729818, -0.648952283581);
\draw[->, red] (2.97080420622, -0.41751930288) -- (2.75417204049, -0.460270323914);
\draw[->, red] (2.9835656861, -0.313585389803) -- (3.62780016563, -0.297775276693);
\draw[->, red] (2.99269215078, -0.209269421232) -- (2.82528433849, -0.462939878866);
\draw[->, red] (2.99817248106, -0.104698490108) -- (2.85837812702, -0.319398346917);
\node[inner sep=0pt] (vortex) at (0.5, 0.5) {\includegraphics[width=2.5cm]{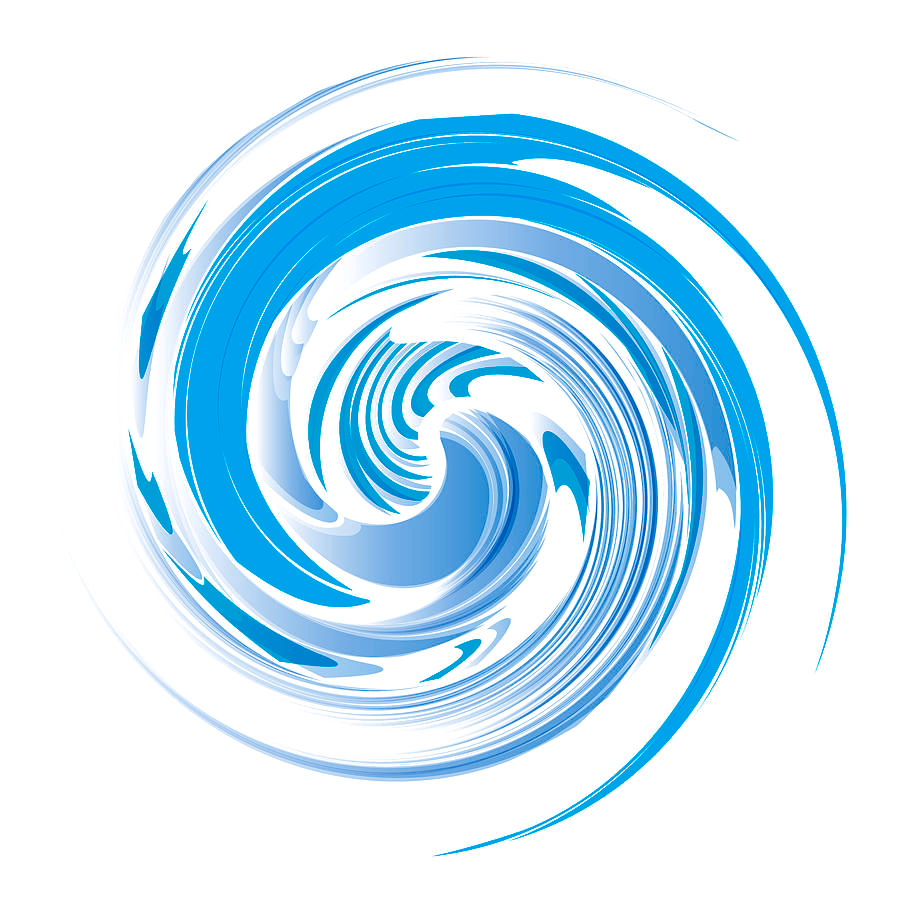}};
\end{tikzpicture}

\end{center}
This picture symbolizes the vortex tube under consideration (blue vortex symbol) surrounded by other remote tubes on a large sphere (orange arrows). The arrows indicate the directions of these remote tubes, which are aligned with the main axis of local strain. We expect the positions and directions of these tubes to be random and uncorrelated when they are far from each other.

If many such bubbles populate space with small but finite density, we would have the "night sky paradox."  The bubbles spread on the far away sphere will compensate the inverse distance squared for a divergent distribution like $\int R^2 d R/R^2$.

This estimate needs to be corrected, as the velocity contributions from various bubbles are uncorrelated, so there is no coherent mean velocity. 

Moreover, a Galilean transformation would remove the finite background velocity, so it does not have any physical effects. 

However, with the strain, there is another story.
Strain coming from remote vortex bubbles
\begin{eqnarray}\label{BSStrain}
    W_{\alpha\beta}(\vec r) = \oh e_{\alpha\mu\gamma}\dbe \dga  \sum_B\int_B d^3 r' \frac{ \omu(\vec r') }{4 \pi | \vec r - \vec r'|} + \left\{ \alpha \leftrightarrow \beta \right\};
\end{eqnarray}
falls off as $1/r^3$, and this time, there could be a mean value $\bar W$, coming from a large number of random terms from various bubbles with distribution $ R^2 d R/R^3 \sim d R/R$.

The space symmetry arguments and Laplace equation $\vec \nabla^2 1/|\vec r| =0$ outside the origin tell us that averaging over the directions of the bubble centers $\vec R = \vec r'-\vec r$ completely cancels this mean value.

The Central Limit Theorem suggests (within our ideal vortex gas model) that such a strain would be a Gaussian tensor variable, satisfying the normal distribution of a symmetric traceless matrix with zero mean value (see Appendix B.)
\begin{eqnarray}\label{GaussTensor}
    d P_\sigma(W) \propto \prod_i d W_{i i} \prod_{i < j} d W _{i j} \delta\left(\sum_i W_{i i}\right)\exp{-\frac{\tr W^2}{2 \sigma^2}}
\end{eqnarray}
The parameter $\sigma$ is related to the mean trace of the square of the random matrix. In $n$ dimensional space
\begin{eqnarray}\label{TraceRel}
\frac{(n+2)(n-1)}{2} \sigma^2 = \VEV{tr W^2}
\end{eqnarray}
The Gaussian random matrices were studied extensively in physics and mathematics. For example, in \cite{RSM}, the distribution of the Gaussian random symmetric matrix (Gaussian Orthogonal Ensemble, $GOE(n)$) is presented.

We achieve the extra condition of zero matrix trace by inserting the delta function of the matrix trace into the invariant measure. This projection preserves the measure's $O(n)$ symmetry as the trace is invariant to orthogonal transformations. We tried to find references for this straightforward extension of the $GOE(n)$ to the space of traceless symmetric matrices.

Separating $SO_3$ rotations $\Omega \in S_2$, we have the measure for eigenvalues $a,b,c$:
\begin{eqnarray}\label{Pab}
    &&d P_\sigma(W) =\frac{1}{4 \pi}d \Omega  d a d b d c \delta(a + b + c) P_\sigma(a,b,c);\\
    && P_\sigma(a,b,c)=  \sqrt{\frac{3}{\pi }} \theta(b-a) \theta(c-b) (b-a)(c-a)(c-b)\nonumber\\
    && \exp{-\frac{a^2 + b^2 + c^2}{2\sigma^2} }
\end{eqnarray}

The moments $\VEV{\left(\frac{a}{\sigma}\right)^m \left(\frac{b}{\sigma}\right)^n}$ of this distribution are calculable analytically (see \cite{MB3}). Here is the table for $m,n =0,1,2,3,4$:
\begin{equation}\label{PabMoms}
\left(
\begin{array}{ccccc}
 1 & 0 & \frac{1}{6} & 0 & \frac{1}{12} \\
 \frac{3 \sqrt{\frac{3}{\pi }}}{2} & -\frac{1}{12} & \frac{3 \sqrt{\frac{3}{\pi }}}{2}-\frac{2 \sqrt{\pi }}{3} & -\frac{1}{24} & 5 \sqrt{\frac{3}{\pi }}-\frac{8 \sqrt{\pi }}{3} \\
 \frac{29}{12} & \frac{2 \sqrt{\pi }}{3}-\frac{3 \sqrt{\frac{3}{\pi }}}{2} & \frac{13}{24} & \frac{8 \sqrt{\pi }}{3}-5 \sqrt{\frac{3}{\pi }} & \frac{49}{144} \\
 \frac{9 \sqrt{\frac{3}{\pi }}}{2} & -\frac{19}{24} & 6 \sqrt{\frac{3}{\pi }}-\frac{8 \sqrt{\pi }}{3} & -\frac{71}{144} & 25 \sqrt{\frac{3}{\pi }}-\frac{40 \sqrt{\pi }}{3} \\
 \frac{209}{24} & \frac{8 \sqrt{\pi }}{3}-7 \sqrt{\frac{3}{\pi }} & \frac{373}{144} & \frac{40 \sqrt{\pi }}{3}-\frac{77}{\sqrt{3 \pi }} & \frac{1747}{864} \\
\end{array}
\right)
\end{equation}

The mean traces of powers of the strain, as well as the powers of its determinant, are also calculable. Odd powers yield zero, and the even powers yield the following tables
\begin{subequations}\label{StrainTraces}
\begin{eqnarray}
   &&\VEV{\tr \left(\frac{\hat W}{\sigma}\right)^{2n}} =\nonumber\\ &&\left\{3,5,\frac{35}{2},\frac{1015}{12},\frac{37345}{72},\frac{185185}{48},\dots\right\};\\
   && \VEV{\left(\det \frac{\hat W}{\sigma}\right)^{2n}} =\nonumber\\ &&\left\{1,\frac{35}{18},\frac{5005}{108},\frac{8083075}{1944},\frac{32534376875}{34992},\frac{29248404810625}{69984}\right\}
\end{eqnarray}
\end{subequations}

The estimate of the variance of the background strain can be done as follows.

\begin{eqnarray}\label{PabTrn}
 \sigma^2 =\VEV{\tr (\hat W^2)} \sim \VEV{\left(\int_{V_B}d^3 \vec r' \vec \omega(\vec r') \right)^2} /\Delta R_B^6 
\end{eqnarray}
where the first factor is the mean value of the variance of volume vorticity of the tube and $\Delta R_B$ is the average distance between tubes.

\subsection{ Energy dissipation at the vortex sheet and its distribution}\label{DissipationCVS}
As we noticed in the previous work \cite{M21c} (see previous sections), the total surface dissipation is conserved on \CVS{} surfaces.
\begin{eqnarray}\label{Etotal}
\mathcal E_{tot} = \sum_{\mathcal S} \mathcal E_{\mathcal S} = \mbox{const}
\end{eqnarray}

Without CVS as a stability condition, the surface dissipation would not be motion integral. The energy would leak from the vortex surfaces and dissipate in the rest of the volume. Thus, the CVS condition is necessary for the vortex sheet turbulence.

While the total dissipation is conserved, the individual contributions to this sum from each tube are not. The long-term interactions between the vortex tubes, arising due to the Gaussian fluctuations of the background strain, lead to the statistical distribution of the energy dissipation of an individual tube.

From analogy with the Gibbs-Boltzmann statistical mechanics, one would expect that the dissipation distribution would come out exponential, with some effective temperature. We put forward this hypothesis in our previous work.

However, the interaction between our tubes is different from that of the Gibbs mechanics. While the background strain is a Gaussian (matrix) variable, the shapes of the tubes and the corresponding dissipation are not. 

These tubes in our incompressible fluid instantly adjust to the realization of the random background strain.
Our exact solutions of the Euler equations with the \CVS{} boundary conditions describe this adjustment.

The general formula \cite{M21c} for the surface dissipation was derived above (\eqref{DissipationStrain}.

For our hyperbolic loop solution, the energy dissipation integral reduces to the following expression (where we restore the implied spatial scale $R_0$ and express the cutoffs in terms of the perimeter $P$ of the loop by \eqref{R0Scaling})
\begin{eqnarray}\label{Dissipation}
  && \frac {\mathcal E} { L \sqrt {\nu }} = 4 \frac{2 a^2 \sqrt{c}}{\sqrt{\pi } }  R_0^3\int_{x_{min}}^{x_{max}} \, d w \, w^2 \abs{1- \i \mu  w^{-\mu -1}}^3 \ra \nonumber\\
  &&P^3 \frac{ a^2 \sqrt{c}}{24\sqrt{\pi }}
\end{eqnarray}
The normalized distribution $W(\zeta)$ for the scaling variable $\zeta = \frac{a^2 \sqrt{c}}{\sigma^{\frac{5}{2}}}$ takes the form \cite{MB3}
\begin{eqnarray}\label{DisPDF}
&& W(\zeta)=2 \zeta^{9/5} \sqrt{\frac{3}{\pi }}\int_{2^{-\ofi}}^{2^{\ofi}}d y \frac{\left(2-y^5\right) \left(y^5+1\right) \left(2 y^5-1\right) }{y^{14}}\nonumber\\
&&\exp{\frac{\left(-y^{10}+y^5-1\right) \zeta^{4/5}}{y^8}}\\
&&\zeta = \frac { 24 \sqrt{\pi} \mathcal E} { L P^3\sqrt {\nu \sigma^5}};
\end{eqnarray}

The expectation value of this scaling variable equals to
\begin{eqnarray}
\bar \zeta = 4.90394
\end{eqnarray}

We show at \pct{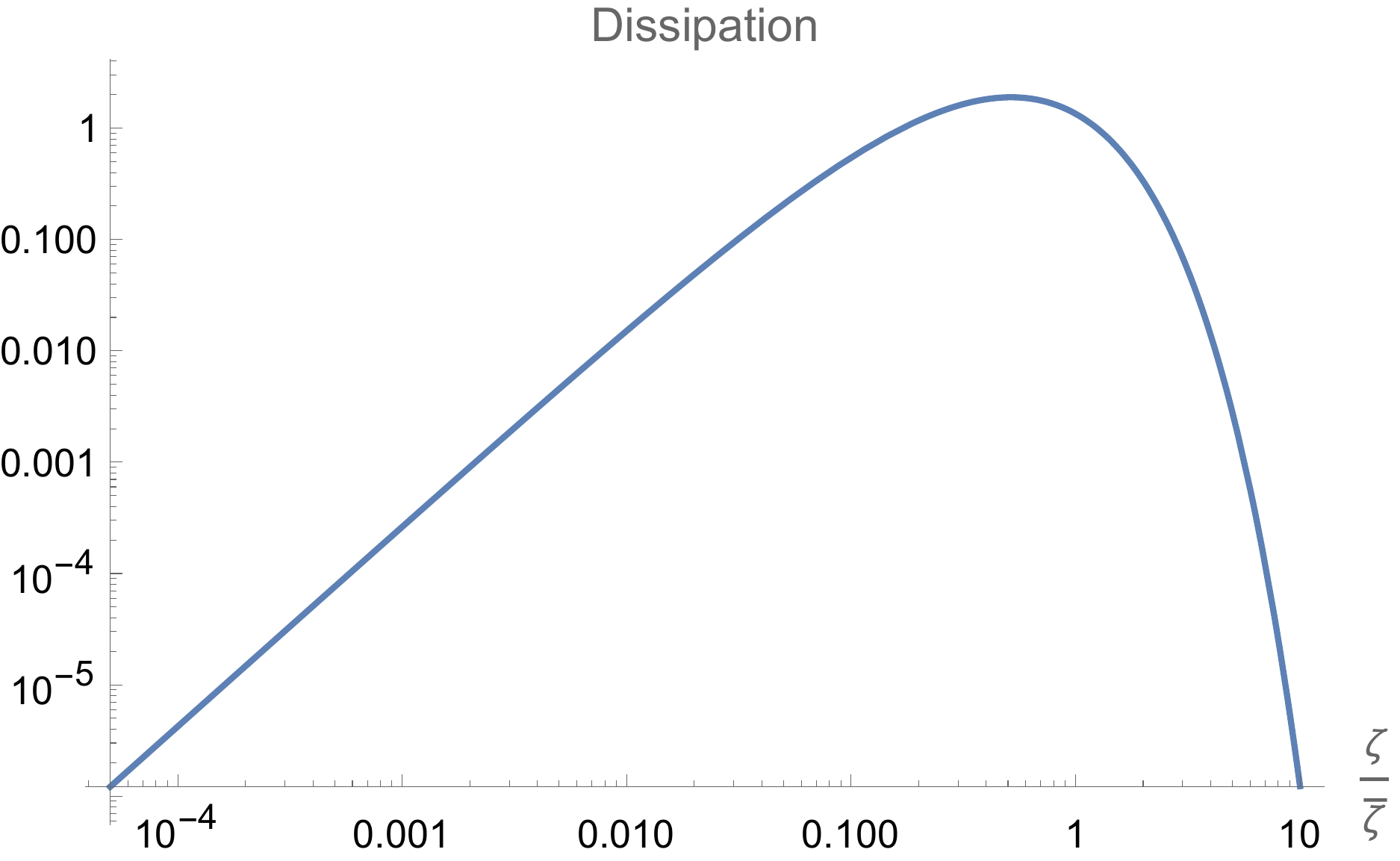}{The energy dissipation PDF (fixed perimeter) in log-log scale} the log-log plot of this distribution.

This $W(\zeta)$ is a completely universal function. We will verify this prediction when the distribution of energy dissipation and tube sizes in numerical or real experiments in the extreme turbulent regime becomes available.

The perimeter $P$ of the loop remains a free parameter of our theory. We need some extra restrictions to find the distribution of these perimeters. This additional restriction of the fixed perimeter of the cross-section makes it quite tedious to compare our distribution of the energy dissipation with numerical simulation.

Moreover, we need to find out if these vortex tubes with non-compact cross-sections, though stable, are indeed realized in a turbulent flow. 

The answer is no. As we shall see in the next sections, the stable solutions are characterized by topological numbers and are more singular than pure vortex sheets. At the core, they match the Burgers vortex with the axis along a local tangent direction to the loop.

These solutions nicely describe the tails of PDF of velocity circulation, which makes them the best candidates for the dominant classical Clebsch fields (Kelvinons).

\section{Vortex Loops}\label{Kelvinon}

The idea behind the Kelvinon theory \cite{M22} is simple: find a solution of the \NS{} equation in the limit of vanishing viscosity such that it conserves the circulation $\Gamma_\alpha$ around a stationary loop $C$.

We call such a hypothetical solution a Kelvinon, honoring Kelvin and his famous theorem of conservation of circulation around a liquid loop. Kelvin theorem applies to arbitrary Euler flow, but Kelvinon flow has a particular stagnant loop. 

To be more precise, this loop is a closed trajectory of a liquid particle, which makes it stationary as a geometric object. This stationarity is sufficient for the Kelvin theorem leading to conserved circulation.

Then, the Wilson loop will be given by the Gaussian average of $\exp{\imath \gamma \Gamma_\alpha}$ over some random velocity background to be specified later:
\begin{eqnarray}\label{GaussAverage}
    &&\Psi[C,\gamma] = \VEV{\exp{\imath \gamma \Gamma_\alpha}};
\end{eqnarray}

Finding such a \NS{} solution is easier said than done, but we have a significant simplification. We are looking for the limit of vanishing viscosity.

\subsection{The Matching Principle}
In this limit, we can use the Euler dynamics everywhere except for some thin boundary regions where a large Laplacian of velocity compensates the viscosity factor in front of this term in the \NS{} equation.

Within the Euler dynamics, boundary regions shrink to lower-dimensional subsets: velocity gap points (shocks) in 1D, point vortexes in 2D, and vortex sheets and vortex lines in 3D.

We have to use full \NS{} equations in these boundary regions. The small thickness of these regions simplifies the geometry of \NS{} equation: we can neglect all the curvature effects.

These flat (or straight line) \NS{} solutions must match the Euler solution outside the boundary regions. Thus, the \NS{} provides the boundary condition to Euler and vice versa. 

This matching principle was suggested in our recent work \cite{M21c, M21d} and applied to the Euler vortex sheets (tangent velocity gaps in purely potential flow).
We discussed it in application to the vortex sheets in the previous section \ref{VortexSheets}.

In general, the matching principle removes the ambiguity of the weak solutions of the Euler dynamics by resolving the singularities at surfaces and lines.

It also lets one select among the various exact solutions of the \NS{} equations in a flat geometry.

The inviscid limit $\nu \ra 0$ in the \NS{} equation does not exist mathematically. At small but finite viscosity, there are anomalies in all physical quantities in the singular Euler flow. These terms in the Hamiltonian grow as the logarithm of the Reynolds number. One could neglect all power corrections at small viscosity, but these logarithmic terms must be summed up at small but finite viscosity.
This situation is analogous to the asymptotic freedom in QCD.

 All studies of weak solutions of the Euler equation, ignoring these anomalies while having a well-defined mathematical meaning,  do not correspond to a physical flow as we see it now.
 
The matching conditions led to so-called \CVS{} equations \eqref{CVSEQ} for the stable vortex sheets (velocity gaps in potential flow). The vortex sheets bounded by a fixed loop do not belong to that category.

As we shall see, a line singularity is also required at the loop, making it a Dirac monopole of the Euler equation. 
This line singularity of the velocity field ($1/r$ pole) corresponds to finite $\vec S(\vec r) \in S_2$ with a particular topology we describe later.

\subsection{Closed monopole line with Burgers vortex at the core}
\pct{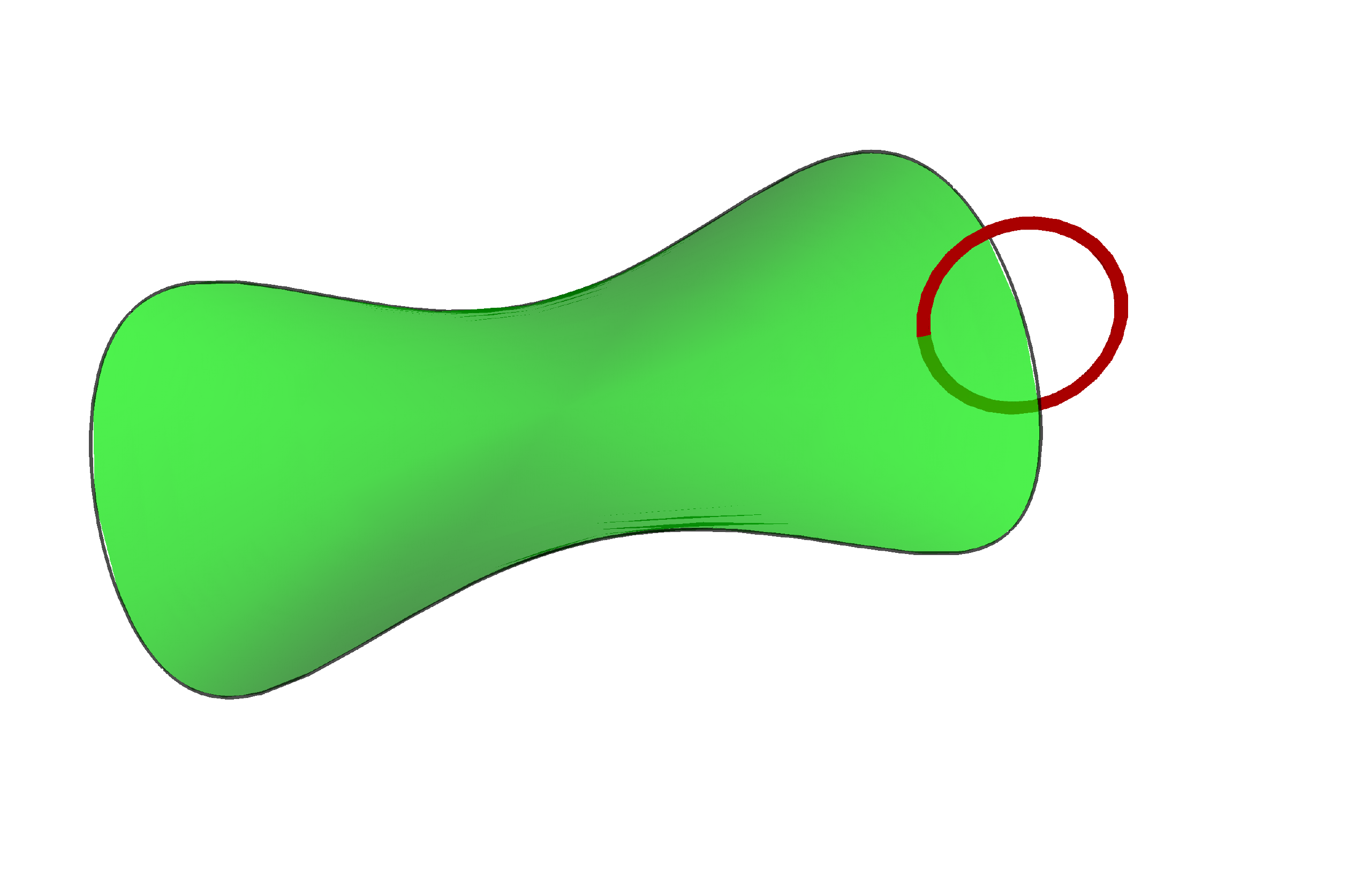}{Kelvinon cycles. The $\beta$ cycle (red) around the $\alpha$ cycle (black) of the vortex sheet (green)}
Let us consider a (yet to be determined)  stationary Euler flow with a singular vortex line at the smooth loop $C$. We surround this loop with a thin tube; its radius varies around the loop and must be much smaller than its local curvature radius;(see Fig.~\ref{fig::GreenPetalRedLoop}.)

Let us first consider a local vicinity of the core, which we assume to be a cylindrical region, given its small radius. 
We assume cylindrical symmetry and take the Burgers vortex solution \cite{BurgersVortex} of the \NS{} equation;
\begin{eqnarray}\label{BurgersVortex}
    && \vec v =\left\{-a x- g(r) y, -b y + g(r) x, c z + d\right\};\\
    && \vec \omega = \left\{0,0,\frac{c  \Gamma_\beta  e^{-\frac{c r^2}{4 \nu }}}{8 \pi  \nu }\right\};\\
    && g(r) = \frac{\left(1-e^{-\frac{c  r^2}{4 \nu }}\right)\Gamma_\beta  }{2 \pi  r^2};\\
    && r = \sqrt{x^2 + y^2};\\
    && \Gamma_\beta =\oint_\beta \vec v \cdot d\vec r;\\
    && a + b = c = \hat t \cdot \hat S \cdot \hat t;\\
    && \hat t = \{0,0,1\}.
\end{eqnarray}

Here $\hat S_{\alpha \beta} =\oh ( \dal \vbe + \dbe \val)$ is the strain tensor far away from the center of the core, and $\vec t$ is the local tangent vector of the loop (the axis of a cylindrical tube).

$\Gamma_\beta$ is the velocity circulation around the cylinder cross-section far away from its axis (the $\beta$ cycle of the tube).

There is an interesting story related to this Burgers vortex. 

The original axial symmetric solution, found by Burgers, was applicable only for axisymmetric strain, $ a = b = \oh c $.

The subsequent studies \cite{MKO94} revealed a more general solution with arbitrary asymmetric strain. 
This solution cannot be found in a closed form. Still, in the turbulent limit $|\Gamma_\beta| \gg \nu$, it reduces to the axial symmetric Burgers solution for the rotational part of velocity, but with the general asymmetric background strain.

These parameters $\Gamma_\beta, a,b, c, d$ have to be matched by an Euler solution at the surface of the tube.

Using \Mathematica{}, we computed the contribution to the Euler Hamiltonian from the circular core of the vortex at $z=0$ 
\begin{eqnarray}
    &&\oh \int_{ x^2 + y^2 < R^2}  d x d y \,\vec v(x,y,0)^2 = \frac{ \pi R^2 d^2}{2} + \frac{\pi  R^4 \left(a^2+b^2\right)}{8} \nonumber\\
    &&+\frac{\left(-E_1\left(\frac{c R^2}{2 \nu }\right)+2 E_1\left(\frac{c R^2}{4 \nu }\right)+\log \left(\frac{c R^2}{8 \nu }\right)+\gamma \right) \Gamma _{\beta }^2}{8\pi }.
\end{eqnarray}

where $E_1(x)$ is an exponential integral function and $\gamma$ is Euler's constant.

In the turbulent limit $\nu \ra 0$ we get
\begin{eqnarray}
   &&\frac{\Gamma_\beta^2}{8\pi} \left(\gamma +\log \left(\frac{c  R^2}{8 \nu }\right)\right)+ \frac{ \pi R^2 d^2}{2} + \frac{\pi  R^4 \left(a^2+b^2\right)}{8} \nonumber\\
   &&+\dots
\end{eqnarray}
where $\dots$ stand for exponentially small terms at $\nu \ra 0$.

The $R^2, R^4$ terms are also negligible in our approximation  $R \ra 0$.
Now, the total contribution from the tube around the loop will be
\begin{eqnarray}
   && \oh \int_{|\vec r-\vec C| < R} d^3 r \vec v^2  \ra \frac{\Gamma_\beta^2}{8\pi}\oint_C \abs{d \vec r}\left(\gamma +\log \frac{c R^2}{8 \nu} \right).
\end{eqnarray}

Adding the integral from the remaining space outside the tube, we get the total Hamiltonian for the Euler flow
\begin{eqnarray}
   &&H_R =\int_{|\vec r-\vec C| > R} d^3 r \frac{\vec v^2}{2}  + \frac{\Gamma_\beta^2}{8\pi} \oint_C \abs{d \vec r}\left(\gamma +\log \frac{c R^2}{8 \nu} \right).
\end{eqnarray}
This Hamiltonian does not depend on $R$ up to the power-like terms at $R\ra 0$; therefore, we can take the limit $ R \ra 0$.

This Hamiltonian must be minimized within the topological class of flows with fixed circulations $\Gamma_\alpha, \Gamma_\beta$.

The circulation $\Gamma_\alpha$ is expressed through the local Burgers parameter $d$ as follows
\begin{eqnarray}
    \Gamma_\alpha = \oint_C  d \vec r \cdot \vec v = \oint_C  d \vec r \cdot \vec t d =  \oint_C |d \vec r| d
\end{eqnarray}

As for the dissipation in the Burgers vortex, it is finite in the turbulent limit.
The vorticity vector is concentrated at the core and exponentially decays outside
so that the enstrophy integral converges at the core
\begin{eqnarray}
   \int_{|\vec r - \vec C| < R} d ^3 \vec r  \vec \omega^2 = \frac{\Gamma_\beta ^2 }{8 \pi  \nu } \oint_C |d \vec r|c \left(1-e^{-\frac{c  R^2}{2\nu }}\right).
\end{eqnarray}

In the turbulent limit, $\nu \ra 0$, we find finite energy dissipation due to our circular Burgers vortex
\begin{eqnarray}
    &&\mathcal E_d =  \nu \int_{|\vec r - \vec C| > R} d ^3 \vec r  \vec \omega^2 + \nu\int_{|\vec r - \vec C| < R} d ^3 \vec r  \vec \omega^2 \nonumber\\
    &&\ra \frac{\Gamma_\beta ^2 }{8 \pi } \oint d \theta c;\\
    && \vec C'(\theta)^2 =1;\\
    && c = \left(\vec C'(\theta )\cdot \vec \nabla\right) \left( \vec v \cdot \vec C'(\theta)\right).
\end{eqnarray}

\subsection{The Kelvinon geometry and topology}

 For the Euler flow, we are considering now, the physical space $\mathcal P$  is the whole space $R_3$ without an infinitesimal tube $\mathcal T$ around the loop $C$ and without the discontinuity surface $S_C$ without its boundary $C$.
 \begin{eqnarray}\label{PhysSpace}
\mathcal P =  R_3\setminus(\mathcal T  \oplus S_C\setminus C)
 \end{eqnarray}   
 At this surface $S_C$, the angular variable $\phi_2$ has a $2 \pi n$ discontinuity, with integer $n$.

The components of the vector $\vec S$ vary continuously in $\mathcal P$,  but the vector $\vec S(\vec r)$ goes 
 $n $ times around a horizontal circle on a sphere when a point $\vec r$ goes around this loop in an infinitesimal circle.

 The velocity field can be written as the \BS{} integral
 \begin{eqnarray}\label{paramsClebsch}
     &&\vec v = Z  \tilde v;\\
     && \vec \omega = Z \tilde \omega;\\
     && \tilde \omega = \oh e_{a b c}  S_a  \vec \nabla S_b \times \vec\nabla S_c;\\
     \label{Vparam}
     && \tilde v = \vec \nabla \Phi  -\vec \nabla \times \vec \Psi;\\ 
     && \vec \Psi(\vec r) = \int d^3 r' \frac{\tilde \omega(\vec r')}{4\pi | \vec r - \vec r'|};\\
     &&\tilde v_n( S_C) =0;
 \end{eqnarray}

 The potential $\Phi$ satisfies the Laplace equation with the Neumann boundary condition at $S_C\setminus C$ to cancel the normal velocity coming from the $\vec \nabla \times \vec \Psi$ term in the \BS{} law. One can prove that the normal derivative of this term is continuous at the surface, so the potential term $\vec \nabla \Phi$ has to cancel the continuous normal velocity at the surface.

This restriction does not affect the rotational part of velocity, which is fixed by the circulation $\Gamma_\beta $.

 There is only a tangent discontinuity of velocity at the surface $S_C\setminus C$,  coming from the delta function in vorticity. The boundary values are
 \begin{eqnarray}
     &&\vec \omega \ra 2 \pi n \delta(z) \vec \sigma \times \vec \nabla \phi_1;\\
     \label{velGap}
     && \vec v^+ - \vec v^- =  2 \pi n \vec \nabla \phi_1;\\
     && S_3^+  = S_3^-;\\
     && \phi_2^+ =\phi_2^- - 2 \pi n;\\
     && \omega^+_n =  \omega^-_n = - \vec \sigma \cdot \vec \nabla \phi_2^\pm \times\vec \nabla \phi_1^\pm
 \end{eqnarray}
 Here $\vec \sigma $ is the local normal vector to the surface, and $z$ is the normal coordinate at this point.

We need to resolve the following paradox. The values of the \CL{} field are continuous across the surface as these values only involve $\cos\phi, \sin\phi$, which are $2\pi$ periodic. How do we get the velocity gap and the vortex sheet with a continuous \CL{} field?

There are two answers to this question. The mathematical answer is that velocity expressed in terms of $\phi,\theta$ is not $2\pi$-periodic
\begin{eqnarray}\label{vphitheta}
    \vec v = -\phi_2 \vec \nabla \phi_1 + \vec \nabla \phi_3
\end{eqnarray}

The last term is a solution to the Poisson equation
\begin{eqnarray}
    &&\vec \nabla^2 \phi_3 = \vec \nabla \cdot \left(\phi_2 \vec \nabla \phi_1\right);\\
    && \phi_3(\vec r) = H(\vec r) + \int d^3 \vec r \frac{\left(\phi_2 \vec \nabla \phi_1\right)  \cdot (\vec r - \vec r')}{4 \pi | \vec r - \vec r'|^3}
\end{eqnarray}

This term is singular but does not have a discontinuity at the surface.

The discontinuity of the first term leads to a velocity gap \eqref{velGap}.

Geometrically, the \CL{} field is a single valued variable on a sphere $\vec S \in S_2$, it maps onto $S_2$ the compactified space $R^3$ without infinitesimal tube $\mathcal G = R^3 \setminus \mathcal T$, which is a mapping of topological solid torus on a sphere $S_2$, described by two integer winding numbers $n, m$.

The velocity field, on the other hand , has a gap on $\mathcal G$ across the surface $\mathcal S$. Therefore, it maps a subspace $ \mathcal P = \mathcal G \setminus \mathcal S$ onto $S_2$. This subspace is a topological ball.

The Hamiltonian in our variational problem for the \CL{} field is expressed as an integral of $\int_G \oh v^2$ plus anomalous terms coming from inside the tube, where Burgers solution applies.

Thus, our Hamiltonian is multivalued on $\mathcal G$ and is single-valued on $\mathcal P$, where the velocity field is single-valued.

The theoretical physics answer is as follows. Let us follow our matching principle and postpone the Euler limit. At fixed viscosity, all the fields are smooth so that we can write for the normal derivative of the \CL{} field at the surface
\begin{eqnarray}
    \d_n \vec S = \left(-\sin \theta \sin\phi, \sin \theta\cos\phi,0\right) \d_n \phi + \textit{ regular terms}
\end{eqnarray}
 With finite viscosity, the $2 \pi n$ gap in $\phi$ will become some smooth regularization of a step function, like the error function in the conventional vortex sheet \eqref{ErrorFunc}.

 So, in a formal Euler limit, when the width of the vortex sheet tends to zero, we get $\d_n \phi \ra 0$ on each surface side.
 However, inside the singular layer of the viscous width $h \sim \sqrt{\nu / (-S_{n n})}$, the normal derivative goes to infinity.

 If we compute this normal derivative in the \NS{} theory and then tend viscosity to zero, we get the anomalous term
 \begin{eqnarray}
     \d_n \vec S \ra  \left(-\sin \theta \sin\phi, \sin \theta \cos\phi,0\right) 2 \pi n \delta(z) + \textit{ regular terms}
 \end{eqnarray}

 This paradox illustrates the main point of our theory: the Euler dynamics is ambiguous and has to be redefined as a limit of the \NS{} dynamics with small viscosity.

 The anomalies are everywhere in the singular Euler flow: in dissipation, in the Hamiltonian, in the helicity, etc., and they are calculable with the matching principle prescription.

 Let us come back to our Kelvinon solution.
In our case, the boundary value $S_3(C)$ is constant around the loop $C$: the velocity gap \eqref{velGap} must vanish at $C$, therefore $\vec \nabla S_3(C) =0$.
The lower side of the surface is mapped on a region at $S_2$ inside the \textit{oriented} curve $\gamma = \vec S(C)$, and the upper side is mapped on the same region at $S_2$ rotated $n$ times around the vertical axis ( $\phi \Ra \phi + 2 \pi n$).

Depending on the signs of winding numbers $m, n$, this oriented region would be either the upper $\Omega_+$ or the lower $\Omega_-$ part of the sphere $S_2$, separated by the horizontal circle $\theta = \lambda$.

Note that both sides of the surface are mapped on the same region  ($\Omega_+$ or $\Omega_-$), with precisely the same area, providing the same circulation.
The choice between $\Omega_\pm$ depends upon the signs of $m, n$. 

There are four distinct possibilities 
\begin{eqnarray}
    &&\sigma_m \equiv\sign{m} = \pm 1;\\
    && \sigma_n \equiv\sign{n} = \pm 1;
\end{eqnarray}

 The circulation at the $\alpha, \beta $ cycles is calculable in terms of these two winding numbers.
 \begin{eqnarray}
    &&\Gamma_\alpha = \oint_\alpha \val d \ral = \int_{S^+_C} d \vec \sigma \cdot \vec \omega^+ =\nonumber\\
    &&\int_{\Omega_+} d \phi_1 \wedge d \phi_2 = 2\pi m Z(1 + \sigma_m\cos\lambda)\\
    &&\Gamma_\beta = \oint_\beta \val d \ral  = \nonumber\\
    &&\int_{\Omega_-} d \phi_1 \wedge d \phi_2=  2 \pi n Z(1 + \sigma_n\cos\lambda);
 \end{eqnarray}

 We took advantage of the constant boundary value 
 \begin{equation}
 S_3(C) = \cos\lambda.
 \end{equation}
Note that changing the sign of each number changes the value of the circulation and its sign.

\pct{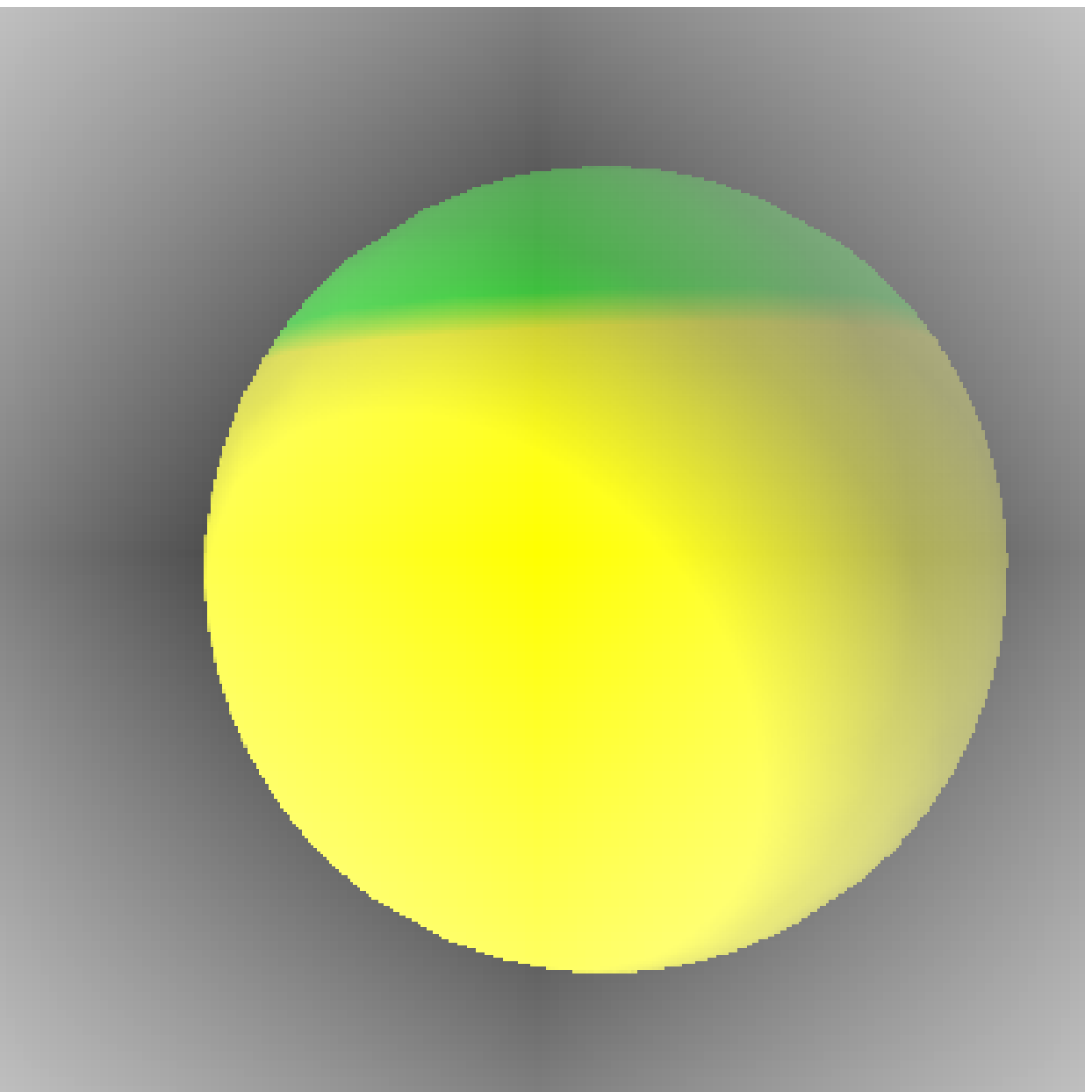}{Circulations are mapped on a yellow/green region $\Omega_\pm$ on $S_2$ for positive/negative sign of the winding number.}
\subsection{Helicity}
The natural question arises: what is a helicity of a Kelvinon?

Let us look at the contribution to helicity integral  \eqref{Helicity}, coming from the discontinuity surface.

 Given our condition of constant $\phi_1(C) $, this contribution to helicity equals zero.

We are left with the anomalous contributions to the helicity integral.
The infinitesimal tube yields:
\begin{eqnarray}\label{tubeHelicity}
    &&\mathcal H_C = \int_{|\vec r - \vec C| < R} \vec v \cdot \vec \omega = \oint_C |d \vec r| v_z \int_{x^2 + y^2 < R^2} d x d y \omega_z(x,y,0) =\nonumber\\
    &&\oint_C |d \vec r| v_z \Gamma _{\beta } \left(1-e^{-\frac{c R^2}{4 \nu }}\right) \ra \Gamma _{\beta } \oint_C |d \vec r| v_z = \nonumber\\
    && \Gamma _{\beta } \oint_C d \vec r \cdot \vec v  = \Gamma_\alpha \Gamma_\beta
\end{eqnarray}

The invariant formula for the turbulent limit of the core vorticity due to the Burgers vortex line is
\begin{eqnarray}\label{omegaLoop}
    \vec \omega(\vec r) \ra \Gamma_\beta \oint  d \vec C(\theta) \delta^3\left( \vec r - \vec C(\theta)\right)
\end{eqnarray}

This formula immediately leads to the same helicity
\begin{eqnarray}
    &&\mathcal H_C = \int d^3 \vec r \vec \omega(\vec r)\cdot \vec v(\vec r) = 
    \nonumber\\
    &&\Gamma_\beta \oint  d \vec C(\theta) \cdot \int d^3 \vec r  \vec v(\vec r) \delta^3\left( \vec r - \vec C(\theta)\right) = \nonumber\\
    &&\Gamma_\beta \oint  d \vec C(\theta) \cdot \vec v(\vec C(\theta)) = \Gamma_\beta \Gamma_\alpha
\end{eqnarray}

This thin tube in $R_3$ is mapped into a finite tube in the \CL{} space, with helicity coming from the spiral motion around this tube;(see Fig.~\ref{fig::WindingTorus}).

\pct{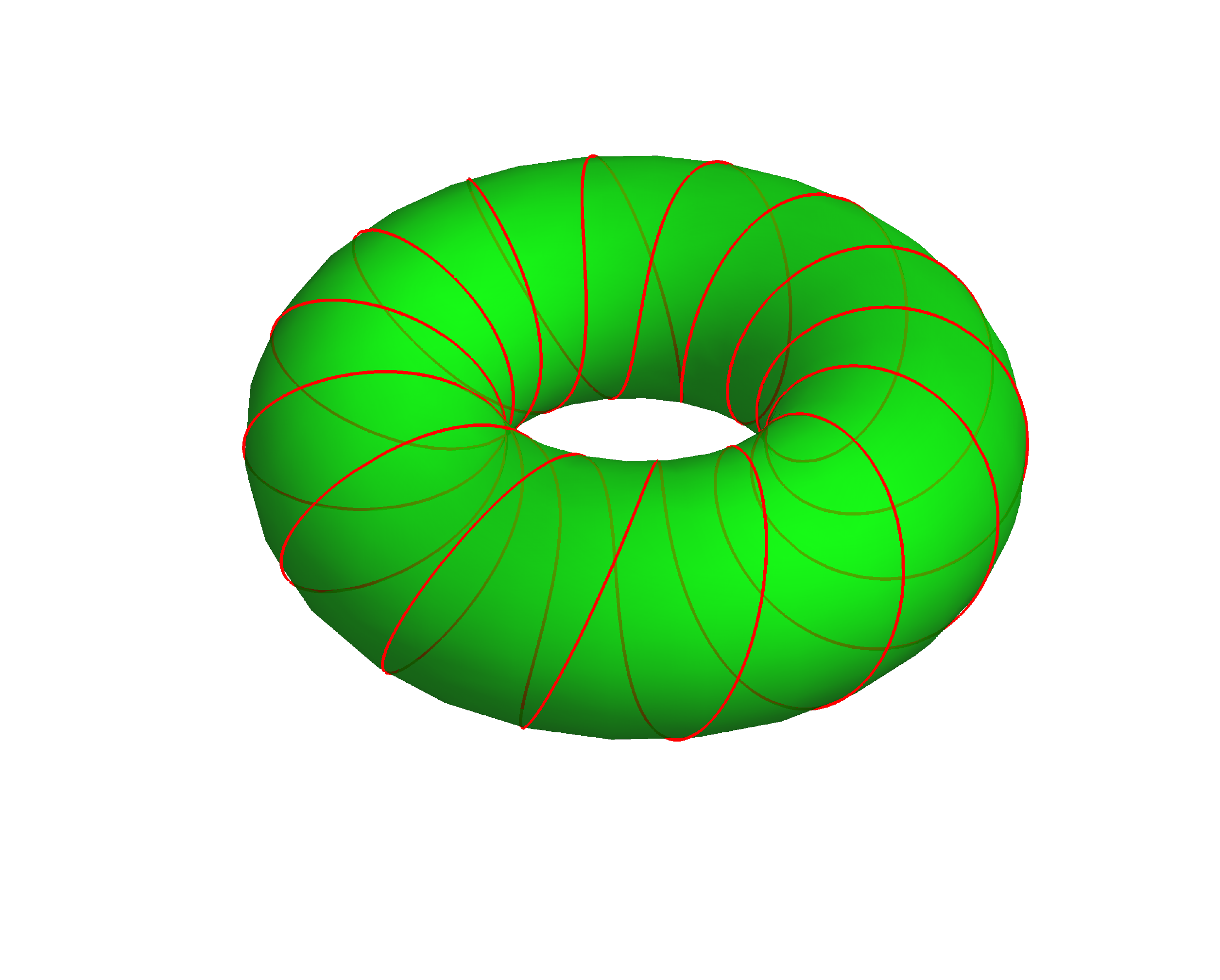}{The spiral motion of liquid particle around the singular vortex line.}

 There is also a  contribution to helicity from the vortex surface.
 
    The global generalization of the Burgers-Townsend formula for the singular vorticity inside the vortex sheet reads
    \begin{eqnarray}
       &&\vec \omega_{S}(\vec r) = 2 \int_{\vec r_1 \in S} d \vec \sigma(\vec r_1) \times \Delta \vec v(\vec r_1) \delta^3(\vec r - \vec r_1);
    \end{eqnarray}

    Substituting this into the helicity integral and using \eqref{velGap}, we get
    \begin{eqnarray}
        &&\mathcal H_{S} = \nonumber\\
        && Z^2\int_{\vec r \in S} d \vec \sigma(\vec r) \times \left(\tilde v(\vec r^+) - \tilde v(\vec r^-)\right) \cdot \left(\tilde v(\vec r^+) + \tilde v(\vec r^-)\right) =\nonumber\\
        && 2Z^2\int_{\vec r \in S} d \vec \sigma(\vec r) \cdot \tilde v(\vec r^+) \times \tilde v(\vec r^-);
    \end{eqnarray}
    
Is such a topology possible with continuous field $\vec S(\vec r) \in S_2$? We have built an example of the \CL{} field with two arbitrary winding numbers and proper boundary conditions at the surface and infinity.

This example is presented in \ref{ExampleClebsch}. It could be used as a zeroth approximation to minimize our renormalized Hamiltonian by relaxation.

\subsection{The energy flow balance and energy minimization}

By definition, the time derivative of any functional of the velocity satisfying the stationary \NS{} equations must vanish.

We regard our physical space as a large ball $\mathcal P$,  with the spherical boundary $\partial \mathcal P =\mathcal B$ separating the turbulent region from the rest of the world. We shall assume that energy flows through this boundary and dissipates inside. 

Let us consider the time derivative of the total Hamiltonian in this space $\mathcal P$ and use the \NS{} equation to reduce it to an enstrophy and the integral of the divergence of an energy current.
Usually, the dissipation is defined as an integral of the square of strain, in which case there will be no $\vec \omega \times \vec v$ term in an energy current. Our formula is equivalent to that, as the square of strain and the square of vorticity differ by gradient terms, which we included in the definition of the energy current.
\begin{eqnarray}
    &&0 = \partial_t H = \int_{\mathcal P} \vec v \cdot \partial_t \vec v = - \mathcal E_d + \int_{\mathcal P}  \vec \nabla \cdot \vec J ;\\
    && \mathcal E_d = \nu \int_{\mathcal P}\vec \omega^2 ;\\
    \label{Ecurr}
    && \vec J =  \left(p + \frac{\vec v^2}{2} \right)\vec v + \nu \vec \omega \times \vec v
\end{eqnarray}

By the Stokes theorem, the integral of $\vec \nabla \cdot \vec J$ reduces to the flux of this current through the external boundary
\begin{eqnarray}
    &&0 =-\mathcal E_d + \int_{ \mathcal B} J_n;
\end{eqnarray}
The flux through this boundary $\mathcal B$ represents the incoming energy flow $\mathcal E_p$ pumping energy from the remaining flow in the infinite volume.

We arrive at the identity
\begin{eqnarray}
    \mathcal E_d =\mathcal E_p
\end{eqnarray}

In the turbulent limit, the dissipation reduces to anomalous dissipation from the singularities at the surfaces $S$ and at the loops $C$ of all the Kelvinons inside the volume $\mathcal P$.

The surface dissipation is
\begin{eqnarray}
    \nu\int_S d S\int_{-\infty}^{\infty} d z \vec \omega_{N S}^2\left(S\setminus C + \vec \sigma z\right)
\end{eqnarray}
where $z$ is the local normal coordinate to the surface, and $\omega_{NS}$ is the \NS{} solution for the vortex sheet vorticity in the viscous layer.

This term was analyzed in previous work \cite{M21c, M21d},  and it vanishes as $\sqrt{\nu}$ in the turbulent limit because the singularity of $\omega^2$ in the vortex sheet yields the factor $1/\sqrt\nu$.

The integral of the Burgers vorticity at the core of the circular line vortex produces the finite result
\begin{eqnarray}\label{AnomalousDissipation}
    \mathcal E_p = \mathcal E_d \equiv \mathcal E = \frac{  \Gamma_\beta^2}{8 \pi }\oint_C |d \vec r| c 
\end{eqnarray}
Finite anomalous dissipation on a singular vortex loop, matching the energy flow from spacial infinity, is what Burgers dreamed up in his pioneering work\cite{BURGERS1948}.

He never finished that work, as the singular topological solutions of the Euler equations, matching his vortex tube, were unknown at the time.

The equation \eqref{AnomalousDissipation} reflects the energy conservation in the stationary solution of the full \NS{} equation.

Expressed in terms of Faddeev variables, velocity and vorticity fields are proportional to $Z$ as in \eqref{Vparam}
\begin{eqnarray}
    && \Gamma_\beta  = 2 \pi n (1 + \sigma_n \cos\lambda)Z;\\
    && c = Z \tilde c;\\
    && \tilde c= \left(\vec t \cdot \vec \nabla \right)\left(\tilde v \cdot \vec t\right);\\
    && \vec t = \frac{d \vec C}{| d \vec C|}
\end{eqnarray}

The vector field $\vec S(\vec r)$ must be found from the minimization of the Hamiltonian at fixed energy flow $\mathcal E$. 

We introduce conventional energy flux per unit volume
\begin{eqnarray}
    &&\eps = \frac{\mathcal E}{V(C)}
\end{eqnarray}
where $ V(C) \sim |C|^3$ is the volume occupied by our Kelvinon, assuming vorticity decays at the scale of distances $r \sim |C|$.

This assumption is valid at the scales $|C|$ below the intrinsic scale $w$ of the Kelvinon. In this limit, there is only one scale $|C|$ left, which justifies the K41 dimensional analysis.

Below, we see that even in this region (lower part of the inertial range), logarithmic terms are modifying the K41 scaling laws.

With these restrictions, we have an estimate:
\begin{eqnarray}\label{Zans}
  &&Z = \left(\frac{\eps V(C) }{\Sigma}\right)^{\ot};\\
   &&\Sigma =   \oh \pi n^2(1 + \sigma_n \cos\lambda)^2 \int_{C} | d \vec r| \tilde S_{t t};\\
   \label{Hren}
   && H_R = Z^2 \int_{|r-C| > R} \frac{\tilde v^2}{2} +\nonumber\\
   && 2\pi Z^2  n^2 (1 + \sigma_n \cos\lambda)^2 \oint_C \abs{d \vec r}\left(\gamma +\log \frac{Z R^2\tilde S_{t t}}{4 \nu}\right);
\end{eqnarray}

The spurious parameter $R$, which formally enters this Hamiltonian, drops from it in the limit $R\ra 0$. The singular terms in the integral over external region $|\vec r - \vec C| > R$ exactly cancel the logarithmic term $\log R$, leaving a finite remainder.

This $R-$independence can be proven by differentiating the reduced Hamiltonian by $\log R$, which selects the lower limit of the radial integration in the $\int_{|r-C| > R} \tilde v^2$. We get velocity square at the surface of the tube, where it is dominated by the singular vortex with circulation $2 \pi n(1+\sigma_n\cos\lambda)$. Adding both terms, we obtain zero in the limit of $R \ra 0$.
\begin{eqnarray}
    &&R \partial_R \left( 4\pi n^2\left(1 + \sigma_n \cos\lambda\right)^2 |C| \log R  +\int_{|r-C| > R} d ^3 \vec r \frac{\tilde v^2}{2}\right) \ra \nonumber\\
    && n^2\left(1 + \sigma_n \cos\lambda\right)^2 \left(4\pi|C|-\oint_C |d \vec r| \int_0^{ 2 \pi} d \theta \frac{2 R^2 }{R^2} \right) =0
\end{eqnarray}

However, in that limit, there is a so-called dimensional transmutation.

The constant term arising after the cancellation of logarithmic terms $\log R$  has a logarithmic dimension. 
The solution for reduced velocity $\tilde v$ has parameters of the dimension of length, and one such parameter takes care of this logarithmic dimension.

This parameter compensates for the logarithmic dimension of the last term $\propto\oint_C \abs{d \vec r}\log \tilde c$ in the reduced Hamiltonian.

Naturally, we would expect the local radius $r_0$ of curvature of the loop to take this role so that the logarithm of strain $\log \tilde c$ would become $\log (\tilde c r_0^2)$ plus dimensionless terms.

Our large parameter is the Reynolds number $ \textbf{Re} = \frac{Z}{\nu}$.

The parameter $Z$ depends on the size $|C|$ of the Kelvinon loop and can be estimated as
\begin{eqnarray}\label{Zeq}
    &&Z \sim (1 + \sigma_n \cos\lambda)^{-\tt}\left(\frac{\eps |C|^3}{|C| |C|^{-2}}\right)^{\ot} \nonumber\\
    &&\sim \eps^{\ot} |C|^{\ft}(1 + \sigma_n \cos\lambda)^{-\tt}
\end{eqnarray}

Here we implicitly assumed that $C$ is the only scale of the Kelvinon solution. There could be another scale $w$ describing the width of the vorticity layer around the discontinuity surface, like in our example in \ref{ExampleClebsch}. In that case, we assume that the minimization of the Euler part of the Hamiltonian relates this width to the size of the loop $C$.

This assumption would not be valid at very large loops, where the dominant surface is expected to be a minimal surface for this loop. In that case, the width may stay finite in the limit of large size $|C|$, which would modify the scaling laws.
\subsection{Asymptotic freedom}

Let us find the asymptotic solution to the minimization problem at large Reynolds number $\frac{Z}{\nu}$.

We are going to bypass the hard problem of finding the spatial shape of the \CL{} field from the passive convection with nonlinear velocity \eqref{Vparam}
\begin{eqnarray}
    && \fbyf{\tilde H_{Euler}}{\phi} = \tilde v \cdot \vec \nabla S_3 =0;\\
    \label{S3eq}
    && \fbyf{\tilde H_{Euler}}{S_3} = -\tilde v \cdot \vec \nabla \phi =0;
\end{eqnarray}
which holds in the region $|\vec r - C| \ge R$ where Euler equation applies.
That includes the matching region $|\vec r - C|= R$ where the Burgers solution also holds.
Here 
\begin{eqnarray}
    \tilde H_{Euler} =  \int_{|\vec r - C| > R} \frac{\tilde v^2}{2}
\end{eqnarray}
The solution to this problem is needed for the subleading terms in the asymptotic expansion in inverse powers of the logarithm of Reynolds number.

The relevant variable in the leading order of this expansion is the constant boundary value $S_3(C) = \cos\lambda$, which has to be determined by the minimization of the Hamiltonian.
Variation of the Euler part $\tilde H_{Euler}$ yields zero by \eqref{S3eq}
\begin{eqnarray}
    \pbyp{\tilde H_{Euler}}{S_3(C)} = \oint_C |d \vec r| \fbyf{\tilde H_{Euler}}{S_3(\vec r)} =0
\end{eqnarray}

We are left with derivatives of $Z$ and derivatives of the anomalous term
\begin{eqnarray}
   && \pbyp{H_R}{S_3(C)} \ra 2 H_R \pbyp{\log Z}{S_3(C)} + \nonumber\\
   &&4\pi Z^2  n^2 \sigma_n(1+ \sigma_n \cos \lambda) |C|\log \frac{Z}{\nu};
\end{eqnarray}
Using
\begin{eqnarray}
    &&\pbyp{\log Z}{S_3(C)} =  -\frac{2}{3} \pbyp{\log(1 + \sigma_n S_3(C))}{S_3(C)} = \nonumber\\
    && \frac{-2\sigma_n}{3(1 + \sigma_n \cos\lambda)};\\
    && H_R \ra Z^2\tilde H_{Euler}  + \nonumber\\
    &&2\pi Z^2  n^2 (1 + \sigma_n \cos\lambda)^2 |C|\log \frac{Z }{\nu}
\end{eqnarray}
we obtain (dropping $O(1)$ terms added to $\log \frac{Z}{\nu}$)
\begin{eqnarray}
     \frac{\tilde H_{Euler}}{1 + \sigma_n \cos\lambda} -  \pi n^2 (1 + \sigma_n \cos\lambda) |C|\log \frac{Z}{\nu} =0
\end{eqnarray}

Finally
\begin{eqnarray}
    && \xi \equiv\frac{1 + \sigma_n \cos\lambda}{\sqrt{ h}} = \frac{2}{\sqrt{3\log \frac{Z}{\nu}}};\\
    && h = \frac{3\tilde H_{Euler}}{4 \pi n^2 |C| }.
\end{eqnarray}

This $\xi$ is our running coupling constant. By dimensional counting $\tilde H_{Euler} \propto |C|$, so that $h = \textit{const}$.
Our running coupling constant decreases as an inverse power of the logarithm of the Reynolds number.

Combining this relation with the relation \eqref{Zeq} we find two implicit equations,
looking like those in the renormalization group.
\begin{eqnarray}\label{RGEQ}
   &&|C| = \textit{const } Z^{\tq}\left(\log Z\right)^{\oq};\\
   \label{xiCeq}
   && \log |C| = \textit{const} + \frac{1 }{\xi^2} - \frac{\log \xi}{2};\\
   \label{RG}
   && \pbyp{\xi}{\log |C|} = \beta(\xi);\\
   && \beta(\xi) = -\frac{\xi^3}{2 +   \xi^2};
\end{eqnarray}

\subsection{Random velocity background}
Up to this point, our theory was purely mathematical: we studied a singular flow with a certain topology and exactly computed anomalous dissipation, Hamiltonian, and helicity in the limit of vanishing viscosity.

We have to make some physical assumptions to explain spontaneous stochastization and compute the probability distribution of the circulation.

These assumptions have little or no meaning for a mathematician, so we suggest taking the final formulas \eqref{Zsol} as a hypothesis to be justified (or disproved) much later.

Our Kelvinon does not live in isolation; there are other Kelvinons in the fluid (or some other vorticity structures). We assume these other Kelvinons to be far away. 

In the leading approximation of our ideal gas of the vortex loops, there are no interactions between these Kelvinons.

However, in the next approximation, we have to account for the small background velocity induced by Kelvinons onto each other.

The total velocity field equals the net \BS{} integral plus a potential term
\begin{eqnarray}
    &&\vec v = \vec \nabla \Phi - \vec \nabla \times \vec \Psi;\\
    && \Psi(\vec r)  =\sum_k \int_{V_k} d^3 r' \frac{\vec \omega(\vec r')}{ 4 \pi | \vec r - \vec r'|}
\end{eqnarray}

In the vicinity of our loop, the contributions to the velocity field from other Kelvinons decrease as a dipole.
\begin{eqnarray}
    \frac{\int_{V_k}d^3 r' \vec \omega(\vec r') \times (\vec r'- \vec C_k)}{4\pi|\vec r-\vec C_{k}|^3}
\end{eqnarray}
where $\vec C_{k}$ is a center of the Kelvinon $k$.
We assume net zero vorticity of each Kelvinon: the contribution from the vortex core \eqref{omegaLoop} to the volume integral of vorticity equals $\oint_C d \vec r =0$.

This $1/r^3$ decay will be compensated by the small but finite density of these Kelvinons in physical space $R_3$. 
As a result, we have the "night sky paradox," the large number of small contributions to the velocity from remaining Kelvinons.

The derivatives of this velocity (the induced strain) are smaller by one more factor of $r$, so we can neglect them at a small density of Kelvinons (i.e., a large mean distance between them).

As long as these Kelvinons are far apart (i.e., their density is small), the total Euler Hamiltonian does not depend on locations or $O(3)$ rotations of these loops.

Thus, these locations and rotations become zero modes, -- arbitrary parameters which do not change the energy.

In principle, these parameters are specified by initial conditions of the \NS{} equations. However, the stationary solution (fixed point) we are studying is degenerate for these parameters (in the small density approximation).

So, with a little stretch of the Central Limit Theorem, we arrive at the random Gaussian background velocity in the turbulent flow, as observed in numerical experiments ( see \cite{S19, S21} and references therein). 

At frozen locations of the other Kelvinons, this background velocity is just a constant vector. 
Later, we average over the Gaussian distribution of this vector.

What is the effect of this small constant shift of the velocity of the Kelvinon?

In the linear approximation, the constant shift does not affect the dissipation, as it cancels in the strain and circulation.

However, our Euler equations are nonlinear, especially with these singular lines and surfaces. There are many sources of change in higher order, in particular, the shift of the Neumann boundary condition for $\vec v \cdot \vec n = Z \tilde v \cdot \vec n$ by the normal component $\vec v_0 \cdot \vec n$ of this background velocity to the surface at some point.

This estimates correction to our dimensionless variables as $\delta S \sim |C| \delta \tilde v \sim \frac{|C|^{\ft}}{Z} \sim ( 1 + \sigma_n \cos\lambda)^{\tt}$. 

At small constant background velocity $\vec v_0$, we expect some second-order local contribution to the free parameter $S_3(C)$.

This contribution should preserve inequality  $-1 < S_3 < 1$, for which purpose it must be a positive quadratic form times $-S_3(C)$
\begin{subequations}\label{Zsol}
\begin{eqnarray}
    &&\delta S_3(C) = -S_3(C)\vec v_0 \cdot \hat q \cdot \vec v_0;\\
    && \hat q \VEV{\vec v_0^2}\propto ( 1 + \sigma_n \cos\lambda)^{\ft};\\
    && \Gamma_\alpha = 2 \pi m Z   \left( 1+\sigma_m \cos\lambda -\sigma_m \cos\lambda\vec v_0 \cdot \hat q \cdot \vec v_0\right);\\
    && \Gamma_\beta = 2 \pi n Z \left( 1 + \sigma_n \cos\lambda  - \sigma_n\cos\lambda \vec v_0 \cdot \hat q \cdot \vec v_0\right)
\end{eqnarray}
The most important effect of these terms is the fluctuations of velocity circulation in the Wilson loop
\begin{eqnarray}
    &&\Psi[C, \gamma] = \nonumber\\
    &&\exp{\i \gamma \tau}\VEV{\exp{-2 \pi \i |m| Z \cos\lambda\gamma \vec v_0 \cdot \hat q \cdot \vec v_0}}_{v_0} ;\\
    && \tau = 2 \pi m Z (1+\sigma_m\cos \lambda);\\
   && \vec v_0 \sim \mathcal N\left(0,\hat \sigma\right);
\end{eqnarray}
\end{subequations}
Computing the  Gaussian integral, we find the Wilson loop:
\begin{eqnarray}\label{WilsonLoop} 
    &&\Psi[C, \gamma] = \frac{\exp{\i \gamma \tau}}{\sqrt{\det {\left(1 - \i \gamma \hat \Lambda\right)}}};\\
    && \hat \Lambda = -4 \pi |m| Z \cos\lambda \sqrt{\hat \sigma} \hat q \sqrt{\hat \sigma}
\end{eqnarray}
This algebraic expression for the Wilson loop looks ridiculously simple.

The complexity hides in the dependence of $Z,\hat q, \lambda$ on the size and shape of the loop, requiring minimization of the Hamiltonian or solving the WKB limit of the loop equations \cite{M21c, M21d}.

Using the asymptotic freedom laws in the previous section, we find two different asymptotic depending on the sign of $m n$
\begin{subequations}
\label{TLxi}
\begin{eqnarray}
\label{Txi}
    &&\tau  \propto |C|^{\ft} \xi^{-\tt} \textit{ if m n <0 };\\
    &&\tau  \propto |C|^{\ft} \xi^{+\ot} \textit{ if m n >0 };\\
\label{Lxi}
    &&  \Lambda \propto |C|^{\ft} \xi^{\tt}
\end{eqnarray}
\end{subequations}

The tip of the distribution and its tails are described by different Kelvinons, which would look in DNS as a regime change or switch of parameters.

So we have three different scales here
\begin{eqnarray}\label{taulambdalog}
   &&\tau   \sim |C|^{\ft} (\log |C|)^{(-\nicefrac{1}{6} \textit{ if m n >0 else }\ot)}; \\
&&\Lambda   \sim |C|^{\ft} (\log |C|)^{-\ot}
\end{eqnarray}

\subsection{PDF tails}

  The probability distribution is given by the Gaussian integral over the quadratic surface
\begin{eqnarray}\label{PDF}
   && P[C,\Gamma] = \nonumber\\
 &&\int \frac{d^3 \vec u}{(2 \pi )^{\frac{3}{2}}} \exp{ - \frac{\vec u^2}{ 2 }} \delta\left(\Gamma - \tau - \oh \vec u \cdot \hat \Lambda \cdot \vec u \right);
\end{eqnarray}

Let us diagonalize $\Lambda$ and assume that all eigenvalues are positive
\begin{eqnarray}
     && \Lambda = \diag{a,b,c}; \\
   &&0 < a < b < c; 
\end{eqnarray}
  
At large positive $\Gamma$, the integral  in \eqref{PDF} factorizes and  tends to
\begin{eqnarray}\label{PDFPos}
   && P[C,\Gamma\ra +\infty] \ra \frac{ \sqrt{2c}\exp{\frac{-\Gamma +\tau}{c}} }{\sqrt{ \pi   (c-a) (c-b)\Gamma}};
\end{eqnarray}

By the derivation, the above formulas apply only for $\sign{\Gamma_\alpha} = \sigma_m$, as both factors $Z, (1+ \sigma_m\cos\lambda)$ are positive.

To obtain the negative $\Gamma_\alpha$, we have to consider the opposite helicity Kelvinon, with the opposite sign of $m$, changing the sign of $\Gamma_\alpha$ but keeping the sign of $\Gamma_\beta$.

The higher terms of the expansion of the PDF at large $\Gamma$ are computed in 

This helicity reflection corresponds to the parity transformation. 
The parity transformation changes only the sign of  $\Gamma_\alpha $, whereas the time-reversion changes the sign of both circulations.
The strain  $\tilde S_{t t}$ changes the sign at time reversion but not at the parity transformation.

Thus, the opposite helicity Kelvinon would still have positive strain projection $\tilde S_{t t}$ in our anomalous Hamiltonian.

The solution for $\lambda$ will be different for the opposite helicity Kelvinon; we get the same exponential law with $-\Gamma$ instead of $\Gamma$ but with different coefficients $ \tau', a',b',c'$.
\begin{eqnarray}\label{PDFNeg}
    P[C,\Gamma\ra -\infty] \ra \frac{ \sqrt{ 2c'}\exp{\frac{\Gamma +\tau'}{c'}} }{\sqrt{ \pi   (c'-a') (c'-b')|\Gamma|}};
\end{eqnarray}

Specifically, with a reflected sign of $m$, the circulation $\Gamma_\alpha$ will change not only the sign but also its magnitude
\begin{eqnarray}
    &&\Gamma_\alpha = - 2 \pi m\left( 1 - \sigma_m \cos\lambda + \sigma_m\cos \lambda \vec v_0 \cdot \hat q \cdot \vec v_0\right);\\
    &&  \tau = -2 \pi m Z (1-\sigma_m\cos \lambda);\\
    && \hat \Lambda = -4 \pi|m| Z \cos\lambda \sqrt{\hat \sigma} \hat q \sqrt{\hat \sigma}
\end{eqnarray}

\section{Comparison with Numerical experiments}

\subsection{Matching the Kelvinon solution to the DNS}\label {KelvDNS}

Let us compare these relations to DNS\cite{S19}, courtesy of Kartik Iyer.

This raw data correspond to $8192^3$  DNS with $R_\lambda = 1300$. 
There are the first $8$ moments of circulation for various Reynolds numbers.

The probability distribution data for velocity circulation corresponds to $r/\eta \approx 140$. 
The circulation is measured in viscosity $\nu$ (here, we restored the correct normalization of circulation, which was lost in \cite{S19} where circulation was measured in some arbitrary units).

This data is the most extensive, with circulation $ -1904.72 < \Gamma < 2048.47 $.
We discarded the extreme values, as the statistics were too small there.

The small absolute values of $\Gamma$ were also discarded, as our theory only applies to large $|\Gamma|
$.
We selected the interval $610 < |\Gamma| < 1426$ corresponding to probabilities $10^{-7} > P > 10^{-14}$ and we fitted

\begin{eqnarray}\label{AsympLaw}
&&\log \left(P(C,\Gamma) \sqrt{|\Gamma|}\right) \approx A_\pm + B_\pm |\Gamma|;\\
&& B_\pm =( -1/c, -1/c');
\end{eqnarray}
\pct{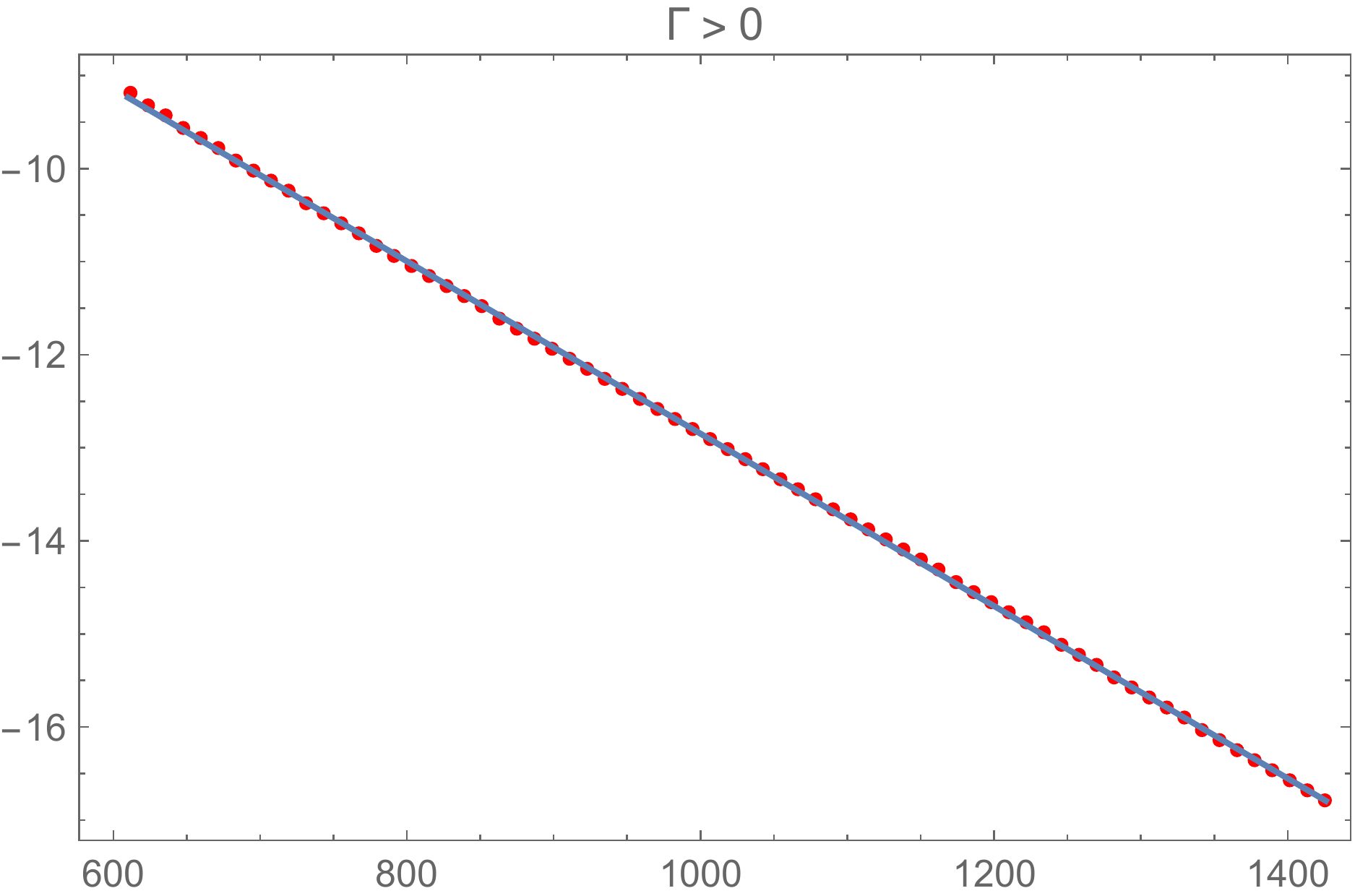}{Fitting \eqref{AsympLaw} for positive $\Gamma$ }

The positive tail has the following fit in the inertial regime (see Fig.~\ref{fig::GammaPos}).
\begin{equation}\label{PosFit}
 \begin{array}{l|llll}
 \text{} & \text{Estimate} & \text{Standard Error} & \text{t-Statistic} & \text{P-Value} \\
\hline
 A_+ & -3.58916 & 0.0105971 & -338.693 & 4.523\times 10^{-110} \\
 B_+ & -0.00925906 & 0.0000101328 & -913.774 & 6.073\times 10^{-139} \\
\end{array}
\end{equation}

\pct{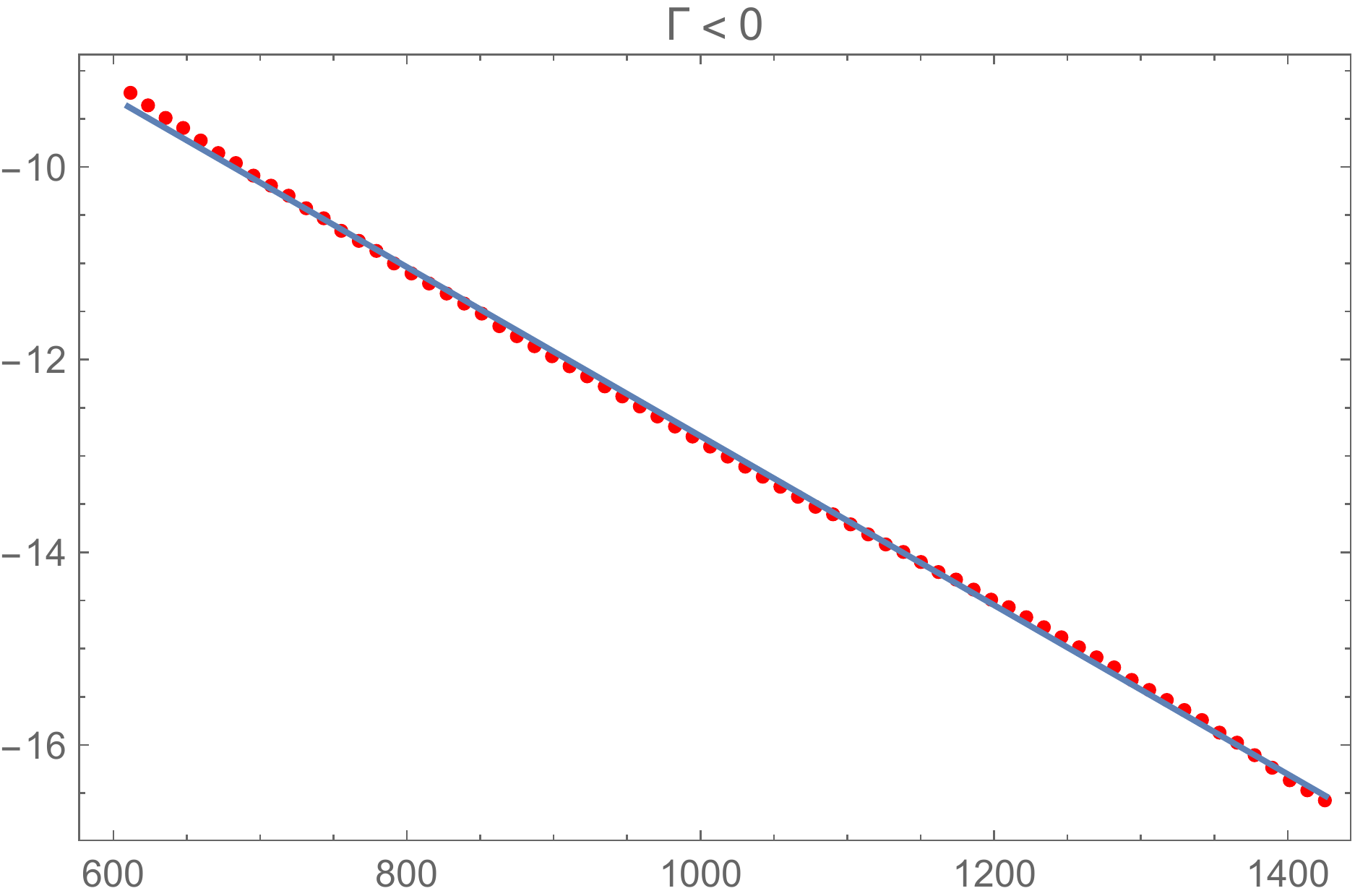}{Fitting \eqref{AsympLaw} for negative $\Gamma$ }

The negative tail  has the following fit in the inertial regime (see Fig.~\ref{fig::GammaNeg})
\begin{equation}\label{NegFit}
 \begin{array}{l|llll}
 \text{} & \text{Estimate} & \text{Standard Error} & \text{t-Statistic} & \text{P-Value} \\
\hline
 A_- & -4.01871 & 0.029290 & -137.201 & 8.071\times 10^{-84} \\
 B_- & -0.00878 & 0.000028 & -313.317 & 8.3194\times 10^{-108} \\
\end{array}
\end{equation}

There is about $5\%$ breaking of time-reversal symmetry $\Gamma \Ra -\Gamma$, which is well beyond the statistical errors $\sim 10^{-3}$.

The ratio 
$\beta = \frac{B_-}{B_+} =0.947737$.

Remarkably, one simple algebraic formula \eqref{WilsonLoop} for the Wilson loop describes all the probability tails for velocity circulation.

It all follows from the Kelvinon solution of the Euler equation with a nontrivial winding number in an approximation of small relative fluctuations of the background velocity in a Kelvinon.

This approximation perfectly describes the tails of circulation probability distribution in DNS of strong isotropic turbulence.

\subsection{Multiple Kelvinon scales and log-multifractal}\label{multifrac}

The numerical experiments \cite{S19,S21} show violation of the K41 scaling law, which for the circulation would read
\begin{eqnarray}\label{K41}
   \textit{K41 scaling:} &&\Gamma_C \sim \eps^{\ot}|C|^{\ft};
\end{eqnarray}

This law was derived from dimensional analysis, assuming $\eps$ to be the only parameter related to time scales.

This analysis does not apply to our theory, as there are logarithmic terms in our Hamiltonian \eqref{Hren}, breaking scale invariance.
Moreover, many Kelvinons are contributing to the Wilson loop, leading to multiple scales.

To be more specific, in the limit of large $\log (Z/\nu)$, for each Kelvinon, we have three relevant scales \eqref{TLxi}, with different powers of a logarithm.

Each Kelvinon contributes to Fourier integral only for one sign of the circulation, as the Wilson loop with positive quadratic form $\hat q$ has singularities in only one semi-plane depending on the sign of $ m n$.
\begin{eqnarray}
    &&\Psi_{m n}[C, \gamma] = \nonumber\\
    &&\frac{\exp{ \i \gamma \tau_{m n}}}{\sqrt{(1 - \i \sigma\gamma a_{m n})(1 - \i\sigma  \gamma b_{m n})(1 - \i \sigma  \gamma c_{m n})} };\\
    && \sigma = \sign{m};\\
    && \Psi[C, \gamma] = \sum \kappa_{m n} \Psi_{m n}[C, \gamma];\\
    && \sum \kappa_{m n} =1;
\end{eqnarray}

We must add contributions from all four signs of $m, n$ to provide both distribution branches. 
\begin{eqnarray}
    &&P_{m n}[C, \Gamma] = \theta\left(\Gamma - \tau_{m n}\right)\nonumber\\
    &&\int_{-\infty}^\infty \Psi_{m n}[C, \gamma]\frac{e^{- \i \gamma\Gamma}}{2 \pi};\\
    &&P_{-m n}[C, \Gamma] = \theta\left(-\Gamma +\tau_{-m n}\right)\nonumber\\
    &&\int_{-\infty}^\infty \Psi_{-m n}[C, \gamma]\frac{e^{- \i \gamma\Gamma}}{2 \pi};
\end{eqnarray}

The moments of circulation, defined as expansion coefficients of the Wilson loop in powers of $\i \gamma$, will become a superposition of powers of these scales, providing complex behavior, difficult to predict.

This superposition is a qualitative explanation of "multifractal," observed in the DNS of\cite{S19}. 

\begin{eqnarray}
    &&\VEV{\Gamma^k} = \sum \kappa_{m n} \VEV{\Gamma^k}_{m n};\\
    && \VEV{\Gamma^k}_{m n} = \left(\pp{x}\right)^k\Psi_{m n}[C, -\i x]_{x =0}
\end{eqnarray}

In particular, the first three moments for each Kelvinon  are (we skipped indexes $a_{m n} \equiv a$, etc. )
\begin{eqnarray}
    &&\VEV{\Gamma}_{m n} = \tau +\frac{1}{2} (a+b+c);\\
    && \VEV{\Gamma^2}_{m n} = \tau ^2+\tau  (a+b+c)+\nonumber\\
    &&\frac{1}{4} \left(3 a^2+2 a (b+c)+3 b^2+2 b c+3 c^2\right);\\
    &&\VEV{\Gamma^3}_{m n} = \tau ^3+\frac{3}{2} \tau ^2 (a+b+c)+\nonumber\\
    &&\tau  \left(\frac{9}{4} (a+b+c)^2-3 (a (b+c)+b c)\right)+\nonumber\\
    &&\frac{15}{8} (a+b+c)^3-\frac{9}{2} (a (b+c)+b c) (a+b+c)+3 a b c;\\
    && \dots
\end{eqnarray}
There is a sum rule related to the third moment, which scales as K41, without anomalies \cite{S19} 
\begin{eqnarray}
    \sum \kappa_{m n} \VEV{\Gamma^3}_{m n} =  -A |C|^4;
\end{eqnarray}

In our theory, the moments of the circulation distribution are the wrong kind of observables, with all three scales \eqref{TLxi} mixed up.

The tails of the probability distribution are better observables.

In general, the three regions of $\Gamma$ overlap so that at the interval $ \tau_{m n} < \Gamma < \tau_{-m n} $ both Kelvinons contribute to the PDF, and at the two remaining tails $ \Gamma < \tau_{m n} $ and $ \Gamma > \tau_{-m n}$ only one of them does. 

The appropriate quantity, with the intermediate range scales eliminated, is the tail CDF
\begin{eqnarray}
    W[C, \Gamma] = \int_{\Gamma}^\infty P[C, \Gamma'] d \Gamma'
\end{eqnarray}
One could invert this relation and consider $\Gamma$ as a function of the probability $w$
\begin{subequations}
    \begin{eqnarray}
    &&\Gamma= F[C,w];\\
    && W[C, \Gamma] = w;\\
    && -w \pbyp{\Gamma}{w} =\frac{w}{P[C, F[C,w]]} \equiv \Lambda[C,w]
\end{eqnarray}
\end{subequations}

In other words, we need to consider the probability density $P[C, \Gamma]$ as a function of probability $w$ to have events with $\Gamma' > \Gamma$.
With the exponential tail of $P[C,\Gamma]$, this $\Lambda$ will be proportional to $c$.

The multifractal scaling laws would correspond to the logarithmic derivative of this $\Lambda[C,w] $ independent of $C$
\begin{eqnarray}
 && \Lambda[C,w] = A(w) |C|^{\eta(w)};\\
   && \pbyp{\log \Lambda[C,w]}{\log |C|} =\eta(w).
\end{eqnarray}

The scaling index $\eta(w)$ is similar to the moments' index $\zeta(n)$, which is assumed to be scale independent in the multifractal model.
It can be related to $\zeta(n)$ at large $n, -\log w$ by using the saddle point method in the integral representation for the moments.

In our theory, this logarithmic derivative does not depend on $w$ for small enough $w(1-w)$, but it does depend on $\log |C|$
\begin{subequations}\label{TailScale}
\begin{eqnarray}
&& \Lambda[C,w] = B(w) |C|^{\ft} \xi^{\tt};\\
    &&\pbyp{\log \Lambda[C,w]}{\log |C|} = \frac{2 }{3} \frac{4 +  \xi^2}{2 + \xi^2}\ra \frac{4 }{3}  - \frac{1}{3\log|C|};
\end{eqnarray}   
\end{subequations}
This scaling law applies to the slopes $c, c'$ in the asymptotic slope of the exponential decay \eqref{AsympLaw}.

In the inner region $\tau_{m n} < \Gamma < \tau_{-m n}$, there is a hot mess of four (or more) Kelvinons, each with its scale.
Three distinct scales have the indexes of \eqref{TLxi}.
The corresponding low moments can imitate multifractal.
It is hard to make quantitative predictions in this inner region until the Kelvinon flow $\vec S(\vec r)$ is computed analytically or numerically.

In the whole region $0 < w <1$, the above logarithmic derivative will depend on $w$ and $\log |C|$.
\begin{eqnarray}
    \eta(w, \log |C|) = \pbyp{\log \Lambda[C,w]}{\log |C|} 
\end{eqnarray}
 This new index is an interesting observable to measure in DNS.
 
 In our theory, it is a function of both variables for $w \sim 0.5$, but near the right tail $ w \ra 0$ and the left tail $w \ra 1$, it becomes independent of $ w$ and approaches the K41 value $\ft$ with negative corrections $O(1/\log |C|)$.

Finally, let us stress that these multifractal laws with logarithmic indexes were derived under the assumption of scale invariance of the energy $\mathcal E[C]$ pumped into the region occupied by the Kelvinon(s). The volume of this region was assumed to scale as $|C|^3$, and the Kelvinon \CL{} field $\vec S(\vec r)$ was assumed to depend only on $\frac{\vec r}{|C|}$.

The size of the loop $C$ is much larger than the viscous scale $h$ (the thickness of the Burgers vortex and the width of the vortex layer in discontinuity surface). However, there may be another, larger intrinsic scale of the Kelvinon, where the one-scale regime changes.

In the case of a large and smooth, almost flat loop with normal deviations $\delta \vec r_\perp \sim L \ll |C|$ there are at least two different scales, and the solution for $\vec S$ would depend on a smaller scale $L$. The volume occupied by a Kelvinon would then scale as $V[C] \sim A_{min}[C]L$ where $A_{min}[C]$ is the minimal surface bounded by $C$ (the plane in this limit).

With this geometry, the above multifractal-log laws would not apply. The replacement is studied in the next Sections using an alternative approach to the turbulent statistics.

\section{The Loop equation and its turbulent limit}\label{SecLoopEq}

We reproduce the lectures in Cargese Summer School and Chernogolovka Summer School in '93.

Many questions raised then are now answered (see previous sections). The Area law was justified (with some restrictions ) theoretically and observed in the DNS 25 years later.

We present these lectures almost the same as they were delivered back in '93, with minimal corrections and occasional comments relating to future work. This theory's natural chronological evolution may be easier to digest than its advanced state.

From the point of view of the theory of vortex sheets and vortex lines presented in the previous Sections, this is an alternative approach that allows one to find the $C-$shape dependence of the Wilson loop and the circulation PDF.

There is much more to this alternative approach. It establishes \textbf{exact} relation between the forced \NS{} equation and certain quantum system with nonlocal dynamical variables ( loops in 3D space).

No assumptions about the small density of the vortex structures or large Reynolds number are involved in this quantum correspondence.

The solution for the Wilson loop (wave function of this quantum system) can be exactly related to a certain one-dimensional nonlinear PDE with initial stochastic data. This representation can be used for analytical and numerical studies.

With advances in quantum computers, this exact quantum correspondence opens the way to large-scale turbulence simulations.

We derive and study this correspondence in this Section and the next ones. The turbulent limit of the \NS{} dynamics corresponds to the WKB limit of this quantum loop system, and it can be studied analytically for large smooth loops.

\subsection{Introduction}

Incompressible fluid dynamics underlies the vast majority of natural phenomena.
It is described by the famous Navier-Stokes equation
which is nonlinear and, therefore, hard to solve.
This nonlinearity makes life more interesting, as it leads to
turbulence. Solving this equation with appropriate initial and boundary
conditions, we expect to obtain the chaotic behavior of the velocity field.

The simplest boundary conditions correspond to infinite space with
vanishing velocity at infinity. We are looking for the translation
invariant probability distribution for the velocity field with an infinite range of
wavelengths. To compensate for the energy dissipation, we add the
usual random force to the \NS{}  equations, with the long wavelength support
corresponding to large-scale energy pumping.

One may attempt to describe this probability distribution by the
Hopf generating functional (the angular bracket denotes time  averaging or
ensemble averaging over realizations of the random forces)
\begin{equation}
Z[J] = \VEV{\exp{\int d^3 r
J_{\alpha}(r)v_{\alpha}(r)}
} \label{eq2}
\end{equation}
which is known to satisfy the linear functional differential equation
\begin{equation}
\dot{Z} = H\left[J,\frac{\delta}{\delta J} \right] Z
\label{eq3}
\end{equation}
similar to the Schrödinger equation for Quantum Field Theory, and
equally hard to solve. Nobody managed to go beyond the Taylor
expansion in source $ J $, which corresponds to the obvious chain of equations
for the equal time correlation functions of the velocity field at various points in space. One could obtain the same equations
directly from Navier-Stokes equations, so the Hopf equation looks
useless.

In this work\footnote{see also \cite{M93} where this approach was initiated and
\cite{TSVS} where its relation with the generalized Hamiltonian dynamics was established.} we argue that one could
significantly simplify the Hopf functional without losing information about
correlation functions.
This simplified functional depends upon the set of 3 periodic
functions of one variable
\begin{equation}
C : r_{\alpha} = C_{\alpha}(\theta)\\;\; 0< \theta< 2\pi
\end{equation}
which set describes the closed loop in coordinate space. The
correlation functions reduce to certain functional derivatives of our loop
functional with respect to $ C(\theta)$ at vanishing loop $ C \rightarrow 0 $.

The properties of the loop functional at large loop $ C $ also have
physical significance. Like the Wilson loops in Gauge Theory, they
describe the statistics of large-scale structures of the vorticity field, which is
analogous to the gauge field strength.

\subsection{The Loop Calculus}\label{LoopCalc}

We suggest using in turbulence the following version of the Hopf functional
\begin{equation}
\Psi \left[C \right] = \VEV{\exp{
    \frac{\i }{\nu}  \oint d C_{\alpha}(\theta)
v_{\alpha}\left(C(\theta)\right) }} \label{eq4}
\end{equation}
which we call the loop functional or the loop field.
It is implied that all angular variable $\theta$ run from $ 0 $ to $ 2\pi$ and
that all the functions of this variable are $ 2\pi$
periodic.\footnote{This parametrization of the loop is a matter of
convention, as the loop functional, is parametric invariant.} The viscosity $
\nu $ was inserted in the denominator in exponential as the only parameter of
proper dimension. 

\textbf{As we shall see below, viscosity plays the role of the
Planck's constant in Quantum mechanics and the turbulence corresponds to the WKB
limit $ \nu \rightarrow 0 $}. It this not an accident that viscosity has the same dimension as Planck's constant!

The ratio of the circulation to viscosity is one definition of the Reynolds number (invariant to Galilean transformations). 

As for the imaginary unit $\i$, there are two reasons to insert it in the exponential. 

First, it makes the motion compact: the phase factor goes around
the unit circle in the evolution of the solution of the \NS{}  equations. So, at large  times one
may expect the ergodicity, with a well-defined average functional  bounded by $1$
by absolute value. \footnote{Otherwise, the loop average would not exists: as we show later (also confirmed by DNS), the PDF for circulation decreases only exponentially.}

Second, with this factor of $\i$, the irreversibility of
the problem is manifest. The time reversal corresponds  to the complex
conjugation of $\Psi$ so that the imaginary part of the  asymptotic value of
$\Psi$ at $t \ra \8$ measures the effects of dissipation.

The analogy with quantum mechanics shows already from a definition of a Wilson loop: this is a
sum over alternative histories of our flow of $\exp{\i \frac{S}{\nu}}$ with the circulation $\Gamma$ playing the role of the Action, and viscosity $\nu$ playing the role of the Planck's constant.

The alternative histories correspond to different realizations of the Gaussian random forces. These random forces drop from the equation for the fixed point in the turbulent limit, but they are implicitly present. The fixed point is degenerate, it is rather a fixed manifold, which we studied in the Kelvinon theory in the previous sections.

Now we have to reconstruct the dynamics of this quantum system using the forced \NS{} equation.

We shall use the field theory notations for the loop integrals,
\begin{equation}
\Psi \left[C \right] = \VEV{\exp{
    \frac{\i }{\nu}  \oint_C d r_{\alpha}v_{\alpha}
 }} \label{eq4'}
\end{equation}
This loop integral can be reduced to the surface
integral of the vorticity field
\begin{equation}
\omega_{\mu\nu} = \d_{\mu}v_{\nu}-\d_{\nu}v_{\mu}
\end{equation}
by the Stokes theorem
\begin{equation}
\Gamma_C[v] \equiv \oint_C dr_{\alpha}v_{\alpha}= \int_{S}  d
\sigma_{\mu\nu} \omega_{\mu\nu} \\;\; \partial S = C
\end{equation}

This integral is the well-known velocity circulation, which measures the net
strength of the vortex lines passing through the loop $ C $. Would we fix
the initial loop $ C $ and let it move with the flow, the loop field would be
conserved by the Euler equation so that only the viscosity effects would be
responsible for its time evolution.

However, this is not what we are trying to
do. We take the Euler rather than Lagrange dynamics so that the loop is fixed
 in space; hence, $\Psi$ is time-dependent already in the Euler equations.
The difference between Euler and \NS{}  equations is the time irreversibility,
which leads to complex average $\Psi$ in \NS{}  dynamics.

It is implied that this field $\Psi\left[C\right]$ is invariant under
translations of the loop $ C(\theta) \rightarrow C(\theta)+ const $. 

The
asymptotic behavior at the large time  with proper random forcing reaches  certain
fixed point, governed by the translation- and scale invariant equations,  which
we derive in this paper.

The Wilson loop corresponds to the Hopf functional (\ref{eq2}) for the
following  imaginary singular source
\begin{equation}
J_{\alpha}(r) =   \frac{\i }{ \nu} \oint_C dr'_{\alpha} \delta^3
\left(r'-r \right) \label{eq8}
\end{equation}

The expansion in powers of $\Gamma_C$ can be written in the exponential using connected correlation functions of the powers
of circulation at equal times.
\begin{equation}
\Psi[C] = \exp{\sum_{n=2}^{\8}\frac{\i^{n}}{n!\,\nu^{n}}\,\VEV{\VEV{
\Gamma_C^n[v]}}}
\end{equation}
This expansion goes in powers of effective Reynolds number, so it diverges in
a turbulent regime. There, the opposite WKB approximation will be used.

Let us come back to the general case of the arbitrary Reynolds number. What
could be the use of such restricted Hopf functional? At first glance, it seems
that we lost most of the information described by the Hopf functional, as the
general Hopf source $J$ depends upon three variables $ x,y,z $, whereas the loop $C$
depends on only one parameter $ \theta $. Still, this information can be
recovered by taking the loops of the singular shape, such as two infinitesimal
loops $R_1, R_2 $,  connected by a couple of wires.

\pct{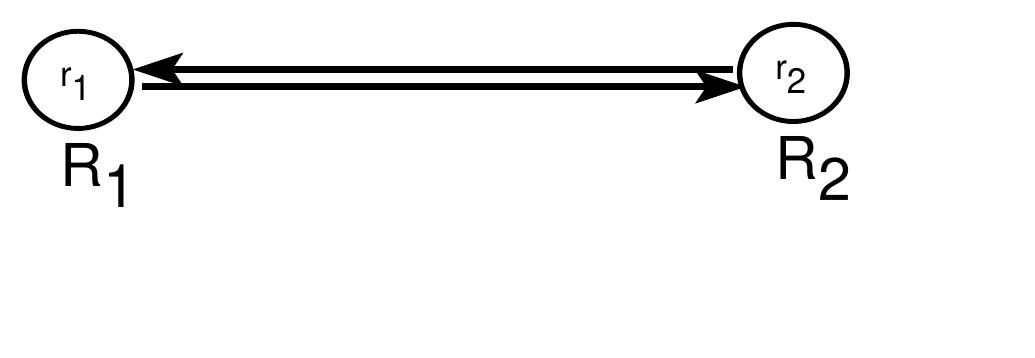}{Wires between two loops }

The loop field, in this case, reduces to
\begin{equation}
\Psi \left[C \right] \rightarrow \left \langle  \exp{
    \frac{\i}{ 2\nu} \Sigma_{\mu\nu}^{R_1}\omega_{\mu\nu}(r_1)
+\frac{\i}{ 2\nu}   \Sigma_{\mu\nu}^{R_2 } \omega_{\mu\nu}(r_2)
} \right \rangle
\end{equation}
where
\begin{equation}
\Sigma_{\mu\nu}^R = \oint_R d r_{\nu}r_{\mu}
\end{equation}
is the tensor area inside the loop $R$. Taking functional derivatives with
respect to the shape of $R_1$ and $R_2$ prior to shrinking them to  points, we
can bring down the product of vorticities at  $r_1$ and $r_2$. Namely, the
variations yield
\begin{equation}
\delta\Sigma_{\mu\nu}^R=  \oint _R\left(d r_{\nu}\delta r_{\mu}+ r_{\mu}d
\delta r_{\nu} \right) =  \oint_R \left( d r_{\nu}\delta r_{\mu} -d
r_{\mu}\delta r_{\nu} \right)
\end{equation}
where integration by parts was used in the second term.

One may introduce the area derivative $\fbyf{}{\sigma_{\mu\nu}(r)}$, which
brings down the vorticity at the given point $ r $ at the loop.
\begin{equation}
-\nu^2 \frac{\delta^2 \Psi \left[C \right]}
{\delta \sigma_{\mu\nu}(r_1)\delta \sigma_{\lambda \rho}(r_2)}
\ra \left \langle \omega_{\mu\nu}(r_1)
\omega_{\lambda \rho}(r_2) \right \rangle
\end{equation}

The careful definition of these area derivatives is of paramount importance
to us. The corresponding loop calculus was developed in\cite{Mig83} in the
context of the gauge theory. Here we rephrase
and further refine the definitions and relations established in that
work.

The basic element of the loop calculus is what we suggest calling the spike
derivative, namely the operator which adds the infinitesimal $
\Lambda $ shaped spike to the loop
\begin{equation}
D_{\alpha}(\theta,\epsilon) = \int_{\theta}^{\theta+2\epsilon}d \phi
\left(
1-\frac{\left|\theta +\epsilon - \phi\right|}{\epsilon }
\right)
    \frac{\delta}{\delta C_{\alpha}(\phi)}
\end{equation}
The finite spike operator
\begin{equation}
\Lambda(r,\theta,\epsilon) =
\exp{r_{\alpha}  D_{\alpha}(\theta,\epsilon) }
\end{equation}
adds the spike of the height $r$. This spike is the straight line from $
C(\theta) $ to $ C(\theta + \epsilon) + r$, followed by another
straight line from $ C(\theta+\epsilon)+r $ to $ C(\theta+2
\epsilon)$,

\pct{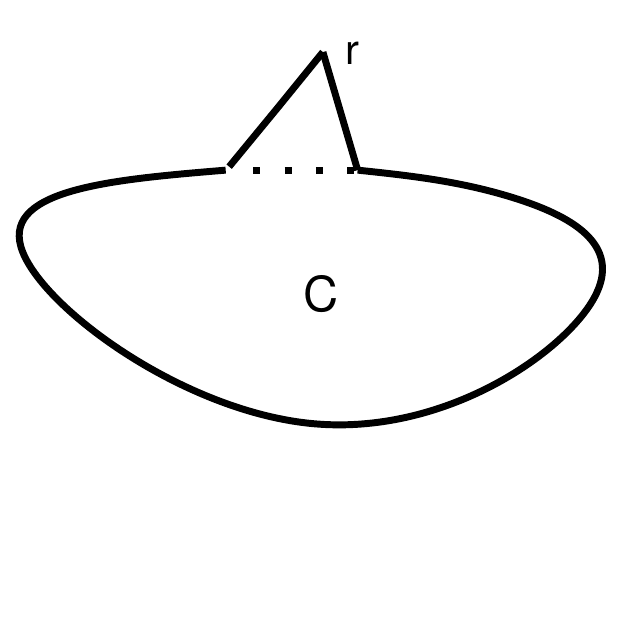}{$\Lambda$ variation of the loop.}
Note that the loop remains
closed, and the slopes remain finite. Only the second derivatives
diverge. The continuity and closure of the loop eliminate the
a potential part of velocity; as we shall see below,
this closure is necessary to obtain the loop equation.

In the limit $ \epsilon \rightarrow 0 $, these spikes are invisible, at
least for the smooth vorticity field, as one can see from the Stokes
theorem (the area inside the spike goes to zero as $ \epsilon $).
However, taking certain derivatives prior to the limit $ \epsilon
\rightarrow 0 $ we can obtain the finite contribution.

Let us consider the operator
\begin{equation}
\Pi \left(r,r',\theta ,\epsilon \right) =
\Lambda  \left(r, \theta,\frac{1}{2} \epsilon \right) \Lambda
\left(r',\theta,\epsilon \right)
\end{equation}
By construction, it inserts the smaller spike on top of a bigger one,
in such a way, a polygon appears (see Fig. \ref{fig::Fig3}).

\pct{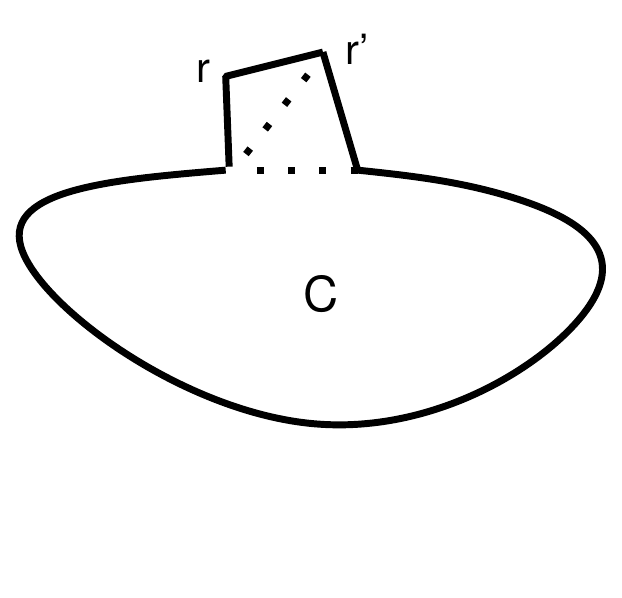}{$\Pi$ variation of the loop}
Taking the derivatives by the  vertices of
this polygon $ r, r' $ , setting $r=r'=0$ and
anti-symmetrizing, we find the tensor operator
\begin{equation}
\Omega_{\alpha\beta}(\theta,\epsilon) =
-\i \nu  D_{\alpha}\left(\theta,\frac{1}{2} \epsilon \right)
D_{\beta}\left(\theta,\epsilon \right) - \{\alpha \leftrightarrow\beta\}
\label{OM}
\end{equation}
which brings down the vorticity when applied to the loop field
\begin{equation}
\Omega_{\alpha\beta}(\theta,\epsilon) \Psi \left[C \right]
\stackrel{\epsilon \rightarrow 0}{\longrightarrow}
\omega_{\alpha\beta}\left(C(\theta)\right)\Psi \left[C \right] \label{eqom}
\end{equation}

The quick  way to check these formulas is to use formal functional
derivatives
\begin{equation}
\frac{\delta \Psi \left[C \right]}{\delta C_{\alpha}(\theta)} =
C'_{\beta}(\theta) \fbyf{\Psi \left[C
\right]}{\sigma_{\alp\bet}\left(C(\theta)\right)}
\end{equation}
Taking one more functional derivative, we find the term with
vorticity times the first derivative of the $ \delta $ function, coming from
the variation of $ C'(\theta) $
\begin{eqnarray}
&&\frac{\delta^2 \Psi [C ]}{\delta C_{\alp}(\theta) \delta
C_{\bet}(\theta')} = \del'(\theta-\theta')\fbyf{\Psi \left[C
\right]}{\sigma_{\alp\bet}\left(C(\theta)\right)} +\nonumber\\
&&
C'_{\gam}(\theta) C'_{\lam}(\theta')\frac{\delta^2\Psi \left[C
\right]}{\delta \sigma_{\alp\gam}\left(C(\theta)\right) \delta
\sigma_{\bet\lam}\left(C(\theta')\right)}
\end{eqnarray}
This term is the only one that survives the limit $ \epsilon
\rightarrow 0 $ in our relation (\ref{eqom}).

So, the area derivative can be defined from the antisymmetric tensor part
of  the second functional derivative as the coefficient in front of $
\delta'(\theta-\theta') $ .  Still, it has all the properties of the first
functional derivative, as it can also be defined from the above first
variation.

The advantage of dealing with spikes is the control over the limit $\eps
\ra 0$, which might be quite singular in applications.

So far, we have managed to insert the vorticity at the loop $ C $ by
variations of the loop field. Later we shall need the vorticity off
the loop, in an arbitrary point in space. This result can be achieved by the
following the sequence of the spike operators
\begin{equation}
\Lambda \left(r,\theta,\epsilon \right) \Pi
\left(r_1,r_2,\theta+\epsilon,\delta \right) \\;\; \delta \ll \epsilon
\end{equation}
This operator inserts the $ \Pi $ shaped little loop at the top of the
bigger spike. In other words, this little loop is translated by a
distance $r$ by the big spike.

Taking derivatives, we find the operator of finite translation of the
vorticity
\begin{equation}
\Lambda \left(r,\theta,\epsilon \right)
\Omega_{\alpha\beta}(\theta+ \epsilon ,\delta)
\end{equation}
and the corresponding infinitesimal translation operator
\begin{equation}
D_{\mu}(\theta,\epsilon)\Omega_{\alpha\beta}(\theta+ \epsilon ,\delta)
\end{equation}
which inserts $ \partial_{\mu} \omega_{\alpha \beta} \left( C(\theta)
\right) $ when applied to the loop field.

Coming back to the correlation function, we are going now to construct
the operator, which would insert two vorticities separated by a distance.
Let us note that the global $ \Lambda $ spike
\begin{equation}
\Lambda \left(r,0,\pi \right) = \oldexp
\left(
    r_{\alpha}\int_{0}^{2\pi}d
\phi  \left(1- \frac{ \left|\phi-\pi \right|}{\pi} \right)
\frac{\delta}{\delta C_{\alpha}(\phi)}\right)
\end{equation}
when applied to a  shrunk loop $ C(\phi) = 0 $ does nothing but
the backtracking from $0$ to $r$ (see Fig. \ref{fig::Fig4}).

\pct{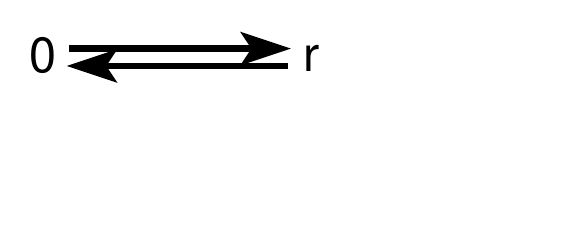}{Backtracking wires}
This backtracking means that the operator
\begin{equation}
\Omega_{\alpha\beta}(0 ,\delta)\Omega_{\lambda \rho}(\pi ,\delta)
\Lambda \left(r,0,\pi \right)
\end{equation}
when applied to the loop field for a shrunk loop yields the vorticity
correlation function
\begin{equation}
\Omega_{\alpha\beta}(0 ,\delta)\Omega_{\lambda \rho}(\pi ,\delta)
\Lambda \left(r,0,\pi \right) \Psi [0] = \left \langle \omega_{\alpha
\beta}(0) \omega_{\lambda \rho}(r) \right \rangle
\end{equation}

The higher correlation functions of vorticities could be constructed in a
similar fashion, using the spike operators. As for the velocity, one
should solve the Poisson equation
\begin{equation}
\partial_{\mu}^2 v_{\alpha}(r) = \partial_{\beta} \omega_{\beta \alpha}(r)
\end{equation}
with the proper boundary conditions, say, $ v=0 $ at infinity.
Formally,
\begin{equation}
v_{\alpha}(r) =
\frac{1}{\partial_{\mu}^{2}}\partial_{\beta} \omega_{\beta \alpha}(r)
\end{equation}

This formula suggests the following definition of  the velocity
operator
\begin{equation}
V_{\alpha}(\theta,\epsilon,\delta) = \frac{1}{D_{\mu}^2(\theta,\epsilon)}
D_{\beta}(\theta,\epsilon) \Omega_{\beta \alpha}(\theta,\delta)\\;\;
\delta \ll \epsilon
\label{VOM}
\end{equation}
\begin{equation}
V_{\alpha}(\theta,\epsilon,\delta)\Psi[C] \stackrel{\delta,\epsilon
\rightarrow 0}{\longrightarrow} v_{\alpha} \left(C(\theta) \right) \Psi[C]
\end{equation}

Another version of this formula is the following integral
\begin{equation}
V_{\alpha}(\theta,\epsilon,\delta)= \int d^3  \rho
\frac{\rho_{\beta}}{4 \pi |\rho|^3}\Lambda \left(\rho,\theta,\epsilon
\right)
\Omega_{\alpha\beta}(\theta+ \epsilon ,\delta)
\end{equation}
where the $ \Lambda $ operator shifts the  $ \Omega $ by a distance $
\rho $ off the original loop at the point $ r = C(\theta + \epsilon)
$, (see Fig. \ref{fig::Fig5}).

\pct{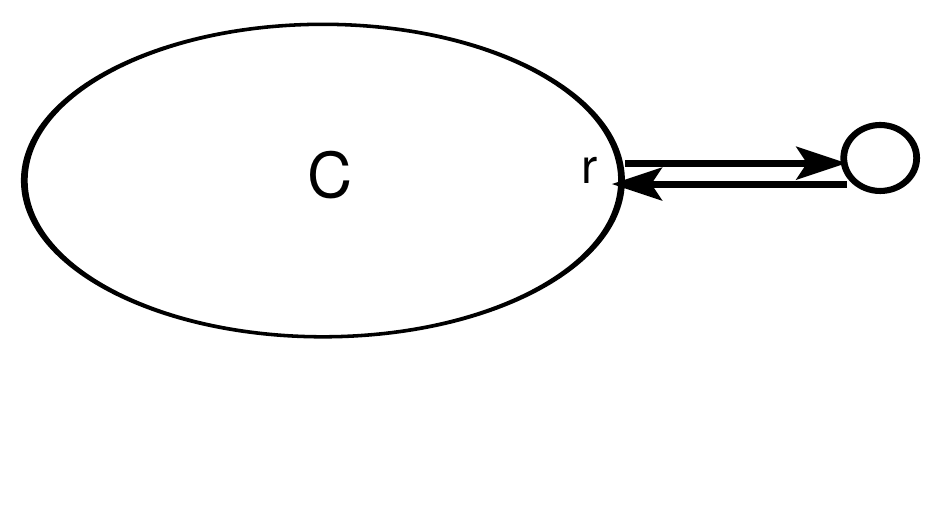}{Area derivative}

\subsection{Loop Equation}\label{LoopEquation}

Let us now derive an exact equation for the loop functional.
Taking the time derivative of the original definition and using the
Navier-Stokes equation we get in front of exponential
\begin{equation}
\oint_C d r_{\alpha}  \frac{\i}{ \nu}
\left(
    \nu \partial_{\beta}^2 v_{\alpha} - v_{\beta}
\partial_{\beta} v_{\alpha} - \partial_{\alpha} p \right)
\end{equation}
The term with the pressure gradient yields zero after integration over
the closed loop, and the velocity gradients in the first two terms
could be expressed in terms of vorticity up to the irrelevant gradient
terms so that we find
\begin{equation}
\oint_C d r_{\alpha}  \frac{\i}{ \nu}
\left(
    \nu \partial_{\beta} \omega_{\beta \alpha} - v_{\beta}
\omega_{\beta \alpha}
\right) \label{Orig}
\end{equation}

Replacing the vorticity and velocity  by the operators discussed in the
the previous Section \ref{LoopCalc}, we find the following loop equation (in explicit
notations)
\bea
-\i \nu\dot{\Psi}[C] = \oint d C_{\alpha}(\theta)
\br
\left(
   \nu D_{\beta}(\theta,\epsilon) \Omega_{\beta \alpha}(\theta,\epsilon) +
\int d^3  \rho
\frac{\rho_{\gam}\left(\Lambda \left(\rho,\theta,\epsilon
\right)
\Omega_{\gam\bet}(\theta+ \epsilon ,\delta)\Omega_{\beta
\alpha}(\theta,\delta)\right)}{4 \pi |\rho|^3} \right) \Psi[C]
\label{PsiC}
\eea

The more compact form of this equation, using the notations of
\cite{Mig83}, reads
\begin{subequations}
\begin{eqnarray}\label{OLD}
 &&  \i\,\nu\dot{\Psi}[C] = {\cal H}_C\Psi ;\\
&&{\cal H}_C =   {\cal H}^{(1)}_C + {\cal H}^{(2)}_C\\
&& {\cal H}^{(1)}_C =
    \nu^2\oint_{C} dr_{\alpha}\partial_{\beta} \frac{\delta }{\delta \sigma_{\beta \alpha}(r)};\\
&&{\cal H}^{(2)}_C =
\nu^2\oint_{C} dr_{\alpha}\int d^3 r'\frac{r'_{\gamma}-r_{\gam}}{4 \pi |r-r'|^3}
\frac{\delta^2}{\delta \sigma_{\beta \alpha}(r)
\delta \sigma_{\beta \gamma}(r')}
\end{eqnarray}
\end{subequations}
We observe viscosity $ \nu $ appearing in front of time and
spatial derivatives, like the Planck constant $\hbar$ in Quantum
mechanics. The first term  ${\cal H}^{(1)}_C$ in our loop Hamiltonian is local in the loop space. 
The second term ${\cal H}^{(2)}_C$ contains the second loop derivatives, acting as a (nonlocal!) kinetic term in loop space.

The first term came from the viscous diffusion term; as we shall see later, it violates the time symmetry.

\subsection{External forces and Dissipation}\label{LoopDissipation}

So far, we have considered so-called decaying turbulence without external
energy source. The  energy
\begin{equation}
E = \int d^3 r \oh \, \val^2
\end{equation}
would eventually all dissipate so that the fluid would stop. In this case
the loop wave function $\Psi$ would asymptotically approach $1$
\begin{equation}
\Psi[C] \stackrel{t \ra \8}\lra 1
\end{equation}

In order to reach the steady state, we  add to the right side of the \NS{} 
equation the usual Gaussian random forces $f_{\alp}(r,t)$ with the space
dependent correlation function
\begin{equation}
\VEV{f_{\alp}(r,t)f_{\bet}(r',t')} = \delta_{\alp\bet}\delta(t-t')F(r-r')
\end{equation}
concentrated at small wavelengths, i.e., slowly varying with $r-r'$.

Using the identity
\begin{equation}
\VEV{f_{\alp}(r,t) \Phi[v(.,t)]} = \oh \int d^3 r' F(r-r')
\fbyf{\Phi[v(.,t)]}{\val(r',t)}
\end{equation}
which is valid for arbitrary functional $\Phi$  we find the following
imaginary potential term in the loop Hamiltonian
\begin{equation}
\delta{\cal H}_C \equiv \i\,U[C]= \oh\frac{\i}{\nu}\,\oint_{C}
dr_{\alpha}\oint_{C} dr'_{\alpha} F(r-r')
\end{equation}

Note that orientation reversal, followed by complex conjugation, changes
the sign of the loop Hamiltonian, as it should. The potential part
involves double loop integration times an imaginary constant. 

The first term in
the kinetic part has one loop integration and one loop derivative times a real constant. 

The second kinetic term has one-loop integration, two-loop derivatives, and a real constant. The left side of the loop equation has
no loop integration and no loop derivatives but has a factor of $\i$.

The relation between the potential and kinetic parts of the loop
Hamiltonian depends on viscosity, or, better to say, it depends upon the
Reynolds number, which is the ratio of the typical circulation to
viscosity. 

When the Reynolds number is small, the
loop wave function is close to $1$. The perturbation expansion in $
\inv{\nu}$ goes in powers of the potential, in the same way, as in Quantum
mechanics. The second (nonlocal) term in the kinetic part of the Hamiltonian
also serves as a small perturbation (it corresponds to the nonlinear term in
the \NS{}  equation).
The first term of this perturbation expansion is just
\begin{equation}
\Psi[C] \ra 1 - \oh\int \frac{d^3
k}{(2\pi)^3}\frac{\tilde{F}(k)}{2\nu^3\,k^2} \left|\oint_C d \ral e^{\i k
r}\right|^2
\end{equation}
with $\tilde{F}(k)$ being the Fourier transform of $F(r)$.
This term is real, as it corresponds to the two-velocity correlation. 
The next term comes from the triple correlation of velocity, and this term is
purely imaginary so that the dissipation shows up.

Direct iterations in the loop space can derive this expansion as
in \cite{Mig83}, inverting the operator in the local part of the kinetic
term in the Hamiltonian. This expansion is discussed in Section \ref{LoopExpansion}. The
results agree with the straightforward iterations of  the \NS{}  equations in
powers of the random force, starting from zero velocity.

So, we have a familiar situation, like in QCD, where the perturbation
theory breaks because of the infrared divergencies. For an arbitrarily small
force in a large system, the region of small $k$ would yield a large
contribution to the terms of the perturbation expansion. Therefore, one
should take the opposite WKB limit $\nu \ra 0$.

In this limit, the wave function should behave as the usual WKB wave
function, i.e., as an exponential\footnote{this is, of course, before averaging over random forces (see below).}
\begin{equation}
\Psi[C] \ra \exp{\frac{\i\,S[C]}{\nu}}
\end{equation}
The effective loop Action $S[C]$ satisfies the loop space Hamilton-Jacobi
equation
\begin{equation}
\dot{S}[C] =-\i U[C] + \oint_{C} dr_{\alpha}
\int d^3 r'\frac{r'_{\gamma}-r_{\gam}}{4 \pi |r-r'|^3}
\frac{\delta S}{\delta \sigma_{\beta \alpha}(r)}
\frac{\delta S}{\delta \sigma_{\beta \gamma}(r')}
\label{SC}
\end{equation}
The imaginary part of $S[C]$ comes from imaginary potential $\i U[C]$, which
distinguishes our theory from reversible Quantum mechanics. The sign
of $\Im S$ must be positive definite since $ |\Psi| <1$. As for the
real part of $S[C]$, it changes the sign under the loop orientation
reversal $C(\theta) \ra C(2\pi-\theta) $.

At finite viscosity, there would be an additional term
\begin{equation}\label{ExtraTerms}
-\nu\oint_{C} dr_{\alpha}\partial_{\beta} \frac{\delta S[C]}{\delta
\sigma_{\beta\alpha}(r)}
-\i\nu \oint_{C} dr_{\alpha}
\int d^3 r' \frac{r'_{\gamma}-r_{\gam}}{4 \pi |r-r'|^3}
\frac{\delta^2 S[C]}{\delta \sigma_{\beta \alpha}(r) \delta \sigma_{\beta
\gamma}(r')}
\end{equation}
on the right of \rf{SC}. As for the term
\begin{equation}
-\oint_{C} dr_{\alpha}
    \i\left(\partial_{\beta}S[C]\right) \frac{\delta S[C]}{\delta
\sigma_{\beta \alpha}(r)}
\end{equation}
which formally arises in the loop equation: this term vanishes, since
$\partial_{\beta}S[C]=0$. This operator inserts backtracking at some point
at the loop without first applying the loop derivative at this point. It was discussed in the previous Section that such backtracking  does not
change the loop functional. This issue was discussed at length in
\ct{Mig83}, where the Leibnitz rule for the operator $ \dal
\fbyf{}{\sigma_{\bet\gam}} $ was established
\begin{equation}
\dal \fbyf{f(g[C])}{\sigma_{\bet\gam}(r)} = f'(g[C])\dal
\fbyf{g[C]}{\sigma_{\bet\gam}(r)}
\end{equation}
In other words, this operator acts as a first-order derivative on the loop
functional with finite area derivative (so-called Stokes type functional).
Then, the above term does not appear.

The Action functional $ S[C] $ describes the distribution of the large-scale vorticity structures; hence, it should not depend on viscosity.
In terms of the connected correlation functions of the circulation
this regime corresponds to the limit when the effective Reynolds number
$\frac{\Gamma_C[v]}{\nu}$ goes to infinity, but the sum of the divergent
series tends to the finite limit. According to the standard picture of
turbulence, the large-scale vorticity structures depend upon the energy
pumping, rather than energy dissipation.

It is understood that  both time $ t $ and the loop
size\footnote{As a measure of the loop size, one may take the square root
of the minimal area inside the
loop.} $ |C| $ should be greater than the viscous scales
\begin{equation}
t \gg t_0 = \nu^{\frac{1}{2}}{\cal E}^{-\frac{1}{2}} \\;\;
|C| \gg r_0 = \nu^{\frac{3}{4}} {\cal E}^{-\frac{1}{4}}
\end{equation}
where $ {\cal E } $ is the energy dissipation rate.

The energy balance equation defines it
\begin{equation}
0 = \d_t\VEV{\oh\,\val^2}= \nu \VEV{\val\d^2 \val} +\VEV{f_{\alp}\val}
\end{equation}
which we can transform to
\begin{equation}
 \oq \nu \VEV{\omega_{\alp\bet}^2} = 3 F(0)
\end{equation}
The left side represents the energy dissipated at a small scale due to
viscosity, and the right side - represents the energy pumped in from a large scale
due to random forces. Their common value is $\et$.

We see that constant $F(r-r')$, i.e., $\tilde{F}(k)\propto \delta(k)$ is
sufficient to provide the necessary energy pumping. However, such forcing
does not produce vorticity, which we readily see in our equation. The
contribution from this constant part to the potential in our loop equation
drops out (this is a  closed loop integral of the total derivative). This reduction of the constant term is
important because it would have the wrong order of magnitude in
the turbulent limit - it would grow as the Reynolds number.

The first term in equation \eqref{ExtraTerms} comes from the viscous diffusion term in the \NS{} equation, and it has two derivatives at the same point. 
This term may have a finite limit at $\nu \ra 0$ in the singular regions, where large Laplacian compensates the factor of $\nu \ra 0$.

As we have seen in the previous Section, this occurred in our Kelvinon solution.

Dropping remaining terms, we arrive at a remarkably simple and universal
functional equation
\begin{eqnarray}
&&\dot{S}[C] =  -\nu\oint_{C} dr_{\alpha}\partial_{\beta} \frac{\delta S[C]}{\delta
\sigma_{\beta\alpha}(r)} + \nonumber\\
&&\oint_{C} dr_{\alpha} \int d^3 r'\frac{r'_{\gamma}-r_{\gam}}{4 \pi |r-r'|^3}
\frac{\delta S}{\delta \sigma_{\beta \alpha}(r)}
\frac{\delta S}{\delta \sigma_{\beta \gamma}(r')}
\label{KIN}
\end{eqnarray}
The steady solution of this equation describes the steady
distribution of the circulation in the strong turbulence. 

\subsection{Loop Equation for the Circulation PDF}\label{LoopPDF}

The loop field could serve as the generating function for the PDF $P_C(\Gamma)
$ for the circulation. The Fourier integral
\begin{equation}
P_C(\Gamma) = \int_{-\8}^{\8} \frac{d g}{2 \pi \nu}
\exp{ \i g/\nu  \left(\C \ral \val(r)  - \Gamma \right)}
\end{equation}
can be analyzed in the same way as the loop field before. The only difference
is that  instead of constant $1/\nu$ we have a variable $g/\nu$ . These factors
can be replaced by
\begin{equation}
g \ra \i \nu \pp{\Gamma}
\end{equation}
acting on $P_C(\Gamma) $.

As a result, we find
\bea
 \pp{\Gamma}\dot{P}_C(\Gamma) = -\oint_{C} dr_{\alpha}
\int d^3 r'\frac{r'_{\gamma}-r_{\gam}}{4 \pi |r-r'|^3}
\frac{\delta^2 P_C(\Gamma)}{\delta \sigma_{\beta \alpha}(r)
\delta \sigma_{\beta \gamma}(r')} \br
 +\nu \pp{\Gamma}
\oint_{C} dr_{\alpha}\partial_{\beta}
     \frac{\delta P_C(\Gamma)}{\delta \sigma_{\beta \alpha}(r)}
- U[C] \frac{\d^3 P_C(\Gamma)}{\d \Gamma^3}
\label{LoopEqForPDF}
\eea
All the imaginary units disappear as they should. The viscosity and
force terms can be neglected in the inertial range in the same way as
before. The only new thing is that one has to assume that $ \Gamma \gg \nu $ is in
the inertial range in addition to the above assumptions about the loop size.

\subsection{Loop Expansion}\label{LoopExpansion}

Let us outline the method of direct iterations of the loop equation. 
A full description of the method can be found in \cite{Mig83}. The idea is to use the following representation of the loop functional
\begin{equation}
\Psi[C] = 1+\sum_{n=2}^{\8} \inv{n} \left\{\oint_C dr_1^{\alp_1} \dots
\oint_C dr_n^{\alp_n}\right\}_{\mbox{cyclic}} W^n_{\alp_1\dots
\alp_n}\left(r_1,\dots r_n\right)
\label{StokesRep}
\end{equation}

This representation holds for every translation invariant functional
with finite area derivatives (so-called Stokes type functional). The
coefficient functions $W$ can be related to these area derivatives. The
normalization $\Psi[0]=1 $ for the shrunk loop is implied.

In general case the integration points $r_1,\dots  r_n$ in \eqref{StokesRep} are
cyclically ordered around the loop $C$. The coefficient functions can be
assumed cyclicly symmetric without loss of generality. However,  in the case
of fluid dynamics, we are dealing with the so-called abelian Stokes
functional. These functionals are characterized by completely symmetric
coefficient functions, in which case the ordering of points can be
removed at the expense of the extra symmetry  factor in the denominator
\begin{equation}
\Psi[C] = 1+\sum_{n=2}^{\8} \inv{n!} \oint_C dr_1^{\alp_1} \dots  \oint_C
dr_n^{\alp_n} W^n_{\alp_1\dots \alp_n}\left(r_1,\dots r_n\right)
\label{ABEL}
\end{equation}
The incompressibility conditions
\begin{equation}
\d_{\alp_k}W^n_{\alp_1\dots \alp_n}\left(r_1,\dots r_n\right)=0
\label{divv}
\end{equation}
does not impose any further restrictions because of the gauge invariance
of the loop functionals. This invariance (nothing to do with the symmetry
of dynamical equations!) follows from the fact that the closed loop
integral of any total derivative vanishes. So, the coefficient functions
are defined modulo such derivative terms. In effect, this means that one
may relax the incompressibility constraints \rf{divv}, without changing
the loop functional.

Let us note that the physical incompressibility
constraints are not neglected to avoid confusion. They are, in fact, present in the loop
equation, where we used the integral representation for the velocity in
terms of vorticity. Still, the longitudinal parts of $W$ drop in the loop
integrals.

The loop calculus for the abelian Stokes functional is especially simple.
The area derivative corresponds to the removal of one loop integration, and
differentiation of the corresponding coefficient function
\begin{equation}
\fbyf{\Psi[C]}{\sigma_{\mu\nu}(r)} = \sum_{n=1}^{\8} \inv{n!} \oint_C
dr_1^{\alp_1} \dots  \oint_C dr_n^{\alp_n}
\hat{V}_{\mu\nu}^{\alp}W^{n+1}_{\alp,\alp_1\dots \alp_n}\left(r,r_1,\dots
r_n\right)
\label{ABEL'}
\end{equation}
where
\begin{equation}
\hat{V}_{\mu\nu}^{\alp} \equiv
\d_{\mu}\delta_{\nu\alp}-\d_{\nu}\delta_{\mu\alp}
\end{equation}
In the nonabelian case, there would also be the contact terms, with $W$ at
coinciding points, coming from the cyclic ordering \ct{Mig83}. In the abelian
case, these terms are absent since $W$ is completely symmetric.

As a next step, let us compute the local kinetic term
\begin{equation}
\hat{L} \Psi[C]  \equiv \oint_C d r_{\nu}
\d_{\mu}\fbyf{\Psi[C]}{\sigma_{\mu\nu}(r)}
\end{equation}
Using the above formula for the loop derivative, we find
\begin{equation}
\hat{L} \Psi[C]  = \sum_{n=1}^{\8} \inv{n!}  \oint_C dr^{\alp}\oint_C
dr_1^{\alp_1} \dots  \oint_C dr_n^{\alp_n} \d^2W^{n+1}_{\alp,\alp_1\dots
\alp_n}\left(r,r_1,\dots r_n\right)
\label{L}
\end{equation}
The net result is the second derivative of $W$ by one
variable. Note that the second term in $\hat{V}_{\mu\nu}^{\alp}$ dropped
as the total derivative in the closed loop integral.

As for the nonlocal kinetic term, it involves the second area derivative
off the loop, at the point, $r'$, integrated over $r'$ with the
corresponding to Green's function. Each area derivative involves the same
operator $\hat{V}$, acting on the coefficient function. Again, the abelian
Stokes functional simplifies the general framework of the loop calculus.
The contribution of the wires cancels here, and the ordering does not
matter so that
\begin{eqnarray}
&&\frac{\delta^2\Psi[C]}
{\delta\sigma_{\mu\nu}(r)\delta\sigma_{\mu'\nu'}(r')}= \sum_{n=0}^{\8}
\inv{n!} \oint_C dr_1^{\alp_1} \dots  \oint_C d r_n^{\alp_n} \nonumber\\
&&
\hat{V}_{\mu\nu}^{\alp}\hat{V'}_{\mu'\nu'}^{\alp'}W^{n+2}_{\alp,\alp',\alp_1\dots
\alp_n}\left(r,r',r_1,\dots r_n\right)
\end{eqnarray}

Using these relations, we can write the steady-state loop equation as
in the figure \ref{fig::Fig6}.

\pct{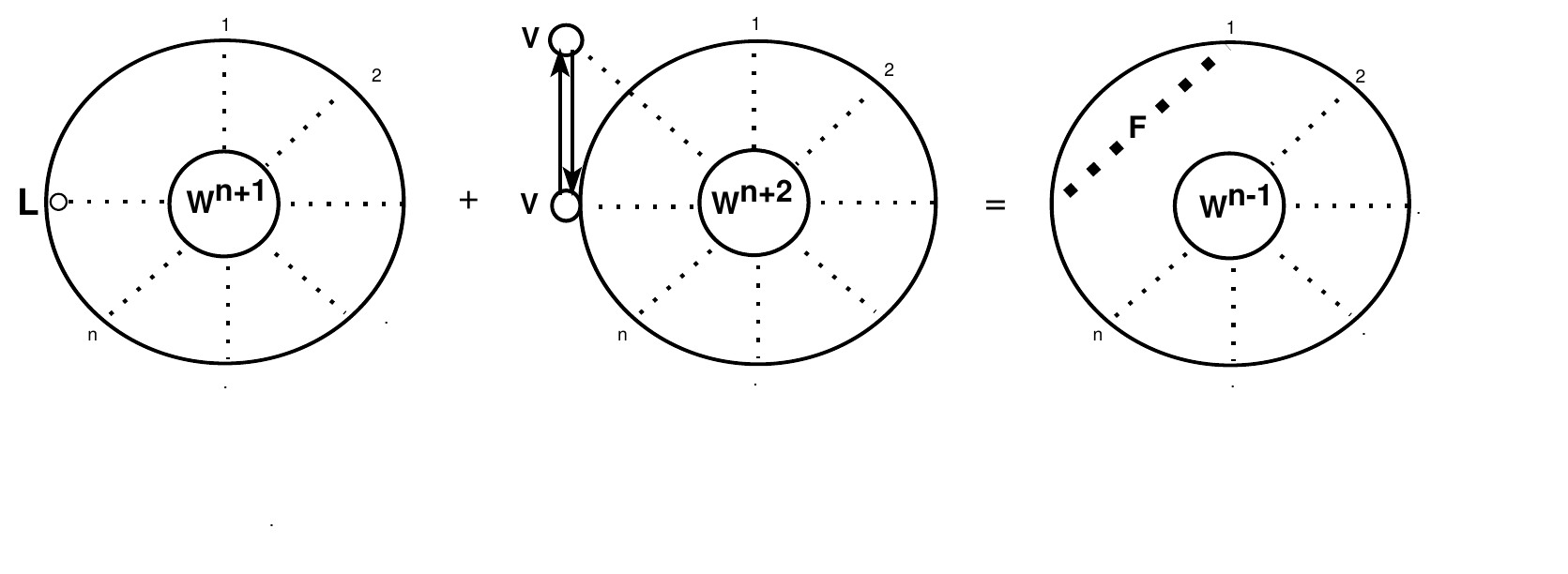}{Loop equation for coefficient functions}
Here the light dotted lines symbolize the arguments $\alp_k, r_k$ of $W$,
the big circle denotes the loop $C$; the tiny circles stand for the loop
derivatives and the pair of lines with the arrow denote the Green's
function. The sum over the tensor indexes and the loop integrations over
$r_k$ are implied.

The first term is the local kinetic term, the second is the nonlocal
kinetic term, and the right is the potential term in the loop
equation. The heavy dotted line in this term stands for the correlation
function $F$ of the random forces. Note that this term is an abelian
Stokes functional as well.

The iterations go in the potential term, starting with $\Psi[C]=1$. In the
next approximation, only the two-loop correction
$W^2_{\alp_1\alp_2}(r_1,r_2)$ is present.  Comparing the terms, we note,
that nonlocal kinetic term reduces to the total derivatives due to the
space symmetry (in the usual terms, it would be $ \VEV{v\omega}$ at
coinciding arguments), so we are left with the local one.

This reasoning yields the equation
\begin{equation}
\nu^3 \d^2 W^2_{\alp\bet}(r-r') = F(r-r')\delta_{\alp\bet}
\end{equation}
modulo derivative terms. The solution is trivial in Fourier space
\begin{equation}
 W^2_{\alp\bet}(r-r') = -\int \frac{d^3 k}{(2\pi)^3} \exp{\i k (r-r')}
\delta_{\alp\bet} \frac{\tilde{F}(k)}{\nu^3 k^2}
\end{equation}
Note that we did not use the transverse tensor
\begin{equation}
P_{\alp\bet}(k) = \delta_{\alp\bet}- \frac{k_{\alp} k_{\bet}}{k^2}
\end{equation}
Though such a tensor is present in the physical velocity correlation, here
we may use $\del_{\alp\bet}$ instead, as the longitudinal terms drop in
the loop integral. This phenomenon is analogous to the Feynman gauge in QED. The
correct correlator corresponds to the Landau gauge.

The potential term generates the four-point correlation $ F \, W^2$, which
matches the disconnected term in the $W^4$ on the left side
\bea
W^4_{\alp_1\alp_2\alp_3\alp_4}\left(r_1,r_2,r_3,r_4\right) \ra
W^2_{\alp_1\alp_2}\left(r_1-r_2\right)W^2_{\alp_3\alp_4}\left(r_3-r_4\right)
+ \br
W^2_{\alp_1\alp_3}\left(r_1-r_3\right)W^2_{\alp_2\alp_4}\left(r_2-r_4\right)
+
W^2_{\alp_1\alp_4}\left(r_1-r_4\right)W^2_{\alp_2\alp_3}\left(r_2-r_3\right)
\eea
The three-point function will appear in the same order in the loop expansion. The corresponding terms in the kinetic part must cancel, as the potential term does not contribute. The local kinetic
term yields the loop integrals of $ \d^2 W^3 $, whereas the  nonlocal one
yields $\hat{V}W^2 \,\hat{V'}W^2$, integrated over $d^3 r'$ with the
Greens's function $ \frac{(r-r')}{4\pi |r-r'|^3}$. The equation has the
structure (see Fig.\ref{fig::Fig7}).

\pct{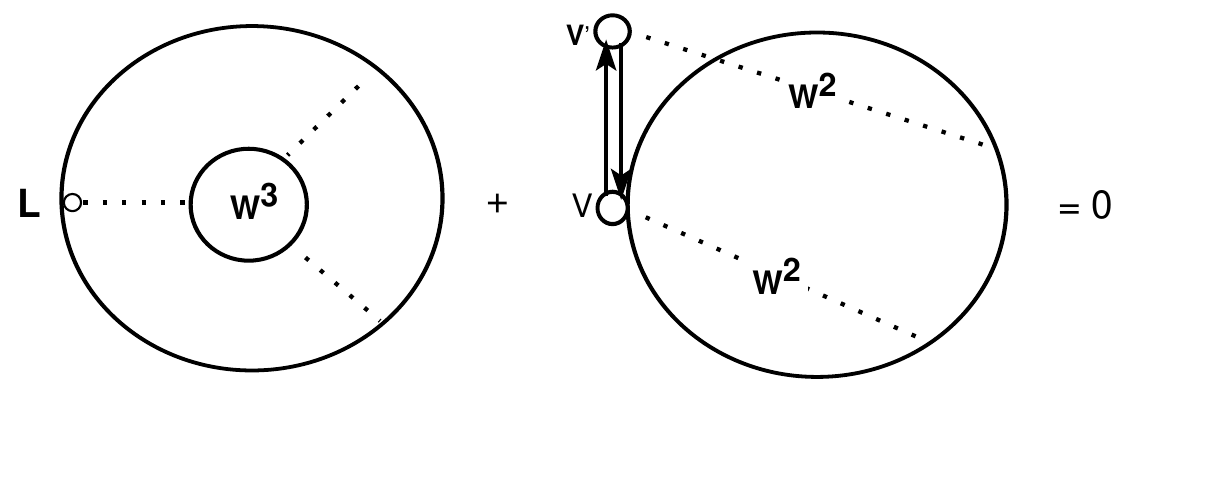}{Relation between third and fourth coefficient functions}

Now it is clear that the solution of this equation for $W^3$ would be the same three-point correlator, which one could obtain (much easier!) by
direct iterations of the \NS{}  equation.

The purpose of this painful exercise was not to give one more method of
developing the expansion in powers of the random force. We
verified  that the loop equations are capable of producing the same
results, as the ordinary chain of the equations for the correlation
functions.

In the above arguments, the loop functional needed to belong to
the class of the abelian Stokes functionals. Let us check that our tensor
area Ansatz
\begin{equation}
\Sigma^C_{\alp\bet}=\oint_C \ral d \rbe
\end{equation}
belongs to the same class. Taking the square, we find
\begin{equation}
\left(\Sigma^C_{\alp\bet}\right)^2 = \oint_C  d \rbe \oint_C  d r'_{\bet}
\ral r'_{\alp} = - \oh  \oint_C  d \rbe \oint_C  d r'_{\bet} (r-r')^2
\end{equation}
where the last transformation follows from the fact that only the cross
term in $ (r-r')^2$ yields nonzero after double loop integration.

Any expansion in terms  of the square of the tensor area, therefore, reduces to the superposition of multiple loop integral of the product of
$(r_i-r_j)^2 $; this is an example of the abelian Stokes functional. 
In the large area limit, this could reduce to a function of some loop integral. An example could be, say
\begin{equation}
\Psi[C] \stackrel{?}= \exp{B\left(1-\left(1+ \frac{\et
\left(\Sigma^C_{\alp\bet}\right)^2}{\nu^3} \right)^{\ot}\right)}
\end{equation}
One could explicitly verify all the properties of the abelian Stokes
functional. This example is unrealistic, though, as it does not have the
odd terms of expansion. In the real world, such terms are present at the
viscous scales. According to our solution, this asymmetry disappears in
the inertial range of loops (which does not apply to velocity correlators in
the inertial range, as those correspond to shrunk loops).

\subsection{How could classical statistics be exactly equivalent to quantum mechanics?}

Let us stop here and think about this remarkable analogy with quantum mechanics.

What is this analogy-- a poetic metaphor or perhaps something deeper, like the Hawking radiation or the ADS/CFT duality?

First of all, this analogy is not a word game -- it is precise mathematical correspondence between two theories: the statistics of classical fluid and the quantum mechanics in loop space.

The wave function of the quantum theory is related to the probability of the classical turbulence by inverse Fourier transformation with the frequency set to inverse viscosity (Planck's constant of our loop space quantum mechanics)
\begin{eqnarray}\label{PDFtoPsi}
   \Psi_C(\Gamma) = \left.\INT{-\8}{\8} \frac{d \omega}{2 \pi}\exp{ \i\omega \Gamma } P_C(\Gamma)\right|_{\omega = \frac{1}{\nu}}
\end{eqnarray}

In the same way as Hawking's interpretation of the period $\beta$ in the black hole solution as inverse temperature, we start with formal relation between two theories by analytical continuation in a complex plane.

The constant $\hbar$ also enters the Schrödinger equation directly, so one cannot use it as a Fourier frequency for the direct transfer. However, with inverse transform, the correspondence involves \textbf{setting} frequency to $1/\nu$, which is a correct mathematical statement.

There are more transformations in our theory -- the \CL{} variables, which look like ordinary fields described as a superposition of planar waves in $R_3 \mapsto R_2$, suddenly wrap up from a plane $R_2$ to a sphere $S_2$. Now they map $R_3 \mapsto S_2$. They are now confined and exist only as internal degrees of freedom, never observable as waves. 

The \CL{} waves were used by Yakhot, and Zakharov \cite{YZ93} in their approximate derivation of the K41 spectrum in the weak turbulence.

We argue that in strong turbulence, these \CL{} fields are confined to a sphere in internal space, unobservable as waves in physical space. The turbulence statistics is related to quantum mechanics in loop space, where the classical configurations of \CL{} field are related to the WKB limit of the wave function.

Still, do these relations represent a formal equivalence, or are there some observable quantum effects in classical turbulence?

As we have seen in previous Sections, there are such quantum effects: quantization of the circulation created by Kelvinons ( the winding numbers of classical solutions for \CL{} fields on a sphere $S_2$).

These Kelvinons are behind the dramatic intermittency effect: exponential decay of the circulation PDF over ten decades in probability. None of the conventional models of turbulence can explain this effect so far.

There are (not yet observed) contributions to PDF with a higher quantized decrement in exponential, like the terms of Planck's radiation, arising as a direct consequence of the quantization of the compact motion.

So, this formal equivalence leads to observable effects, confirmed in DNS.

\subsection{Matrix Model}\label{MatrixModel}

The Navier-Stokes equation represents a very special case of nonlinear
PDE. There is a well-known Galilean invariance
\begin{equation}
v_{\alpha}(r,t) \rightarrow v_{\alpha}(r-u t,t) + u_{\alpha}
\end{equation}
which relates the magnitude of the velocity field with the scales of time
and space. \footnote{At the same time, it tells us that the constant part of
velocity is frame-dependent, so it better be eliminated if we
would like to have a smooth limit at large times. Most notorious
large-scale divergencies in turbulence are due to this unphysical
constant part.} Let us make this relation more explicit.

First, let us introduce the vorticity field
\begin{equation}
\omega_{\mu\nu} = \partial_{\mu} v_{\nu} -\partial_{\nu} v_{\mu}
\end{equation}
and rewrite the Navier-Stokes equation as follows
\begin{equation}
\dot{v}_{\alpha} = \nu \partial_{\beta} \omega_{\beta \alpha} -
v_{\beta}\omega_{\beta \alpha} - \partial_{\alpha} w \\;\;
w = p + \frac{v^2}{2}
\end{equation}

This $ w $ is the well-known enthalpy density to be found from the
incompressibility condition $ div v = 0 $, i.e.
\begin{equation}
\partial^2 w = \partial_{\alpha}v_{\beta}\omega_{\beta \alpha}
\end{equation}

As a next step, let us introduce the "covariant derivative" operator
\begin{equation}
D_{\alpha} = \nu \partial_{\alpha} - \frac{1}{2}v_{\alpha}
\end{equation}
and observe that
\begin{equation}
2 \left[D_{\alpha} D_{\beta} \right] = \nu \omega_{\beta \alpha}
\end{equation}
\begin{equation}
2  D_{\beta}\left[ D_{\alpha}D_{\beta} \right] + {\it h.c.}=
\nu \partial_{\beta} \omega_{\beta \alpha} -
v_{\beta}\omega_{\beta \alpha}
\end{equation}
where $ {\it h.c.}$ stands for hermitian conjugate.

These identities allow us to write down the following dynamical equation for the covariant derivative operator
\begin{equation}
\dot{D}_{\alpha} =  D_{\beta}\left[ D_{\alpha}D_{\beta} \right]
- D_{\alpha} W + {\it h.c.}
\end{equation}

As for the incompressibility condition, we can write it as follows
\begin{equation}
\left[D_{\alpha} D^{\dagger}_{\alpha} \right] =0
\end{equation}
The enthalpy operator $ W = \frac{w}{\nu}$ is to be determined from
this condition, or, equivalently
\begin{equation}
\left[D_{\alpha} \left[D_{\alpha} W \right] \right] =
    \left[D_{\alpha}, D_{\beta}\left[ D_{\alpha}D_{\beta} \right]\right]
\end{equation}

We see that the viscosity disappeared from these equations. This
paradox is resolved by extra degeneracy of this dynamics: the
anti-hermitian part of the $ D $ operator is conserved. Its value at
the initial time is proportional to viscosity.

The operator equations are invariant by the time
independent unitary transformations
\begin{equation}
D_{\alpha} \rightarrow S^{\dagger}D_{\alpha}S\\;\; S^{\dagger}S = 1
\end{equation}
and, in addition, to the time-dependent unitary
transformations with
\begin{equation}
S(t) = \exp{ \frac{1}{2\nu} t u_\beta \left(D_{\beta} -
D_{\beta}^{\dagger} \right)}
\end{equation}
corresponding to the Galilean transformations.

One could view the operator $ D_{\alpha} $ as the  matrix
\begin{equation}
\left\langle i | D_{\alpha} | j \right \rangle =
\int d^3r \psi_i^{\star}(r) \nu \d_{\alp} \psi_j(r)-
\frac{1}{2} \psi_i^{\star}(r)v_{\alpha}(r) \psi_j(r)
\end{equation}
where the functions $ \psi_j(r) $ are the Fourier of Chebyshev
functions depending upon the geometry of the problem.

The finite mode approximation would correspond to the truncation of this
infinite size matrix to finite size $ N $. This approximation is not the same
as leaving $ N $ terms in the mode expansion of the velocity field.
The number of independent parameters here is $ O(N^2) $ rather than $
O(N)$. Is the unitary symmetry worth paying
such a high price in numerical simulations?

The matrix model of the Navier-Stokes equation has some theoretical beauty.
Moreover, it raises hopes of a simple asymptotic probability distribution. The
ensemble of random hermitian matrices was recently applied to the
the problem of Quantum Gravity \cite{QG}, which led to a genuine
breakthrough in the field.

Unfortunately, the model of several coupled random matrices, which is
the case here, is much more complicated than the one-matrix model
studied in Quantum Gravity. The dynamics of the eigenvalues are coupled
to the dynamics of the "angular" variables, i.e., the unitary matrices
$ S $ in the above relations. We could not directly apply the technique of
orthogonal polynomials, which was so successful in the one matrix
problem.

Another technique that proved to be successful in QCD and Quantum
Gravity is the loop equation. This method, which we discuss
at length in this paper, works in field theory problems with hidden
geometric meanings. The turbulence is an ideal case, much simpler
than QCD or Quantum Gravity.

\subsection{The Reduced Dynamics}\label{ReducedDynamics}

Let us now try to  reproduce the dynamics of the loop field in a simpler
Ansatz
\begin{equation}
  \Psi[C] = \left \langle \oldexp
	\left(
	   \frac{\i}{\nu}\oint d C_{\alpha}(\theta) P_{\alpha}(\theta)
	\right) \right \rangle \label{Reduced}
\end{equation}
The difference with the original definition (\ref{eq4}) is that our new
function $ P_{\alpha}(\theta) $ depends directly on $ \theta $ rather
then through the function $ v_{\alpha}(r) $ taken at $ r_{\alpha} =
C_{\alpha}(\theta) $. This transformation is the $ d \Ra 1 $ dimensional
reduction we mentioned above. From the point of view of the loop
functional, there is no need to deal with field $ v(r) $; one could
take a shortcut.

The reduced dynamics  must be fitted to the Navier-Stokes
dynamics of the original field. With the loop calculus developed above, we
have all the necessary tools to build these reduced dynamics.

Let us assume some unknown dynamics for the $P $ field
\begin{equation}
  \dot{P}_{\alpha}(\theta) = F_{\alpha}\left(\theta,[P] \right).
\end{equation}
Furthermore, compare the time derivatives of the original and reduced Ansatz. We
find in (\ref{Reduced}) instead of (\ref{Orig})
\begin{equation}
     \frac{\i}{\nu}\oint d C_{\alpha}(\theta)
	F_{\alpha}\left(\theta,[P]\right)
\end{equation}

Now we observe that $P'$ could be replaced by the functional
derivative, acting on the exponential in (\ref{Reduced}) as follows
\begin{equation}
  \frac{\delta}{\delta C_{\alpha}(\theta)}
	\leftrightarrow  -\i\nu P'_{\alpha}(\theta)
\end{equation}
This equation means that one could take the operators of the
Section 2 which are expressing velocity and vorticity in terms of the spike
operator, and replace the functional derivative as above.
This transformation yields the following formula for the spike derivative
\begin{eqnarray}
  &&D_{\alpha}(\theta,\epsilon) = \nonumber\\
  &&-\i\nu \int_{\theta}^{\theta+2
\epsilon} d \phi
	\left(
	1- \frac{\left|\theta + \epsilon - \phi \right|}{\epsilon}
	\right) P'_{\alpha}(\phi) =\nonumber\\
	&&-\i\nu
 \int_{-1}^{1}d \mu
	 \mbox{ sgn}(\mu)
	 P_{\alpha} \left(\theta + \epsilon (1+ \mu) \right)
	 \label{DP}
\end{eqnarray}
We obtained the weighted discontinuity of the function $ P(\theta) $,
which, in the naive limit $ \epsilon \rightarrow 0 $, would become the
true discontinuity. However, the function $ P(\theta) $ has in general
the stronger singularities than discontinuity, so that this limit
cannot be taken yet.

Thus, we arrive at the dynamical equation for the $P$ field
\begin{equation}
  \dot{P}_{\alpha} = \nu D_{\beta} \Omega_{\beta \alpha} - V_{\beta}
\Omega_{\beta \alpha} \label{Pdot}
\end{equation}
where the operators $ V , D, \Omega $ of the Section \ref{LoopCalc} should
be regarded as the ordinary numbers, with  definition (\ref{DP}) of $D$ in
terms of $P$.

All the functional derivatives are gone! We needed them only to
prove equivalence of reduced dynamics to the Navier-Stokes dynamics.

The function $ P_{\alpha}(\theta) $ would become complex now,
as the right side of the reduced dynamical
equation is complex for real $ P_{\alpha}(\theta) $.

Let us discuss this puzzling issue in more detail. The origin of
imaginary units was the factor of $ \imath $ in the exponential of the loop field. We had to insert this factor to make the
loop field decrease at large loops due to oscillations of
the phase factors. Later this factor propagated to the definition of
the $ P $ field.

Our spike derivative $ D $ is purely imaginary for real $ P $, and so
is our $ \Omega, $ operator. As a result, the velocity operator $ V $ is real.
Therefore the $ D \Omega $ term in the
reduced equation (\ref{Pdot}) is real for real $ P $ whereas the
$ V \Omega $ term is purely imaginary.

This relation does not contradict the moment equations, as we saw before. The
terms with even/odd numbers of velocity fields in the loop functional are
real/imaginary, but the moments are real, as they should be. The complex
dynamics of $ P $ doubles the number of independent variables.

There is one serious problem, however. Inverting the spike operator $
D_{\alpha} $ we implicitly assumed that it was anti-hermitian and
could be regularized by adding an infinitesimal negative constant to $
D_{\alpha}^2 $ in denominator. Indeed, this prescription works perturbatively in
each term of expansion in time or size of the loop, as we checked.
However, beyond this expansion, there would be a problem of
singularities, which arise when $ D_{\alpha}^2(\theta) $ vanishes at
some  $ \theta $.

In general, this would occur for complex $ \theta $ when the
imaginary and real part of $ D_{\alpha}^2(\theta) $ simultaneously
vanish. One could introduce the complex variable
\begin{equation}
  	e^{\imath \theta}=z\\;\;
	e^{-\imath \theta}= \frac{1}{z}\\;\;
	 \oint d \theta = \oint
	\frac{dz}{\imath z}
\end{equation}
where the contour of $z $ integration encircles the origin around the
unit circle. Later, with time evolution, these contours must
be deformed, to avoid  complex roots of $ D_{\alpha}^2(\theta) $.

\subsection{Initial Data}\label{Loopinitial}
Let us study the relation between the initial data for the original
and reduced dynamics. Let us assume that the initial field is distributed
according to some translation invariant probability density,
so that the initial value of the loop field does not depend on the
constant part of $C(\theta)$.

One can expand the translation invariant loop field in functional Fourier
transform
\begin{equation}
  \Psi[C] = \int DQ\delta^3 \left(\oint d \phi Q(\phi) \right)
	 W[Q] \oldexp
	\left(
	\imath \oint d \theta C_{\alpha}(\theta) Q_{\alpha}(\theta)
	\right)
\end{equation}
which can be inverted as follows
\begin{equation}
  \delta^3 \left( \oint d \phi Q(\phi)\right) W[Q] =
	\int DC\Psi[C]\oldexp
	\left(
	-\imath \oint d \theta C_{\alpha}(\theta) Q_{\alpha}(\theta)
	\right)
\end{equation}

Let us take a closer look at these formal transformations. The
functional measure for these integrations is defined according to the
scalar product
\begin{equation}
  (A, B) = \oint \frac{d \theta}{2 \pi} A(\theta) B(\theta)
\end{equation}
which diagonalizes in the Fourier representation
\begin{eqnarray}
 && A(\theta) = \sum_{-\infty}^{+\infty} A_n e^{\imath n \theta};\;A_{-n} = A_n^{\star}\nonumber\\
  &&(A,B) =  \sum_{-\infty}^{+\infty} A_n B_{-n} =\nonumber\\
  &&
A_0 B_0 + \sum_{1}^{\infty} a'_n b'_n + a''_n b''_n\nonumber\\
&&a'_n = \sqrt{2} \Re A_n,\; a''_n = \sqrt{2} \Im A_n
\end{eqnarray}

The corresponding measure is given by an infinite product of the
Euclidean measures for the imaginary and real parts of each Fourier
component
\begin{equation}
  DQ = d^3 Q_0 \prod_{1}^{\infty} d^3 q'_n d^3 q''_n
 \end{equation}
The orthogonality of Fourier transformation could now be explicitly
checked, as
\begin{eqnarray}
  &&\int DC \exp{ \imath \int d \theta C_{\alpha}(\theta)
	\left(
	 A_{\alpha}(\theta) - B_{\alpha}(\theta)
	\right) }\\ \nonumber
	&&= \int d^3 C_0 \prod_{1}^{\infty} d^3 c'_n d^3 c''_n \\ \nonumber
  &&\oldexp
\left( 2 \pi \imath
	\left(
	 C_0 \left(A_0-B_0 \right) +
	\sum_{1}^{\infty} c'_n\left(a'_n - b'_n \right)+
	 c''_n\left(a''_n - b''_n \right)
	\right)
\right)\\ \nonumber
&&=\delta^3\left(A_0-B_0 \right)
\prod_{1}^{\infty} \delta^3\left(a'_n - b'_n \right)
\delta^3\left(a''_n - b''_n \right)
\end{eqnarray}

Let us now check the parametric invariance
\begin{equation}
  \theta \rightarrow f(\theta)\\;\; f(2\pi) -f(0)
= 2\pi \\;\; f'(\theta) >0
\end{equation}
The functions $ C(\theta) $ and $ P(\theta) $ have
zero dimension in the sense that only their argument transforms
\begin{equation}
  C(\theta) \rightarrow C \left( f(\theta) \right) \\;\;
 P(\theta) \rightarrow P\left( f(\theta) \right)
\end{equation}
The functions $ Q(\theta) $ and $ P'(\theta) $ in above transformation
have dimension one
\begin{equation}
  P'(\theta) \rightarrow  f'(\theta) P'\left( f(\theta) \right)\\;\;
Q(\theta) \rightarrow  f'(\theta) Q \left( f(\theta) \right)
\end{equation}
so that the constraint on $ Q $ remains invariant
\begin{equation}
  \oint d \theta Q(\theta) = \oint df(\theta) Q\left( f(\theta)
\right)
\end{equation}

The invariance of the measure is easy to check for infinitesimal
reparametrization
\begin{equation}
  f(\theta) = \theta + \epsilon(\theta)\\;\; \epsilon(2\pi) = \epsilon(0)
\end{equation}
which changes $C$ and the $L_2$ norm $(C,C)$ as follows
\begin{eqnarray}
  &&\delta C(\theta) = \epsilon(\theta) C'(\theta) \\
&& \delta (C,C) = \oint \frac{d \theta}{2\pi}
\epsilon(\theta) 2 C_{\alpha}(\theta) C'_{\alpha}(\theta) \\\nonumber
&&=-\oint \frac{d \theta}{2\pi}\epsilon'(\theta)C_{\alpha}^2(\theta)
\end{eqnarray}
The corresponding Jacobian reduces to
\begin{equation}
  1 - \oint d \theta \epsilon'(\theta) =1
\end{equation}
in virtue of periodicity.

This relation proves the parametric invariance of the functional Fourier
transformations. Using these transformations, we could find the
probability distribution for the initial data of
\begin{equation}
 P_{\alpha}(\theta) = - \nu\int_{0}^{\theta} d \phi Q_{\alpha}(\phi)
\end{equation}

The simplest but still meaningful distribution of the initial velocity field is the Gaussian one, with energy concentrated in the macroscopic
motions. The corresponding loop field reads
\begin{equation}
  \Psi_0[C] = \oldexp
	\left(
	 -\frac{1}{2} \oint dC_{\alpha}(\theta)
	\oint dC_{\alpha}(\theta') f\left(C(\theta)-C(\theta')\right)
	\right)
\end{equation}
where $ f(r-r') $ is the velocity correlation function
\begin{equation}
  \left \langle v_{\alpha}(r) v_{\beta}(r') \right \rangle =
\left(\delta_{\alpha \beta}- \partial_{\alpha} \partial_{\beta}
\partial_{\mu}^{-2} \right) f(r-r')
\end{equation}
The potential part drops out in the closed loop integral.

The correlation function varies at the macroscopic scale, which means that
we could expand it in the Taylor series
\begin{equation}
  f(r-r') \rightarrow f_0 - f_1 (r-r')^2 + \dots \label{Taylor}
\end{equation}
The first term $ f_0 $ is proportional to initial energy density,
\begin{equation}
  \frac{1}{2} \left \langle v_{\alpha}^2 \right \rangle =\frac{d-1}{2}
f_0
\end{equation}
and the second one is proportional to initial energy dissipation
rate
\begin{equation}
 {\cal E}_{0} = -\nu  \left \langle  v_{\alpha} \partial_{\beta}^2
v_{\alpha} \right \rangle = 2 d(d-1) \nu f_1
\end{equation}
where $ d=3 $ is the dimension of space.

The constant term in (\ref{Taylor}) as well as $ r^2 + r'^2 $ terms
drop from the closed
loop integral, so we are left with the cross-term $ r r' $
\begin{equation}
  \Psi_0[C] \rightarrow  \oldexp
	\left(
	 - f_1 \oint dC_{\alpha}(\theta)
	\oint dC_{\alpha}(\theta') C_{\beta}(\theta)C_{\beta}(\theta')
	\right)
\end{equation}
This distribution is almost Gaussian: it reduces to Gaussian one by
extra integration
\begin{equation}
  \Psi_0[C] \rightarrow  {\rm const }\int d^3 \omega \oldexp
	\left(
	 -\omega_{\alpha \beta}^2
	\right)
	\oldexp
	\left(
	 2\imath \sqrt{f_1}
	 \omega_{\mu\nu} \oint dC_{\mu}(\theta) C_{\nu}(\theta)
	\right)
\end{equation}
The integration here goes
over all $ \frac{d(d-1)}{2} =3 $ independent $ \alpha < \beta $
components of the antisymmetric tensor $ \omega_{\alpha \beta} $.
Note that this is ordinary integration, not the
functional one. The physical meaning of this $ \omega $ is the random
constant vorticity at the initial moment.

At fixed $ \omega $ the Gaussian functional integration over $ C $
\begin{equation}
  \int DC \oldexp
	\left(
	 \imath  \oint d \theta
	\left(\frac{1}{\nu}
	  C_{\beta}(\theta) P'_{\beta}(\theta)
	+2 \sqrt{f_1}
	\omega_{\alpha \beta} C'_{\alpha}(\theta)C_{\beta}(\theta)
	\right)
	\right)
\end{equation}
can be performed explicitly, it reduces to the solution of the saddle
point equation
\begin{equation}
  P'_{\beta}(\theta) = 4\nu\sqrt{f_1}\omega_{\beta \alpha}
C'_{\alpha}(\theta)
\end{equation}
which is trivial for constant $ \omega $
\begin{equation}
  C_{\alpha}(\theta) =
\frac{1}{4\nu\sqrt{f_1}} \omega^{-1}_{\alpha \beta} P_{\beta}(\theta)
\end{equation}
The inverse matrix is not unique in odd dimensions since $ \mbox{Det }
\omega_{\alpha\beta} = 0 $.  However, the resulting pdf for $ P $ is unique.
This probability distribution  is  Gaussian, with the correlator
\begin{equation}
  \left \langle P_{\alpha}(\theta) P_{\beta}(\theta') \right \rangle =
2\imath \nu\sqrt{f_1} \omega_{\alpha \beta} {\rm sign}(\theta'-\theta)
\label{Corr}
\end{equation}

Note that antisymmetry of $ \omega $ compensates that of the sign
function so that this correlation function is symmetric, as it should
be. However, it is anti-hermitian, which corresponds to purely
imaginary eigenvalues. The corresponding realization of the $ P$
functions are complex!

Let us study this phenomenon for the Fourier components.
Differentiating the last equation by $ \theta $ and
Fourier transforming, we find
\begin{equation}
  \left \langle P_{\alpha,n} P_{\beta,m}  \right \rangle
= \frac{4\nu}{m} \delta_{-n m} \sqrt{f_1}\omega_{\alpha \beta}
\end{equation}

This relation cannot be realized with complex conjugate Fourier components $
P_{\alpha,-n} = P_{\alpha,n}^{\star} $ but we could take
$\bar{P}_{\alpha,n} \equiv P_{\alpha,-n} $ and $ P_{\alpha,n} $ as
real random  variables, with the correlation function
\begin{equation}
  \left \langle \bar{P}_{\alpha,n}P_{\beta,m} \right \rangle
= \frac{4\nu}{m}\delta_{n m}\sqrt{f_1} \omega_{\alpha \beta} \\;\; n>0
\end{equation}
The trivial realization is
\begin{equation}
   \bar{P}_{\alpha,n} =\frac{4\nu}{n} \sqrt{f_1}\omega_{\alpha \beta}
P_{\beta,n}
\end{equation}
with $P_{\beta,n} $ being Gaussian random numbers with unit
dispersion.

As for the constant part $ P_{\alpha,0} $ of $ P_{\alpha}(\theta) $ ,
it is not defined, but it drops from equations by translational
invariance.

\subsection{Possible Numerical Implementation}\label{Numerical}

The above general scheme is abstract and complicated. Could it
lead to any practical computation method? It would depend upon the
success of the discrete approximations of the singular equations of
reduced dynamics.

The most obvious approximation would be the truncation of Fourier
expansion at some large number $ N $. With Fourier components
decreasing only as powers of $ n $, this approximation is doubtful.
In addition, such truncation violates the parametric invariance, which
looks dangerous.

It is safer to approximate $ P(\theta) $ by a
sum of step functions so that it is piecewise constant. The
parametric transformations vary the lengths of intervals of
constant $ P(\theta) $ but leave invariant these constant values.
The corresponding representation reads
\begin{equation}
  P_{\alpha}(\theta) = \sum_{l=0}^{N} \left(p_{\alpha}(l+1)-p_{\alpha}(l)
\right)
 \Theta \left(\theta-\theta_l \right)\\;\; p(N+1) = p(1),\; p(0) = 0
\label{Thetas}
\end{equation}
It is implied that $ \theta_0 =0 < \theta_1 < \theta_2 \dots < \theta_N <
2\pi $.
By construction, the function $ P(\theta) $ takes value
$p(l)$ at the interval $ \theta_{l-1} < \theta < \theta_{l}
$.

We could take $ \dot{P}(\theta) $ at the middle of this interval as
approximation to $ \dot{p}(l) $.
\begin{equation}
	\dot{p}(l) \approx \dot{P}(\bar{\theta}_l)\\;\;
   \bar{\theta}_l = \frac{1}{2}\left(\theta_{l-1} + \theta_l \right)
\end{equation}
As for the time evolution of angles
$ \theta_l $ , one could differentiate (\ref{Thetas}) in time
and find
\begin{equation}
 \dot{P}_{\alpha}(\theta) =
\sum_{l=0}^{N} \left(\dot{p}_{\alpha}(l+1)-\dot{p}_{\alpha}(l) \right)
 \Theta \left(\theta-\theta_l \right) -
\sum_{l=0}^{N} \left(p_{\alpha}(l+1)-p_{\alpha}(l) \right)
 \delta(\theta-\theta_l)\dot{\theta_l}
\end{equation}
from which one could derive the following approximation
\begin{equation}
\dot{\theta_l} \approx \frac{\left(p_{\alpha}(l)-p_{\alpha}(l+1)
\right)}{ \left(p_{\mu}(l+1)-p_{\mu}(l)\right)^2}
\int_{\bar{\theta}_l}^{\bar{\theta}_{l+1}} d \theta
\dot{P}_{\alpha}(\theta)
\end{equation}

The extra advantage of this approximation is its simplicity. All the
integrals involved in the definition of the spike derivative
(\ref{DP}) are trivial for the piecewise constant $ P(\theta) $. So,
this approximation can be, in principle, implemented on the computer.
This formidable task exceeds the scope of the present Section,
which we view as purely theoretical.

\subsection{The loop equation in the WKB limit and singular Euler solutions}\label{SteadyUnsteady}

The loop equation's classical ($\nu\ra 0$) limit can be related to the singular  Euler flow, regularized as a Burgers vortex. 

If one takes any steady solution $\vec v_{cl}$ of the \NS{}  equation and  computes
\begin{eqnarray}
    &&P_{cl}[C;\Gamma] = \VEV{ \delta\left(\Gamma - \oint_C \vec v_{cl} \cdot d \vec r\right)} =\nonumber\\
    &&\INT{-\8}{\8}\frac{d g}{2 \pi} \exp{-\i g\Gamma }
    \VEV{\exp{\i g\oint_C \vec v_{cl} \cdot d \vec r}}
\end{eqnarray}
it will satisfy the steady loop equation. 
In the limit $\nu \ra 0$, the right side of the time derivative of the circulation becomes 
\begin{eqnarray}\label{EulerCirculation}
\oint_C (\vec\omega_{cl}\times \vec v_{cl} ) \cdot  d \vec r 
\end{eqnarray}

This integral will vanish for any classical steady solution of the \textbf{Euler} equation, which corresponds to \GBF{} ($\vec \nabla \times \vec v \times \vec \omega =0$).

However, we need a weaker condition: only the contour integral must vanish.

In case the velocity field satisfies the boundary condition
\begin{eqnarray}\label{vtangent}
\vec  v_{cl}(\vec r = \vec C(\theta))  \times \vec C'(\theta) =0
\end{eqnarray}
the circulation derivative \eqref{EulerCirculation}  will also vanish.

Another possibility would be vorticity aligned with the tangent vector
\begin{eqnarray}\label{omegatangent}
\vec  \omega_{cl}(\vec r = \vec C(\theta))  \times \vec C'(\theta) =0
\end{eqnarray}
We will discuss these boundary conditions in more detail later.

There is also an anomaly term coming from the viscosity term in the \NS{} equation.
\begin{eqnarray}
    \nu \VEV{\oint_C  d \vec r \cdot \left(\vec \nabla \times \vec\omega_{cl}\right)}  
\end{eqnarray}

The anomaly in the dissipation and the Hamiltonian (see previous Section) comes from the Burgers vortex in the vicinity of the loop, with $\vec r$ being the vector to the nearest point at the loop and $\Gamma_\beta$ being velocity circulation around the dual loop $\tilde C$ encircling $C$ at a given point. The size of this loop is much smaller than $C$ but much larger than a viscous width $h = \sqrt{\nu/c}$.

The Burgers solution we studied above, in \eqref{BurgersVortex}, has a finite circulation$\Gamma_\beta$.
\begin{eqnarray}
    &&\vec \omega_B(\vec r) \ra \vec t \; \frac{c  \Gamma_\beta  e^{-\frac{c \vec \rho^2}{4 \nu }}}{8 \pi  \nu };\\
    && \vec \rho = \vec r-  (\vec r \cdot \vec t)\vec t;
\end{eqnarray}
The curl of this vortex 
\begin{eqnarray}
    \vec \nabla \times \vec \omega_B(\vec r)  = \vec t \times \vec  \rho\; \frac{c^2  \Gamma_\beta  e^{-\frac{c \vec \rho^2}{4 \nu }}}{16 \pi  \nu^2 }
\end{eqnarray}
is orthogonal to the loop tangent vector $\vec t$.
Therefore,
\begin{eqnarray}
    \oint d \vec r \cdot \left(\vec \nabla \times \vec \omega_B(\vec r) \right)=0;
\end{eqnarray}

Consequently, we can set $\nu =0$ and neglect this term, as there is no anomaly here.

One can prove the absence of an anomaly using a simpler argument. The Euler derivative \eqref{EulerCirculation} of the circulation vanishes by itself on a Burgers vortex solution of the \NS{} equation. The remaining term, the anomaly, also must vanish as the net derivative vanishes on any stationary solution of the \NS{} equation.

Another possibility would be for these two terms to cancel each other, in which case we would have to keep the anomaly term in the loop equation.

Note also that the Burgers vortex satisfies both stability equations \eqref{vtangent}, \eqref{omegatangent} as well as the \GBF{} condition as $\vec v(C)$ is also directed along the local tangent vector $\vec t$ of the loop.

In the general case, we can use any steady vortex sheet providing finite circulation for this solution.

The steady vortex sheet requires the boundary value of velocity to be aligned with the boundary, plus it must be tangent to the vortex sheet inside the boundary.

The circulation will be conserved, as the steady loop equation requires.

This conservation is a manifestation of the Kelvin theorem. In the case of the steady boundary of the vortex sheet, the liquid contour along this boundary does not move. Thus, the Kelvin theorem says the circulation is conserved.

As we argued above, the domain wall bounded by a singular vortex line, which we discussed in the section \ref{Kelvinon}, could be a realization of a Kelvinon.

\textit{Note added in proof:} Note that a fixed point of the loop equation  \textbf{does not require} a stationary solution of the \NS{} or Euler equation. It is sufficient to have a conserved circulation. Thus, even a non-stationary solution of the Euler equation, regularized as a Burgers vortex in a tube surrounding a fixed loop $C$, would provide a fixed point, i.e., a steady probability distribution for circulation around a given loop in space. In other words, it suffices to have a time-dependent solution of the \NS{} with an extra restriction of a given circulation $\Gamma$ around a fixed loop in space. The PDF is then defined as a time average for this solution as a function of the value of the circulation.

\subsection{Tensor Area law}

The Wilson loop in QCD decreases as the exponential of the minimal area
encircled by the loop. This decay leads to quark confinement. What is the
similar asymptotic law in turbulence? The physical mechanisms leading to
the area law in QCD are absent here. Moreover, there is no guarantee that
$\Psi[C]$ always decreases with the size of the loop.

This observation makes it possible to look for the simple Anzatz, which was not
acceptable in QCD, namely
\begin{equation}
P[C, \Gamma] = s\left(\Sigma_{\mu\nu}^C, \Gamma\right)
\end{equation}
where
\begin{equation}
\Sigma_{\mu\nu}^C= \oint_C r_{\mu} d r_{\nu}
\end{equation}
is the tensor area encircled by the loop $C$. The difference between this
area and the scalar area is the positivity property. The scalar area
vanishes only for the loop, which can be contracted to a point by removal
of all the backtracking. As for the tensor area,  it vanishes 
for the $8$ shaped loop, with the opposite orientation of petals.

Thus, there are some large contours with vanishing tensor areas, for which
there would be no decrease in the $\Psi$ functional.
In QCD, the Wilson loops must always decrease at large distances due to
the finite mass gap. Here, large-scale correlations 
play a central role in a turbulent flow. So, at this point, we see no
convincing arguments to reject the tensor area Ansatz.

This Ansatz in QCD was not only unphysical, but it also failed to reproduce the
correct short-distance singularities in the loop equation. In turbulence,
there are no such singularities. Instead, there are the  large-distance
singularities, which all should cancel in the loop equation.

For this Anzatz, the (turbulent limit of the) loop equation is automatically satisfied without further restrictions.
Let us verify this important property. The first area derivative yields
\begin{equation}
\omega_{\mu\nu}^C(r)=\fbyf{S}{\sigma_{\mu\nu}(r)} = 2\pbyp{s}{
\Sigma_{\mu\nu}^C}
\end{equation}
The factor of $2$ comes from the second term in the variation
\begin{equation}
\fbyf{\Sigma^C_{\alp\bet}}{\sigma_{\mu\nu}(r)}=
\del_{\alp\mu}\del_{\bet\nu}-\del_{\alp\nu}\del_{\bet\mu}
\end{equation}
Note that the right side does not depend on $r$. Moreover, one can shift
$r$ aside from the base loop $C$ with proper wires inserted. The area
derivative would stay the same as the contribution of wires drops.

This observation implies that the corresponding vorticity $\omega_{\mu\nu}^C(r) $ is
space independent, it only depends upon the loop itself. 

The velocity can
be reconstructed from vorticity up to irrelevant constant terms
\begin{equation}
\vbe^C(r) = \oh\,\ral\,\omega_{\alp\bet}^C
\end{equation}
One can formally derive this relation  from the above integral representation
\begin{equation}
\vbe^C(r) =\int d^3 r'\frac{\ral-r'_{\alp}}{4 \pi
|r-r'|^3}\omega_{\alp\bet}^C
\label{INTG}
\end{equation}
as a residue from the infinite sphere $ R = |r'| \ra \8$. One may insert
the regularizing factor $ |r'|^{-\epsilon}$ in $\omega$, compute the
convolution integral in Fourier space and check that in the limit $
\epsilon  \ra 0^+$, the above linear term arises. So, one can use the above
form of the loop equation, with the analytic regularization prescription.

Now, the $v\,\omega$ term in the loop equation reads
\begin{equation}
\oint _C d r_{\gam} \,\vbe^C(r)\,\omega_{\bet\gam}^C \propto
\Sigma_{\gam\alp}^C\,\omega_{\alp\bet}^C\,\omega_{\bet\gam}^C
\end{equation}
This tensor trace vanishes because the first tensor is antisymmetric, and
the product of the last two antisymmetric tensors is symmetric with
respect to $\alp\gam$.

So, the positive and negative terms cancel each other in our loop
equation, like the "income" and "outcome" terms in the usual kinetic
equation. We see that there is an equilibrium in our loop space kinetics.

{}From the point of view of the notorious infrared divergencies in
turbulence, the above calculation explicitly demonstrates how they cancel.
By naive dimensional counting, these terms were linearly divergent. The
space isotropy lowered this to logarithmic divergence in \rf{INTG}, which
reduced to finite terms at closer inspection. Then, the explicit form of
these terms was such that they were all canceled.

This cancellation originates from the angular momentum conservation in
fluid mechanics. The large loop $C$  creates the macroscopic eddy with
constant vorticity  $\omega_{\alp\bet}^C$ and linear velocity $ v^C(r)
\propto r$. This eddy is a well-known static solution of the \NS{}  equation.

The
eddy is conserved due to the angular momentum conservation. The only
nontrivial thing is the eddy vorticity's functional dependence upon the loop's shape and size $C$. This expression is a function of the tensor area
$\Sigma_{\mu\nu}^C$, rather than a general functional of the loop.

Combining this Anzatz with the Kolmogorov scaling
law, we arrive at the tensor area law
\begin{equation}
P[C, \Gamma] \propto F\left(|\Gamma|\et^{-\ot}\abs{\Sigma^C_{\alp\bet}}^{-\tt} \right)
\label{AREA}
\end{equation}
Note, however, that we did not prove this law. The absence of decay for large
twisted loops with zero tensor area is unphysical. Also, the spatial picture differs from what we expect in turbulence. The uniform vorticity, even a
random one, as in this solution, contrasts the observed intermittent
distribution. Besides, there must be corrections to the asymptotic law,
whereas the tensor area law is {\em exact}. This solution is far too simple: it is a trivial steady Euler flow rather than a flow with a steady loop we are seeking.

We
discussed this unphysical solution mostly as a test of the loop technology.

\subsection{Scalar Area law}

Let us now study the scalar area law, which is a valid Anzatz for the
asymptotic decay of the circulation PDF.
We summarized in Appendix A the equations (the Weierstrass theory) for the minimal surface 
All we need here is the following representation
\begin{equation}
A \ra \inv{2L^2_{\Gamma}}\,
\int \int d \sigma_{\mu\nu}(x) d \sigma_{\mu\nu}(y)
\exp{-\pi\frac{(x-y)^2}{L^2_{\Gamma}}}
\end{equation}
where $ L_{\Gamma} =|\Gamma|^{\tq} \et^{-\oq} $.
The distance $ (x-y)^2$ is measured in 3-space, and integration goes
along the minimal surface. We implied that  its size is much larger than
$ L_{\Gamma} $.

In this limit, one can perform the integration over, say, $ y $  along the
local tangent plane at $ x $ in small vicinity $ y-x \sim L_{\Gamma} $ .
Afterward, $ L_{\Gamma} $ factors cancel. We are left then with the
ordinary scalar area integral
\begin{equation}
 A \ra \oh \int d\sigma_{\mu\nu}(x)
 d \sigma_{\mu\nu}(y) \delta^2(x-y) \ra \int d^2x \sqrt{g}
\end{equation}

In the previous, regularized form, the area represents the so-called Stokes
functional\ct{Mig83}, which can be substituted into the loop equation.
The area derivative of the area reads
\begin{equation}
\fbyf{A}{\sigma_{\mu\nu}(x)} =
\inv{L^2_{\Gamma}}\, \int  d \sigma_{\mu\nu}(y)
\exp{-\pi\frac{(x-y)^2}{L^2_{\Gamma}}}
\end{equation}
In the local limit, this reduces to the tangent tensor
\begin{equation}
\fbyf{A}{\sigma_{\mu\nu}(x)} \ra \int  d \sigma_{\mu\nu}(y) \delta^2(x-y)
= t_{\mu\nu}(x)
\end{equation}
We implied that the point $x$ approaches the contour from inside
the surface so that the tangent tensor is well defined
\begin{equation}
t_{\mu\nu}(x) = t_{\mu}n_{\nu} - t_{\nu} n_{\mu}
\end{equation}
Here $ t_{\mu}$ is the local tangent vector of the loop, and $ n_{\nu} $
is the inside normal to the loop along the surface.

The second area derivative of the regularized area in this limit is just
the exponential
\begin{equation}
\frac{\delta^2
A}{\delta\sigma_{\alpha\beta}(x)\delta\sigma_{\gamma\delta}(y)}=
\frac{\delta_{\alpha\gamma} \delta_{\beta\delta} -\delta_{\alpha\delta} \delta_{\beta\gamma}}{L^2_{\Gamma}}\, \exp{-\pi\frac{(x-y)^2}{L^2_{\Gamma}}}
\end{equation}
Should we look for the higher terms of the asymptotic expansion at a large area, we would have to take into account the shape of the minimal
surface, but in the thermodynamical limit, we could neglect the curvature of
the loop and use the planar disk.

Let us use the general WKB form of PDF
\begin{equation}
P_C(\Gamma) = \inv{\Gamma}
\exp{-Q\left(\frac{A}{t^{2\kappa}},\frac{\Gamma}{t^{2 \kappa-1}} \right)}
\end{equation}
We shall skip the arguments of effective Action $ Q $.
We find on the left side of the loop equation
\begin{equation}
\d_t Q \d_{\Gamma} Q - \d_t \d_{\Gamma} Q
\end{equation}
On the right side, we find  the following integrand
\begin{equation}
\left(\left(\d_{A}Q\right)^2-\d^2_{A}Q\right)
\,\fbyf{A}{\sigma_{\alpha\beta}(r)}
\fbyf{A}{\sigma_{\gamma\beta}(r')} -
\d_{A}Q \,\frac{\delta^2 A}{\delta\sigma_{\alpha\beta}(r)
\delta\sigma_{\gamma\beta}(r')}
\end{equation}

The last term drops after the  $r' $ integration by symmetry.
The leading terms in the WKB approximation on both sides are those with
the first derivatives. We find
\begin{equation}
\d_t Q \d_{\Gamma}Q =
\left(\d_{A}Q\right)^2 \C \ral \fbyf{A}{\sigma_{\alpha\beta}(r)}
\int d^3 r' \frac{\rga-r'_{\gamma}}{4\pi|r-r'|^3}\,
\fbyf{A}{\sigma_{\gamma\beta}(r')}
\end{equation}

In the last integral, we substitute the above explicit form of the
area derivatives and perform the $ d^3r' $ integration first. In the
thermodynamic limit only the small vicinity $ r'-y \sim L_{\Gamma} $
contributes, and we find
\begin{equation}
\int d^3 r' \frac{\rga-r'_{\gamma}}{4\pi|r-r'|^3}\,
\fbyf{A}{\sigma_{\gamma\beta}(r')} \ra
L_{\Gamma}^2 \,\int d \sigma_{\gamma\beta}(y)
\frac{\rga-y_{\gamma}}{4\pi|r-y|^3}
\end{equation}

This integral logarithmically diverges. We compute it with
the logarithmic accuracy with the following result
\begin{equation}
\int d \sigma_{\gamma\beta}(y)
\frac{\rga-y_{\gamma}}{4\pi|r-y|^3}
\propto \frac{t_{\beta}}{ \pi} \ln \frac{L^2_{\Gamma}}{A}
\end{equation}
The meaning of this integral is the average velocity in the WKB
approximation. This velocity is tangent to the loop, up to the next
correction terms at a large area.

Now, the emerging loop integral vanishes due to symmetry
\begin{equation}
\C \ral t_{\beta} t_{\alpha\beta} =0
\end{equation}
as the line element $ d \ral $ is directed along the tangent vector $
t_{\alpha} $, and the tangent tensor $ t_{\alpha\beta} $ is antisymmetric.
We used a similar mechanism in the tensor area solution; only there the
cancellations emerged at the global level after the closed loop
integration. 

Here the right side of the loop equation vanishes locally
at every point of the loop. Thus, we see that the scalar area 
represents the steady solution of the loop equation in the leading WKB
approximation.

However, this steady solution no longer corresponds to a steady Euler flow-- the circulation is conserved only for this specific loop.

The area conservation is the same phenomenon we discussed in the previous section \ref{SteadyUnsteady}. The loop equation is satisfied under the boundary condition for the velocity field: it is directed along the boundary, making this boundary steady.

It might be instructive to compare this solution with another known, exact
solution of the Euler dynamics, namely the Gibbs solution
\begin{equation}
P[v] = \exp{-\beta \int d^3 r \oh \val^2 }
\end{equation}
For the loop functional it reads
\begin{equation}
\Y = \exp{-\frac{\gamma^2}{2\beta}\C \ral \C r'_{\beta} \delta^3(r-r')}
\end{equation}
The integral  diverges, and it corresponds to the perimeter law
\begin{equation}
\C \ral \C r'_{\beta} \delta^3(r-r') \ra r_0^{-2} \oint_C |dr|
\end{equation}
where $ r_0 $ is a small distance cutoff. For the PDF it yields
\begin{equation}
P(\Gamma) \propto \exp{-\frac{\Gamma^2\beta r_0^2}{2 \oint_C |dr|}}
\end{equation}

We observe the same thing when substituting the Gibbs solution into the loop equation. Average velocity is tangent to the loop, which leads to
a vanishing integrand in the loop equation. The difference is that in our
case, this is true only asymptotically; there are the next corrections.

This equation does not fix the shape of the function $ Q $ in the leading
WKB approximation. In a scale-invariant theory, it is natural to expect the
power law
\begin{equation}
Q\left(\frac{A}{t^{2\kappa}},\frac{\Gamma}{t^{2 \kappa-1}} \right) \ra
\mbox{const } \left(\Gamma^{2\kappa} A^{1-2\kappa}\right)^{\mu}
\label{MuLaw}
\end{equation}
There is one more arbitrary index $ \mu$ involved. Even for the Kolmogorov law
$ \kappa = \frac{3}{2} $ the $ \Gamma $ dependence remains unknown.

As we have seen, the K41 laws are broken for the Kelvinon solution $ \mu = \frac{1}{2 \kappa}$.
\subsection{Discussion of Area Law}

So, we found two asymptotic solutions of the loop equation in the
turbulent limit, not counting the Gibbs solution. 

The tensor area solution is
mathematically cleaner, but its physical meaning contradicts the intermittency
paradigm. It corresponds to the uniform vorticity with random magnitude and
random direction rather than the regions of high vorticity interlaced with
regions of low vorticity observed in turbulent flows.

The recent numerical experiments\cite{S19,S21} favor the scalar area rather than
the tensor one. Also, violations of the Kolmogorov scaling were observed in these experiments.

The scalar area law was observed only for the exponential tails for large circulations; 
the tip of the PDF has more complex laws, rather than the WKB limit of the loop equation.

The scalar area is less trivial than the tensor one. The minimal area as a
functional of the loop cannot be represented as any explicit contour integral
of the Stokes type. Therefore it corresponds to an infinite number of higher
correlation functions present. Moreover, there could be several minimal
surfaces for the same loop, as the equations for the minimal surface are
nonlinear. The one with the least area should be accepted.

The condition of minimality of the area geometrically means that the normal vector is divergence-free in 3 dimensions. The surface variation can be viewed as the difference between the area of the surface $\delta_+ S $ with a little bump in the positive normal direction and the surface $\delta_- S $ with the same bump in the negative normal direction.

This difference is a surface integral of the normal vector $n_\mu$ over the closed surface $\delta S = \delta_+ S - \delta_- S$
\begin{eqnarray}
\delta A = \int_{\delta S} d \sigma_\mu n_\mu 
\end{eqnarray}
which, by the Stokes theorem, is equal to the volume integral of the divergence of vector $ N_\mu$ with boundary value $n_\mu$ at $\delta S$.
\begin{eqnarray}
&&\delta A = \int_{\delta V} \d_\mu N_\mu ;\\
&& \d \delta V = \delta S;\\
&& N_\mu(\delta S) = n_\mu
\end{eqnarray}

The condition $\d_\mu N_\mu =0$ is equivalent to the usual requirement that the trace of external curvature tensor equals zero. We discuss this in the next sections in more detail.

In our case, this normal vector represents the vorticity in the local neighborhood of the surface. Thus, the vorticity is divergent-less.

\subsection{Minimal surfaces}\label{Minimal}

Let us present the modern theoretical physicist's view of the classical Weierstrass theory of minimal surfaces. 
One can describe the minimal surface by a parametric equation
\begin{equation}
S: \ral = X_{\alpha}\left(\xi_1,\xi_2\right)
\end{equation}
The function $ X_{\alpha}(\xi) $ should provide the minimum to the
area functional
\begin{equation}
A[X] = \int_S \sqrt{d \sigma_{\mu\nu}^2} = \int d^2 \xi \sqrt{\mbox{Det } G}
\end{equation}
where
\begin{equation}
G_{ab} = \d_a X_{\mu}\d_b X_{\mu},
\end{equation}
is the induced metric.
For general studies, it is sometimes convenient to introduce the unit
tangent tensor as an independent field and minimize
\begin{equation}
A\left[X,t,\lambda\right] = \int d^2 \xi  \left(\,e_{ab} \d_a X_{\mu}
\d_b X_{\nu} \, t_{\mu\nu} + \lambda \left(1-  t_{\mu\nu}^2 \right) \right)
\end{equation}
{}From the classical equations, we will find then
\begin{equation}
t_{\mu\nu} = \frac{e_{ab}}{2\lambda}  \d_a X_{\mu} \d_b X_{\nu}\;;
t_{\mu\nu}^2 = 1,
\end{equation}
which shows equivalence to the old definition.

For the actual computation of the minimal area, it is convenient to introduce
the auxiliary internal metric  $ g_{ab} $
\begin{equation}
A\left[X,g\right] = \oh \int_S d^2 \xi  \, \tr{ g^{-1} G} \,\sqrt{\mbox{Det }
g}.
\label{gG}
\end{equation}
The straightforward minimization by $ g_{ab} $ yields
\begin{equation}
 g_{ab} \,\tr{ g^{-1} G} =  2 G_{ab},
\end{equation}
which has a family of solutions
\begin{equation}
g_{ab} = \lambda G_{ab}.
\end{equation}
The local scale factor $ \lambda $ drops from the area functional, and we
recover the original definition. So, we could first minimize the quadratic
functional \rf{gG} of $ X(\xi) $ (the linear problem) and
then minimize by $ g_{ab} $ (the nonlinear problem).

The crucial observation is the possibility of choosing conformal
coordinates, with the diagonal metric tensor
\begin{equation}
g_{ab} = \delta_{ab} \rho,\; g^{-1}_{ab} = \frac{\delta_{ab}}{\rho},\;
\sqrt{\mbox{Det } g} = \rho;
\end{equation}
after which the local scale factor $\rho$ drops from the integral
\begin{equation}
A[X,\rho] = \oh \int_S d^2 \xi \d_a X_{\mu} \d_a X_{\mu}.
\end{equation}
However, the $ \rho $ field is implicitly present in the problem,
through the boundary conditions.

Namely, one has to allow an arbitrary parametrization of the boundary curve
$ C $. We shall use the upper half plane of $ \xi$ for our surface, so the
boundary curve corresponds to the real axis $ \xi_2 = 0 $. The Boundary
condition will be
\begin{equation}
X_{\mu}(\xi_1,+0) = C\left(f(\xi_1)\right),
\label{Bcon}
\end{equation}
where the unknown function $ f(t) $ is related to the boundary value of $
\rho$ by the boundary condition for the metric
\begin{equation}
g_{11}  = \rho =  G_{11} = \left(\d_1 X_{\mu}\right)^2 = C_{\mu}'^2 f'^2
\end{equation}
As it follows from the initial formulation of the problem, one should now
solve the linear problem for the $ X $ field, compute the area and
minimize it as a functional of $ f(.) $. As we shall see below, the
minimization condition coincides with the diagonality of the metric at the
boundary
\begin{equation}
\left[\d_1X_{\mu} \d_2 X_{\mu}\right]_{\xi_2=+0} = 0
\label{Diag}
\end{equation}

The linear problem is nothing but the Laplace equation $ \d^2 X = 0 $ in the
upper half
plane with the Dirichlet boundary condition \rf{Bcon}. The solution is
well known
\begin{equation}
X_{\mu}(\xi) = \int_{-\8}^{+\8} \frac{d t}{\pi} \frac{C_{\mu}\left(f(t)\right)
\,
\xi_2}{\left(\xi_1-t\right)^2 + \xi_2^2}
\end{equation}
The area functional can be reduced to the boundary terms by virtue of the
Laplace equation
\begin{equation}
A[f]  = \oh \int d^2 \xi \d_a \left(X_{\mu} \d_a X_{\mu}\right) = -\oh
\int_{-\8}^{+\8} d \xi_1 \left[X_{\mu} \d_2  X_{\mu}\right]_{\xi_2 = +0}
\end{equation}
Substituting here the solution for $ X $, we find
\begin{equation}
A[f] = -\frac{1}{2\pi}\, \Re \int_{-\8}^{+\8}d t \int_{-\8}^{+\8} d t'
\frac{C_{\mu}(f(t))\, C_{\mu}(f(t'))}{(t-t'-\i0)^2}
\end{equation}
This formula can be rewritten in a nonsingular form
\begin{equation}
A[f] = \frac{1}{4\pi}\,  \int_{-\8}^{+\8}d t \int_{-\8}^{+\8} d t'
\frac{\left(C_{\mu}(f(t))- C_{\mu}(f(t'))\right)^2}{(t-t')^2}
\end{equation}
which is manifestly positive.

Another nice form can be obtained by integrating by-parts
\begin{equation}
A[f] = \frac{1}{2\pi}\,  \int_{-\8}^{+\8}d t f'(t) \int_{-\8}^{+\8} d t'
f'(t') C'_{\mu}(f(t))\, C'_{\mu}(f(t')) \log | t-t'|
\end{equation}
This form allows one to switch to the inverse function $ \tau(f) $, which is more convenient for optimization
\begin{equation}
A[\tau] = \frac{1}{2\pi}\,  \int_{-\8}^{+\8}d f \int_{-\8}^{+\8} d f'
 C'_{\mu}(f)\, C'_{\mu}(f') \log |\tau(f)-\tau(f')|
\end{equation}

In the above formulas, it was implied that $ C(\8) = 0 $. One could switch
to more traditional circular parametrization by mapping the upper half
plane inside the unit circle
\begin{equation}
\xi_1 + \i \xi_2 = \i \frac{1-\omega}{1+\omega} \;; \omega = r e^{\i
\alpha}\; ; r \le 1.
\end{equation}
The real axis is mapped at the unit circle. Changing variables in above
integral we find
\begin{equation}
X_{\mu}(r,\alpha) =  \Re \int_{-\pi}^{\pi}  \frac{d \theta}{\pi}
C_{\mu}(\phi(\theta))
\left(\frac{1}{1- r\exp{\i\alpha- \i\theta}}- \frac{1}{1 +
\exp{-\i\theta}}\right)
\end{equation}
Here
\begin{equation}
\phi(\theta) = f\left(\tan{\frac{\theta}{2}}\right).
\end{equation}
The last term represents an irrelevant translation of the surface, so it
can be dropped. The resulting formula for the area reads
\begin{equation}
A[\phi] = \frac{1}{4 \pi} \int_{-\pi}^{\pi} d \theta \int_{-\pi}^{\pi} d
\theta'
\frac{\left(C_{\mu}(\phi(\theta))-
C_{\mu}(\phi(\theta'))\right)^2}{\left| e^{\i \theta} - e^{\i \theta'}
\right|^2}
\end{equation}
or, after integration by parts and inverting parametrization
\begin{equation}
A[\theta] = \frac{1}{2 \pi} \int_{-\pi}^{\pi} d \phi \int_{-\pi}^{\pi} d \phi'
C'_{\mu}(\phi)\,C'_{\mu}(\phi') \log \left| \sin \frac{\theta(\phi) -
\theta(\phi')}{2} \right|
\end{equation}

Let us now minimize the area as a functional of the boundary
parametrization $ f(t) $ (we shall stick to the upper half plane). The
straightforward variation yields
\begin{equation}
0 = \Re \int_{-\8}^{+\8} d t' \frac{C_{\mu}(f(t'))\,
C_{\mu}'(f(t))}{(t-t'+\i 0)^2}
\label{NL}
\end{equation}
which duplicates the above diagonality condition \rf{Diag}.
Note that in virtue of this condition, the normal vector $ n_{\mu}(x) $ is
directed towards $ \d_2 X_{\mu} $ at the boundary. Explicit formula reads
\begin{equation}
n_{\mu}\left(C(f(t))\right) \propto \Re \int_{-\8}^{+\8} d t'
 \frac{C_{\mu}(f(t'))}{(t-t'+\i 0)^2}
\end{equation}

Let us have a closer look at the remaining nonlinear integral equation
\rf{NL}. In terms of inverse parametrization, it reads
\begin{equation}
0 = \Re\int_{-\8}^{+\8} d f \frac{ C'_{\mu}(f) C'_{\mu}(f')}{\tau(f)
-\tau(f') + \i 0}
\end{equation}
Introduce the vector set of  analytic   functions
\begin{equation}
F_{\mu}(z) =
\int_{-\8}^{+\8}\frac{d f}{\pi} \frac{C'_{\mu}(f)}{\tau(f)-z}
\end{equation}
which decrease as $ z^{-2} $ at infinity. The discontinuity in the real
axis
\begin{equation}
\Im F_{\mu}(\tau + \i 0) = C_{\mu}'(f) f'(\tau)
\end{equation}
Which provides the implicit equation for the parametrization $ f(\tau) $
\begin{equation}
\int d \tau \Im F_{\mu}(\tau +\i0) = C_{\mu}(f)
\end{equation}
We see that the imaginary part points in the tangent direction at the
boundary. As for the boundary value of the real part of $ F_{\mu}(\tau) $ it
points in the normal direction along the surface
\begin{equation}
\Re F_{\mu} \propto n_{\mu}
\end{equation}
Inside the surface, there is no direct relation between the derivatives of $
X_{\mu}(\xi) $ and $ F_{\mu}(\xi) $.

The integral  equation \rf{NL} reduces to the trivial boundary condition
\begin{equation}
 F_{\mu}^2(t+\i0) =  F_{\mu}^2(t-\i0)
\end{equation}
In other words, there should be no discontinuity of $ F_{\mu}^2 $ at the real
axis.
Here is the solution, which is analytic in the upper half plane and decreases at infinity as $
z^{-2} $:
\begin{equation}
F_{\mu}^2(z) = (1+ \omega)^4\,P(\omega);\; \omega = \frac{\i - z}{\i + z}
\end{equation}
where $ P(\omega) $ defined by a series, convergent at $ |\omega| \le 1 $.
In particular, this could be a polynomial.
The coefficients of this series should be found from an algebraic
minimization problem, which cannot be pursued forward in the general case.

The flat loops are trivial, however.
In this case, the problem reduces to the conformal transformation mapping
the loop onto the unit circle. For the unit circle, we have 
\begin{equation}
C_1 + \i C_2 = \omega;\; F_1 = \i F_2 = - \frac{(1+ \omega)^2}{2};\; P = 0.
\end{equation}
Small perturbations around  the circle or any other flat loop can be
treated systematically by a perturbation theory.

\section{Exact Scaling Index}\label{ExactIndex}
This Section is based on a recent paper published in the '20-ties, after the DNS \cite{S19} confirmed area law.
\subsection{Introduction}
Let us summarize the main equations of the loop dynamics as viewed from the 21st Century.
The basic variable in the Loop Equations is circulation around a closed loop in coordinate space.

\begin{equation}
    \Gamma_C = \oint_C \vec v d\vec r 
\end{equation}

The PDF  for  velocity circulation as a functional of the loop 
\begin{equation}
    P\left ( C,\Gamma\right) =\left < \delta\left(\Gamma - \oint_C \vec v d\vec r\right)\right>
\end{equation}
with brackets \begin{math}< > \end{math} corresponding to time average or average over random forces, was shown to satisfy certain functional equations (loop equation). 

\begin{equation}\label{LoopEqPDF}
\frac{\partial}{\partial \Gamma} \frac{\partial}{\partial t}  P\left ( C,\Gamma\right) 
=\oint_C d r_i \int d^3\rho\frac{ \rho_j }{4\pi|\vec \rho|^3}\frac{\delta^2 P(C,\Gamma)}{\delta \sigma_k(r) \delta \sigma_l(r + \rho)}\left(\delta_{i j}\delta_{k l} - \delta_{j k}\delta_{i l}\right)
\end{equation}

The area derivative is defined using the difference between $P(C+\delta C,\Gamma)- P(C,\Gamma)$ where an infinitesimal loop $\delta C$ around the 3d point $r$ is added as an extra connected component of $C$. In other words, let us assume that the loop $C$ consists of an arbitrary number of connected components $C = \sum C_k$. 
We add one more infinitesimal loop at some point away from all $C_k$. 
In virtue of the Stokes theorem, the difference comes from the circulation $\oint_{\delta C} \vec v d\vec r$, which reduces to vorticity at $r$
\begin{equation}
  P(C+\delta C,\Gamma)-P(C,\Gamma) =  d\sigma_i(r)    \left <\omega_i(r) \delta'\left(\Gamma - \oint_C \vec v d\vec r\right)\right>
\end{equation}
where
\begin{equation}
    d\sigma_k(r) = \oint_{\delta C} e_{i j k} r_i d r_j,
\end{equation}
is an infinitesimal vector area element inside $\delta C$.
In general, for the Stokes type functional, by definition:
\begin{equation}
    U[C+\delta C]-U[C] = d\sigma_i(r) \frac{\delta U[C]}{\delta \sigma_i(r)}
\end{equation}
The Stokes condition $\oint_{\delta S} d \sigma_i \omega_i =0$ for  any closed surface $\delta S$ translates into
\begin{equation}\label{StokesSurf}
   \oint_{\delta S} d \sigma_i \frac{\delta U[C]}{\delta \sigma_i(r)} =0
\end{equation}

The fixed point of the chain of the loop equations (\ref{SecLoopEq}) was shown to have solutions corresponding to two known distributions: Gibbs distribution and (trivial) global random rotation distribution. In addition, we found the third, nontrivial solution, which is an arbitrary function of minimal area \begin{math} A_C \end{math} bounded by $C$.
\begin{equation}
    P(C,\Gamma) = F\left(A_C,\Gamma\right)
\end{equation}
The Minimal area can be reduced to the Stokes functional by the following regularization
\begin{equation}\label{Regularized}
  A_C =\min_{S_C} \int_{S_C} d \sigma_{i}(r_1) \int_{S_C} d \sigma_{j}(r_2) \delta_{i j} \Delta(r_1-r_2)
\end{equation}
with 
\begin{equation}
\Delta(r) = \frac{1}{r_0^2} \exp{-\pi \frac{r^2}{r_0^2}}; r_0 \rightarrow 0
\end{equation}
representing two-dimensional delta function, and integration goes over minimized surface $S_C$ (see Fig.\ref{fig::SimpleMinimal}, created with \Mathematica{}, \cite{MinSurfaceX}).
\pct{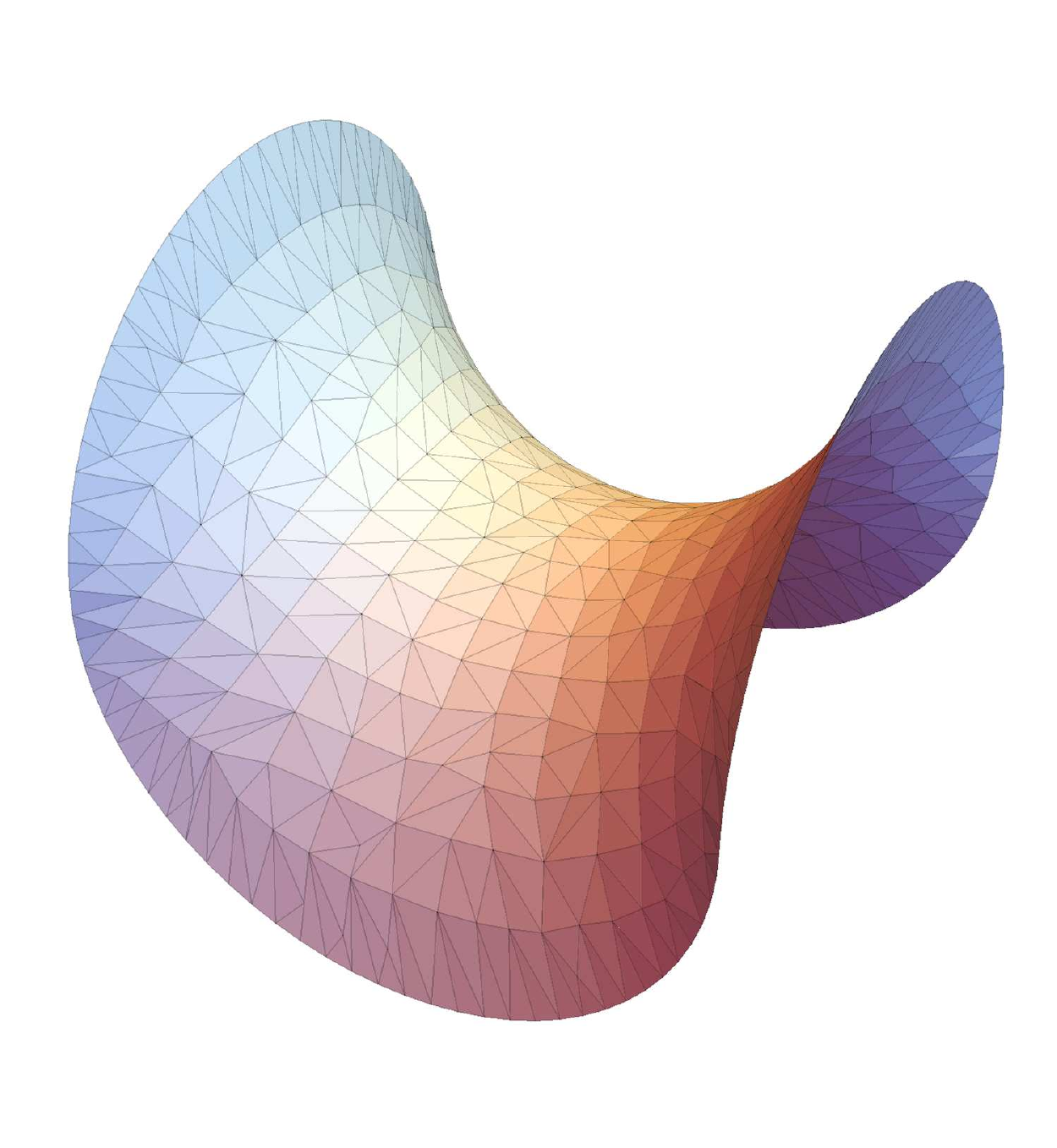}{Minimal surface, topologically equivalent to a disk (sphere with one hole), bounded by curved loop $C$.}

In the general case, the loop $C$ consists of $N$ closed pieces $C = \sum_{k=1}^N C_k$ and the surface $S_C$ must connect them all so that it is topologically equivalent to a sphere with $N$ holes and no handles (see Fig.\ref{fig::MultiLoopMinimal} created with \Mathematica \cite{MinSurfaceX}).
\pct{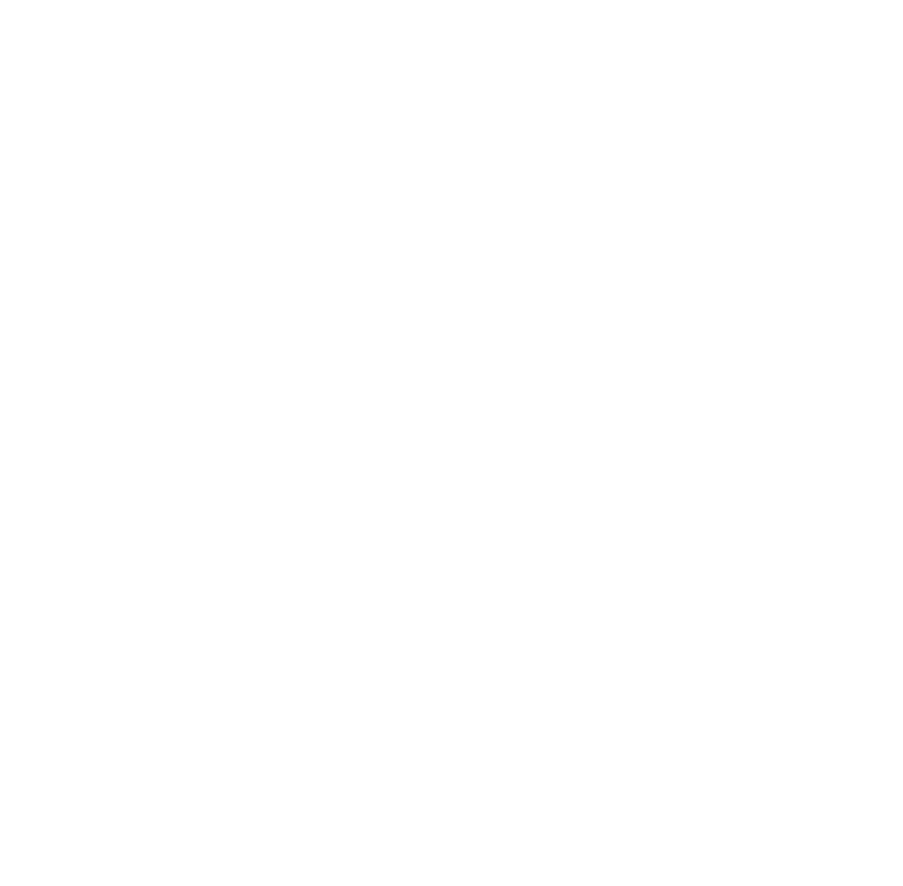}{Minimal Surface, topologically equivalent to a sphere with $N=6$ holes, bounded by loop $C= \sum_{k=1}^6 C_k$.}
 
If some pieces are far from others, the minimal surface will make thin tubes reach from one closed loop $C_k$ to another via some central hub where all the tubes grow out of the sphere.
In our case, we need only two extra little loops, both close to the initial contour $C$, but for completeness, we must assume an arbitrary number of closed pieces in $C$.

In real world this $r_0$ would be the viscous scale $\left(\frac{\nu^3}{\cal E}\right)^{\nicefrac{1}{4}}$.
This is a positive definite functional of the surface, as one can easily verify using spectral representation:
\begin{eqnarray}
    &&\int_{S_C} d \sigma_{i}(r_1) \int_{S_C} d \sigma_{j}(r_2) \delta_{i j} \Delta(r_1-r_2) \nonumber\\
    &&\propto \int d^3k \exp{-\frac{k^2r_0^2}{4\pi}} \left| \int_{S_C} d \sigma_{i}(r)e^{i k r} \right|^2
\end{eqnarray}
In the limit $r_0 \rightarrow 0$, this definition reduces to the ordinary area:
\begin{equation}
  A_C \rightarrow \min_{S_C} \int_{S_C} d^2 \xi \sqrt{g} 
\end{equation}

The Stokes condition (\eqref{StokesSurf}) follows from the extremum condition. When the surface changes into $S'$, the linear variation reduces to the surface integral (\eqref{StokesSurf}), with $\delta S = S'-S$ being the infinitesimal closed surface between $S'$ and $S$. This linear variation must vanish by the definition of the minimal surface for the regularized area and its local limit.

The area derivative of the Minimal area in regularized form, then, as before, reduces to the elimination of one integration (see \eqref{vorticity})

 With this regularization, the area derivative is defined everywhere in space but exponentially decreases away from the surface. 

Precisely at the surface, it reduces to twice the unit normal vector\footnote{As for Stokes condition $\partial_i \frac{\delta A_C}{\delta \sigma_i(r)} =0$ one can readily check in a local coordinate frame where $\vec r = (x,y,z) $ and the surface equation is $ z = \frac{1}{2} \left(k_1 x^2 + k_2 y^2\right)$, that at $r_0 \rightarrow 0$ the Stokes condition reduces to $\int_{-\infty}^{\infty}d x \int_{-\infty}^{\infty} d y \partial_z \Delta (\vec r) \propto (k_1 + k_2)=0 $ which is the well-known equation of vanishing mean curvature at a minimal surface.}.

Should we go to the limit $r_0 \rightarrow 0$ first, we would have to consider the minimal surface connecting the original loop $C$ and infinitesimal loop $\delta C$. Such a minimal surface would have a thin tube connecting the point $r$ to  $\Tilde{r}$ at the original minimal surface along its local normal $\vec n$ (see Fig.\ref{fig::Peak}).
\pct{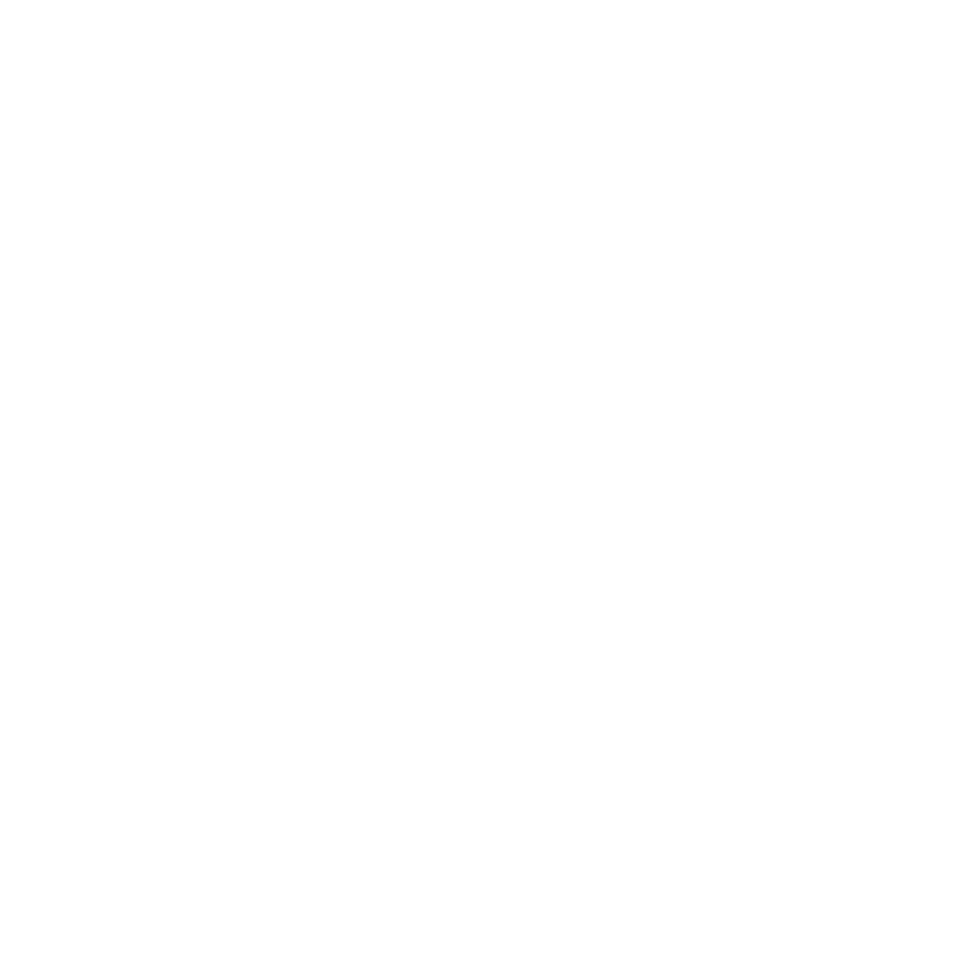}{Minimal Surface, stretched to reach a remote point (infinitesimal loop) }

We are not going to investigate this complex problem here -- with the regularized area, we have an explicit formula, and we need this formula only at the boundary in leading log approximation (see \cite{M19a}).

As we argued in the old paper \cite{M93} we expect scale-invariant solutions, depending on $\gamma =\Gamma A_C^{-\alpha}$ in our scale invariant equations, with some critical index $\alpha$, yet to be determined.
Let us stress again that the Kolmogorov value of the scaling index $\alpha_{K} = \frac{2}{3}$ does \textbf{not} follow from the loop equations; this is an additional assumption based on dimensional counting and Kolmogorov anomaly \cite{M93} for the third moment of velocity. As we stressed in that paper, the Kolmogorov anomaly poses no restrictions on the vorticity correlations and cannot be used to determine our scaling index. 

The Area law is expected to be an asymptotic solution at large enough circulations and areas where PDF is small. Such PDF tails are usually interpreted as Kelvinons or classical solutions in some variables \cite{FKLM}. In our variables, this is the minimal area as a functional of its boundary loops  $C = \sum_k C_k$. 

This Universal Area Law was confirmed in numerical experiments \cite{S19} with Reynolds up to $10^4$ with very high accuracy over the whole inertial range of circulations and areas with PDF from 1 down to $10^{-8}$. See \cite{M19a} for analysis of numerical results and their correspondence with Area Law.

Let us present a more conventional physical interpretation of this Area Law.

Let us consider the vorticity field (\eqref{vorticity}) generated by a minimal surface with thickness $r_0$. It rapidly decreases outside the surface and equals twice the normal vector at the surface.
The corresponding velocity field will be defined everywhere in space by the integral
\begin{equation}
    v_i(r) \propto  e_{i j k}\int d^3 r'\frac{r'_j -r_j}{|\vec r - \vec r'|^3} n_k(\Tilde{r}') \Delta(r-r')
\end{equation}

In particular, directly at the surface
\begin{equation}
    v_i(r) \propto r_0 e_{i j k}\int_{S_C} d^2 r'\frac{r'_j -r_j}{|\vec r - \vec r'|^3} n_k(r')
\end{equation}
If $r$ approaches the edge, this integral logarithmically diverges (we use the frame where the surface normal is directed to $z$, the local tangent to $C$ goes along $x$, and the inside tangent vector of the surface at its boundary $\d S = C$ goes along $y$, and $\rho =x,y$ are local coordinates on the surface near its edge.
\begin{eqnarray}
   && v_a(C - \epsilon) \propto r_0 e_{a b }\int_{-\infty}^{\infty} d x \int_0^\infty d y \rho_b \left((y+\epsilon)^2 + x^2\right)^{-\frac{3}{2}} \propto \nonumber\\
   &&r_0 \delta_{a 1 } \int_0^\infty d y \frac{y}{(y+\epsilon)^2} \propto r_0 \delta_{a 1 }\ln \epsilon
\end{eqnarray}
Now, as is well known (Kelvin theorem, see also \cite{M19a})
\begin{equation}
    \partial_t \Gamma \propto e_{i j k} \int_C d r_i v_j \omega_k 
\end{equation}
In our coordinate frame, using the fact that $\vec \omega \propto \vec n$ is directed along $z$
\begin{equation}
    \partial_t \Gamma \propto  \int_C d x v_2   
\end{equation}
which vanished as $v_2=0$. 
So, the circulation would be conserved for this particular vorticity distributed in a thin layer around a minimal surface.

The above computations do not apply to the vorticity field, which is spread in space away from the surface, which was the case for the Kelvinon in the previous Section.

The only way to reconcile the Kelvinon solution with the minimal area solution is to consider the case of the loop $C$, which is much larger than the width $w$ of the Kelvinon velocity and vorticity fields around the discontinuity surface $\mathcal S(C)$.

In this limit, the effective volume occupied by Kelvinon will grow as area $V \propto A[\mathcal S(C)] w$ and minimization of the Hamiltonian by the shape of $S$ would result in a minimal surface $\mathcal S(C) = \mathcal S_{min}(C)$.

\subsection{Equation for Scaling Index}

Let us start without the assumption of scaling law, with a weaker assumption of some unknown function of the minimal area as the scale of $\Gamma$
\begin{equation}\label{AutoModel}
     P(C,\Gamma)\rightarrow G(\ln A_C)\Pi\left(\Gamma G(\ln A_C)\right) 
\end{equation}
The factor of $G(\ln A_C)$ in front of scaling function $\Pi(\gamma)$ follows from the fact that $\Gamma P(C,\Gamma)$ must be scale-invariant, regardless of how effective scale $G$ depends on $\ln A_C$.

Let us derive the self-consistency equation for $G(\ln A)$.

On the one hand:
\begin{eqnarray}
    &&\left< \int_{S_C} d\vec \sigma \cdot \vec  \omega \delta'\left(\Gamma - \int_{S_C} d\vec \sigma \cdot \vec \omega\right) \right>\nonumber\\
    &&= \partial_\Gamma\left(\left< \int_{S_C} d\vec \sigma \cdot \vec  \omega \delta\left(\Gamma - \int_{S_C} d\vec \sigma \cdot \vec \omega\right)\right> \right)\nonumber\\ 
    &&=\partial_\Gamma \left(\Gamma\left< \delta\left(\Gamma - \int_{S_C} d\vec \sigma \cdot \vec \omega\right)\right>\right)\nonumber\\
    &&=\partial_\Gamma \left(\Gamma G(\ln A_C) \Pi\left(\Gamma G(\ln A_C)\right) \right) \label{One}
\end{eqnarray}
On another hand:
\begin{eqnarray}
    &&\left< \int_{S_C} d\vec \sigma\cdot \vec  \omega \delta'\left(\Gamma - \int_{S_C} d\vec \sigma \cdot \vec \omega\right) \right>\nonumber\\
    &&= -\left<\int_{S_C} d\vec \sigma \cdot  \frac{\delta}{\delta \vec \sigma} \delta\left(\Gamma - \int_{S_C} d\vec \sigma \cdot \vec \omega\right)\right>\nonumber\\
   &&= -\int_{S_C} d\vec \sigma \cdot \frac{\delta}{\delta \vec \sigma}\left< \delta\left(\Gamma - \int_{S_C} d\vec \sigma \omega\right)\right>\nonumber\\
    &&= -\int_{S_C} d\vec \sigma \cdot  \frac{\delta}{\delta \vec \sigma}\left( G(\ln A_C) \Pi\left(\Gamma G(\ln A_C)\right) \right)\label{Second}
\end{eqnarray}
Recall that at the minimal surface
\begin{align}
    \frac{\delta A_C}{\delta \vec \sigma} &= 2 \vec n\\
    \int_{S_C} d\vec \sigma \vec n &= A_C
\end{align}

Moreover, we find in (\ref{One}):
\begin{equation}
    G \left( \Pi(\Gamma G) + \Gamma G \Pi'(\Gamma G)\right)
\end{equation}
and in (\ref{Second}):
\begin{equation}
   -2 \frac{\partial}{\partial \ln A_C} \left( G(\ln A_C) \Pi\left(\Gamma G(\ln A_C)\right) \right)=
   -2G'\left( \Pi(\Gamma G) + \Gamma G \Pi'(\Gamma G)\right)
\end{equation}
Comparing these two expressions, we find the differential equation
\begin{equation}
    G' = -\frac{1}{2} G
\end{equation}
with solution
\begin{equation}
    G \propto A_C^{-\frac{1}{2}}
\end{equation}
Note that there is no restriction on the PDF scaling function $\Pi(\gamma)$.

Note also that as $\Gamma$ changes sign on time reversal, we expect the dissipation reflects itself in the asymmetry of PDF. Indeed, as measured in \cite{S19} and discussed in the \ref{KelvDNS}, the left tail of the PDF differs from the right tail.

\subsection{Higher Correlations}\label{HiCor}
Now let us check that this remarkable solution is compatible with the higher correlations.

Let us recall  the results  \cite{M19a}:
\begin{equation}\label{MultiNormal}
    \left < \vec \omega_1\dots \vec\omega_k \delta\left(\Gamma - \oint_C \vec v d\vec r\right)\right >  = \vec n_1 \dots \vec n_k A_C^{-\alpha - k(1-\alpha)}\Omega_k\left(\Gamma A_C^{-\alpha}\right)
\end{equation}
The scaling functions $\Omega_k(\gamma)$ with $\Omega_0(\gamma) = \Pi(\gamma)$ being scaling PDF, satisfy recurrent equations :
\begin{equation}\label{Recurrent}
    \Omega_{k+1}\left(\gamma\right)= 2 \alpha \gamma\Omega_{k}\left(\gamma\right)  -2(1-\alpha)k\int_\gamma^{\pm \infty}\Omega_{k}(y) \,d y
\end{equation}
 \begin{equation}
   \left < \delta\left(\Gamma - \oint_C \vec v d\vec r\right)\right> = A_C^{-\alpha}\Pi\left(\frac{\Gamma}{A_C^{\alpha}}\right)
 \end{equation}
 \begin{equation} 
     \left < \vec \omega \delta\left(\Gamma - \oint_C \vec v d\vec r\right)\right> =2\alpha\vec n \frac{\Gamma}{A_C} \Pi\left(\frac{\Gamma}{A_C^{\alpha}}\right)
 \end{equation}
 
 From the original derivation using the area derivative, it follows that this equation holds on \textbf{the whole surface} as well as at its edge $C$.
 Therefore, as pointed out in \cite{M19a} we can integrate this equation over the surface:
 \begin{equation}
   \left <  \int_{S_C} d \vec \sigma(r) \vec \omega(r) \delta\left(\Gamma - \oint_C \vec v d\vec r\right)\right> =2\alpha\int_{S_C} d \vec \sigma(r)\vec n(r) \frac{\Gamma}{A_C} \Pi\left(\frac{\Gamma}{A_C^{\alpha}}\right)
 \end{equation}
 
 On the left side we obtain $\int_{S_C} d \vec \sigma(r) \vec \omega(r) = \Gamma$ in virtue of the $\delta$ function, and on the right side we get
 $\int_{S_C} d \vec \sigma(r)\vec n(r) = A_C$. As a result, after canceling $\Gamma$, we obtain the equation:
 \begin{equation}
     \Pi\left(\frac{\Gamma}{A_C^{\alpha}}\right) = 2 \alpha \Pi\left(\frac{\Gamma}{A_C^{\alpha}}\right)
 \end{equation}
from which we conclude that
\begin{equation}
    \alpha = \frac{1}{2}
\end{equation}
as we already found in the previous section \ref{ExactIndex}.

With the next equation, though, things are getting tricky, and this is where we got stuck in \cite{M19a}:
 \begin{equation} 
     \left < \omega_1 \omega_2 \delta\left(\Gamma - \oint_C \vec v d\vec r\right)\right> = n_1  n_2  A_C^{-\frac{3}{2}}\Omega_2\left(\Gamma A_C^{-\frac{1}{2}}\right)
 \end{equation}
with 
\begin{equation}\label{Omega2}
    \Omega_2(\gamma) = \gamma^2\Pi(\gamma) -\int_{ \gamma}^{\pm\infty} \Pi(y)y \,d y 
\end{equation}
Integrating over the surface, we get, as before, after going to scaling variables and setting $\alpha=\frac{1}{2}$, on the left :
\begin{equation}
    \gamma^2\Pi(\gamma)
\end{equation}
and on the right
\begin{equation}
    \gamma^2\Pi(\gamma)  -\int_{ \gamma}^{\pm\infty} \Pi(y)y \,d y 
\end{equation}
Everything stops! The left part does not match the right part.

Here is what we should have included: the contact terms. The correlations of vorticity were obtained, assuming the points do not coincide. In general, we can expect extra contact term:
\begin{equation} \label{ContactTerm}
     \left < \omega_1 \omega_2 \delta\left(\Gamma - \oint_C \vec v d\vec r\right)\right> = n_1  n_2  A_C^{-\frac{3}{2}}\Omega_2\left(\Gamma A_C^{-\frac{1}{2}}\right) + X \delta_{1 2} \delta^2(1-2)
 \end{equation}
where $X$ is to be determined, $\delta_{1 2}$ is Kronecker delta for vector indexes, and $\delta^2(1-2)$ is an invariant delta function on the surface. These contact terms display themselves only in the integral relations we have here and do not change correlations at far away points (which was assumed in \cite{M19a}).
With this term present, we get a perfect match if
\begin{equation}
    X = \int_{ \gamma}^{\pm\infty} \Pi(y)y \,d y 
\end{equation}
What could be the origin of such a term? Let us consider the second derivative of (\ref{ContactTerm}) by $\Gamma$. On the left, we find 
\begin{eqnarray}
    &&\left < \omega_1 \omega_2 \delta''\left(\Gamma - \oint_C \vec v d\vec r\right)\right>=
    \frac{\delta^2}{\delta \sigma(1)\delta \sigma(2)}\left < \delta\left(\Gamma - \oint_C \vec v d\vec r\right)\right> 
    =\nonumber\\
    &&\frac{\delta^2}{\delta \sigma(1)\delta \sigma(2)} A_C^{-\frac{1}{2}} \Pi\left(\frac{\Gamma}{A_C^{\frac{1}{2}}}\right)
\end{eqnarray}
The contact term precisely of the form we need comes from the second functional derivative of $A_C$
\begin{equation}
    \frac{\delta^2 A_C}{\delta \sigma(1)\delta \sigma(2)}\partial_{A} A^{-\frac{1}{2}} \Pi\left(\Gamma A^{-\frac{1}{2}}\right) =  2\delta_{1 2} \delta^2(1-2)\partial_{A} A^{-\frac{1}{2}} \Pi\left(\Gamma A^{-\frac{1}{2}}\right)
\end{equation}

The rest of the terms were accounted for in recurrent equations for vorticity expectation values in \cite{M19a}, but this one was missed (or, better to say, ignored as we assumed all points were separate).
The derivative of the $X$ term matches this one as
\begin{equation}
    \partial_\gamma^2\int_{ \gamma}^{\pm\infty} \Pi(y)y \,d y  = - \left(\Pi(\gamma)+ \gamma \Pi'(\gamma)\right)
\end{equation}
and
\begin{equation}
  2 \partial_{A} A^{-\frac{1}{2}} \Pi\left(\Gamma A^{-\frac{1}{2}}\right)\propto -\left(\Pi(\gamma)+ \gamma \Pi'(\gamma)\right)
\end{equation}
Here we dropped factors of $A_C$ as they are known to match by dimensional counting.
So, the contact terms from the second functional derivative precisely match the missing terms in the self-consistency relation for the two-point function.

Let us present the result for the next correlation functions with corrected coefficients at $\alpha= \frac{1}{2}$, which were created in \Mathematica\ using symbolic integration by parts:
\begin{align}\label{OmegaRelations}
\Omega_{1}(\gamma) &= \gamma  \Pi(\gamma)\\
\Omega_{2}(\gamma) &= \gamma ^2 \Pi(\gamma)-\int_{\gamma }^{\pm \infty } y \Pi (y) \, d y\\
\Omega_{3}(\gamma) &= \gamma ^3 \Pi(\gamma)-3 \int_{\gamma }^{\pm \infty } \gamma  y \Pi (y) \, d y\\
\Omega_{4}(\gamma) &= \gamma ^4 \Pi(\gamma)+\frac{3}{2} \int_{\gamma }^{\pm \infty } y \left(y^2-5 \gamma ^2\right) \Pi (y) \, d y\\
\Omega_{5}(\gamma) &= \gamma ^5 \Pi(\gamma)+\frac{5}{2} \int_{\gamma }^{\pm \infty } \gamma  y \left(3 y^2-7 \gamma ^2\right) \Pi (y) \, d y\\
\Omega_{6}(\gamma) &= \gamma ^6 \Pi(\gamma)-\frac{15}{8} \int_{\gamma }^{\pm \infty } y \left(21 \gamma ^4+y^4-14 \gamma ^2 y^2\right) \Pi (y) \, d y\\
\Omega_{7}(\gamma) &= \gamma ^7 \Pi(\gamma)-\frac{21}{8} \int_{\gamma }^{\pm \infty } \gamma  y \left(33 \gamma ^4+5 y^4-30 \gamma ^2 y^2\right) \Pi (y) \, d y\\
&\dots
\end{align}

\subsection{Matching Kelvinon with Area Law and multifractals}

The Kelvinon solution \ref{Kelvinon} for the turbulent statistics is complementary to the above Area law solution.

With the Kelvinon, we established the asymptotic exponential decay of the PDF, with pre-exponential factor $1/\sqrt{|\Gamma|}$ and understood the origin of the asymmetry (i.e. time-reversal violation in anomalous Euler Hamiltonian). The $C$ -dependence in that approach was established only for small loops compared with intrinsic scale $L$, where the K41 scaling laws apply, modified by the powers of the logarithm of scale.

With the Loop equation, in the opposite limit of large smooth loops, we can establish the $C-$ dependence but not the $\Gamma-$ dependence.

Matching the two solutions, we arrive at the asymptotic expansion of the scaling function $\Pi(\gamma)$ of the previous Section:
\begin{eqnarray}\label{matchKL}
    &&\Pi(\gamma) \propto \frac{\exp{ \gamma_c-\gamma}}{\sqrt{\gamma-\gamma_c}} \sum_0^\infty F_n \left(\frac{1}{\gamma-\gamma_c}\right)^n
\end{eqnarray}
The coefficients $F_n$ are computed in \ref{AsympPDF}, starting with
\begin{eqnarray}
   && F_0 =1,\\
   && F_1 =\frac{ c (a+b) -2 a b}{4 (c-a) (c-b)}.
\end{eqnarray}

This formula applies to asymptotic tails at both $\gamma, |C|$ large. Large values mean the Reynolds number $|\Gamma|/\nu$ above a certain critical threshold (see \ref{multifrac}).

The change of regime was observed in \cite{S19,S21} and was interpreted first as bi-fractal, then as a new phase of turbulence.

In the Kelvinon solution, this change of regime corresponds to a single Kelvinons dominating the asymptotic decay of the PDF. As we saw in \ref{multifrac}, each Kelvinon has a critical value of the circulation, below which value its contribution to the PDF tail vanishes.

This critical value $\tau$ has a different dependence on a scale compared to the decrement of the exponential (which we have normalized to $1$ in this solution).

The parameter $\gamma_c$ here  equals the ratio of these two scales $\gamma_c = \frac{\tau}{c}$. In the asymptotically free region of small loops, this ratio is proportional to some power of the logarithm of scale.

Now we can substitute our solution into the general relation \eqref{OmegaRelations}. The integrals reduce to the error function, but only the asymptotic limits are relevant here. We find these limits by expanding in $y-\gamma$ and solving recurrent equation for $\Omega_n$:
\begin{eqnarray}\label{LKasympt}
\Omega_n \ra \gamma^{n - \oh}e^{\gamma_c-\gamma} \left( 1 + \frac{2 F_1 + \gamma_c - n(n-1)}{2\gamma}\right);
\end{eqnarray}

We also can compute an effective scaling index for the circulation in the region where the area law holds
\begin{equation}\label{zeta}
    \zeta(p) = \frac{d \log \left<|c \gamma|^p\right>}{d \log \sqrt{A_C}}
\end{equation}

Using exponential tail \eqref{matchKL} and computing the integral for moments by the saddle point at large $p$ we find the following expansion in inverse powers of  $p$
\begin{eqnarray}\label{zetaexp}
    &&\zeta(p) \sim  p \pbyp{\log c}{\log \sqrt{A_C}} + \left(1 + \frac{1}{2 p}\right) \pbyp{\gamma_c}{\log \sqrt{A_C}} + \O{1/p^2}\ra\nonumber\\
    && p + f(\log A_C)\left(1 + \frac{1}{2 p}\right) + \O{1/p^2}
\end{eqnarray}

The first term is equal to $p$ provided the decrement $c$ scales as $\sqrt{A_C}$. The next term represents a correction to our scaling law at large $p$.
The function $f(\log A_C)$ is known in the perturbative region, where it decreases as some negative power of $\log |C|$ (see \eqref{taulambdalog}).
\begin{eqnarray}
    &&f(\log |C|) \sim \pbyp{(\log |C|)^\mu}{\log |C|} \propto (\log |C|)^{\mu -1};\\
    &&\mu = \frac{1}{6} \textit{ if } m n > 0 \textit{ else } \tt
\end{eqnarray}

The asymptotic dependence $\zeta(p) \ra p + \textit{const}$   can be verified directly by studying the ratio of higher moments (related to the tails of the PDF). 
We compared the raw data from \cite{S19} with this prediction. We took the ratio of the moments $M_p = \VEV{\Gamma^p}$ at largest available $p$ and defined the circulation scale as 
\begin{equation}
    S = \sqrt{\frac{M_8}{M_6}} \sim \left(\sqrt{A_C}\right)^{\frac{\zeta(8)-\zeta(6)}{2}}. 
\end{equation}

We fitted using \Mathematica{} $S(r)$ as a function of the size $r = \frac{a}{\eta}$ of the square loop measured in the Kolmogorov scale $\eta$.
The quality of a linear fit was very high with adjusted $R_2=0.9996$. 
The so-called ANOVA table summarizes this fit as follows
\begin{equation}\label{LinearFitR}
\begin{array}{cccccc}
 \text{} & \text{DF} & \text{SS} & \text{MS} & \text{F-Statistic} & \text{P-Value} \\
 R & 1 & 2.30567 & 2.30567 & 22984.4 & \text{4.008150477989548$\grave{ }$*${}^{\wedge}$-15} \\
 \text{Error} & 8 & 0.000802519 & 0.000100315 & \text{Null} & \text{Null} \\
 \text{Total} & 9 & 2.30648 & \text{Null} & \text{Null} & \text{Null} \\
\end{array}
\end{equation}
The linear fit is shown in Fig.\ref{fig::FitSqrtArea}
\pct{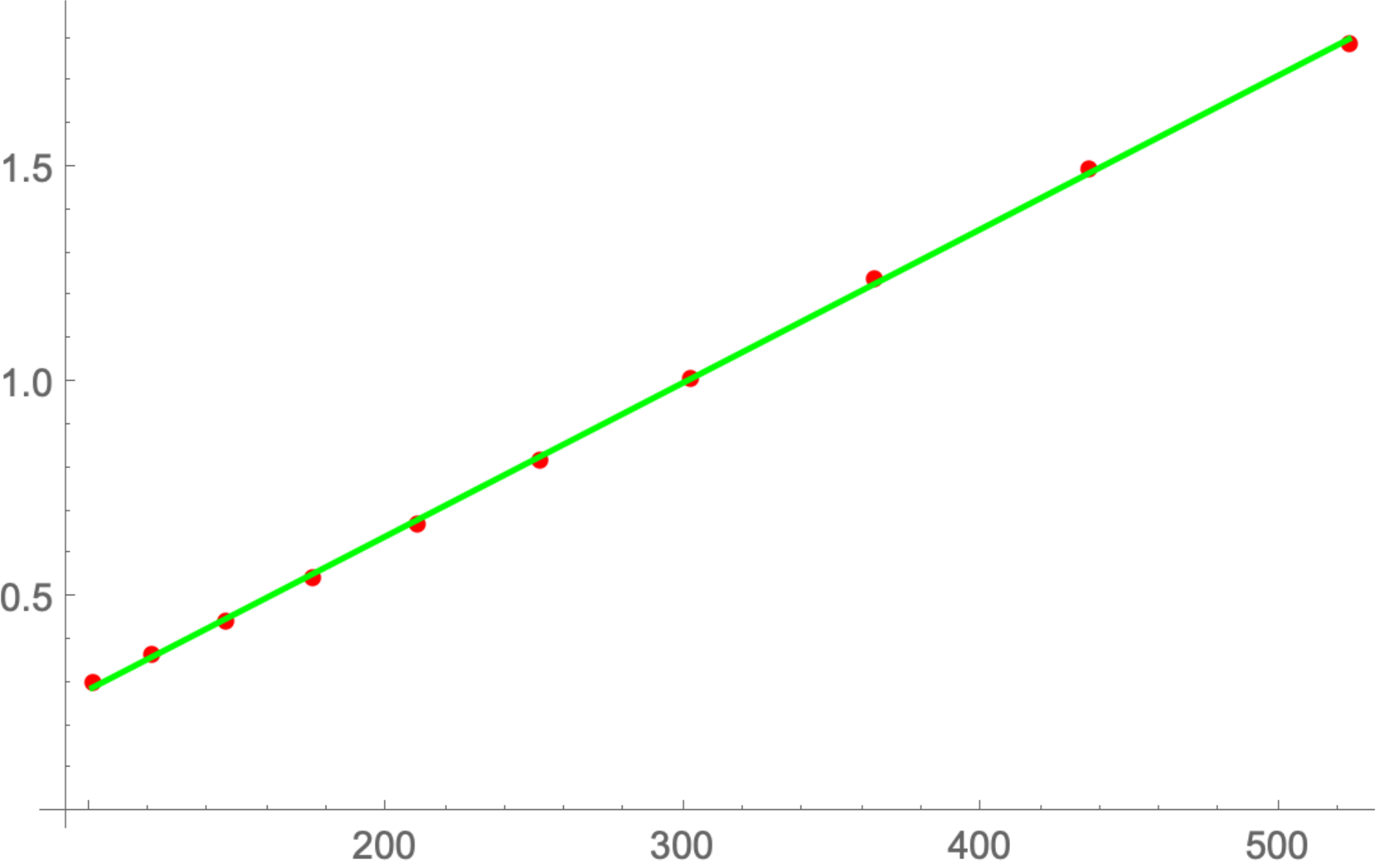}{Linear fit of the circulation scale $S = \sqrt{\frac{M_8}{M_6}}$ (with $M_p = \VEV{\Gamma^p}$) as a function of the $a/\eta$ for inertial range $ 100 \le a/\eta \le 500$. Here $a$ is the side of the square loop $C$, and $\eta$ is a \KO{} scale. The linear fit $S = -0.073404 + 0.00357739 a/\eta$ is almost perfect:  adjusted $R_2 = 0.999609$}.

The errors are likely artifacts of random forcing at an $8K$ cubic lattice.

This is not to say that some other nonlinear formulas cannot fit this data equally well or maybe even better; for example, fitting $\log S$ by $\log R$ would produce a very good linear fit with $\zeta(8) -\zeta(6)\approx 2.2  $ instead of our prediction $2$. 

Data fitting cannot derive the physical laws -- it can only verify them against some null hypothesis, especially in the presence of a few percent of systematic errors related to finite size effects and harmonic quasi-random forcing. We believe that distinguishing between $2.2$ and $2$ is an over-fit in such a case.

\subsection{Discussion}
The main result of this Section is an exact computation of the asymptotic scaling law  $\Gamma \sim \sqrt{Area}$ from self-consistency of the Minimal Area solution of the Loop Equations. With correct recurrent equations (\ref{Recurrent}), the contact terms needed for consistency of surface integrals of the vorticity correlation functions naturally arise from second functional derivatives of the regularized area (\eqref{Regularized}).

The ends start meeting in this exotic solution, which initially raised so much confusion 26 years ago.

If we start believing the Minimal surface, we have to try and see how the correlation between two circulations $\Gamma_1, \Gamma_2$ around planar loops in the $x y$ plane depends upon the separation between these loops in the orthogonal direction $z$. (see Fig.\ref{fig::Tube}).
\pct{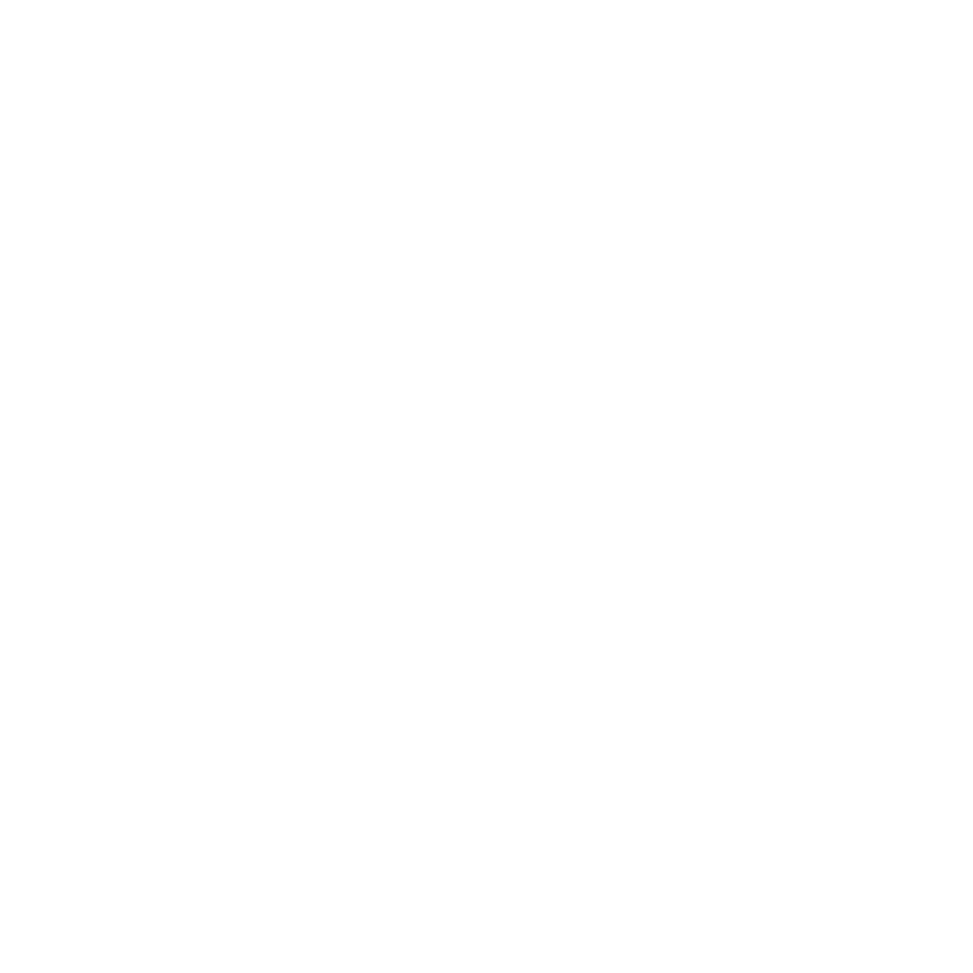}{Tube-like minimal surface connecting two separated loops.}

When the small loop starts inside the big one and moves in the $z$ direction, the minimal surface will grow like a tower between these two loops. In the limit of a small loop much less than the big one (see Fig.\ref{fig::Peak}), we must approach the vorticity correlation as a function of the normal distance to the minimal surface for the large square (see Fig. \ref{fig::Peak}). How does that correlation depend on $z$? 

Also, the "soccer gate" loop (made of two perpendicular squares touching one side) makes an even more interesting test than we suggested before. One could verify that mean vorticity is directed along the normal to the minimal surface anywhere at this surface, not just at the edge. Here is the minimal Surface for Soccer Gates loop, created with \Mathematica\ package \cite{MinSurfaceX}(see Fig. \ref{fig::SoccerGates}).
\pct{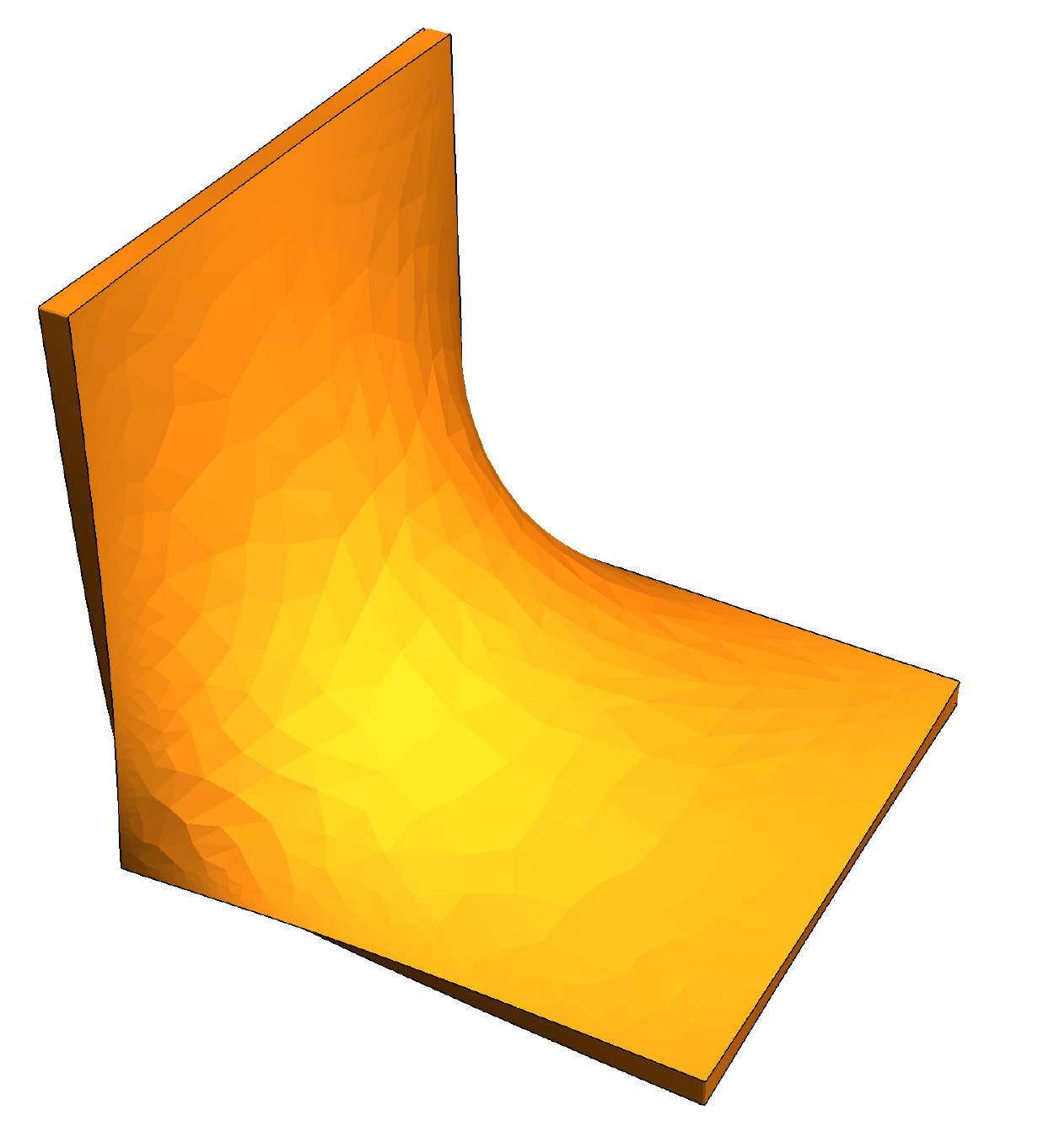}{Minimal surface bounded by soccer gates. Artificial thickness imitates viscosity effects.}

In the next Section, we analyze  the minimal area solution for minimal curved surfaces.
\section{Area Law for curved surfaces}
This Section is based on a recent paper published in the '20-ties, after the DNS \cite{S19} confirmed area law.
\subsection{Introduction}
In the previous two papers \cite{M19a},\cite{M19b}, we reviewed and advanced the Minimal Area Solution \cite{M93} for the Loop Equation in turbulence, which was recently verified experimentally \cite{S19}.
Let us repeat this theory's latest revision before advancing it further.

The basic variable in the Loop Equations is circulation around a closed loop in coordinate space $\oint_C \val d \ral$.

The results of this Section only apply to large smooth loops $C$ compared to an effective width of the stationary solution (Kelvinon) for the Euler vorticity surrounding the discontinuity surface. In this limit, the variational estimates say that the surface has a shape of a minimal surface bounded by $C$.

The PDF  for  velocity circulation as a functional of the loop 
\begin{equation}
    P\left ( C,\Gamma\right) =\left < \delta\left(\Gamma - \oint_C \vec v d\vec r\right)\right>
\end{equation}
with brackets \begin{math}< > \end{math} corresponding to time average or average over random forces, was shown to satisfy certain functional equations (loop equation). 

\begin{equation}\label{LoopEq0}
\frac{\partial}{\partial \Gamma} \frac{\partial}{\partial t}  P\left ( C,\Gamma\right) 
=\oint_C d r_i \int d^3\rho\frac{ \rho_j }{4\pi|\vec \rho|^3}\frac{\delta^2 P(C,\Gamma)}{\delta \sigma_k(r) \delta \sigma_l(r + \rho)}\left(\delta_{i j}\delta_{k l} - \delta_{j k}\delta_{i l}\right)
\end{equation}

The area derivative is defined using the difference between $P(C+\delta C,\Gamma)- P(C,\Gamma)$ where an infinitesimal loop $\delta C$ around the 3d point $r$ is added as an extra connected component of $C$. In other words, let us assume that the loop $C$ consists of an arbitrary number of connected components $C = \sum C_k$. 
We add one more infinitesimal loop at some point away from all $C_k$. 
In virtue of the Stokes theorem, the difference comes from the circulation $\oint_{\delta C} \vec v d\vec r$, which reduces to vorticity at $r$
\begin{equation}
  P(C+\delta C,\Gamma)-P(C,\Gamma) =  d\sigma_i(r)    \left <\omega_i(r) \delta'\left(\Gamma - \oint_C \vec v d\vec r\right)\right>
\end{equation}
where
\begin{equation}
    d\sigma_k(r) = \frac{1}{2}\oint_{\delta C} e_{i j k} r_i d r_j
\end{equation}
is an infinitesimal vector area element inside $\delta C$.
In general, for the Stokes type functional, by definition:
\begin{equation}
    U[C+\delta C]-U[C] = d\sigma_i(r) \frac{\delta U[C]}{\delta \sigma_i(r)}
\end{equation}
The Stokes condition $\partial_i \omega_i(r) =0$ translates into
\begin{equation}\label{StokesCond}
   \partial_i \frac{\delta U[C]}{\delta \sigma_i(r)} =0
\end{equation}
where $\partial_i = \frac{\partial}{\partial r_i}$ is an ordinary spatial derivative rather than a singular loop derivative introduced in the Non-Abelian Loop Equations.

The fixed point of the chain of the loop equations (\ref{LoopEq0}) was shown to have a solution which is an arbitrary function of minimal area \begin{math} A_C \end{math} bounded by $C$.
\begin{equation}
    P(C,\Gamma) = F\left(A_C,\Gamma\right)
\end{equation}
The Minimal area can be reduced to the Stokes functional by the regularization \eqref{Regularized}.

In the real world, this $r_0$ would be the effective width of the Kelvinon solution for the vorticity layer around the discontinuity surface.

This is a positive definite functional of the surface, as one can easily verify using spectral representation:
\begin{eqnarray}
   && \int_{S_C} d \sigma_{i}(r_1) \int_{S_C} d \sigma_{j}(r_2) \delta_{i j} \Delta(r_1-r_2) \propto \nonumber\\
   &&\int d^3k \exp{-\frac{k^2r_0^2}{4\pi}} \left| \int_{S_C} d \sigma_{i}(r)e^{i k r} \right|^2
\end{eqnarray}
In the limit $r_0 \rightarrow 0$, this definition reduces to the ordinary area:
\begin{equation}
  A_C \rightarrow \min_{S_C} \int_{S_C} d^2 \xi \sqrt{g} 
\end{equation}

\subsection{Stokes Condition and Mean Curvature}
Let us study this amazing duality of Minimal Surface to turbulent flow. First of all, why a minimal surface? No string theory would require this minimal surface to arise as a classical solution (or at least we do not know any well-defined string theory equivalent to turbulence despite some interesting observations \cite{TSVS}).

The Stokes condition (\eqref{StokesCond}) is satisfied by the minimum condition. When the surface changes into $S'$, the linear variation reduces by volume Stokes theorem to 
\begin{equation}
    \oint_{S'-S} d \sigma_i(r) \frac{\delta U[C]}{\delta \sigma_i(r)} = \int_{\delta V} d^3 r \partial_i \frac{\delta U[C]}{\delta \sigma_i(r)}
\end{equation}
with $\delta V$ being infinitesimal volume between $S'$ and $S$. This linear variation must vanish by the definition of the minimal surface for the regularized area and its local limit.

The area derivative of the Minimal area in regularized form, then, as before, reduces to the elimination of one integration
\begin{equation}\label{vorticity}
  \frac{\delta A_C}{\delta \sigma_i(r)} = 2 \int_{S_C} d \sigma_{i}(\rho) \Delta(r-\rho) \rightarrow 2 n_i(\Tilde{r})\exp{-\pi \frac{r_{\perp}^2}{r_0^2}}
\end{equation}
Where $n_i(\Tilde{r}) $ is the local normal vector to the minimal surface at the nearest surface point $\Tilde{r}$ to the 3d point $r$, and $r_{\perp}$ is the component normal to the surface at $\Tilde{r}$. With this regularization, the area derivative is defined everywhere in space but exponentially decreases away from the surface. Exactly at the surface, it reduces to twice the normal unit vector.

Let us investigate this issue in more detail.

Let us choose $x,y$ coordinates in a local tangent plane to the minimal surface at some point which we set as an origin. In the quadratic approximation (which will be enough for our purpose), the equation of the surface reads:
\begin{align}
     z &= \frac{1}{2} \left(K_1 x^2  + K_2 y^2\right)\\
     n &= \frac{\left[ -K_1 x, -K_2 y, 1\right]}{\sqrt{1 +  K_1^2 x^2 + K_2^2 y^2}}\label{Normal2}
\end{align}
where $K_1, K_2$ are main curvatures in planes $y=0, x=0$.

Now, at $r_0 \rightarrow 0$, the Stokes condition reduces to :
\begin{align}
     &\int_{-\infty}^{\infty}d x \int_{-\infty}^{\infty} d y \partial_z \Delta (\vec r)\\
     &\propto
     \int_{-\infty}^{\infty}d x \int_{-\infty}^{\infty} d y \frac{z}{r_0^4} \exp{-\pi \frac{x^2+ y^2}{r_0^2}}\\ &\propto
     \int_{-\infty}^{\infty}d x \int_{-\infty}^{\infty} d y \frac{K_1 x^2 + K_2 y^2}{r_0^4} \oldexp\left(-\pi \frac{x^2+ y^2}{r_0^2}\right) \\
     &\propto K_1 + K_2 =0\label{MeanCurvature}
\end{align}
 
This sum is the mean curvature. So, the Stokes condition is \textbf{equivalent} to the equation for the minimal surface. It is nice to know!

One may check that any rotational invariant representation of the delta function
\begin{eqnarray}
    &&\Delta (\vec r) = \eps^{-3} f\left(\frac{\vec r^2}{\eps^2}\right);\\
    &&\int d^3 r f(\vec r^2) =1;
\end{eqnarray}
in the limit $\eps \ra 0$ leads to the same equation $K_1 + K_2 =0$.
Therefore, this relation of the minimal surface with the Stokes condition does not depend upon the regularization method.
\subsection{Loop Equation Beyond Logarithmic Approximation}

An important unanswered question in our recent papers \cite{M19a,M19b} is whether the loop equation was satisfied beyond the leading logarithmic approximation.

Within the Area law Anzatz, the steady solution of the loop equation (\ref{LoopEq0}) reduces in the limit $r_0 \rightarrow 0$ to:
\begin{equation}\label{LE1}
    r_0 \oint_C d r_i \int_{S_C} d \sigma(r') \frac{ (r'_i - r_i)n_k(r)n_k(r') - n_i(r')(r'_k - r_k) n_k(r) }{|\vec r' - \vec r |^3}=0
\end{equation} 
Here $d\sigma(r') = \left| d \vec \sigma(r')\right|$ is the scalar area element for the point $\vec r'$ at the surface. The distance $r_i - r'_i$ is measured in Euclidean space rather than along the surface.
The factor of $r_0$ implies that the time derivative of our PDF is very small in the limit when viscosity goes to zero. In the $d-$ dimensional problem, the factor would be $r_0^{d-2}$, and in particular, in two dimensions, there would be no factor.
By renormalizing time:
\begin{equation}
    t = \tau T_0; \, T_0 = \frac{A_C}{|\Gamma|} \left(\frac{A_C}{r_0^2}\right)^{\frac{d-2}{2}}
\end{equation}
we eliminate this factor from the loop equation altogether. However, remember that reaching the equilibrium PDF we are investigating would take a large time  $T_0$  at $d >2 $. ( we reinserted here missing dimensional factors ).

After dropping the factor $r_0$ in 3 dimensions:
\begin{equation}\label{LoopEq1}
    \oint_C d r_i \int_{S_C} d \sigma(r') \frac{ (r'_i - r_i)n_k(r)n_k(r') - n_i(r')(r'_k - r_k) n_k(r) }{|\vec r' - \vec r |^3}=0
\end{equation}
This formula is the final form of the Loop Equation for steady PDF, depending on the minimal area. 

\subsection{Exact Solution for Flat Loop}

Here is the biggest news of this paper: the minimal surface \textbf{exactly} solves the loop equation for a flat loop.

The thing is that at a flat loop (in the $x,y $ plane), the minimal surface is flat as well so that both $\vec n(r) = \vec n(r') = (0,0,1)$ which makes the second term in (\ref{LoopEq1}) zero. As for the first term, it reduces to the gradient and vanishes after integration over a closed loop:
\begin{equation}
    \oint_C d r_i \int_{S_C} d \sigma(r')\frac{ (r'_i - r_i) }{|\vec r' - \vec r |^3} \propto  \int_{S_C} d \sigma(r')\oint_C d r_i \partial_{r_i}\frac{ 1}{|\vec r' - \vec r |} =0
\end{equation}

So, we claim that this is not just an asymptotic solution for the tails of the PDF; this is the exact solution.

In the case of 2D turbulence, where all loops are planar, vorticity is pseudoscalar, and the normal vector is $n =\pm 1$ depending on the orientation of the loop, the loop equation for the minimal surface Anzatz reads
\begin{equation}
    \oint_C d r_i \int_{S_C} d \sigma(r')\frac{ (r'_i - r_i) }{|\vec r' - \vec r |^2} \propto  \int_{S_C} d \sigma(r')\oint_C d r_i \partial_{r_i}\ln |\vec r' - \vec r | =0
\end{equation}

Let us return to the loop equation's exact solution in three dimensions for a planar loop.
In that case, it must apply to the moments of the circulation, in particular, to the second moment
\begin{equation}
    \left<\Gamma^2\right> = A_C \int_{-\infty}^{\infty} d \gamma \gamma^2 \Pi(\gamma)
\end{equation}
This formula raised objections\footnote{Sasha Polyakov, private communication.}: one can explicitly compute this double integral
\begin{equation}\label{Moment2VV}
    \left<\Gamma^2\right> = \oint_C d r_i \oint_C d r'_j \left< v_i(r) v_j(r')\right>
\end{equation}
taking the scaling law $\left< v_i(r) v_j(r')\right> \propto \delta_{i j} |r-r'|^{2 \alpha -1}$ where $\alpha$ is the scaling index of circulation in terms of the area. Integral looks nothing like an area, and, say, for the rectangle, it shows manifest dependence upon the aspect ratio at a fixed area.

Our answer is very simple: this is so for Kolmogorov index $\alpha = \frac{2}{3}$ as well as any other index except our prediction\footnote{Note that the argument of \cite{M19b} applies to an arbitrary dimension of space, as long as the vorticity surface was 2-dimensional.} $\alpha = \frac{1}{2}$. In this exceptional case, the second moment can be directly proven equal to the area.

Namely, in case $2\alpha = 1$, the velocity has zero dimension, so its correlator is proportional to $\ln |\vec r-\vec r'|$.
The double loop integral by Stokes theorem reduces to a double area integral
\begin{align}
    &\left<\Gamma^2\right> = \oint_C d r_i \oint_C d r'_j \left< v_i(r) v_j(r')\right>=\\
    &\int_{S_C} d \sigma(r) \int_{S_C} d \sigma(r')\left< \omega_3(r) \omega_3(r')\right> \propto \\
    &\int_{S_C} d \sigma(r) \int_{S_C} d \sigma(r') \nabla^2 \ln |\vec r-\vec r'| \propto \\
    &\int_{S_C} d \sigma(r) \int_{S_C} d \sigma(r') \delta(\vec r-\vec r')  = A_C
\end{align}
Q.E.D.

We also claim that for our solution, all higher moments are proportional to the powers of the area for the flat loop. However, this is hard to verify directly, as we have yet to determine the exact form of higher-order velocity correlations.

Note, however, that the same logarithmic velocity correlator in 3D space, as required to compute the second moment for a non-planar loop, will no longer produce $\delta$ functions for vorticity correlations. We may use the Stokes theorem and integrate it over some curved surface, but the vorticity correlator
\begin{equation}
    \left< \omega_i(r) \omega_j(r')\right> \propto \left(\delta_{i j} \partial^2 - \partial_i \partial_j \right)   \ln |\vec r - \vec r'|  
\end{equation}
will have long term tails $\propto \frac{1}{(\vec r - \vec r')^2} $ unless taken on a flat surface.
Therefore, our solution does not imply a short-range correlation of vorticity, just that it has a scaling dimension $-1$.

An interesting property of our solution is that it leaves the scaling function arbitrary as long as it depends upon the minimal area. The dependence of higher vorticity correlations in a circulation background is uniquely expressed in the basic PDF scaling function, but this function remains arbitrary.

We shall accept the solution for the planar loop and try to generalize it for the non-planar one.

\subsection{Loop Equation in Quadratic approximation}\label{LoopEqQuadratic}

Let us introduce a local quadratic approximation to a surface, as before (with $K_1 = K, K_2= -K$), and $n(r')$ given by (\ref{Normal2}), with $z$ direction being the local normal vector at the contour. The curvature $K$ refers to the integration point $r$. 

As we found the minimal surface in Appendix H in \cite{M93} (see the previous Section), the conformal coordinates at the boundary are consistent with these $x,y$ in quadratic approximation: $x$ goes along the local tangent direction of the loop, and $y$ goes inside the surface. As for $z$ it goes along the normal to the surface $n(r)$. We take $r$ as an origin. The loop $C $ expands as a piece of parabola in $x y $ plane:
\begin{align}
    C_1 &= x \\
    C_2 &= b x^2\\
    C_3 &=0
\end{align}

Therefore the surface integral in the quadratic approximation of the integrand in $\oint_C d r_i $ becomes:
\begin{equation}
    \int_{-\infty}^{\infty} d y
    \int_{-\infty}^{\infty} x\,d x \,
    \theta\left(y - b x^2\right)
     \frac{
     1 + \frac{K^2}{2}\left(x^2-y^2\right)
     }
     {
     \sqrt{
     \left(x^2 + y^2\right)^3\left(1 + K^2 \left(x^2 + y^2\right)\right)
     }
     }
\end{equation}

This integral vanishes by reflection symmetry $x\rightarrow -x$.

However, there is no reason to expect this reflection symmetry to hold beyond quadratic approximation. Cubic terms in the contour equation alone would destroy this reflection symmetry.

\subsection{Minimal Area In Conformal Metric Field}\label{MinAreaConfMetric}
The computations in the previous Section \ref{LoopEqQuadratic}  suggest the following "conformal" Ansatz
\begin{equation}\label{Conformal}
  A_C[\phi] =\min_{S_C} \int_{S_C} d \sigma_{i}(r_1) \int_{S_C} d \sigma_{j}(r_2) \delta_{i j} \Delta(r_1-r_2) \oldexp\left(\frac{1}{2}( \phi(r_1) + \phi(r_2))\right)
\end{equation}
where conformal metric $\phi(r)$ is some external field defined in all $R_3$, not just on the surface. 

The local limit of this functional $A_C[\phi]$ tends to be the area in the external conformal metric
\begin{align}
    &A_C[\phi] \underset{r_0\rightarrow 0}{\longrightarrow} \int_{S_C} d \sigma(r) \oldexp\left(\phi\left(R(x)\right)\right)\\\label{LocalArea}
   &d \sigma(r) = \sqrt{\left(d \sigma_i(r)\right)^2} = d^2 x \sqrt{\det G}
\end{align}
with induced metric tensor, corresponding to parametric equation $\vec r = \vec R(x), x = (u,v)$ of the surface $S_C$
\begin{equation}
    G_{a b}(x) = \partial_a R_\mu(x) \partial_b R_\mu(x)
\end{equation}
We derive an equation for $\phi$ later but now consider this area in the external field $A_C[\phi]$ as a functional of the surface at fixed external field $\phi(r)$.
First of all, one can verify that this is a positive definite functional, just as before
\begin{align}
    &\int_{S_C} d \sigma_{i}(r_1) \int_{S_C} d \sigma_{j}(r_2) \delta_{i j} \Delta(r_1-r_2)\oldexp\left(\frac{1}{2}( \phi(r_1) + \phi(r_2))\right)\\
    &\propto \int d^3k \oldexp\left(-\frac{k^2 r_0^2}{4\pi}\right) \left| \int_{S_C} d \sigma_{i}(r)\oldexp\left(i k r + \frac{1}{2}\phi(r)\right) \right|^2
\end{align}

Now, the Stokes condition will, as before, be satisfied by minimality. Still, it will be interesting to derive a replacement of the mean curvature equation for the ordinary minimal surface.
Repeating the above computations in the local tangent plane, we find here:
\begin{equation}\label{MeanCurv}
    K_1 + K_2 = n_i(x) \partial_i \phi(r)_{\vec r =\vec R(x)}
\end{equation}
On the left, we have mean curvature at the surface, and on the right, we have the normal derivative of the conformal metric field projected at the surface. 

This property means that we can keep the ordinary minimal surface, with $ K_1 + K_2 =0$, and ensure that the conformal metric varies only along the surface but not in the normal direction. In the local tangent plane, this means that $\phi(x,y,z) = \phi(x,y,0)$, so it changes only in a local tangent plane but not in the normal direction.

Remember, this is just the Stokes condition, not yet the loop equation, but the net result is that we keep the minimal surface in the sense of being the surface of the minimal area in ordinary induced metric. However, the Extended Area $A_C[\phi]$ has an extra factor $\oldexp(\phi(r))$, which varies along the surface.

For the loop equation to hold, the following condition must be valid:
\begin{eqnarray}\label{SelfCons}
    &&\oint_C d r_i \oldexp\left(\phi(r)\right)\int_{S_C} d \sigma(r')\oldexp\left(\phi(r')\right) \nonumber\\
    &&\frac{ (r'_i - r_i)n_k(r)n_k(r') - n_i(r')(r'_k - r_k) n_k(r) }{|\vec r' - \vec r |^3}=0
\end{eqnarray}
We call it the self-consistency relation for the metric field. At a given surface $S_C$, this is an equation for the metric field on the surface.
Thus we get a closed set of integrodifferential equations for the parametric equation of surface $r_i =R_i(u,v)$  and the external conformal metric $\phi(u,v)$, which is a two-dimensional vector field. So, we have a two-dimensional surface embedded in four-dimensional space $x,y,z,\phi$.

Note also that there would now extra terms in the loop equation of the structure
\begin{equation}
    \int d^3 \rho \frac{\rho_j}{|\vec \rho|^3}\Delta(\vec \rho)
    \oldexp\left(\frac{1}{2}\phi(\vec r+\vec \rho)+ \frac{1}{2}\phi(\vec r)\right)
\end{equation}
coming from the second area derivative of generalized area $A_{S_C}[\phi]$. These terms no longer vanish by space symmetry in the presence of a conformal metric field, but the leading term at $r_0 \rightarrow 0$ will be a gradient which vanishes after loop integration\footnote{Note also that extra condition $n_k \partial_k n(r)=0$ was not used here, as $n_k(r) d r_k =0$.}
\begin{eqnarray}
 &&  \oint_C d r_i \int d^3 \rho \frac{\rho_i}{|\vec \rho|^3}\Delta(\vec \rho)\oldexp\left(\frac{1}{2}\phi(\vec r+\vec \rho) + \frac{1}{2}\phi(\vec r)\right) \underset{r0\rightarrow 0}{\propto}  \nonumber\\
 &&\oint_C d r_i\partial_i \phi(\vec r)\exp{\phi(\vec r)} = 0
\end{eqnarray}
The next terms with $\partial^3 \phi, \left(\partial^2 \phi\right) \partial \phi, \left(\partial \phi\right)^3$ will already have $r_0^2$ in front of them so that they will be negligible compared to the leading $O(1)$ term in the loop equation.

Let us integrate the first term in (\ref{SelfCons}) by parts
\begin{eqnarray}
    &&\oint_C d r_i \oldexp\left(\phi(r)\right)\int_{S_C} d \sigma(r')\oldexp\left(\phi(r')\right) \frac{ n_i(r')(r'_k - r_k) n_k(r) }{|\vec r' - \vec r |^3}=\nonumber\\
    &&\oint_C d r_i \oldexp\left(\phi(r)\right)\int_{S_C} d \sigma(r')\oldexp\left(\phi(r')\right) n_k(r)n_k(r')\partial_{r_i}\frac{1}{|\vec r' - \vec r |}=\nonumber\\
    &&-\oint_C d r_i \oldexp\left(\phi(r)\right)\left(\partial_{i}n_k(r) + n_k(r)\partial_{i}\phi(r)\right)\nonumber\\
    &&\int_{S_C} d \sigma(r')\oldexp\left(\phi(r')\right)\frac{ n_k(r')}{|\vec r' - \vec r |}
\end{eqnarray}
Moving term with $\partial_i \phi(r)$ to the left and all remaining terms to the right, we get
\begin{eqnarray}\label{LoopEqConf}
    &&\oint_C d r_i \oldexp\left(\phi(r)\right)\partial_{i}\phi(r)\int_{S_C} d \sigma(r')\oldexp\left(\phi(r')\right)\frac{ n_k(r)n_k(r')}{|\vec r' - \vec r |}=\nonumber\\
    &&-\oint_C d r_i \oldexp\left(\phi(r)\right)\int_{S_C} d \sigma(r')\oldexp\left(\phi(r')\right) \nonumber\\
    &&\left(\frac{ \left(n_i(r')-n_i(r)\right)(r'_k - r_k) n_k(r) }{|\vec r' - \vec r |^3} + \frac{\partial_{i}n_k(r)  n_k(r')}{|\vec r' - \vec r |}\right)
\end{eqnarray}
We used the fact $ d r_i n_i(r) =0$ to subtract $n_i(r)$ from $n_i(r')$ and remove spurious singularity in the surface integral.
At this point exact solution seems impossible - even the mean curvature equation is a problem that can only be tackled by numerical minimization.

But we never know.

\subsection{Vorticity Correlations and Equation for Scaling Index}\label{HicorWithRho}

Let us now repeat the computation of vorticity correlations. The area derivatives are modified trivially (because the regularized area is a quadratic functional of $ d \sigma_i$, we get a factor of 2):
\begin{align}
    \frac{\delta A_C[\phi]}{\delta \sigma_i(r)} &= 2 n_i(r) \exp{ \phi(r)}\\
    \int_{S_C} d \sigma_i(r) \frac{\delta A_C[\phi]}{\delta \sigma_i(r)} &= 2\int_{S_C} d \sigma_i(r) n_i(r)\exp{ \phi(r)}\\
    &= 2  A_C[\phi]
\end{align}
So, we get the same equations for vorticity correlations, including the self-consistency equation for $\alpha$:
\begin{equation}
    \alpha = \frac{1}{2}
\end{equation}
All the vorticity correlations will acquire extra factors of $\exp{ \phi}$:
\begin{eqnarray}\label{MultiNormalPhi}
   && \left < \vec \omega_1\dots \vec\omega_k \delta\left(\Gamma - \oint_C \vec v d\vec r\right)\right >  =\nonumber\\
   &&\left|A_C[\phi]\right|^{-\frac{1}{2}}\vec n_1 \oldexp( \phi_1)\dots \vec n_k \oldexp( \phi_k)\nonumber\\
   &&A_C[\phi]^{-\frac{k}{2}}\Omega_k\left(\Gamma A_C[\phi]^{-\frac{1}{2}}\right)
\end{eqnarray}

So, the formulas look the same, and the scaling functions $\Omega_k\left(\Gamma A_C[\phi]^{-\frac{1}{2}}\right)$ satisfy the same recurrent equations and results expressing them in terms of integrals of $\Omega_0(\gamma) = \Pi(\gamma)$ stay the same as in \eqref{Recurrent},\eqref{OmegaRelations}.
The only thing which changed is the extra dependence of the coordinates provided by conformal metric factor $e^\phi$.
So, we cannot predict the spatial dependence of vorticity correlations until we solve these heavy equations for $\phi$ and $\vec R(u,v)$.
\subsection{Sum over topologies and quantum equivalence}

Here is an important observation. The loop equation is a \textbf{linear} Schrödinger equation in loop space, even though the original \NS{} and Euler equations are nonlinear.

There is no superposition principle in fluid dynamics, but there is such a principle in quantum mechanics in loop space.

We already used this superposition principle when we took the averaging over the random forces out of the loop functional. The average of the loop functionals, each satisfying the loop equation in virtue of the classical Euler equation with fixed force $\vec f$, also represented a solution to the loop equation.

Now we apply the same argument to the topological classes.

In virtue of the superposition principle for the linear equation, the general solution would be a superposition for the loop functional as well as for its Fourier transform, -- the PDF
\begin{eqnarray}
   &&\Psi[C,\gamma] = \sum_{n, m \in \Z}   \kappa_{ n,m} \Psi_{n m}[C, \gamma]=\nonumber\\
   && \sum_{n, m \in \Z} \frac{\kappa_{n m}}{\sqrt{\det{\left(1 - \i\gamma \hat \Lambda\right)}}};\\
   && \Pi[C,\Gamma] =\sum_{n, m \in \Z}   \kappa_{ n,m} \Pi_{n m}[C, \Gamma]
\end{eqnarray}

What are the weights $\kappa_{n m}$? At the moment, we cannot say, other than demand, that these weights must add up to 1. This problem remains for future research.

Here are some general considerations.

The solution of the \NS{} equation is supposedly unique in the absence of forcing and at fixed initial data. There is even a Millennium Prize offered for proving or (for the sake of caution) disproving this widely accepted conjecture.

There is, though, an instability at a large enough Reynolds number (small enough viscosity compared to a typical velocity circulation).

In this case of infinitesimal viscosity, infinitesimal random forces at the boundary are known to lead to stochastic behavior, making the whole question of the uniqueness of the solution of the \NS{} equation moot.
The real turbulence problem is: what manifold is covered by this stochastic motion, and what is the nature of spontaneous breaking of time reversal invariance in the Euler equation?

Coming back to the sum over topologies, let us consider the time evolution of the \NS{} flow.
There was a deterministic flow at the initial moment, far from any fixed points. According to our scenario, it approaches one of the stable fixed points at some time, corresponding to certain topological charges of the \CL{} field.

By definition of the winding numbers, no continuous deformation can change them.
However, in our system, we assumed some random external boundary forces $\vec f$, supplying the energy dissipated at large velocity gradients.

With a certain probability, these random forces can change topology as they violate the continuity of the solution. Moreover, an enhancement effect caused by large vorticity in vortex sheets leads to finite results of infinitesimal forcing at infinitesimal viscosity.

In that case, the flow will jump between topological classes, approach the fixed point in each class and cover the vicinity of this fixed point (which is, in fact, a fixed manifold).

Another interpretation of the averaging over topological classes is volume averaging.
Our solution is non-uniform; there are singular vortex structures scattered in place. It is impossible to predict exactly the positions and locations of these structures.

However, any measurement in "isotropic" turbulence would involve averaging over the volume.
The vortex structures in the infinite volume would come in various shapes, positions, orientations, and \textit{topological classes}.

Therefore, the volume average enforces averaging over topological classes and creates an illusion of strong isotropic turbulence.
Any snapshot of a strong turbulent flow is anisotropic and inhomogeneous, but we can predict only volume averages.

In that picture, the coefficients $\kappa_{m n}$ are time (or volume) frequencies of \CL{} field getting into the corresponding topological class as a result of Gaussian random forcing.

Naturally, there is no sum over histories in our classical theory. The \NS{}  flow does not go over various alternatives with the weight $\exp{ \i \frac{S}{\nu}}$ at the same moment.

Instead, in the Kelvinon solution, we have an average of $\exp{\i \frac{\Gamma}{\nu}} $ over the Gaussian background velocity, on top of averaging over topological classes of this solution.

There are some dynamic instabilities triggered by external forcing, resulting in a weighted sum of contributions from various topological classes to the time average of the PDF over infinite time (and infinite space).

The turbulent flow spends some fraction of time in each of these classes, jumping from one to another, triggered by (unlikely) large fluctuations of the random force.

From that interpretation, it follows that all $\kappa_{m n} >0$, unlike quantum mechanics, where the eigenfunctions could add to wave function with complex or negative coefficients.

\section{Conclusions and Discussion}

Here is how our theory answers the questions posed in the Prologue.

\begin{itemize}
\item \textit{Is there an analogy between turbulent velocity and vector potential in the gauge field theory?}

     No, this analogy is spurious and misleading. There is no invariance of adding $\vec \nabla \phi$ to the velocity field. The incompressibility condition $\vec \nabla \cdot \vec v =0$ is not a gauge condition but rather a physical restriction on a flow, leading to long-range correlations, unlike the gauge theory, where the gauge condition does not affect physics.
     
     However, there is a \textbf{different kind of gauge invariance} related to the true gauge potentials, namely the \CL{} fields.
     This gauge invariance leads to the confinement of these \CL{} fields in strong turbulence.

     The observable velocity and vorticity are gauge-invariant functionals of these gauge potentials.

\item \textit{Are there some "fundamental particles" hidden inside?}

     This negative remark by Feynman was a prophecy. He subconsciously reminded us about the fundamental particles, and he was right!
     There are confined quarks in turbulence, and they are called Faddeev variables or spherical \CL{} fields. 
     
     The vorticity and velocity are invariant under local gauge (symplectic) transformations \eqref{GaugeTranClebsch} of coordinates on the \CL{} target space. The sphere is a particular gauge choice in that gauge theory.
     Their topology is what makes the Kelvinon stable. There are remarkable "quantum" effects in classical turbulence (quantization of circulations $\Gamma_\alpha, \Gamma_\beta$) related to the compactness of the classical motion of these fields in their target space $S_2$.
     
     One cannot observe spherical \CL{} as particle excitations, but one can study the structure of the vorticity field made from the \CL{} field\footnote{The thin vortex tubes of the type of the Burgers vortex were observed in DNS \cite{Tubes19}}.
     
     Why do we call these variables fundamental? They are canonical Hamiltonian variables with \PB{} of elementary rigid rotators or spins.
     All other variables (velocity, vorticity) are the gauge-invariant nonlinear functionals of these \CL{} fields, like electromagnetic current made from quarks in QCD.
     
\item \textit{How is inviscid limit of the \NS{} dynamics different from the Euler dynamics?}

 The inviscid limit of the \NS{} dynamics does not exist. At arbitrary small viscosity, there are logarithmic terms in the Hamiltonian $\log \frac {Z}{\nu} $ where the constant $Z$ measured the scale of the circulation.

 However, at small viscosity, the Euler dynamics apply in most physical space, except for singular vortex lines and surfaces. Inside thin layers and tubes surrounding these singularities of the Euler equation, the \NS{} equation applies. 
 
 The Euler solution outside can be matched with the \NS{} solution inside (Burgers cylindrical vortex and Burgers-Townsend vortex sheet).

 The ambiguity of weak solutions of the Euler equation is resolved by matching with these \NS{} vortex solutions.
 As a result, the anomalous terms appear, which stay finite (Helicity, Dissipation) or grow logarithmically (Hamiltonian) as viscosity goes to zero.
 
    \item  \textit{What is the origin of the randomness of the circulating fluid? }
    
        The origin of randomness is a large number of vortex tubes (Kelvinons) scattered far away from each other. 
        These tubes contribute to the background velocity leading to fluctuations in each one.

        The \BS{} contributions of each tube to the background velocity decay as the third power of distance, but with a finite density of these tubes in infinite space, the random terms add up to finite Gaussian background velocity $\vec v_0$.
        
        The fluctuations of the velocity circulation in virtue of parity and Galilean invariance can be approximated as a quadratic form in background velocity.
        
        This dependence leads to our equilibrium distribution of the normalization factor $Z$ and, consequently, circulation.
        
        \item \textit{Is it spontaneous, and what makes it irreversible?}
    
        Yes, this randomness comes from inner vorticity distribution; in that sense, it is spontaneous. It does not need external forcing; it is self-consistent, like a mean field in the phase transition theory. 
        
        Ultimately, the energy comes from external forces at the boundary of the turbulent flow in an infinite volume. But the randomness influencing our isolated Kelvinon is the accumulated random background velocity contributing to the energy flux into this tube. This random background velocity was measured in DNS and was shown to obey Gaussian distribution (see \cite{S19, S21} and references within).
        
        Irreversibility (time-reflection breaking) comes from the \NS{} anomalies. For arbitrary small viscosity, there is a finite contribution from the singular line of the Kelvinon vortex, regularized by a Burgers vortex at the core. In a limit, there are finite extra terms in the Euler Hamiltonian, helicity, and the energy dissipation from singular vortexes. 
        
        The dissipation is linear with the tangent component of the strain along the loop;
        the anomalous contribution to the Hamiltonian involves the logarithm of the same tangent component of the strain; therefore, this component must be positive. This positivity requirement explicitly breaks time reversal, as the strain is time-odd.
        
        In addition, there is a spontaneous breaking of time reversal by the quantized circulation $\Gamma_\alpha$ of our Kelvinon. The winding number $m$ is time-odd and can have two signs. A particular solution explicitly breaks this symmetry, leading to an asymmetric circulation PDF.
        
  \item \textit{What are the properties of the fixed manifold covered by this random motion?}
  
        This fixed manifold in our Kelvinon solution for each sign of the circulation is a parabolic surface $ \Gamma = \tau +\vec v_0 \cdot \hat \Lambda \cdot \vec v_0, \vec v_0 \in R_3$ describing the velocity circulation $\Gamma$ as a function of a vector of background velocity $\vec v_0$. The measure on this surface is Gaussian.
        
        Another way to describe this manifold is to consider the solution for the \CL{} field $\vec S(\vec r, \vec v_0)$.
        As a function of $\vec r$, it is a map $R_3\mapsto S_2$, but this map also depends on a Gaussian random vector $\vec v_0$ covering the above parabolic surface.

        On top of that, there is another degeneracy involved. The solution for $\vec S$ is defined up to the symplectic transformation of coordinates in the target space, changing its metric while preserving the area and the topology. So, this solution represents a point on gauge orbit. There could be a time evolution around this orbit, corresponding to arbitrary time-dependent symplectic transformation, leaving invariant the velocity field.
        
\item \textit{What is the microscopic mechanism of the observed multi-fractal scaling laws?}

    There are no scaling laws in our theory. The K41 scaling laws are modified, like in QCD, by powers of the logarithm of the scale, not by the anomalous dimensions, like in conformal field theory. There is no conformal symmetry in the anomalous Hamiltonian we have found, but there are explicit violations of K41 scaling by logarithmic terms in the anomalous part of the Hamiltonian.
    
    However, the moments of circulation we can compute in a certain range of scales (small enough loops $C$) are slowly varying functions of the scale (superposition of terms with K41 power of scale times different powers of the logarithm of scale). 
    
    Such functions would be indistinguishable from anomalous scaling dimensions at the modest Reynolds numbers in existing DNS. A much larger inertial range and much more precise measurements would be needed to verify or disprove our log-scaling laws for the moments of circulation.

    However, the physical mechanism for anomalous scaling is determined- it originates in anomalous terms in the turbulent Hamiltonian, coming from the Burgers core of the singular vortex lines.

\end{itemize}
\section{New game for smart kids}
\begin{itemize}
     \item \textit{Dear field theorists, topologists, and mathematical physicists!} 
     
     Stop looking inside the black holes and elementary particles -- there is beauty and mystery right under your nose, in every river, every ocean, and even in the glass of wine, should you swirl it a little.

     This work just scratched the surface of singular topological solutions in turbulence. The target space could have a higher topology than $S_2$, the dimension of physical space could be higher than $3$, etc. The existence of Kelvinon as a minimum of a bounded Hamiltonian is plausible but not proven. There may be an exact solution, like an instanton in non-abelian gauge theory.

     \item \textit{Dear numerical experimentalists!}
     
     This theory can only be finished with your input, like any theory needing experimental data. There are so many predictions to verify besides the exponential decay of PDF. 
     
     The spatial structure of the Kelvinon can be measured, and the predictions \eqref{Recurrent},\eqref{LKasympt} for correlation functions with
     \begin{equation}
         \Omega_0(\gamma) = a/\sqrt{|\gamma|} \exp{- b |\gamma|} 
     \end{equation}
     on the vortex surface could be compared to more accurate numerical experiments. 
     
     Also, the moments of background strain $\hat S$: \eqref{PabMoms}, \eqref{StrainTraces} can be easily verified in DNS (one has to take averages over regions in space where $\hat S^2 \ll \VEV{\hat S^2}$, to exclude dissipation domain).

     \item \textit{Finally, the new breed of quantum computer enthusiasts!}
     
     The loop equation is exact and describes a quantum system, which can be discretized in the same way as the lattice loop equation and mapped directly to qubits. The loop equation on the cubic lattice was investigated for the Lattice gauge theory. Each loop is described as a collection of steps in each of $2 d$ directions on a $d-$ dimensional cubic lattice, summing up to zero. These steps would become discrete dynamic variables in quantum mechanics.
     
     Mapping them into qubits would open the way to simulate extremely strong turbulence. 
\end{itemize}
\section*{Acknowledgments}

I am grateful to Theo Drivas, Dennis Sullivan, and other Stony Brook University Einstein seminar participants for stimulating discussions of this work, which helped me find errors in earlier versions. 

Deep discussions at the seminars in IAS were both inspiring and elucidating; I would like to thank Peter Sarnak, Ed Witten, and other participants of the IAS seminars for their questions and comments.

I reported this work at various stages at Tel Aviv University, Haifa University, New York University, Max Planck Institute, Nankai symposium, London Royal Institution seminar, Bangalore Turbulence Worksop 2023, and some others. I am grateful to the organizers of these talks and seminars, especially Grisha Falkovich, Marek Karliner, Joshua Feinberg, Yang-Hui He, Uriel Frisch and Kostya Khanin, for their attention and help.

At all stages of this long study, I greatly benefited from suggestions and criticism from my colleagues Sasha Polyakov, Pavel Wiegmann,  Nikita Nekrasov, Sasha Zamolodchikov,  and Nigel Goldenfeld.

Grigory Volovik, Camillo De Lellis, and Elia Bru\'e added significant pieces to my research and allowed me to publish these pieces here as Appendices. They should be credited for that, though they do not share responsibility for my conjectures.

Special thanks to Victor Yakhot and Katepalli Sreenivasan, who discussed, criticized, and supported my work all these years.

Arthur Migdal's help with \Mathematica is also greatly appreciated.

I also appreciate the thoughtful comments of my referee in IJMPA.

This research is supported by a Simons Foundation award ID $686282$ at NYU Abu Dhabi.
I am very grateful to Jim for his constant help and inspiration for my work, extending far beyond the financial support. 

This project was completed while I was working in ADIA. The atmosphere of creative freedom and academic research in the SPD department in ADIA, encouraged at that time by our Executive Director Majed Alromaithi, was essential in this final stage of my work. Thank you, Majed.

\newpage
\bibliographystyle{elsarticle-num} 
\bibliography{bibliography}
\newpage
\appendix{

 %%%%%%%%%%%%%%%%%%%%%%%Grammarly down from here

\section{The Cylindrical CVS equations}

Let us re-derive the cylindrical \CVS{} equations of \cite{M21c} without unnecessary assumption of real and equal functions $d \Gamma_\pm$  in \eqref{Vgamma}.

The normal vector in complex notation
\begin{eqnarray}
 \sigma =  \i C'(\theta)
\end{eqnarray}
The normal projection of velocity field
\begin{equation}
     v_x \sigma_x +  v_y \sigma_y = \Re (v_x - \i v_y) \sigma =
   \frac{\Im \left( (v_x - \i v_y) C'(\theta) \right)}{|C'(\theta)|}
\end{equation}

The first \CVS{} equation (vanishing normal velocity on each side) becomes
\begin{eqnarray}\label{Neumann}
   \Im\left( V_\pm(C(\theta)) C'(\theta)\right) =0;\forall \theta
\end{eqnarray}

The second and third \CVS{} equations require the computation of the strain related to the complex velocity $F_\pm = \phi'_\pm $. 
The strain on each side is a $3 \times 3$ matrix 
\begin{eqnarray}
  \hat S_\pm(\eta) = \left(
\begin{array}{ccc}
 \Re F'_\pm(\eta)+a & - \Im F'_\pm(\eta) & 0 \\
 -\Im F'_\pm(\eta) & - \Re F'_\pm(\eta) +b & 0\\
  0         &0                 &c\\
\end{array}
\right)
\end{eqnarray}

Let us introduce notations
\begin{eqnarray}
&&F(\eta) = \oh(F_+(\eta) + F_-(\eta));\\
&&\Delta F(\eta) = F_+(\eta) - F_-(\eta);\\
&&p = (a+b)/2;\\
&&q =(a-b)/2
\end{eqnarray}

The null vector equation $(\hat S_+(\eta) + \hat S_-(\eta)) \cdot \Delta \vec v =0$ provides the following complex equation
\begin{eqnarray}\label{eq1}
   && \Delta F^*(\eta)(F'(\eta) + q )  +  p  \Delta F(\eta) =0;\forall \eta \in C ;
\end{eqnarray}

The product 
\begin{eqnarray}
\Gamma(\theta) = C'(\theta) \Delta F(C(\theta))
\end{eqnarray}
is real, in virtue of the Neumann boundary conditions.
The difference between the two Neumann equations \eqref{Neumann} reduces to $\Im \Gamma =0$.

Thus, multiplying \eqref{eq1} by $C'(\theta)C'^*(\theta)$ and using the fact that $\Gamma(\theta)$ is real  we find
\begin{eqnarray}\label{NullVector}
   && C'(\theta) \left(F'(C(\theta)) + q \right)  +  p  C'^*(\theta) =0;\forall \eta \in C ;
\end{eqnarray}

This equation is simpler than it looks: it is reduced to the total derivative.
\begin{eqnarray}
   \dd{\theta} \left( F(C(\theta))+ q C(\theta) + p C^*(\theta) \right) =0 
\end{eqnarray}

The generic solution is
\begin{eqnarray}\label{Ceq}
    F(C(\theta))+ q C(\theta) + p C^*(\theta) = A
\end{eqnarray}
with some complex constant $A$. 

Plugging it back to the \eqref{Neumann} we have everything cancel except one term
\begin{eqnarray}
   \Im\left(A C'(\theta)\right) =0;\forall \eta =C(\xi)
\end{eqnarray}

The  nontrivial solution for $C(\theta)$ would correspond to $A =0$.

This solution is equivalent to the equation \eqref{CVSCyl} we stated in the text.
\section{Symmetric Traceless Random Matrices}\label{RandomStrain}
 
 Let us investigate the general properties of the Gaussian random matrix measure. First, the trace of the matrix is invariant against orthogonal transformations
 \begin{eqnarray}
 &&W \Ra O^T \cdot W \cdot O;\\
 && O^T = O^{-1};\\
 && \tr W \Ra \tr W
 \end{eqnarray}
 therefore this measure stays $O(n)$ invariant after insertion of the delta function of the trace.
 
 The mean value of  $\tr W^2$ can be computed by the following method. Consider the normalization integral
 \begin{eqnarray}
 Z_n(\sigma) = \int d P_\sigma(W)
 \end{eqnarray}
 By rescaling the matrix elements $W_{i j} = \sigma w_{i j}$ we find the property
 \begin{eqnarray}\label{Zint}
 Z_n(\sigma) = \sigma^{\frac{(n+2)(n-1)}{2}} Z_n(1)
 \end{eqnarray}
 
 Taking the logarithmic derivative of the original integral, we get the identity
 \begin{eqnarray}
 \frac{\sigma Z_n'(\sigma)}{Z_n(\sigma)} = \frac{\int d P_\sigma(W) \tr W^2}{\sigma^2 Z_n(\sigma)} = \frac{\VEV{\tr W^2}}{\sigma^2}
 \end{eqnarray}

 On the other hand, taking the logarithmic derivative from \eqref{Zint}, we find 
 \begin{eqnarray}
 \frac{\sigma Z_n'(\sigma)}{Z_n(\sigma)} = \frac{(n+2)(n-1)}{2}
 \end{eqnarray}
 This formula produces the trace relation \eqref{TraceRel}.

\section{Spherical Gauge}\label{SphericalGauge}
 
 The symmetric metric tensor $g_{i j}$ in 2 dimensions has three independent components: two diagonal values $g_{1 1}, g_{2 2}$ and one off-diagonal value $g_{1 2} = g_{2 1}$. 
 
 We take stereographic coordinates $z = z_1 + \i z_2 = \tan\frac{\theta}{2}  e^{\i \varphi}$
 \begin{subequations}
\begin{eqnarray}
    && g_{i j} = \delta_{i j} \rho;\\
   &&\rho = \frac{1}{\left(1 + |z|^2\right)^2};\\
   && z_a = \frac{S_a}{1 + S_3};\\
   && S_a = \frac{2 z_a}{1 + |z|^2};\\
   && S_3 = \frac{1-|z|^2}{1 + |z|^2};\\
   && d^2 S = d z_1  d z_2 \rho
\end{eqnarray}
\end{subequations}

The $O(3)$ rotation in these coordinates reads (with $I,J,K = 1,2,3, a,b,c...= 1,2$)
\begin{subequations}
 \begin{eqnarray}
 &&\delta S_I = e_{I J K} S_J \alpha_k; \\
 && \delta z_a = \alpha_3 e_{a b} z_b -\oh(1-|z|^2)\tilde \alpha_a  - \tilde \alpha_b z_b z_a;\\
 && \tilde \alpha_b  = e_{b c} \alpha_c;
  \end{eqnarray}
\end{subequations}
The $O(3)$ transformation of the metric tensor involves the matrix $R_{i j} = \d_j \delta z_i$
\begin{eqnarray}
    &&R_{i j} =  \alpha_3 e_{i j} + \tilde \alpha_i z_j - \tilde \alpha_j z_i - \delta_{i j}\tilde \alpha z ;\\
    && \delta_{O(3)} g_{i j} = R_{a i} g_{a j} +  R_{a j} g_{a i}
\end{eqnarray}

Computing the variation $\delta_{O(3)} g_{i j}$ of the conformal metric $g^c_{i j} = \rho \delta_{i j}$ we find
\begin{subequations}
\begin{eqnarray}
    &&\delta_{O(3)} g^c_{1 2} = 0;\\ 
    &&\delta_{O(3)} g^c_{1 1} = -2 \rho \left(z_1 \alpha_2 - z_2 \alpha_1 \right);\\ 
     &&\delta_{O(3)} g^c_{2 2} = -2 \rho \left(z_1 \alpha_2 - z_2 \alpha_1 \right);
\end{eqnarray}
\end{subequations}
 The gauge transformation of conformal metrics produces
 \begin{subequations}
 \begin{eqnarray}
    && \delta_{gauge} g^c_{1 2} = (h_{2 2} - h_{1 1})\rho;\\
    && \delta_{gauge} g^c_{1 1} = 2 h_{1 2} \rho;\\
    && \delta_{gauge} g^c_{2 2} = -2 h_{1 2} \rho;\\
    && h_{i j} = \d_i\d_j h(z_1, z_2)
 \end{eqnarray}
\end{subequations}

Now, we do not want to break the rotational invariance of the spherical metric. 
Therefore, any $h(z)$ satisfying the equations
\begin{eqnarray}
    && h_{2 2} - h_{1 1} = 0;\\
    && h_{1 2}  = -(z_1 \alpha_2 - z_2 \alpha_1) ;\\
    && - h_{1 2}  = -( z_1 \alpha_2 - z_2 \alpha_1) 
\end{eqnarray}
with some finite constant $\alpha_1, \alpha_2$ should not be restricted by our gauge conditions. Adding the last two equations, we immediately see that no such gauge functions could imitate the $O(3)$ rotations.

The independent conditions $h_{1 2} =0, h_{1 1} = h_{2 2}$ combine into one complex equation
\begin{equation}
    \hat L h = \rho \frac{\d^2}{\d \bar z^2}h =0
\end{equation}
These gauge conditions leave out arbitrary linear function $h = A + B_i z_i$, corresponding to constant shifts of \CL{} field. These constant shifts can be fixed by placing the origin at the South Pole, which we did.

For remaining nontrivial \SYM{} we have the gauge fixing Gaussian integral 
\begin{subequations}
\begin{eqnarray}
   &&\int D \lambda D \mu D h \exp{\i \int d^2 S \lambda \hat L_1 h+ \mu \hat L_2 h} ;\\
   && \hat L_1 h = \rho \Re \hat L h;\\
   && \hat L_2 h = \rho \Im \hat L h;
\end{eqnarray}
\end{subequations}
The regularized determinant $\det{\hat L}$ is a universal number, which does not depend on our dynamical variables.

This operator being non-Hermitian, we are not sure how to regularize and compute this determinant, but this is immaterial, as it does not depend on dynamic variables and thus drops from the measure.
 
 The symmetric metric tensor $g_{i j}$ in 2 dimensions has three independent components: two diagonal values $g_{1 1}, g_{2 2}$ and one off-diagonal value $g_{1 2} = g_{2 1}$. 
 
 We take stereographic coordinates $z = z_1 + \i z_2 = \tan\frac{\theta}{2}  e^{\i \varphi}$
 \begin{subequations}
\begin{eqnarray}
    && g_{i j} = \delta_{i j} \rho;\\
   &&\rho = \frac{1}{\left(1 + |z|^2\right)^2};\\
   && z_a = \frac{S_a}{1 + S_3};\\
   && S_a = \frac{2 z_a}{1 + |z|^2};\\
   && S_3 = \frac{1-|z|^2}{1 + |z|^2};\\
   && d^2 S = d z_1  d z_2 \rho
\end{eqnarray}
\end{subequations}

The $O(3)$ rotation in these coordinates reads (with $I,J,K = 1,2,3, a,b,c...= 1,2$)
\begin{subequations}
 \begin{eqnarray}
 &&\delta S_I = e_{I J K} S_J \alpha_k; \\
 && \delta z_a = \alpha_3 e_{a b} z_b -\oh(1-|z|^2)\tilde \alpha_a  - \tilde \alpha_b z_b z_a;\\
 && \tilde \alpha_b  = e_{b c} \alpha_c;
  \end{eqnarray}
\end{subequations}
The $O(3)$ transformation of the metric tensor involves the matrix $R_{i j} = \d_j \delta z_i$
\begin{eqnarray}
    &&R_{i j} =  \alpha_3 e_{i j} + \tilde \alpha_i z_j - \tilde \alpha_j z_i - \delta_{i j}\tilde \alpha z ;\\
    && \delta_{O(3)} g_{i j} = R_{a i} g_{a j} +  R_{a j} g_{a i}
\end{eqnarray}

Computing the variation $\delta_{O(3)} g_{i j}$ of the conformal metric $g^c_{i j} = \rho \delta_{i j}$ we find
\begin{subequations}
\begin{eqnarray}
    &&\delta_{O(3)} g^c_{1 2} = 0;\\ 
    &&\delta_{O(3)} g^c_{1 1} = -2 \rho \left(z_1 \alpha_2 - z_2 \alpha_1 \right);\\ 
     &&\delta_{O(3)} g^c_{2 2} = -2 \rho \left(z_1 \alpha_2 - z_2 \alpha_1 \right);
\end{eqnarray}
\end{subequations}
 The gauge transformation of conformal metrics produces
 \begin{subequations}
 \begin{eqnarray}
    && \delta_{gauge} g^c_{1 2} = (h_{2 2} - h_{1 1})\rho;\\
    && \delta_{gauge} g^c_{1 1} = 2 h_{1 2} \rho;\\
    && \delta_{gauge} g^c_{2 2} = -2 h_{1 2} \rho;\\
    && h_{i j} = \d_i\d_j h(z_1, z_2)
 \end{eqnarray}
\end{subequations}

Now, we do not want to break the rotational invariance of the spherical metric. 
Therefore,  any $h(z)$ satisfying the equations
\begin{eqnarray}
    && h_{2 2} - h_{1 1} = 0;\\
    && h_{1 2}  = -(z_1 \alpha_2 - z_2 \alpha_1) ;\\
    && - h_{1 2}  = -( z_1 \alpha_2 - z_2 \alpha_1) 
\end{eqnarray}
with some finite constant $\alpha_1, \alpha_2$ should not be restricted by our gauge conditions. Adding the last two equations, we immediately see that no such gauge functions could imitate the $O(3)$ rotations.

The independent conditions $h_{1 2} =0, h_{1 1} = h_{2 2}$ combine into one complex equation
\begin{equation}
    \hat L h = \rho \frac{\d^2}{\d \bar z^2}h =0
\end{equation}
These gauge conditions leave out arbitrary linear function $h = A + B_i z_i$, corresponding to constant shifts of \CL{} field. These constant shifts can be fixed by placing the origin at the South Pole, which we did.

For remaining nontrivial \SYM{} we have the gauge fixing Gaussian integral 
\begin{subequations}
\begin{eqnarray}
   &&\int D \lambda D \mu D h \exp{\i \int d^2 S \lambda \hat L_1 h+ \mu \hat L_2 h} ;\\
   && \hat L_1 h = \rho \Re \hat L h;\\
   && \hat L_2 h = \rho \Im \hat L h;
\end{eqnarray}
\end{subequations}
The regularized determinant $\det{\hat L}$ is a universal number, which does not depend on our dynamical variables.

This operator being non-Hermitian, we are not sure how to regularize and compute this determinant, but this is immaterial, as it does not depend on dynamic variables and thus drops from the measure.

\section{Burgers Vortex contribution to the Hamiltonian of the Kelvinon}\label{LoopSing}

Let us compute the contribution of small radial coordinates to the Hamiltonian of the Kelvinon.
 We assume the loop to be a straight line in this approximation and take the Burgers vortex solution \cite{BurgersVortex} for the \NS{} equation in this region.
\begin{eqnarray}
    && \vec v =\left\{-\alpha x- g(r) y, -\alpha  y + g(r) x, 2 \alpha z\right\};\\
    && g(r) = \frac{\left(1-e^{-\frac{\alpha  r^2}{2 \nu }}\right)\Gamma  }{2 \pi  r^2};\\
    && r = \sqrt{x^2 + y^2};\\
    && \Gamma = 4 \pi n;\\
    && \alpha = \oh \hat t \cdot \hat S \cdot \hat t;\\
    && \hat t = \{0,0,1\} 
\end{eqnarray}

The contribution to the Euler Hamiltonian from the circular core of the vortex at $z=0$ was computed using \Mathematica{}
\begin{eqnarray}
    &&\oh \int_{ x^2 + y^2 < R^2}  d x d y \,\vec v(x,y,0)^2 = \nonumber\\
    &&2 \pi n^2 \left(\gamma +\log \left(\frac{\alpha  R^2}{4 \nu }\right)\right) + \frac{1}{4}\pi\alpha ^2 R^4 + \nonumber\\
    &&2 \pi  n^2 \text{Ei}\left(-\frac{R^2 \alpha }{\nu }\right)-4 \pi  n^2 \text{Ei}\left(-\frac{R^2 \alpha }{2 \nu }\right)
\end{eqnarray}

where $\text{Ei}(x)$ is an exponential integral function.

In the turbulent limit $\nu \ra 0$ we get
\begin{eqnarray}
   2 \pi n^2 \left(\gamma +\log \left(\frac{\alpha  R^2}{4 \nu }\right)\right) + \frac{1}{4}\pi\alpha ^2 R^4 + O\left(\exp{-\frac{R^2 \alpha }{2 \nu }}\right)
\end{eqnarray}
\newpage
\section{Initial data for the Kelvinon}\label{ExampleClebsch}

Let us present an explicit example of the \CL{} field with the required topology, which could serve as initial data for the Hamiltonian minimization by relaxation.

We introduce a surface of the minimal area bounded by our loop $C$
\begin{eqnarray}
   S_{min}(C) = \argmin_{S: \partial S = C} \int_S  d S 
\end{eqnarray}
For every point $\vec r \in R_3$, there is the nearest point $\vec r_1$ at the minimal surface $S_{min}(C)$. 
\begin{eqnarray}
   \vec r_1 = \argmin_{ \vec r' \in S_{min}(C)} (\vec r - \vec r')^2;
\end{eqnarray}

For this point $\vec r_1 $ at the surface there is also a nearest point $\vec r_0$ at its edge $C$, minimizing the geodesic distance  $d( a,  b)$ along the surface from $\vec r_1$ to the edge
\begin{eqnarray}
   &&s_0 = \argmin_{s} d(\vec r_1, \vec C(s));\\
   && \vec r_0 = \vec C(s_0);
\end{eqnarray}

Let us also introduce the local frame with vectors $\vec t(s), \vec n(s), \vec \sigma(s)$ at the loop:
\begin{subequations}
\begin{eqnarray}
&& \vec C'(s)^2 =1 ;\\
&&\vec t(s) = \vec C'(s);\\
&& \vec n(s) = \frac{ \vec C''(s)}{|\vec C''(s)|}\\
&&\vec \sigma(s) =\vec t(s) \times \vec n(s);
\end{eqnarray}
\end{subequations}\label{frame}

Our field is then defined as follows:
\begin{subequations}\label{alphabetarho}
\begin{eqnarray}
&& \alpha = \frac{2 \pi s_0}{ \oint | d \vec C|};\\
&& \beta = \arg \left((\vec r - \vec r_0)\cdot \left(\vec n(s_0) + \imath\vec \sigma(s_0)\right) \right);\\
&& \rho = \sqrt{(\vec r - \vec r_1)^2 + d(\vec r_1, \vec r_0)^2} ;\\
&& \theta =\frac{\lambda w^2 + \pi \rho^2 }{w^2 + \rho^2};\\
&& \phi =  m \alpha +  n \beta;\\
&&\vec S =  \left(\sin\theta \cos\phi, \sin\theta \sin\phi, \cos\theta\right) ;\\
\end{eqnarray}
\end{subequations}
The parameter $w$ plays the role of the size of the Kelvinon in physical space $R_3$.

When the point $\vec r $ approaches the nearest point $\vec r_1$ at the minimal surface, the difference $ \vec \eta = \vec r - \vec r_1$ is normal to this surface.

When the point $\vec r_1$ approaches the nearest point $\vec r_0$ at the edge $C$ of the surface, the geodesic becomes a straight line in $R_3$, tangent to the surface and $\rho$ becomes Euclidean distance to the loop
\begin{eqnarray}
  &&d(\vec r_1, \vec r_0) \ra |\vec r_1 - \vec r_0|;\\
   &&\rho^2 \ra  |\vec r - \vec r_1|^2 + |\vec r_1 - \vec r_0|^2 = |\vec r - \vec r_0|^2 
\end{eqnarray}

Note that all variables $s_0, \vec r_0, \vec r_1, \alpha, \beta,\rho,\theta, \phi$ depend on $\vec r \in R_3$ through the minimization of the distance to the surface and the loop. By construction, $\rho= |\vec r - \vec r_0|$ away from the surface, when $\vec r_1 = \vec r_0, d(\vec r_1, \vec r_0) =0$.

The Euler angles $\theta, \phi$ for the \CL{} field take the boundary values at the loop :
\begin{eqnarray}
&& \phi(\vec r \ra C) \ra  m \alpha + n \beta;\\
&&\theta(\vec r \ra C) =  \lambda + O\left((\vec r-C)^2\right);
\end{eqnarray}
and $\theta(\infty) = \pi$. 

One may estimate the decay rate of $\vec \nabla \cos\theta \sim 1/|\vec r|^3, \vec \nabla \phi\sim 1/|\vec r|$, which corresponds to vorticity decaying as $1/|\vec r|^4$.

This decay is sufficient for the convergence of the enstrophy integral at infinity.

Let us move $\vec r$ along the normal from the surface at $\vec r_1$. Our parametrization of $\theta$ does not change in the first order in normal shift $\vec \eta = \vec r - \vec r_1$, as $\rho^2$ has only quadratic terms in $\vec \eta$.

 We conclude that the normal derivative of the \CL{} field $\phi_1 = Z(1+ \cos\theta)$ vanishes
\begin{eqnarray}
\partial_n \phi_1=0;
\end{eqnarray}

 We requested vanishing normal velocity at the discontinuity surface for this surface to be stationary.

 In terms of the \CL{} parametrization, the normal velocity would vanish provided
 \begin{eqnarray}\label{dnS3}
\partial_n \Phi = \left(\vec \nabla \times \Psi\right)_n; \; \vec r \in S_C\setminus C
\end{eqnarray}

As for the angular field $\phi$ in \eqref{alphabetarho}, its normal derivative does not vanish in the general case. The angle $\alpha$ does not change when the point $\vec r$ moves in normal direction  $\vec N(\vec r_1)$ from the surface projection $\vec r_1$ by infinitesimal shift $\vec \eta = \eps \vec N(\vec r_1) $. However, another angle $\beta$ changes in the linear order in $\eps$ as
\begin{eqnarray}
   \vec N(\vec r_1) \cdot \left(\vec n(s_0) + \imath\vec \sigma(s_0)\right) \neq 0
\end{eqnarray}

\section{Asymptotic expansion of the Kelvinon probability tail}\label{AsympPDF}

The probability distribution \eqref{PDF} does not reduce to a known special function.  However, the terms of asymptotic expansion at large $\Gamma$ can be computed analytically.

Integrating out the third component of the vector $\xi$, we find (up to irrelevant constant normalization)
\begin{eqnarray}
   && \Pi(\gamma) \propto \frac{\exp{- \frac{\mu}{c}}}{\sqrt{c \mu}} \int d^2 \xi \frac{\exp{-\frac{1}{2}\xi_i^2 +\frac{\xi_i\Lambda_{i j} \xi_j}{2 c}} }{\sqrt{\left(1 - \frac{\xi_i \Lambda_{i j} \xi_j}{2\mu}\right)}};\\
   && \mu = \gamma - \tau;
\end{eqnarray}

Now, we expand this square root in Taylor expansion in $1/\mu$
\begin{eqnarray}
    \frac{1}{\sqrt{\left(1 - \frac{\xi_i \Lambda_{i j} \xi_j}{2\mu}\right)}} = \sum_0^\infty \frac{\Gamma(n + \oh)}{\sqrt{\pi} n! \mu^n} \left(\frac{\xi_i \Lambda_{i j} \xi_j}{2}\right)^n
\end{eqnarray}
Interchanging the summation and integration and representing
\begin{eqnarray}
   && \left(\frac{\xi_i \Lambda_{i j} \xi_j}{2}\right)^n \exp{\frac{\xi_i\Lambda_{i j} \xi_j}{2 c}} =\nonumber\\
   && \left(\d_z\right)^n \left.\exp{z\frac{\xi_i\Lambda_{i j} \xi_j}{2 }}\right|_{z = 1/c}
\end{eqnarray}
we use \Mathematica{}, and get the following asymptotic expansion
\begin{eqnarray}
    &&\Pi(\gamma) \sim \frac{\exp{- \frac{\gamma-\tau}{c}}\sqrt{c}}{\sqrt{(c-a)(c-b) (\gamma-\tau)}} \sum_0^\infty F_n\frac{c^n}{(\gamma-\tau)^{n}};\\
    && F_n = \frac{\Gamma \left(n+\frac{1}{2}\right) \left(\frac{a b}{(c-a) (c-b)}\right)^{n/2} 
    P_n\left(\frac{(a+b) c-2 a b}{2 \sqrt{a b (c-a)(c-b)}}\right)}{\sqrt{\pi }}
\end{eqnarray}
where $P_n(z)$ are Legendre polynomials.

These coefficients $F_n$ are rational functions of the ratios of eigenvalues $a,b,c$.
The first four coefficients
\begin{eqnarray}
   && F_0 =1,\\
   && F_1 =\frac{1}{4} \left(\frac{a}{c-a}+\frac{b}{c-b}\right),\\
   && F_2 = \frac{3 \left(c^2 \left(3 a^2+2 a b+3 b^2\right)+8 a^2 b^2-8 a b c (a+b)\right)}{32 (a-c)^2 (b-c)^2},\\
   && F_3 = 
   \frac{15 (2 a b-c (a+b)) }{128 (a-c)^3 (c-b)^3}*\nonumber\\
   &&\left(c^2 \left(5 a^2-2 a b+5 b^2\right)+8 a^2 b^2-8 a b c (a+b)\right);\\
   &&\dots
\end{eqnarray}
\section{Geometric structure of constrained vortex surfaces
 ({by Camillo De Lellis, Elia Bru\'e})}\label{deLellis}
 {
 \begin{center}
    \parbox{0.8\textwidth}{
This short note derives some geometric conditions that Migdal's constrained vortex surfaces have to satisfy.
}
\end{center}

\subsection{CVS conditions: geometric interpretation and instability in the compact case}

Let $\mathcal S \subset \mathbb R^3$ be a (sufficiently smooth) complete connected surface. We denote by $\mathcal{S}_-$ the region enclosed by $\mathcal S$, and set $\mathcal{S}_+ := \mathbb R^3 \setminus (\mathcal S \cup \mathcal{S}_-)$.

According to Migdal's definition, a CVS (constrained vortex surface) solution to the Euler equations is a pair of (sufficiently smooth) velocity fields $(v_-, v_+)$ defined on $\mathcal{S}_-$ and $\mathcal{S}_+$, respectively, such that
\begin{align}
	& {\rm div}\, v_+ = {\rm div}\, v_-  = 0 \label{eq:divfree}\\
	&{\rm curl}\, v_+ = {\rm curl}\, v_- = 0 \label{eq:curlfree}\\
	v_+\, , v_- \quad & \mbox{are tangent to $\mathcal S$}  \label{eq:tangent}
	\\
	(Dv_+ + Dv_-)&\cdot (v_+-v_-) = 0 \label{eq:strain}
	\\
	|v_+|^2 - |v_-|^2 & \mbox{ is locally constant on $\mathcal S$.}\label{eq: Bernoulli}
\end{align}
Moreover, the constrained vortex surface has to satisfy the following stability condition
\begin{equation}\label{e:stability}
	\langle D(v_+ + v_-)\cdot n\, , n \rangle < 0\, ,
\end{equation}
where $n$ is the exterior normal to $\mathcal S$ and $\langle a,b\rangle$ denotes the scalar product of the vectors $a$ and $b$.

We have the following geometric characterization of \eqref{eq:tangent}, \eqref{eq:strain} and \eqref{eq: Bernoulli}.

\begin{theorem}\label{t:identity-curvature}
	Conditions \eqref{eq:tangent}, \eqref{eq:strain}, \eqref{eq: Bernoulli} are satisfied iff
	\begin{equation}\label{eq: geometric}
		\begin{cases}
			A(v_-, v_-) = A(v_+, v_+)
			\\
			[v_+, v_-] = 0 \, ,
		\end{cases}
	\end{equation}
	where $A$ is the second fundamental form of $\mathcal S$, and $[v_+, v_-]$ is the Lie bracket, i.e.
	\begin{equation*}
		[v_+, v_-] = Dv_+ \cdot v_- - Dv_- \cdot v_+\, .
	\end{equation*}
\end{theorem}

\begin{remark}
	Recall that the Lie Bracket $[v_+, v_-] = (Dv_+ \cdot v_- - Dv_- \cdot v_+)|_{\mathcal S}$ is tangent to $\mathcal{S}$ because $v_+$ and $v_-$ are tangent to $\mathcal{S}$.
\end{remark}

\begin{proof}
	
	Condition \eqref{eq: Bernoulli} is equivalent to
	\begin{equation}\label{eq:Bernoulli2}
		(D v_+ \cdot v_+ - D v_- \cdot v_-)|_{\mathcal S}
		\quad
		\mbox{is parallel to $n$}\, .
	\end{equation}
	We rewrite \eqref{eq:strain} as $I + II = 0$, where
	\begin{align*}
		I & = D v_+ \cdot v_+ - D v_- \cdot v_-
		\\
		II & = - Dv_+ \cdot v_- + Dv_- \cdot v_+ \, ,
	\end{align*}
	and observe that $I$ is normal to $\mathcal S$, as a consequence of \eqref{eq:Bernoulli2}, while $II$ is tangent to $\mathcal S$. Indeed,
	\begin{equation*}
		\langle II, n \rangle = A(v_+, v_-) - A(v_-, v_+) = 0\, .
	\end{equation*}
	In particular $I + II =0$ iff $I=0$ and $II=0$. 
	It is immediate to rewrite the two conditions as \eqref{eq: geometric}.
	
\end{proof}

Regarding the stability condition \eqref{e:stability}, we can prove the following rigidity result.

\begin{theorem}\label{t:nonexistence}
	If $\mathcal S \subset \mathbb R^3$ is a smooth compact closed surface and $v_+, v_-$ a pair of vector fields that satisfy the conditions \eqref{eq:divfree} and \eqref{eq:tangent}, then
	\begin{equation}
%		\label{e:average0}
		\int_{\mathcal S} \langle D(v_+ + v_-)\cdot n\, , n\rangle = 0\, .
	\end{equation}
	In particular, \eqref{e:stability} cannot hold.
\end{theorem}

The same argument allows for a similar conclusion when $\mathcal{S}$ is a smooth compact surface with boundary $\partial \mathcal{S}$ and $v^++v^-$ is tangent to the boundary.

\begin{theorem}\label{t:nonexistence-2}
Assume $\mathcal S \subset \mathbb R^3$ is a smooth compact surface with smooth boundary $\partial \mathcal{S}$ and $v_+, v_-$ a pair of vector fields that satisfy the conditions \eqref{eq:divfree} and \eqref{eq:tangent}. If in addition $v^++v^-$ is tangent to $\partial \mathcal{S}$, then
	\begin{equation}\label{e:average0}
		\int_{\mathcal S} \langle D(v_+ + v_-)\cdot n\, , n\rangle = 0\, .
	\end{equation}
	In particular, \eqref{e:stability} cannot hold.
\end{theorem}

\begin{remark} Clearly Theorem \ref{t:nonexistence} can be thought as the particular case of Theorem \ref{t:nonexistence-2} has empty boundary.  
	The smoothness needed by the proof given below is that the surface and its boundary are both $C^1$ (i.e. they have a tangent at every point, which in turn varies continuosly), and that the vector fields are $C^1$, i.e. continuously differentiable.
\end{remark}

\begin{proof}[Proof of Theorem \ref{t:nonexistence-2}]
	Fix a point $p\in \mathcal S$ and choose a local orthonormal frame around $p$ consisting of $e_1, e_2, n$, where the vectors $e_1$ and $e_2$ are tangent to $\mathcal S$. Note that, since both $v_+$ and $v_-$ are divergence free we have that $D (v_+ + v_-)$ is a trace-free matrix, and thus
	\begin{equation}\label{e:trace-free}
		\langle D (v_+ + v_-)\cdot n, n \rangle = - \langle D (v_+ + v_-) \cdot e_1, e_1 \rangle -  \langle D (v_+ + v_-) \cdot e_2, e_2 \rangle\, .
	\end{equation}
	Let $g$ be the Riemannian scalar product induced on $\mathcal S$ by the Euclidean one, and denote by $\nabla^\mathcal{S}$ the corresponding Levi-Civita connection. Recall that the latter coincides with the connection induced by the Euclidean connection. This amounts to say that, if $X, Y$ and $Z$ are tangent vector fields to $\mathcal S$, then
	\[
	g (\nabla^{\mathcal S}_Y X, Z) = \langle DX \cdot Y, Z\rangle\, .
	\]
	If we apply this to $Y=Z= e_i$ and $X = v_+ + v_-$, \eqref{e:trace-free} can be rewritten as 
	\begin{equation}\label{eq:divergence}
		g (\nabla^\mathcal{S}_{e_1} (v_+ + v_-), e_1) + g (\nabla^\mathcal{S}_{e_2} (v_+ + v_-), e_2) = -  \langle D(v_+ + v_-)\cdot n\, , n\rangle\, .
	\end{equation}
	Notice now that the left hand side of \eqref{eq:divergence} is the divergence in $\mathcal S$ of the tangent vector field $v_+ + v_-$, namely we have
	\begin{equation}\label{e:divergence}
		{\rm div}_\mathcal{S}\, (v_+ + v_-) = -  \langle D(v_+ + v_-)\cdot n\, , n\rangle\, .
	\end{equation}
	Next, by Gauss theorem, the fact that $\mathcal{S}$ is has a smooth boundary implies that
	\begin{equation}\label{e:by-parts}
	\int_\mathcal{S} {\rm div}_\mathcal{S}\, (v_+ + v_-) = - \int_{\partial \mathcal{S}} (v_++v_-) \cdot \nu\, ,
	\end{equation}
	where we denote by $\nu$ the smooth unit vector field on $\partial \mathcal{S}$ which is tangent to $\mathcal{S}$, orthogonal to $\partial \mathcal{S}$ and points ``inwards'', i.e. towards $\mathcal{S}$.
Since $v^++v^-$ is parallel to $\partial \mathcal{S}$, the integrand in the right hand side of \eqref{e:by-parts} vanishes identically, which allows us to conclude \eqref{e:average0}.
\end{proof}

\subsection{Rigidity of the CVS condition for closed surfaces}

We prove rigidity results for closed connected CVS surfaces without imposing the stability condition \eqref{e:stability}. We prove that
\begin{itemize}
	\item[(A)] there are no CVS solutions with $\mathcal{S}$ homeomorphic to the sphere;

	\item[(B)] if $\mathcal{S}$ has genus bigger than $1$, then $\mathcal{S}$ cannot be real analytic;

	\item[(C)] if $\mathcal{S}$ has genus $1$, then there are no axisymmetric solutions.
\end{itemize}
The precise formulation of (A) is the following:

\begin{theorem}\label{thm:sphere}
	If $(v_-, v_+)$ satisfies the CVS conditions \eqref{eq:divfree}, \eqref{eq:curlfree}, \eqref{eq:tangent}, \eqref{eq:strain}, \eqref{eq: Bernoulli}, and $\mathcal{S}$ is homeomorphic to the sphere, then $v_+ = 0$ and  $v_- = 0$.
\end{theorem}

Point (B) is a consequence of a more general fact which we state in the following theorem.

\begin{theorem}\label{t:main}
	The following holds:
	\begin{itemize}
		\item[(i)] If $\mathcal S$ is a smooth closed connected CVS surface with genus different than $1$, then $|v_+|^2 = |v_-|^2$.
		\item[(ii)] If $\mathcal S$ is any smooth CVS surface and $|v_+|^2 = |v_-|^2$, then for any point $q\in \mathcal S$ one of the following conditions must necessarily hold:
		\begin{itemize}
			\item[(a)] The Gauss curvature of $\mathcal S$ is $0$ at $q$;
			\item[(b)] $v_+ (q)= v_- (q)$;
			\item[(c)] $v_+ (q)=- v_- (q)$.
		\end{itemize}
	\end{itemize}
\end{theorem}

From the above theorem we draw the following simple 

\begin{corollary}\label{c:absence}
	Assume $\mathcal S$ is a closed smooth connected CVS surface with genus strictly higher than $1$. Then either $v_+ = v_-$ on some nontrivial open subset of $\mathcal S$, or $v_+=-v_-$ on some nontrivial open subset of $\mathcal S$. 
\end{corollary}

\begin{proof}
	In fact, if the sets $\{v_+ = v_-\}$ and $\{v_+=-v_-\}$ contain no interior points, then the Gauss curvature would vanish on a dense set, and by continuity it must vanish anywhere. But it is well known that a surface with vanishing Gauss curvature cannot be closed.	
\end{proof}

 The condition is even more stringent if $\mathcal S$ is real analytic, i.e. it can be described as the graph of a function with a converging Taylor series around any point, up to rotation of the coordinates. 

\begin{corollary}\label{c:absence2}
	Assume $\mathcal S$ is a closed real analytic connected CVS surface with genus strictly higher than $1$. Then either $v_+=v_-$ on $\mathcal S$, or $v_+=-v_-$ on $\mathcal S$. 
\end{corollary}

\begin{proof}
	Since both $v_+$ and $v_-$ are locally gradients of harmonic functions which satisfy the Neumann boundary condition, it turns out that their restrictions to $\mathcal S$ are real analytic as well. But real analytic functions which vanish on a nontrivial open set, must vanish identically because $\mathcal S$ is connected.
\end{proof}

We finally detail the non existence result (C) for axisymmetric solutions of genus $1$.
Let us consider a closed simple curve $[0,1] \ni t \to \gamma(t) = (\gamma_r(t), \gamma_z(t))\in (0,\infty)^2$, and the associated torus of rotation $\mathcal{S} \subset \mathbb R^3$, parametrized by
\begin{equation}\label{eq:torus}
	(t,\theta) \to (\gamma_r(t) \cos \theta, \gamma_r(t) \sin \theta, \gamma_z(t))\, .
\end{equation}
We use polar coordinates $(x,y,z)=(r\cos\theta, r\sin\theta,z)$. 
Recall that $\frac{\partial}{\partial r}$, $\frac{1}{r}\frac{\partial}{\partial \theta}$, $\frac{\partial}{\partial z}$ is an orthonormal frame.

We look for CVS solutions with axisymmetry. The general Ansatz is the following.
\begin{align*}
	v_{-} & :=  \alpha \frac{1}{r^2}\frac{\partial}{\partial \theta } = \nabla \theta 
	\\
	v_+ & := a(r,z) \frac{\partial }{\partial r} + b(r,z) \frac{1}{r} \frac{\partial }{\partial \theta} + c(r,z) \frac{\partial }{\partial z}\, ,
\end{align*}
for some $\alpha\in \mathbb{R}$ and $C^1$ functions $a$, $b$ and $c$.

\begin{theorem}\label{thm:axisymm torus}
	Let $\mathcal{S}$ be given by \eqref{eq:torus}, for some simple closed curve $\gamma=(\gamma_r,\gamma_z)$ of class $C^2$.
	If $(v_-,v_+)$ is an axisymmetric pair that satisfies the CVS conditions \eqref{eq:divfree}, \eqref{eq:curlfree}, \eqref{eq:tangent}, \eqref{eq:strain}, \eqref{eq: Bernoulli}, then either $v_+=v_-$ on $\mathcal{S}$, or $v_+ = - v_-$ on $\mathcal{S}$.
\end{theorem}

\subsection{Proof of Theorem \ref{thm:sphere}}
Since $\mathcal{S}_-$ is simply connected we can write $v_- = \nabla \Phi_-$ for some harmonic function $\Phi_-$, whose gradient is tangent to $\mathcal{S}$. A simple integration by parts gives
\begin{equation}
  \int_{\mathcal{S}_-} |v_-|^2 \, dx
  =
  \int_{\mathcal{S}_-} |\nabla \Phi_-|^2 \, dx 
  =
  \int_{\mathcal{S}} \Phi_- \nabla \Phi_- \cdot n \, dx 
  =
  0\, ,
\end{equation} 
where $n$ denotes the exterior normal to $\mathcal{S}$.
We deduce that $v_- = 0$. Thanks to Theorem \ref{t:main}(i) we deduce that $v_+=0$ on $\mathcal{S}$.
Since $v_+$ is harmonic, the unique continuation principle implies that $v_+$ vanishes identically.

\subsection{Proof of Theorem \ref{t:main}} 

Let us begin by proving (i).
A well-known theorem in topology implies that any tangent vector field to a closed surface $\mathcal S$ with genus different than $1$ must necessarily vanish at some point. This excludes that the constant in \eqref{eq: Bernoulli} is positive (because then $v_+$ would never vanish) or negative (because then $v_-$ would never vanish). 

\medskip

Let us pass to the proof of (ii).
The key ingredient is the following lemma, whose proof is postponed at the end of this section.

\begin{lemma}\label{lemma:rigidity}
	Let $\Sigma \subset \mathbb R^3$ be a smooth surface and let $U\subset \Sigma$ be an open set. Assume the existence of smooth velocity fields $u,w$ defined on $U$ and tangent to $\Sigma$. If
	\begin{itemize}
		\item[(i)] $u(p),w(p) \neq 0$ and $g_p( v, w) = 0$ for any $p\in U$, where $g$ is the metric induced by the ambient space $\mathbb R^3$;
		\item[(ii)] $[u,w]=0$ in $U$;
		\item[(iii)] $u$ and $v$ are gradients in $U$, i.e. there exist $\alpha,\beta:U \to \mathbb R$ such that $g_p(u,\cdot) = {\rm d}_p \alpha$ and $g_p(v,\cdot) = {\rm d}_p \beta$ for any $p\in U$.
	\end{itemize}
	Then, the Gaussian curvature of $\Sigma$ is zero in $U$.
\end{lemma}

Let us explain how to prove Theorem \ref{t:main}(ii) given Lemma \ref{lemma:rigidity}. Fix $q$ such that $v_+ (q) \neq v_- (q)$ and $v_+ (q) \neq v_- (q)$ and let $U$ be a neighborhood of $q$ where $v_+ - v_-$ and $v_+ + v_-$ never vanish. 
Since $|v_+|^2 = |v_-|^2$, we have $g_p(u,w)=0$ for any $p\in U$. Theorem \ref{t:identity-curvature} implies that $[v_+, v_-]=0$, hence $[u,w]=0$ in $U$. Moreover, \eqref{eq:curlfree} imply that $v_+$ and $v_-$ are irrotational in $\mathcal{S}_+$ and $\mathcal{S}_-$, respectively, and \eqref{eq:tangent} says that $v_+$ and $v_-$ are tangent to $\mathcal S$. This implies that, $v_+$ and $v_-$ are locally gradients in $\mathcal S$. In particular $u$ and $w$ are locally gradients in $U$. We are in position to apply Lemma \ref{lemma:rigidity} with $\Sigma = \mathcal{S}$, which implies that $K =0$ in $U$, where $K$ denotes the Gaussian curvature of $\mathcal S$. This implies in particular that the Gauss curvature of $\mathcal S$ vanishes at $q$.

\subsubsection{Proof of Lemma \ref{lemma:rigidity}} Fix $p\in U$. We claim that there exists local coordinates around $p$ 
\begin{equation*}
	X: W\subset \mathbb R^2 \to U\, , \quad
	(x,y) \to X(x,y) \in U\, ,
\end{equation*}
such that 
\begin{equation*}
	\frac{\partial}{\partial x} = u \, , \quad
	\frac{\partial}{\partial y} = w \, .
\end{equation*}
The claim follows from (ii), indeed we can define
\begin{equation*}
	X(x,y) = \phi_u^x \circ \phi_w^y(p)\in U\, ,
\end{equation*}
where $\phi_u$ and $\phi_w$ are the flow maps associated to $u$ and $v$, respectively. Observe that $X(0,0)=p$. Condition (ii) implies that the flow maps commute, hence
\begin{equation}\label{e:X}
	\frac{\partial}{\partial x} X(x,y) = u\circ X(x,y)\, ,
	\quad
	\frac{\partial }{\partial y} X(x,y) = w\circ X(x,y)\, ,
\end{equation}
in particular (i), along with the implicit function theorem, says that $X$ is a local parametrization in a neighborhood of $p$. 

\medskip

Let $h$ be the metric in the new coordinates $(x,y)$. It turns out that
\begin{equation}
	h_{x,y} = h_{y,x} 
	= g\left(\frac{\partial}{\partial x}, \frac{\partial}{\partial y}\right)
	= g(u,w)
	= 0
\end{equation}
as a consequence of the previous claim and condition (i). We claim that
\begin{equation}\label{eq:main claim h}
	\frac{\partial}{\partial y} h_{x,x} = \frac{\partial }{\partial x} h_{y,y} = 0\, .
\end{equation}
It immediately implies that $\Sigma$ is flat in a neighborhood of $p$. Indeed, as a consequence of \eqref{eq:main claim h} we can write the metric as
\begin{equation*}
	h = a(x) dx^2 + b(y) dy^2 \, ,
\end{equation*}
which is clearly flat.

Let us prove \eqref{eq:main claim h}. As a consequence of (iii) we get
\begin{align}
	\frac{\partial}{\partial y} h_{x,x} &= \frac{\partial}{\partial y} g\left( \frac{\partial}{\partial x}, \frac{\partial}{\partial x} \right)
	= 2g\left( \nabla_{\frac{\partial}{\partial y}} \frac{\partial}{\partial x}, \frac{\partial}{\partial x} \right)\nonumber\\
	&= 2 {\rm Hess}\, \alpha \left( \frac{\partial}{\partial y}, \frac{\partial}{\partial x} \right)\label{z500}
\end{align}
the symmetry of the Hessian implies
\begin{equation}\label{z501}
	{\rm Hess}\, \alpha \left( \frac{\partial}{\partial y}, \frac{\partial}{\partial x} \right)
	= {\rm Hess}\, \alpha \left( \frac{\partial}{\partial x}, \frac{\partial}{\partial y} \right)
	= g\left( \nabla_{\frac{\partial}{\partial x}} \frac{\partial}{\partial x}, \frac{\partial}{\partial y} \right)\, ,
\end{equation}
on the other hand we know that $g\left(\frac{\partial}{\partial x},\frac{\partial}{\partial y} \right)=0$, hence
\begin{align}
	g\left( \nabla_{\frac{\partial}{\partial x}} \frac{\partial}{\partial x}, \frac{\partial}{\partial y} \right)
	& = \frac{\partial}{\partial x} g\left(\frac{\partial}{\partial x},\frac{\partial}{\partial y} \right)
	- g\left(  \frac{\partial}{\partial x}, \nabla_{\frac{\partial}{\partial x}} \frac{\partial}{\partial y} \right)\nonumber\\
	& = - g\left(  \frac{\partial}{\partial x}, \nabla_{\frac{\partial}{\partial y}} \frac{\partial}{\partial x} \right)\, ,\label{z502}
\end{align}
where in the last step we used that $\left[ \frac{\partial}{\partial x}, \frac{\partial}{\partial y} \right]=0$. By collection \eqref{z500}, \eqref{z501}, and \eqref{z502} we deduce
\begin{equation*}
	\frac{\partial}{\partial y} h_{x,x} = - \frac{\partial}{\partial y} h_{x,x}\, ,
\end{equation*}
which implies our claim. We argue in the same way to show that $\frac{\partial}{\partial x} h_{y,y}=0$.

\subsection{Proof of Theorem \ref{thm:axisymm torus}}
Without loss of generality we assume that $\gamma :[0,L]\to (0,\infty)^2$ is parametrized by arclength, where $L$ is the length of $\gamma$.
Recall that we look for solutions of the form
\begin{align*}
	v_{-} & :=  \alpha \frac{1}{r^2}\frac{\partial}{\partial \theta }
	\\
	v_+ & := a(r,z) \frac{\partial }{\partial r} + b(r,z) \frac{1}{r} \frac{\partial }{\partial \theta} + c(r,z) \frac{\partial }{\partial z}\, ,
\end{align*}
where $\alpha\in \mathbb{R}$ and $a$,$b$,$c$ are $C^1$ functions.

\begin{lemma}\label{lem:curl}
	We have that
	${\rm curl}\, v_+=0$ if and only if $b(r,z)r$ is constant and
	\begin{equation}\label{eq:curl a b}
		{\rm curl}\, \left(a(r,z) \frac{\partial }{\partial r}  + c(r,z) \frac{\partial }{\partial z}\right) = 0\, .
	\end{equation}
\end{lemma}

\begin{proof}
	We compute
	\begin{equation*}
		{\rm curl}\, v_+ = 
		\left(  \frac{1}{r}\frac{\partial c}{\partial \theta} - \frac{\partial b}{\partial z} \right) \frac{\partial}{\partial r}
		+ 
		\left(  \frac{\partial a}{\partial z} - \frac{\partial c}{\partial r} \right) \frac{1}{r}\frac{\partial}{\partial \theta}
		+
		\frac{1}{r}\left( \frac{\partial (r b)}{\partial r} - \frac{\partial a}{\partial \theta} \right) \frac{\partial}{\partial z}\, .
	\end{equation*}
	Since $v_+$ is axisymmetric, we have that ${\rm curl}\, v_+ = 0$ if and only if
	\begin{equation}
		\begin{cases}
			\frac{\partial b}{\partial z} = 0
			\\
			\frac{\partial (rb)}{\partial r} = 0
			\\
			{\rm curl} \left(   a \frac{\partial }{\partial r} + c \frac{\partial}{\partial z}  \right) = 0\, .
		\end{cases}
	\end{equation}
	The first two conditions amount to $br = C $, for some constant $C\in \mathbb R$. The latter amounts to \eqref{eq:curl a b}.
\end{proof}

Let us begin by considering the case $\alpha\neq 0$. Without loss of generality we can assume $\alpha=1$.
After imposing ${\rm curl}\,  v_+ =0$, we can write
\begin{equation}
	v_+ = a(r,z) \frac{\partial }{\partial r}  + c(r,z) \frac{\partial }{\partial z} + C v_- =: w_+ + C v_-\, .
\end{equation}
To satisfy the CVS conditions we need to impose the following properties on $w_+$:
\begin{itemize}
	\item[(1)] ${\rm div}\, w_+ = {\rm curl}\, w_+ = 0$
	\item[(2)] $w_+$ is tangent to $\mathcal{S}$
	\item[(3)] $|w_+|^2 + \frac{C^2-1}{r^2} = \ell$ in $\mathcal{S}$, for some $\ell \in \mathbb R$
	\item[(4)] $[w_+, v_-] = 0$ in $\mathcal{S}$
	\item[(5)] $A(w_+ + C v_- , w_+ + C v_-) = A(v_-,v_-)$
\end{itemize}
We show that (2), (3) and (4) imply that $\gamma_r$ is constant.

\medskip

{\em Condition (4).}
We compute
\begin{align*}
	[w_+, v_-]  & =  \nabla_{w_+} v_- - \nabla_{v_-} w_+
	\\&
	= \nabla_{a\frac{\partial}{\partial r} + c \frac{\partial}{\partial z}} \left(\frac{1}{r^2}\frac{\partial}{\partial \theta}\right) 
	- \frac{1}{r^2} \nabla_{\frac{\partial}{\partial \theta}} \left(  a \frac{\partial}{\partial r} + c\frac{\partial}{\partial z}\right)
	= - \frac{2}{r^3} a \frac{\partial}{\partial \theta}\, .
\end{align*}
Hence, we have to impose 
\begin{equation}\label{eq: a=0}
	a(\gamma_r(t),\gamma_z(t)) = 0\, , \quad \text{for any $t\in [0,L]$}\, .
\end{equation}

\medskip

{\em Condition (2) and (3).} Recall that 
\begin{equation*}
	n = (\dot \gamma_z(t) \cos \theta, \dot \gamma_z(t) \sin \theta, -\dot \gamma_r(t))\, ,
\end{equation*}
where $n$ denotes the exterior normal to $\mathcal{S}$.
By using \eqref{eq: a=0}, we deduce
\begin{equation}\label{eq:w_+ after comm}
	w_+ = c(r,z)\frac{\partial }{\partial z} \, ,
	\quad \text{on $\mathcal S$}\, ,
\end{equation}
hence
\begin{equation*}
	0 = w_+ \cdot n = - c(\gamma_r(t),\gamma_z(t))\, \dot \gamma_r(t)\, .
\end{equation*}
In particular, by employing \eqref{eq:w_+ after comm} and (3) we get
\begin{align*}
	0 &= |c(\gamma_r(t),\gamma_z(t))|^2 |\dot \gamma_r(r)|^2 
	= |w_+|^2 |\dot \gamma_r(t)|^2\\
	& = \left(\ell + \frac{1-C^2}{\gamma_r^2(t)}\right) |\dot \gamma_r(t)|^2\, .
\end{align*}
If either $C^2\neq 1$ or $\ell \neq 0$, then $\gamma_r = {\rm const}$. This is impossible because $\gamma$ is a closed simple curved.

If $C^2=1$ and $\ell = 0$, then (3) implies that $|w_+|^2 = 0$ on $\mathcal{S}$. It amounts to $v_+ = v_-$ when $C=1$ and $v_+ = - v_-$ when $C=-1$. 

\medskip

{\em The case $\alpha = 0$.}
Let us now assume that $\alpha=0$.
By applying Lemma \ref{lem:curl} we deduce
\begin{equation*}
	v_+ = a(r,z) \frac{\partial}{\partial r} + c(r,z) \frac{\partial }{\partial z} + C \frac{1}{r^2} \frac{\partial}{\partial \theta}
	=: w_+ + C \frac{1}{r^2} \frac{\partial}{\partial \theta}\, .
\end{equation*}
In this case, to satisfy the CVS conditions we need to impose the following properties on $w_+$:
\begin{itemize}
	\item[(1')] ${\rm div}\, w_+ = {\rm curl}\, w_+ = 0$
	\item[(2')] $w_+$ is tangent to $\mathcal{S}$
	\item[(3')] $|w_+|^2 + \frac{C^2}{r^2} = \ell$ in $\mathcal{S}$, for some $\ell \in \mathbb R$
	\item[(4')] $A(w_+ + C \frac{1}{r^2}\frac{\partial}{\partial \theta} , w_+ + C \frac{1}{r^2}\frac{\partial}{\partial \theta}) = 0$.
\end{itemize}
We show that (4') forces $C=0$. 
Since $w_+$ is tangent to $\mathcal{S}$, there exists $\lambda: \mathcal{S} \to \mathbb{R}$ such that
\begin{equation}\label{z}
	w_+(\gamma) = \lambda \left(\dot \gamma_r \frac{\partial}{\partial r} + \dot \gamma_z \frac{\partial}{\partial z}\right)
	\, .
\end{equation}
We use the well-known identities
\begin{align*}
	A\left(\dot \gamma_r \frac{\partial}{\partial r} + \dot \gamma_z \frac{\partial}{\partial z},\dot \gamma_r \frac{\partial}{\partial r} + \dot \gamma_z \frac{\partial}{\partial z}\right) & = \kappa
	\\
	A\left( \frac{\partial}{\partial \theta}, \frac{\partial}{\partial \theta}\right) & = \gamma_r \dot \gamma_z
	\\
	A\left(\dot \gamma_r \frac{\partial}{\partial r} + \dot \gamma_z \frac{\partial}{\partial z}, \frac{\partial}{\partial \theta}\right) & = 0\, ,
\end{align*}
where $\kappa= \ddot\gamma_z \dot\gamma_r - \ddot\gamma_r \dot\gamma_z$ is the curvature of $\gamma$.
We deduce
\begin{equation}\label{eq:main}
	0 = A \left(w_+ + C \frac{1}{r^2}\frac{\partial}{\partial \theta} , w_+ + C \frac{1}{r^2}\frac{\partial}{\partial \theta}\right) = \lambda^2\kappa + \frac{C^2}{\gamma_r^3} \dot \gamma_z
\end{equation}
Let us consider $t_0\in [0,L]$, a minimum point for $\gamma_r$. Since $\dot \gamma_r(t_0) = 0$ and $\gamma$ is parametrized by arclength, we deduce that $\dot \gamma_z(t_0)\neq 0$. Hence, \eqref{eq:main} implies that $\lambda(t_0)\neq 0$. So, in a small neighborhood of $t_0$ we can rewrite \eqref{eq:main} as
\begin{equation*}
     \ddot\gamma_z \dot\gamma_r - \ddot\gamma_r \dot\gamma_z
     = \kappa
     = \frac{C^2}{\lambda^2\gamma_r^3} \dot \gamma_z\, .
\end{equation*}
We multiply the latter by $\dot \gamma_z$, and use the identity $\ddot \gamma_z \dot \gamma_z = - \ddot \gamma_r \dot \gamma_r$ (which is a consequence of $(\dot \gamma_r)^2 + (\dot \gamma_z)^2=1$), to get
\begin{equation}\label{eq:main2}
	\ddot \gamma_r = - \frac{C^2}{\lambda^2 \gamma_r^3}(1-(\dot \gamma_r)^2)\, .
\end{equation}
Using that $\ddot \gamma_r(t_0)\ge 0$ and $\dot \gamma_r(t_0) = 0$ we conclude that $C=0$. 

We now use (3') to deduce that $v_+ = 0$. Indeed, $\lambda^2 = \ell$, since $\lambda$ is continuous we deduce that $\lambda$ is a constant. If $\lambda = 0$ then $v_+=0$.
If $\lambda \neq 0$, then \eqref{eq:main2} gives $\kappa = 0$, which contradicts the fact that $\gamma$ is closed.

\subsection{Solutions with cylindrical symmetry}

In \cite{M21d} Migdal finds stable solutions to the CVS equations with linear growth and
vorticity concentrated on a cylinder $\mathcal S \subset \mathbb{R}^3$.
It turns out that the cross section of $\mathcal S$ is noncompact. Given the cylindrical symmetry and the linear growth at infinity, that is the best one can hope for. Below we show that there are {\em no} solutions to the CVS equations (irrespectively of the stability condition) with cylindrical symmetry, compact cross section and linear growth.

Consider:
\begin{itemize}
	\item[(i)] a smooth simple closed curve $\sigma \subset \mathbb R^2$;

	\item[(ii)] the bounded simply connected  domain $\Omega_-$ bounded by $\sigma$;

	\item[(iii)] the unbounded domain $\Omega_+:= \mathbb R^2\setminus (\sigma \cup \Omega_-)$;

	\item[(iv)] the surface $\mathcal S \subset \mathbb R^3$ given by $\sigma \times \mathbb R$;

	\item [(v)] the cylindrical domains $\mathcal{S}_\pm :=  \Omega_\pm \times \mathbb{R}$.
\end{itemize}
We are looking for two bounded vector fields $v_\pm : \Omega_\pm \to \mathbb R^2$ and a quadratic function $q: \mathbb R^3 \to \mathbb R$ with the following properties:
\begin{itemize}
	\item[(a)] $q$ is harmonic;
	\item[(b)] the maps $u_\pm := (v_\pm,0) + \nabla q : \mathcal{S}_\pm \to \mathbb R^3$ are divergence free, curl-free and tangent to $\mathcal S$;
	\item[(c)] for any point $p\in \mathcal S$, the vector $u_+ (p) - u_- (p)$ belongs to the kernel of $Du_+ (p) + Du_- (p)$.
\end{itemize} 

We claim the following.

\begin{theorem}\label{t:cylinder}
	$v_\pm$ and $q$ must vanish identically.
\end{theorem}

We can in fact consider the following more general situation:
\begin{itemize}
\item[(i')] $\sigma_i$, $i\in \{1, \ldots, N_0\}$, is an arbitrary finite collection of simple closed curves which are pairwise disjoint;
\item[(ii')] $\{\Omega_j\}$, $j\in \{1, \ldots, N_0+1\}$ are the connected components of $\mathbb R^2\setminus \bigcup_i \sigma_i$;
\item[(iii')] The surface $\mathcal{S}\subset \mathbb R^3$ is the union of the cylinders $\mathcal{S}_i := \sigma_i \times \mathbb R$;
\item[(iv')] The cylindrical domains are given by $\Lambda_j:= \Omega_j \times \mathbb R$.
\end{itemize}
Under these more general assumptions we are looking for $N_0+1$ bounded vector fields $v_j : \Omega_j\to \mathbb R^2$ and a quadratic function $q: \mathbb R^3 \to \mathbb R$ with the following properties:
\begin{itemize}
\item[(a')] $q$ is harmonic;
\item[(b')] The maps $u_j:= (v_j, 0) +\nabla q: \Lambda_j \to \mathbb R^3$ are divergence-free, curl-free, and tangent to $\partial \Lambda_j \subset \mathcal{S}$;
\item[(c')] If $\Lambda_i$ and $\Lambda_j$ have a common boundary $\mathcal{S}_k$, then for any point $p\in \mathcal{S}_k$ the vector $u_i (p)-u_j (p)$ belongs to the kernel of $Du_i (p)+ Du_j (p)$. 
\end{itemize}

Under these assumptions Theorem \ref{t:cylinder} can be generalized to

\begin{theorem}\label{t:cylinders}
	$v_i$ and $q$ must vanish identically.
\end{theorem}

\begin{proof}[Proof of Theorem \ref{t:cylinder}]
	Consider the function $\varphi (x,y) = q (x,y,0)$ and the vector field $\xi_- (x,y) = v_- (x,y) + \nabla \varphi (x,y)$. Observe that $v_-$ is curl-free and, since $\Omega_-$ is simply connected, there is a potential $\zeta_-: \Omega_- \to \mathbb R$ for $\xi_-$. Now, 
	\begin{equation}\label{e:potential-1}
	\frac{\partial \zeta_-}{\partial \nu}=0 \qquad \mbox{on $\sigma = \partial \Omega_-$.}
	\end{equation}
	On the other hand, since $\varphi$ is quadratic, $\Delta \zeta_-$ is a constant. Observe that, therefore, from 
	\begin{equation}\label{e:potential-2}
	\int_{\Omega_-} \Delta \zeta_- = \int_{\sigma} \frac{\partial \zeta_-}{\partial \nu} = 0\, ,
	\end{equation}
	it turns out that 
	\begin{equation}\label{e:potential-3}
	\Delta \zeta_- = \Delta \varphi = 0\, .
	\end{equation}
	But then we can integrate by parts to conclude 
	\begin{equation}\label{e:potential-4}
	\int_{\Omega_-} |\nabla \zeta_-|^2 = \int_\sigma \zeta_- \frac{\partial \zeta_-}{\partial \nu} = 0\, .
	\end{equation}
	In particular $\xi_-$ vanishes identically. 
	
	Define next $\xi_+ (x,y) = v_+ (x,y) + \nabla \varphi (x,y)$. (c) implies that $\xi_+ (x,y)$ is in the kernel of $D\xi_+ (x,y)$ for every $(x,y)\in \sigma$. Assume $\zeta_+$ is a potential for $\xi_+$ in some simply connected domain $U\cap \Omega_+$, where $U$ is the neighborhood of some point $(x,y)\in \sigma$. Then the latter condition can be rewritten as 
	\[
	\frac{1}{2} \nabla |\nabla \zeta_+|^2 = 0
	\quad \text{on $\sigma \cap U$} 
	\, . 
	\]
	If $|\nabla \zeta_+|=0$ on $\sigma\cap U$, then $\zeta_+$ is a constant over $\sigma \cap U$ and if we extend $\zeta_+$ to $\Omega_-\cap U$ by setting it equal to the latter constant, we immediately see that $\zeta_+$ is $C^1$ on $U$ and weakly harmonic, hence harmonic. So $\zeta_+$ must vanish by unique continuation for harmonic functions. If $|\nabla \zeta_+|= c >0$, we conclude that $D^2 \zeta_+(p)$ has a nontrivial kernel for every $p\in \sigma \cap U$, but since the trace of the two-dimensional matrix $D^2 \zeta_+ (p)$ is zero, we must conclude that $D^2 \zeta_+$ vanishes identically on $\sigma\cap U$. But this means that $\frac{\partial \zeta_+}{\partial x}$ and $\frac{\partial \zeta _+}{\partial y}$ are both locally constant over $\sigma\cap U$. The assumption that $\frac{\partial \zeta_+}{\partial \nu}=0$ on $\sigma\cap U$ implies therefore that $\nabla \zeta_+$ must vanish identically on $\sigma \cap U$.
	
	Having concluded that both $\xi_-$ and $\xi_+$ vanish identically, we immediately conclude that actually $u_+=u_-$ on $\mathcal S$ and that the corresponding function $u$ given by defining $u = u_\pm$ on $\mathcal{S}_\pm$ is the gradient of a quadratic harmonic function. 
	Since $u$ must be tangent to $\mathcal S$, we see right away that $u$ must vanish identically. 
\end{proof}

\begin{proof}[Proof of Theorem \ref{t:cylinders}]
The proof is by induction over the number $N_0$ of curves. The start of the the induction, namely $N_0=1$, is in fact Theorem \ref{t:cylinder}. Consider therefore an arbitrary $N_0>1$ and assume that the theorem is correct when $N_0$ is substituted by $N_0-1$. Fix a collection of curves $\sigma_i$ as in (i') above. Each $\sigma_i$ bounds a unique simply connected domain $\Xi_i$ in $\mathbb R^2$, and given that the curves are pairwise disjoint, each $\sigma_j$ with $j\neq i$ is either contained in $\Xi_i$ or in the interior of its complement. Since the curves are finitely many it is obvious that one of them is an ``innermost'' curve, namely there is a $\Xi_i$ which does not contain any curve $\sigma_j$. Without loss of generality we can assume $i=1$ and observe that $\Xi_1$ must be one of the domains $\Omega_i$: again without loss of generality we can assume it is $\Omega_1$. We then denote by $\Omega_2$ the only other connected component of $\mathbb R^2\setminus \bigcup_i \sigma_i$ whose boundary intersects $\sigma_1$ (in fact we must have $\sigma_1 \subset \partial \Omega_2$, but observe that the inclusion might be strict). If we set $v_- := v_1$ and $v_+ := v_2$, we can now repeat the argument of Theorem \ref{t:cylinder}. First of all the potential $\zeta_-$ exists in our case as well because $\Omega_1$ is simply connected, and hence the conclusions \eqref{e:potential-1}-\eqref{e:potential-2} can all be drawn in our case as well. The subsequent argument leads to the conclusion that $v^+$ is in fact a smooth continuation of $v^-$ across $\sigma_1\cap U$ in any simply connected neighborhood $U$ of $p\in \sigma= \sigma_1$: the argument can be taken verbatim in our case as long as $U$ does not intersect any other curve $\sigma_j$ with $j>1$. In particular we conclude that $\sigma_1$ could actually be eliminated from the collection of curves because the function $\tilde{v}$ defined to be $v_1$ in $\Omega_1$ and $v_2$ in $\Omega_2$ is in fact smooth across $\sigma_1$. Having reduced the number of curves by $1$ we can apply the inductive assumption and conclude the validity of the theorem. 
\end{proof}\label{Camillo}
}
 \section{Topological defects in Turbulent flows,
 (by Grigory Volovik)}\label{Volovik}
 {
 % v_8
% \documentstyle[12pt]{article}
% \documentclass[prb,
%  twocolumn,
%  superscriptaddress,showpacs,amsmath,amssymb]{revtex4}
% \usepackage{amsfonts}
% \usepackage{bm}
% \usepackage{verbatim}

% \usepackage{graphicx}

% \begin{document}

% \title{On Migdal vortex sheet turbulence and monopole terminating Dirac strings}

% \author{G.E.~Volovik}
% \affiliation{Low Temperature Laboratory, Aalto University,  P.O. Box 15100, FI-00076 Aalto, Finland}
% \affiliation{Landau Institute for Theoretical Physics, acad. Semyonov av., 1a, 142432,
% Chernogolovka, Russia}

% \date{\today}

% \begin{abstract}
% {}
% \end{abstract}
% \pacs{
% }

% \maketitle
\begin{center}
\parbox{0.8\textwidth}{
    Let us consider the Migdal view on classical turbulence (see Refs.\cite{M20c,M21b})  and look for the connections between the classical hydrodynamics and turbulence in liquids and quantum hydrodynamics and turbulence in fermionic superfluids.
    }
\end{center}

\subsection{Classical vs quantum turbulence}

Of course, there are important differences between the turbulence in quantum and classical systems.
In classical liquids, the viscous term has more derivatives that nonlinear Euler term. In fermionic superfluids both terms have the same number of derivatives,
see Eq.(2) in Ref. \cite{Finne2003}. The corresponding Reynolds number is the ratio of two dimensionless parameters -- the internal parameters of the liquid, reactive coefficient $1-\alpha'$ and the dissipative one $\alpha$.  
This Reynolds number does not depend on velocity: it depends only on such internal characteristic of the liquid as temperature $T$. By changing  $T$,  the transition to quantum turbulence is experimentally observed, when the ratio of the reactive and dissipative parameters crosses unity. 

The possible relation of viscosity anomaly to chiral anomaly may follow from the observation that the parameters $\alpha$ and  $\alpha'$ in Eq.(2) of  Ref.\cite{Finne2003} are related to the Adler-Bell-Jackiw (ABJ) chiral anomaly. The vortex has chirality, and its motion produces the spectral flow in the vortex cores, which is similar to that in ABJ anomaly and which in some limit case is described by the ABJ equation.\cite{Volovik2003} In that limit, which takes place for vorticity in chiral superfluid $^3$He-A, the ABJ equation with the proper prefactor has been experimentally confirmed.\cite{Bevan1997}

\subsection{Clebsch hydrodynamics vs Heisenberg ferromagnets}

The velocity field in Clebsch variable $(\phi_1,\phi_2,\phi_3)$  is $v_i= -\phi_2\nabla_i \phi_ 1 + \nabla_i \phi_3$. In the Migdal theory, velocity and vorticity are determined in the compact space $S^2$, see Eq.(43) for vorticity in terms of the unit vector $\hat{\bf S}$: 
\begin{equation}
\omega_{ik}=(\nabla \times {\bf v})_{ik} =  Z(\hat{\bf  S} \cdot (\nabla_i \hat{\bf  S} \times \nabla_k \hat{\bf  S}))\ \,.
\label{MerminHo1}
\end{equation}
The polar and azimuthal  components of the unit vector $\hat{\bf S}$ are
$\phi_1/Z= \cos \theta$, $\phi_1 =\phi$, while $Z$ is adiabatic invariant in the absence of viscosity.

Eq.(\ref{MerminHo1}) also takes place in chiral superfluid $^3$He-A, where vorticity is expressed in terms of the unit vector of the orbital angular momentum $\hat{\bf l} \equiv \hat{\bf  S}$. 
In this superfluid the prefactor $Z$ is fixed,  $Z=\hbar/4m_3$, where $m_3$ is the mass of the $^3$He atom, \cite{VollhardtWolfle,Volovik2003} and this equation is known as the Mermin-Ho relation. Since in the chiral superfluid $\kappa=\pi \hbar/m_3$ is the  the quantum of circulation, the prefactor $Z$ in Migdal hydrodynamics plays the role of the circulation quantum, see Sec.\ref{quantization}. 

However, the Clebsch fields are not well determined, since $\phi_2$ is not defined at $\theta=0$. Actually these  Clebsch variables are exactly the same as used in ferromagnets, where the magnetization is ${\bf M}=M\hat{\bf m}\equiv Z\hat{\bf S}$, $M-M_z=M(1- \cos \beta) \equiv Z(1- \cos \beta)=\phi_1$, and $\phi \equiv \phi_2$. Here $\beta$ is the angle of deflection of magnetization from the $z$-axis, and $\phi$ is the azimuthal angle of magnetization.  The role of vorticity is played by the effective gauge field  $F_{ik} \equiv \omega_{ik}$ acting on electrons from magnetization.\cite{Volovik1987} The pseudoelectric field $F_{i0}\equiv \omega_{i0}$ gives rise to the so-called spin-motive force acting on electrons. These gauge fields are related to the Berry phase. 

In ferromagnets, $\phi$ and $M_z$ (or $M-M_z$, since $M$ is the dynamical invariant) are canonically conjugate variables. But since $\phi\equiv \phi_2$ is not well determined, the $3+1$ action $\int dt dV M_z\dot\phi$ is substituted by the $4+1$ Wess-Zumino term, $\int dt d\tau dV F_{0\tau}$, see Eq.(19) in Ref. \cite{Volovik1987}. 
Probably such approach is to be used in Migdal theory.

One may say that the big difference with ferromagnets is that there the magnetization is observable, while in hydrodynamics the Clebsch variables are not. However, if the spin-orbit interaction is neglected, the  magnetization becomes unobservable.
But the effective gauge invariant gauge field $F_{\mu\nu}$ acting on fermions in ferromagnets will be observable in the same way as its analog -- gauge invariant vorticity $\omega_{\mu\nu}$, which is expressed in terms of the unit vector $\hat{\bf S}$ of the analog of magnetization.

\subsection{Topological objects: hedgehog, instanton, domain wall,  quantum of  circulation}
\label{topology}

Since $M\equiv Z$ is the dynamical invariant in reversible dynamics, the reversible hydrodynamics is fully determined by the unit vector of magnetization $\hat{\bf m}\equiv \hat{\bf S}$. It has two topological invariants $\pi_2$ and $\pi_3$. 

The $\pi_2$ invariant is $N= \int \, dS_i \omega^i /Z$. It is proportional to the velocity circulation along the moving loop, which is conserved in moving liquid in the absence of dissipation. For the closed surface, the invariant $N$ should be integer describing the  topological defects -- the hedgehogs or monopoles. In ferromagnets, where $\hat{\bf m}$ is the original variable and vorticity (gauge field $\omega_{ik}$) is the secondary variable, the hedgehog in $\hat{\bf m}$ field looks as the Dirac monopole in the Berry phase, with the unobservable Dirac line, at which the Berry phase  changes by $2\pi$. 

In hydrodynamics, where vorticity $\omega_{ik}$ is the original variable and $\hat{\bf S}$ is the secondary one, the hedgehogs in the $\hat{\bf S}$ field are impossible as separate objects. The Dirac line attached to such monopole is singular and is observable. The same is in the chiral superfluids, where the Dirac line is the singular doubly quantized vortex, $m=2$ (or $m=2N$ in general case, where $N$ is the topological charge of the monopole). Nevertheless, in Migdal theory the integer $m$ is important, because the circulation $\Gamma = Zm$, and thus the prefactor $Z$ plays the role of circulation quantum, and $m$ is the number of quanta.

The $\pi_3$ invariant is the Hopf invariant, which is proportional to the integral over helicity: $N_3= \int d^3r \,{\bf v}\cdot{\boldsymbol\omega}/Z^2$. The corresponding topological defect is instanton, which describes the process of creation of helicity from the "vacuum". 

In Heisenberg ferromagnets there are no domain walls, because $\pi_0=0$. In hydrodynamics the surfaces are possible at which the prefactor $Z$ changes sign. Probably the vortex sheets in Migdal theory -- the thin pancake-like regions of
increasing vorticity\cite{Kuznetsov2018} are the analogous objects.

\subsection{Dynamic invariants in hydrodynamics (circulation and helicity) and their  quantization in superfluids}
\label{quantization}

In Migdal theory, the circulation $\Gamma = Zm$ is the dynamic invariant. In superfluids,  the quantity $m$ becomes the  integer number, i.e. the circulation is quantized: $Z=2\pi \hbar /m_4$ and $Z=\pi \hbar /m_3$ is the quantum of circulations in Bose superfluid $^4$He with mass of atom $m_4$ and  in Fermi superfluid $^3$He with mass of atom $m_3$ correspondingly, and $m$ is the number circulation quanta. 
 Thus in Migdal theory, the prefactor $Z$ plays the role of circulation quantum, and $m$ plays the role of the number of quanta, which becomes integer in the quantum case.

In the same manner, in superfluids the integral of helicity is quantized, $H= Z^2 N_3$ with integer $N_3$. The integer $N_3$ in superfluids is the topological invariant: it is the knot invariant in Bose superfluid $^4$He and in Fermi superfluid $^3$He-B, and the Hopf invariant $\pi_3(S^2)$ in the chiral superfluid $^3$He-A. In both Fermi superfluids, the quantum of helicity is  $Z^2\propto (\hbar/m_3)^2$. In classical hydrodynamics, the prefactor $Z^2\propto (\hbar/m_3)^2$ serves as the "quantum of helicity", and $N_3$ serves as  the "number of quanta", which becomes the integer topological invariant in the quantum case of superfluids. 

The quantization of helicity  and of circulation in superfluids provide two examples, which illustrate the  main property of the dynamical invariants in classical systems. The dynamic invariant in classical systems become quantized in integer numbers in the quantum systems, $N_3$ and $N$ correspondingly.

Similar situation is in Heisenberg ferromagnets. In classical equations $M\equiv Z$ is the dynamic invariant, which is conserved in the Landau–Lifshitz equation without dissipation. In ferromagnets the total magnetic moment $\int d^3x M$ becomes quantized  in terms of $\hbar/2$, i.e.
 $\int d^3x M=k \hbar/2$, where $k$ is integer. This quantization follows from the Wess-Zumino term in action.\cite{Volovik1986}

\subsection{Hydrodynamics in terms of tetrads}
\label{tetrads}

Let us introduce the orthogonal triad of unit length:
\begin{equation}
{\bf  e}_1\,\,,\,  {\bf  e}_2\,\,,\,  \hat{\bf  S}={\bf  e}_1\times {\bf  e}_2  \,. 
% \label{tetrads} % duplicate with section label
\end{equation}
The velocity field:
\begin{equation}
v_i = Z({\bf  e}_1\cdot \nabla_i {\bf  e}_2 - {\bf  e}_2\cdot \nabla_i {\bf  e}_1)\ \,, 
\label{velocity}
\end{equation}
where $Z$ is the dynamical invariant.
The vorticity is expressed in terms $\hat{\bf  S}$ field (according to the Mermin-Ho relation) in the same way as in Migdal theory in Eq.(43) of Ref.\cite{M20c}:
\begin{equation}
\omega_{ik}=(\nabla \times {\bf v})_{ik} =  Z(\hat{\bf  S} \cdot (\nabla_i \hat{\bf  S} \times \nabla_k \hat{\bf  S}))\ \,.
\label{MerminHo}
\end{equation}

The triad field variables correspond to three Clebsch fields, but as distinct from the Clebsch fields the triad is well determined everywhere, except for the real topological singularities.

The $SO(3)$ space of triad, and $S^2$ space of  $\hat{\bf  S}$ provide the geometry, including the effective metric, and topology.
The topological invariants are well determined in terms of $\hat{\bf  S}$, or in terms of rotation of the dyad about $\hat{\bf  S}$-axis. They are integer for special geometries, while the circulation and helicity are represented by these invariants multiplied by the dynamical invariants $Z$ and $Z^2$ correspondingly. 

\subsection{Dirac monopole with observable Dirac strings}
\label{monopole}

The hydrodynamics of incompressible fluid has been discussed in terms of the ${\bf S}$-vector in Ref. \cite{Kuznetsov1980}.
The unit vector description allows the topological objects $\pi_3(S^2)=Z$, which describe the Hopf structure, and also the instantons. It also allows the defects of the  group $\pi_2(S^2)=Z$,  the hedgehogs, or monopoles, in which the  ${\bf S}$-vector wraps $N$ times around a sphere. In tetrads, where the target space is $SO(3)$, the Hopf structures are allowed, since $\pi_3(SO(3))=Z$. But the hedgehogs do not exist as separate objects, since $\pi_2(SO(3))=0$. That is why the monopole in ${\bf S}$-field has singular Dirac string (the vortex singularity)
attached to the monopole, which is observable.\cite{Blaha1976,VolovikMineev1976,Volovik2000}

 The simplest monopole is the radial hedgehog in the $\hat{\bf S}$ field:  $\hat{\bf S}=\hat{\bf r}$. 
 Its topological $\pi_2$ charge is $N=1$. Let us consider two configurations of $N=1$ monopole: (i) with one Dirac string with winding number $m=2N=2$;  and (ii) with two strings, each with  winding number $m=1$.
 
 \subsection{Dirac monopole with one Dirac string}
\label{monopole1}

 $(z,\rho,\phi)$ are cylindrical coordinates and $(r,\theta,\phi)$ are spherical coordinates.
 The velocity field is
\begin{equation}
{\bf v}({\bf r}) = Z{\hat{\boldsymbol\phi}} \frac{1-\cos\theta}{r\sin\theta} \,\,, \,\, \nabla\cdot{\bf v}=0 \,,
\label{velocityMonopoleOneString}
\end{equation}
and vorticity field:
\begin{equation}
{\boldsymbol\omega}({\bf r}) = Z \frac{\hat{\bf r}}{r^2} + 2Z\theta(-z){\hat{\bf  z}} \delta_2({\boldsymbol\rho}) \,\,,\,\, \nabla\cdot {\boldsymbol\omega}=0\,.
\label{vorticityMonopoleOneString}
\end{equation}
The vorticity is equivalent to magnetic field of Dirac monopole with charge $Z$, i.e. ${\bf B} \equiv {\boldsymbol\omega}({\bf r})$. The magnetic flux is conserved,  $\nabla\cdot{\bf B}=0$. The $4\pi Z$ flux from the monopole is compensated by the magnetic flux along the singular flux tube (observable Dirac string) to the monopole. Here there is the single Dirac string on the lower half-axis, $z<0$.  It is the vortex line with circulation $2Z$, or winding number $m=2$ in Migdal notations.
The triads are 
\begin{equation}
\hat{\bf S}({\bf r}) =\hat{\bf r} \,\,,\,  {\bf e}_1 + i{\bf e}_2= e^{i\phi}({\hat{\boldsymbol\theta}} +i {\hat{\boldsymbol\phi}})  \,.
% \label{triadMonopoleTwoStrings} % doubly defined label
\end{equation}

This topological structure with vortex terminated by monopole may have some connection to tornado, see Figs. \ref{TornadoMonopoleFig} and \ref{TornadoMonopole2}. Here the funnel cloud moves towards the ground and when it reaches the ground it forms tornado in Fig.\ref{TornadoFig}. Before the tornado is formed the funnel cloud is the vortex with the end point. The end point represents the monopole-hedgehog in the $\hat{\bf S}$-field and the funnel cloud represents its Dirac string. Probably the monopole may exist also at the final stage of the tornado: when tornado is dying the end of the vortex leaves the ground.

\subsection{Dirac monopole with two Dirac strings}
\label{monopole2}

 In this case the velocity field is:
\begin{equation}
{\bf v}({\bf r}) = -Z{\hat{\boldsymbol\phi}} \frac{\cos\theta}{r\sin\theta} \,,
\label{velocityMonopoleTwoStrings}
\end{equation}
and vorticity field:
\begin{equation}
{\boldsymbol\omega}({\bf r}) = Z \frac{\hat{\bf r}}{r^2} +  Z\,\sign(z)\,\hat{\bf  z} \delta_2({\boldsymbol\rho}) \,.
\label{vorticityMonopoleTwoStrings}
\end{equation}
Now the flux from the monopole is compensated by the magnetic fluxes which enter the monopole along two Dirac strings. 
Each string (vortex line) has circulation $Z$, or winding number $m=1$ in Migdal notations.
The triads are 
\begin{equation}
\hat{\bf S}({\bf r}) =\hat{\bf r} \,\,,\,  {\bf e}_1 + i{\bf e}_2= {\hat{\boldsymbol\theta}} +i {\hat{\boldsymbol\phi}}  \,.
\label{triadMonopoleTwoStrings}
\end{equation}

\subsection{General case with two Dirac strings and monopole with charge N}
\label{monopoleN}

The monopole in ${\bf S}$-field with topological charge $N$, has the Dirac strings with the total winding number $m=2N$.
The triads can be chosen as extension of $m=1$ monopole in Eq.(\ref{triadMonopoleTwoStrings}):
\begin{equation}
 {\bf e}_1 + i{\bf e}_2= ({\hat{\boldsymbol\theta}} +i {\hat{\boldsymbol\phi}})^N  \,.
\label{NMonopole}
\end{equation}
There are two Dirac strings, at $z<0$ and at $z>0$, each with winding number $m=N$ (the total winding number is $m=2N$).
It is not clear whether the monopoles with $|N|>1$ could be realized. 

\subsection{Monopole in spherical layer}
\label{SphericalLayer}

The monopole configuration with $N=1$ with two strings, each with $m=1$, is realized   for hydrodynamics  of ideal incompressible fluid in a thin spherical layer.\cite{Chefranov2017} Here ${\bf S}=\hat{\bf r}$, and thus the topological charge of the monopole is $N=1$.
The first term in the RHS of Eq.(\ref{vorticityMonopoleTwoStrings}) can be neglected in the limit of large radius of the sphere. Then the velocity field satisfies both stationary equations for incompressible fluid:
\begin{equation}
\nabla\cdot{\bf v}=0 \,\,,\,\, \nabla\times({\bf v}\times{\boldsymbol\omega})=0 \,.
\label{HydroRquations}
\end{equation}
This is the monopole, which core is represented by the inner sphere. 

%%%%%%%%%%%%%%%%%%%%%%%%%%%%%%%%%%%%%%%%%%%%%%%%%%%%%%%%%%%%%
%%%%%%%%%%%%%%%%%%%%%%%%%%%%%%%%%%%%%%%%%%%%%%%%%%%%%%%%%%%%%
\begin{figure}%[htt]
 \includegraphics[width=0.5\textwidth]{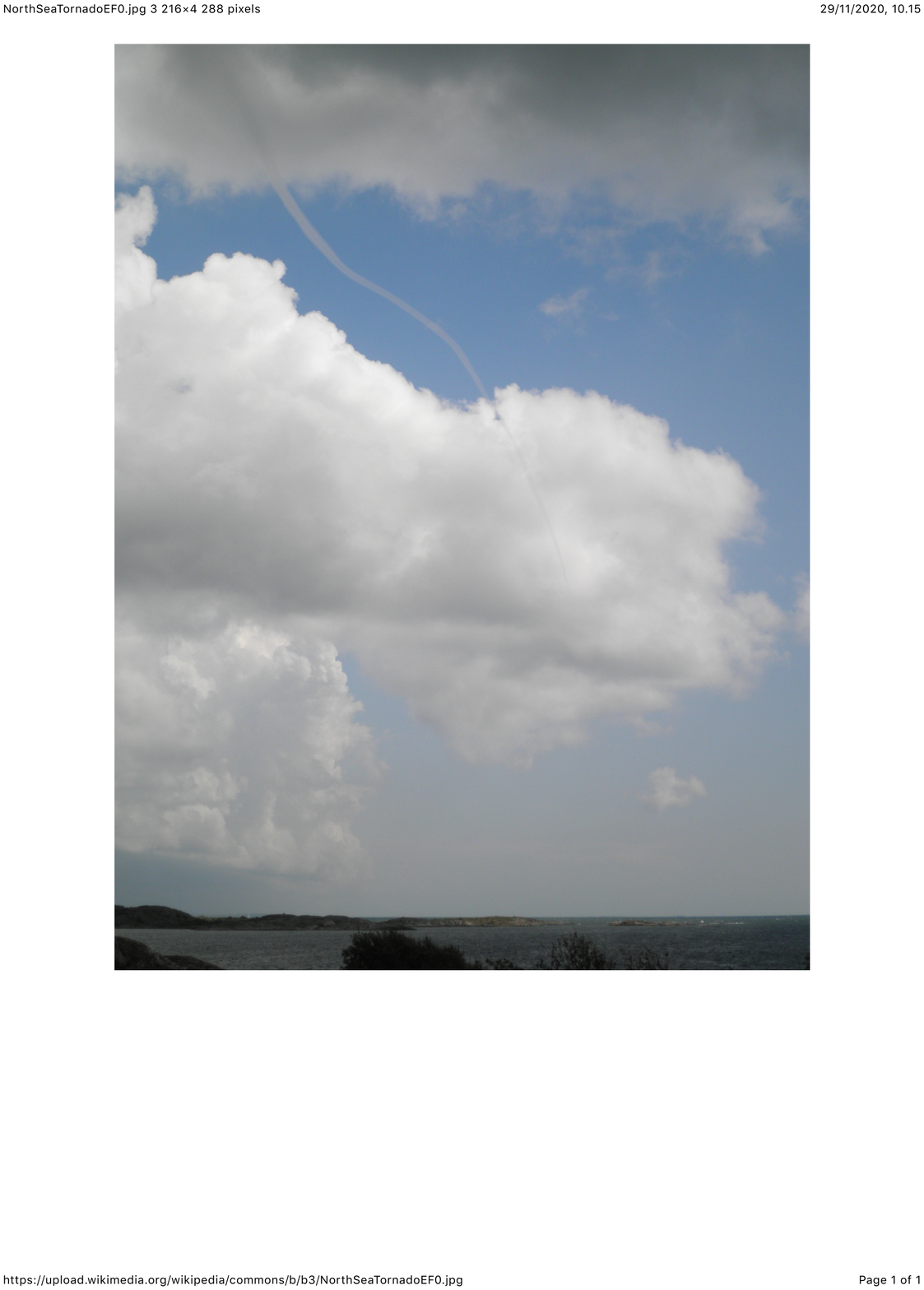}
 \caption{Since the end point of the vortex is not seen, this is probably the funnel cloud, which is the vortex (Dirac string) terminated  by the Dirac monopole.}
 \label{TornadoMonopoleFig}
\end{figure}
%%%%%%%%%%%%%%%%%%%%%%%%%%%%%%%%%%%%%%%%%%%%%%%%%%%%%%%%%%%%%
%%%%%%%%%%%%%%%%%%%%%%%%%%%%%%%%%%%%%%%%%%%%%%%%%%%%%%%%%%%%%

%%%%%%%%%%%%%%%%%%%%%%%%%%%%%%%%%%%%%%%%%%%%%%%%%%%%%%%%%%%%%
%%%%%%%%%%%%%%%%%%%%%%%%%%%%%%%%%%%%%%%%%%%%%%%%%%%%%%%%%%%%%
\begin{figure}%[htt]
 \includegraphics[width=0.5\textwidth]{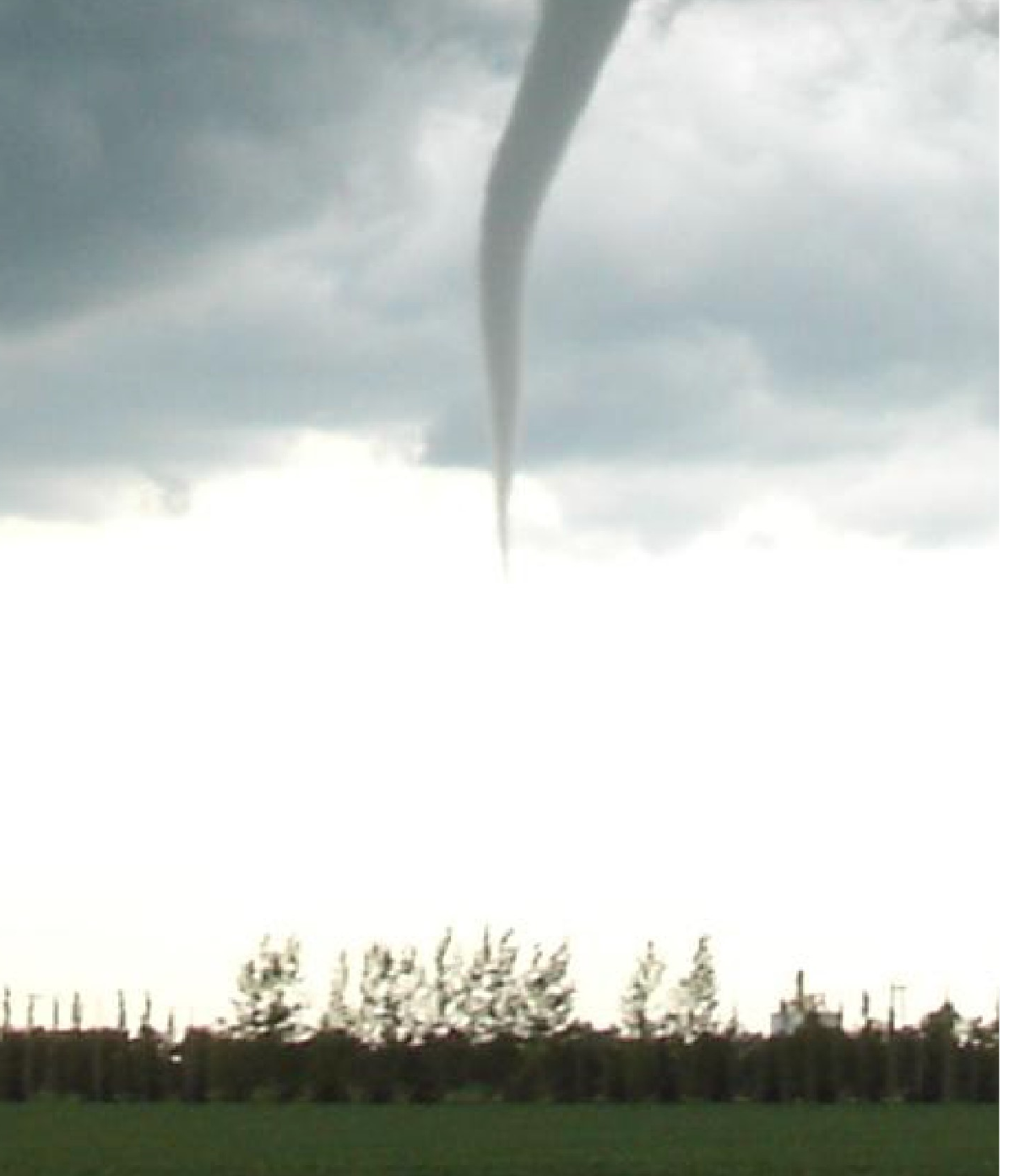}
 \caption{One more monopole.}
 \label{TornadoMonopole2}
\end{figure}
%%%%%%%%%%%%%%%%%%%%%%%%%%%%%%%%%%%%%%%%%%%%%%%%%%%%%%%%%%%%%
%%%%%%%%%%%%%%%%%%%%%%%%%%%%%%%%%%%%%%%%%%%%%%%%%%%%%%%%%%%%%

%%%%%%%%%%%%%%%%%%%%%%%%%%%%%%%%%%%%%%%%%%%%%%%%%%%%%%%%%%%%%
%%%%%%%%%%%%%%%%%%%%%%%%%%%%%%%%%%%%%%%%%%%%%%%%%%%%%%%%%%%%%
\begin{figure}%[htt]
 \includegraphics[width=0.5\textwidth]{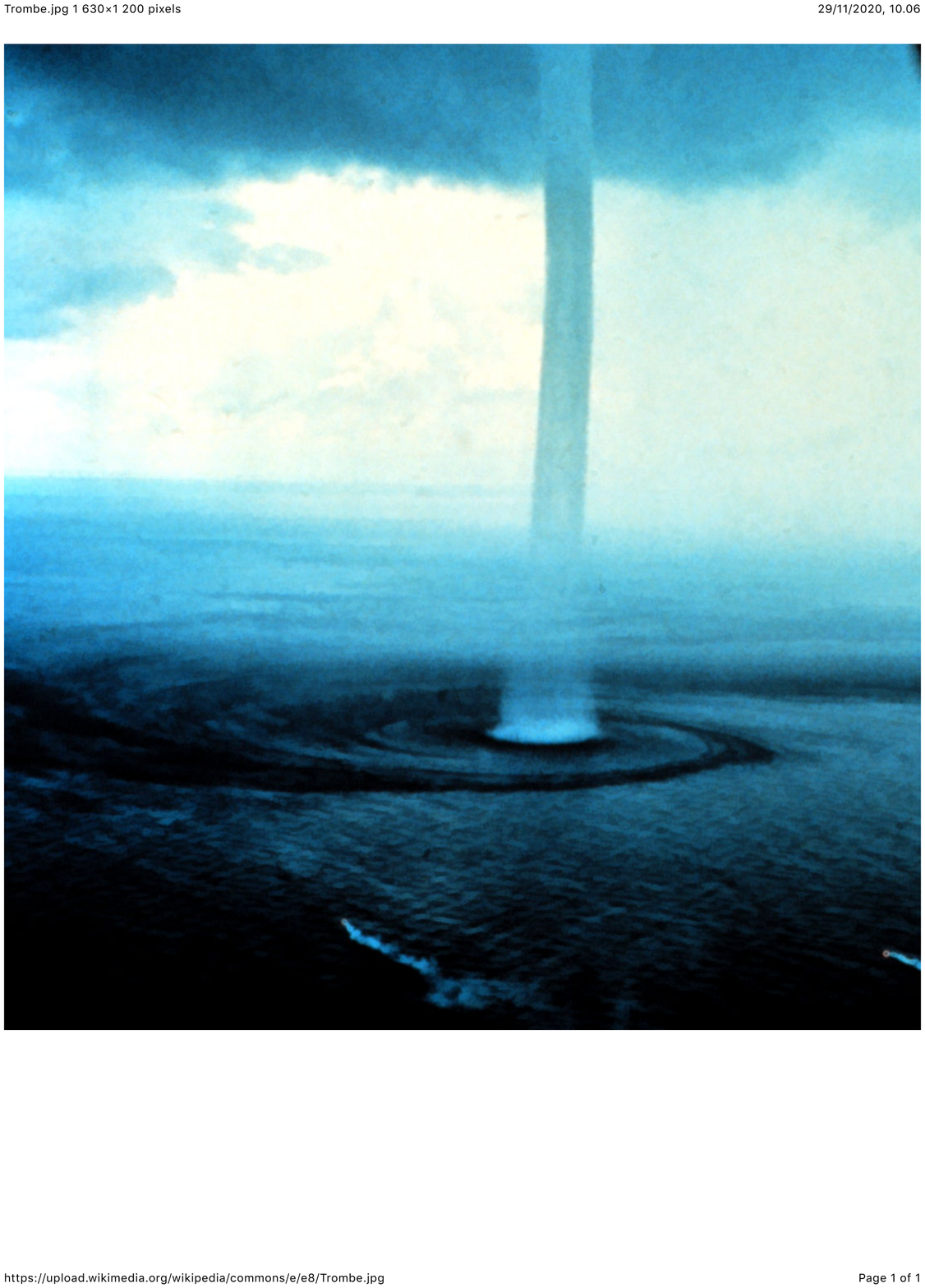}
 \caption{Tornado can be represented as Dirac string, while the spherical core of the Dirac monopole is represented by the Earth.}
 \label{TornadoFig}
\end{figure}
%%%%%%%%%%%%%%%%%%%%%%%%%%%%%%%%%%%%%%%%%%%%%%%%%%%%%%%%%%%%%
%%%%%%%%%%%%%%%%%%%%%%%%%%%%%%%%%%%%%%%%%%%%%%%%%%%%%%%%%%%%%

The simplest realization is tornado in Figs. \ref{TornadoFig} and \ref{Tornado2Fig}, where the Earth represents the core of the monopole.

%%%%%%%%%%%%%%%%%%%%%%%%%%%%%%%%%%%%%%%%%%%%%%%%%%%%%%%%%%%%%
%%%%%%%%%%%%%%%%%%%%%%%%%%%%%%%%%%%%%%%%%%%%%%%%%%%%%%%%%%%%%
\begin{figure}%[htt]
 \includegraphics[width=0.5\textwidth]{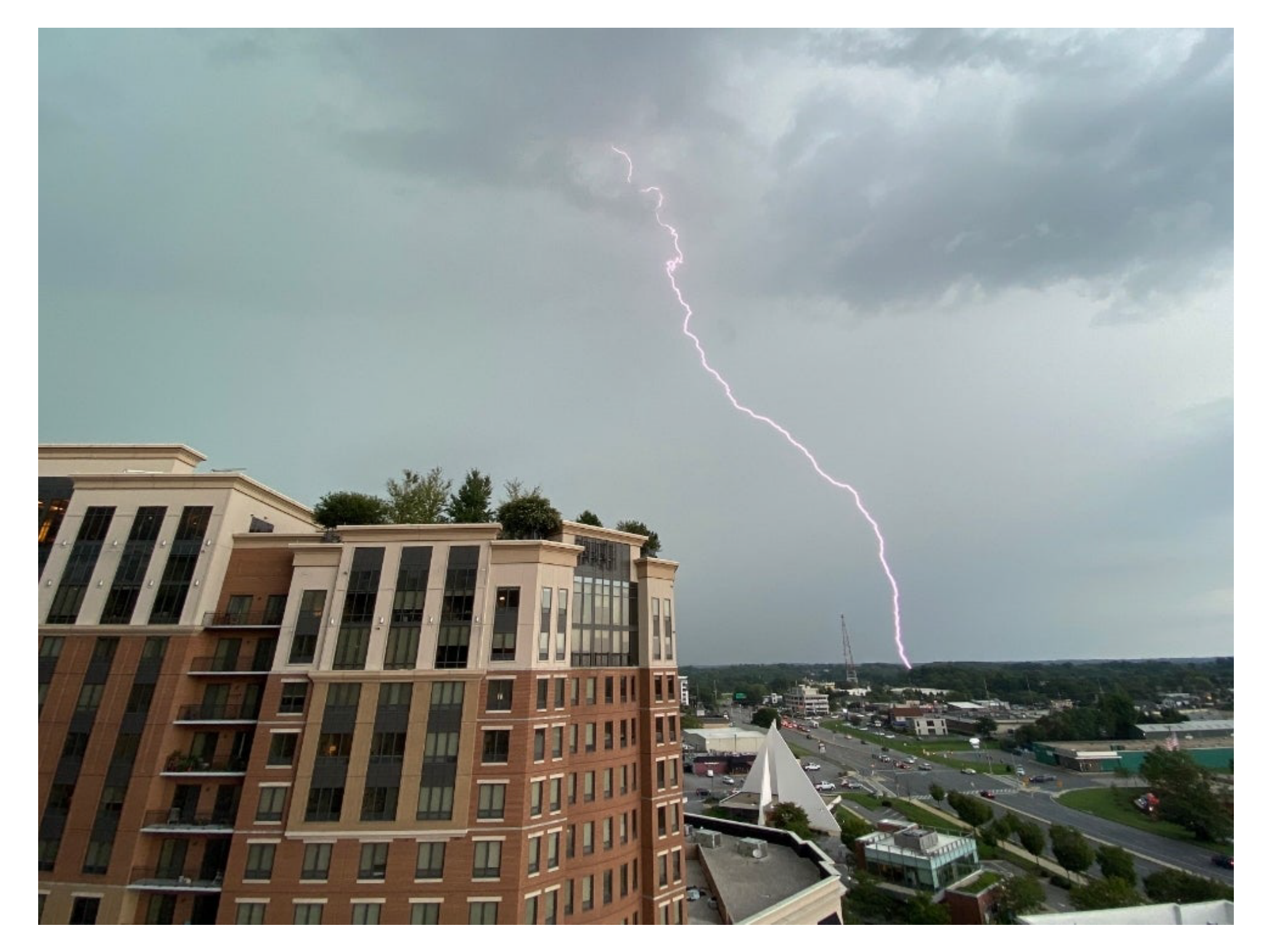}
 \caption{One more Dirac string}
 \label{Tornado2Fig}
\end{figure}
%%%%%%%%%%%%%%%%%%%%%%%%%%%%%%%%%%%%%%%%%%%%%%%%%%%%%%%%%%%%%
%%%%%%%%%%%%%%%%%%%%%%%%%%%%%%%%%%%%%%%%%%%%%%%%%%%%%%%%%%%%%

%%%%%%%%%%%%%%%%%%%%%%%%%%%%%%%%%%%%%%%%%%%%%%%%%%%%%%%%%%%%%
%%%%%%%%%%%%%%%%%%%%%%%%%%%%%%%%%%%%%%%%%%%%%%%%%%%%%%%%%%%%%
\begin{figure}%[htt]
 \includegraphics[width=0.5\textwidth]{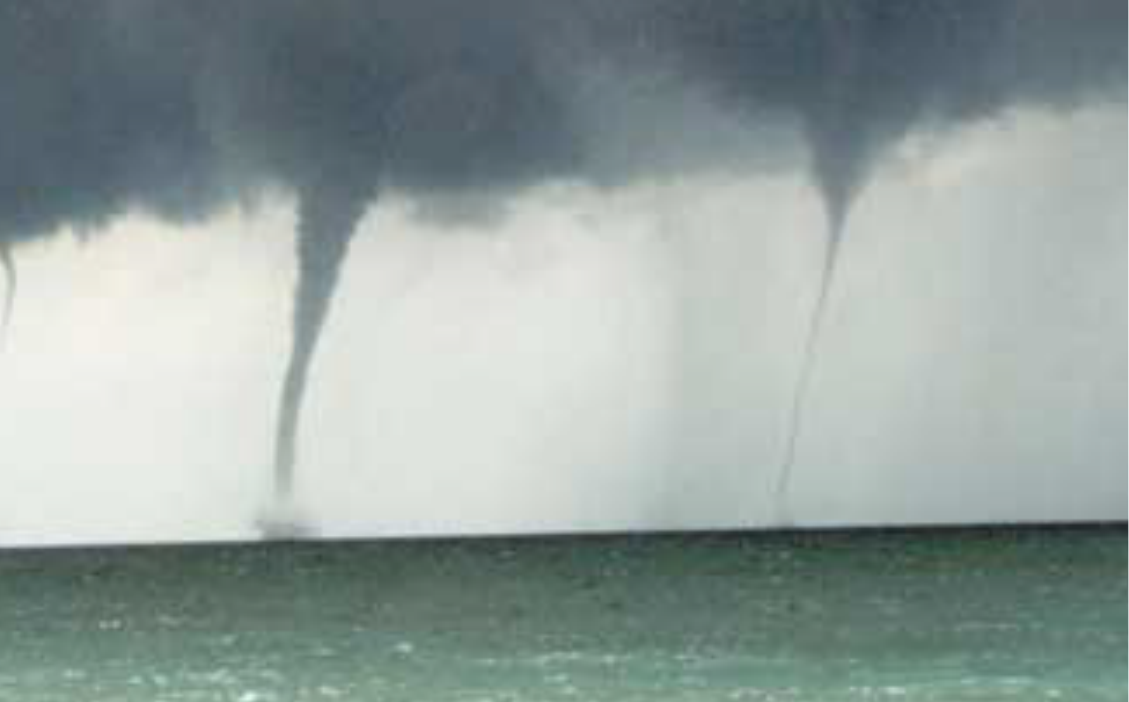}
 \caption{Dirac strings again}
 \label{Tornado3Fig}
\end{figure}
%%%%%%%%%%%%%%%%%%%%%%%%%%%%%%%%%%%%%%%%%%%%%%%%%%%%%%%%%%%%%
%%%%%%%%%%%%%%%%%%%%%%%%%%%%%%%%%%%%%%%%%%%%%%%%%%%%%%%%%%%%%

Viscosity resolves the singularity on Dirac line. This configuration can be also described in terms of $\hat{\bf S}$-vector:
$\hat{\bf S}({\bf r})$ is vertical outside the tornado and becomes radial on the axis.
The simplest model for that is the Rankine vortex, which has constant vorticity inside the tube, ${\boldsymbol\omega}={\rm const}\hat{\bf z}$, and zero vorticity outside the tube, ${\boldsymbol\omega}=0$.  The cylindrical surface of the tube experiences singularity, the jump in vorticity.

\subsection{From monopole to pancake vortex sheet}
\label{MonopoleSheet}

Now we can adiabatically deform the inner sphere, so that the topological invariant $N=1$ is conserved. Then the flow will be automatically adjusted  to the new configuration of the core.

There are several interesting deformations with the conservation of axial symmetry. 

One may adiabatically decrease the radius of the core. In the limit of small radius the configuration will represent the monopole with the singular core. 

One may deform the inner sphere to the disk of small thickness without violating the axial symmetry and the discrete symmetry 
with respect to inversion: ${\bf S}({\bf r})=-{\bf S}(-{\bf r})$, ${\bf v}({\bf r})={\bf v}(-{\bf r})$, ${\boldsymbol\omega}({\bf r})=-{\boldsymbol\omega}(-{\bf r})$. This would result in the Migdal vortex sheet.

So, let us consider the monopole structure obtained when the inner sphere in Sec.\ref{SphericalLayer} is adiabatically (i.e. with conservation of topological invariant $N=1$) deformed to the thin disk. In the vicinity of the disk one has  $S_z(z)=|S_z|\, \sign(z)$,  i.e. the disk represents the domain wall, where $S_z$ changes sign, but 
$|S_z(\rho)|$ is continuous across the sheet. Velocity is ${\bf v}= ZS_z\nabla\phi_2({\bf r})$. Let us first consider 
$\phi_2({\bf r})=\phi_2(\phi)=\phi$, the azimuthal angle.  In this case the flow is circulating in the opposite directions on two sides of the disk:
\begin{equation}
{\bf v}= ZS_z\nabla\phi =Z\frac{|S_z(\rho)|}{\rho}\sign(z) {\hat{\boldsymbol\phi}}\,.
\label{SheetRotated}
\end{equation}
The vorticity has the $\delta(z)$ singularity at the disk and $\delta_2({\boldsymbol\rho})$ singularity on the axis:
\begin{equation}
{\boldsymbol\omega}= {\hat{\boldsymbol\rho}}\, \delta(z) \frac{Z|S_z(\rho)|}{\rho}+
Z\hat{\bf z}\,  \sign{z} \,\left( \delta_2({\boldsymbol\rho})+ \frac{1}{\rho}  \partial_\rho| S_z(\rho)|\right)\,.
\label{SheetOmegaRotated}
\end{equation}

This structure represents the monopole with $N=1$  and two Dirac strings, each with winding number $m=1$. The core of the monopole is represented by the disk. 
On two sides of the disk one has the flow circulating in opposite directions.  The singular contribution $\delta_2({\boldsymbol\rho})$ to vorticity along $z$ axis represents two Dirac strings with opposite circulations. These Dirac strings enter the monopole (vortex sheet disk) from two sides.

% In Migdal picture of the vortex sheet $\phi_2$ depends on $\rho$, i.e. $\phi_2(\rho)$, and velocity is radial on two sides of the sheet:
% \begin{equation}
% {\bf v}= ZS_z(\rho)\nabla\phi_2(\rho) =Z |S_z(\rho)|\frac{d\phi_2}{d\rho}\sign(z) {\hat{\boldsymbol\rho}}\,.
% \label{SheetMigdal}
% \end{equation}
% The incompressibility condition requires the dependence of velocity also on $z$, i.e. $\phi_2(z,\rho)$ with $v(z)\propto z$ inside the sheet.
 
\subsection{Wall bounded by string}
\label{KLS}

In the present scenario, the disk itself represents the $Z_2$ domain wall at which $S_z$ changes sign. In the body of the disk, the vector  $\hat{\bf S}_\perp$  has $2\pi n$ rotations across the disk. In the Migdal scenario the vector $\vec S_\perp$ changes sign., and it makes half-integer number of rotations $(2 n +1)\pi$, or, a reflection in addition to $2\pi n$ rotations.\cite{M20c,M21b} 

In both cases the edge of the disk represents the circular string (ring), which terminates the domain wall. This string has either the $\pi_1$ winding number $n$, or the winding number $n+1/2$.
In the first case it is the analog of the soliton terminated by strings,\cite{MineyevVolovik1978} while in the second case it is the analog of  Kibble-Lazarides-Shafi (KLS) wall\cite{Kibble1982,Kibble1982b} bounded by Alice strings.
The similar configuration, the domain wall bounded by Alice string (half-quantum vortex), has been experimentally studied in the polar phase of superfluid $^3$He.\cite{Makinen2019} The general topology of the KLS walls bounded by string was discussed in Ref. \cite{VolovikZhang2020}. In both cases, solitons terminated by strings and domain walls terminated  by strings, the classification is in terms of the relative homotopy groups. The winding number $n$ or $n+1/2$ is the element of topology of such combined objects.

These combined objects are described by the combination of topologies. This includes  $\pi_2(S^2)=Z$  (the topological charge  $N=2m$ of the hedhehog-monopole and the winding number of Dirac strings, which enter the monopole);
$\pi_1(S^1) = Z$ (the winding number $n$ of the soliton and of the circular ring, which terminates the soliton); $\pi_0(Z_2) = Z_2$  (the domain wall).

% The vortex sheet geometry discussed by Migdal\cite{M20c,M21b} is in Eq.(\ref{SheetMigdal}).

\subsection{ SO(3)  hydrodynamics of incompressible liquid}
\label{SO3hydro}

Let us introduce instead of the triads with unit length ${\bf e}_a$, the more general triads ${\bf E}_a$:
 \begin{equation}
{\bf E}_1 =A{\bf e}_1 \,,\, {\bf E}_2 =A{\bf e}_2  \,,\, {\bf L}={\bf E_3}= {\bf E}_1 \times {\bf E}_2=Z{\bf e}_3\equiv Z{\bf S} \,,
\label{tetradMotion}
\end{equation}
\begin{equation}
{\bf E}_1^2={\bf E}_2^2=A^2=Z \,\,, \,\, {\bf E}_1\cdot {\bf E}_2= 0\,. 
\label{Z}
\end{equation}

The velocity field is:
\begin{equation}
v_i = {\bf  E}_1\cdot \nabla_i {\bf  E}_2   \,. 
\label{velocity2}
\end{equation}
Note that this is the general form of velocity, which is not violated by the condition of incompressibility. The latter condition is obtained by gauge transformation in Eq.(\ref{GaugeTransformation}). 

We have the following PB for canonically conjugate variables:
\begin{equation}
\{E_1^i({\bf r}_1), E_2^k({\bf r}_2)\}= \delta^{ik}\delta({\bf r}_2- {\bf r}_1)\,.
\label{EE}
\end{equation}
Eq.(\ref{EE}) gives:
\begin{eqnarray}
\{L^i({\bf r}_1), E^k_a({\bf r}_2)\}= e^{ikl}E^l_a\delta({\bf r}_2- {\bf r}_1)\,,
\label{LE}
\\
\{L^i({\bf r}_1), L^k({\bf r}_2)\}= e^{ikl}L^l\delta({\bf r}_2- {\bf r}_1)\,,
\label{LL}
\\
\{v_i({\bf r}_1), E_k^a({\bf r}_2)\}= \nabla_i   E_k^a \, \delta({\bf r}_2- {\bf r}_1) \,\,,\, a=1,2,3\,.
\label{VE}
\end{eqnarray}
The Eq.(\ref{VE}) gives the following  motion equations for the tetrad fields ${\bf E}_a$, with $a=1,2,3$: 
 \begin{equation}
\dot{\bf E}_a + ({\bf v}\cdot\nabla){\bf E}_a=0\,,
% \label{tetradMotion} % duplicate label
\end{equation}
Multiplying each of 3 equations to the corresponding ${\bf E}_a$ one obtains that each of three equation gives the equation for $Z$:
 \begin{equation}
\dot Z + ({\bf v}\cdot\nabla)Z=0\,.
\label{ZMotion}
\end{equation}
This demonstrates that $Z$ can be chosen as constant in spacetime.

The PB for velocity field are:
\begin{equation}
\{v_i({\bf r}_1), v_k({\bf r}_2)\}= e_{ikl} \omega^l({\bf r}_1) \delta({\bf r}_2- {\bf r}_1)\,.
\label{vv}
\end{equation}

Gauge transformation which does not change the distribution of vector ${\bf L}$
 (and the vorticity, which is expressed in terms of  ${\bf L}$), is:
\begin{equation}
 {\bf E}_1 + i{\bf E}_2 \rightarrow ( {\bf E}_1 + i{\bf E}_2)  e^{i\Phi}   \,\,, \,\, {\bf v} \rightarrow {\bf v} + \nabla \Phi \,.
\label{GaugeTransformation}
\end{equation}
This is the rotation of the dyad ${\bf E}_1$ and ${\bf E}_2$ by angle $\Phi$ about the local direction of  ${\bf L}$-vector.
The gauge transformation is needed to satisfy the incompressibility condition $\nabla\cdot{\bf v}=0$. It is analogous to transformation from 3 Clebsch variables to 2. But as distinct from Clebsch variables,  triads correctly deal with with such structures, where Clebsch variables have unphysical singularities (for example in the configuration with monopole and Dirac strings in Sec.\ref{monopole}).

\subsection{Spin connection and torsion}
\label{SpinConnection}

The metric, which follows from tetrads, is flat:
\begin{equation}
g^{ik}=\delta^{ab}E_a^i E_b^k \,.
% \label{metric} % duplicate label, collision in paper_base.tex, but it's not used here, so.
\end{equation}
Spin connection and torsion are  
\begin{equation}
C^{12}_i = -C^{21}_i = {\bf  E}_1\cdot \nabla_i {\bf  E}_2= - {\bf  E}_2\cdot \nabla_i {\bf  E}_1 =  v_i\ \,.
\label{SpinConnectionEq}
\end{equation}
\begin{equation}
T^a_{ik}=\partial_i    E^a_k -\partial_k   E^a_i  \ \,.
\label{Torsion}
\end{equation}

\subsection{Conclusion}
\label{CConclusion}

It looks that not only the relativistic quantum field theories, but also the quantum hydrodynamics of topological superfluids can be useful for study the important structures emerging in classical liquids.

{\bf Acknowledgements}. This work has been supported by the European Research Council (ERC) under the European Union's Horizon 2020 research and innovation programme (Grant Agreement No. 694248).  

 }
\end{document}